\documentclass[a4paper,11pt]{book} 
\usepackage[utf8]{inputenc}
\usepackage{xspace}
\usepackage[french, english]{babel}
\usepackage[T1]{fontenc}
\usepackage{lmodern}
\usepackage[margin=28mm,includeheadfoot,bindingoffset=5mm]{geometry}
\usepackage{graphicx,color} 
\usepackage[svgnames]{xcolor} 

\usepackage{subcaption}
\usepackage{tikz}
\usetikzlibrary{arrows.meta,bending}

\usepackage{etoolbox}          
\usepackage{calc}              
\usepackage{datetime}                
\usepackage{eso-pic}                 
\usepackage{enumitem} 
 
\usepackage{cite} 

\newcommand{\doidoi}[2]{\href{http://dx.doi.org/#1}{#2}}
\newcommand{\arxiv}[1]{\href{https://arxiv.org/abs/#1}{arXiv:#1}}

\usepackage[toc,page]{appendix}

\usepackage{epigraph}

\usepackage{tocloft}
\usepackage{color} 
\definecolor{Blue}{rgb}{0.00, 0.00, 1.00}
\definecolor{Red}{rgb}{1.00, 0.00, 0.00}

\newcommand{\myeqref}[1]{Eq.~\eqref{#1}}
\newcommand{\subsubref}[2]{\mbox{{\color{DarkBlue}#2} \ref{#1}}}

\newenvironment{itquote}
  {\begin{quote}\itshape}
  {\end{quote}\ignorespacesafterend}

\usepackage[pagebackref]{hyperref}
\hypersetup{colorlinks,
citecolor=DarkRed,
urlcolor=DarkBlue,linkcolor=DarkBlue,anchorcolor=DarkRed,
pdfdisplaydoctitle=true,pdfpagemode=UseOutlines,%
bookmarksnumbered=true,bookmarksopen=true
}
\newcounter{reqcount}
\newcommand{\pubitem}[1]{%
  \item[(\ref*{#1})] \refstepcounter{reqcount}\label{#1}
}
\makeatletter
\newcommand{\pubref}[1]{%
  {\def\@linkcolor{orange}\hyperref[#1]{(\ref*{#1})}}%
}
\makeatother 

\usepackage{setspace} 

\usepackage{amsmath, mathtools}
\usepackage{amsfonts,bm,bbm}     
\usepackage{upgreek}             
\usepackage{hhline} 
\usepackage{amssymb}
\usepackage{physics}

\newcommand{\bea}{\begin{eqnarray}}
\newcommand{\eea}{\end{eqnarray}}
\newcommand{\be}{\begin{equation}}
\newcommand{\ee}{\end{equation}}
\newcommand{\bee}{\begin{equation*}}
\newcommand{\eee}{\end{equation*}}

\newcommand{\HH}{\mathbf{H}}
\newcommand{\OO}{\mathbf{O}}
\newcommand{\UU}{\mathbf{U}}
\newcommand{\Ss}{\mathbf{S}}
\newcommand{\LL}{\mathbf{\Lambda}}
\newcommand{\WW}{\mathbf{W}}
\newcommand{\InverseW}{\mathbin{\rotatebox[origin=c]{180}{$\WW$}}}
\newcommand{\XX}{\mathbf{X}}
\newcommand{\JJ}{\mathbf{J}}
\newcommand{\AAA}{\mathbf{A}}
\newcommand{\BB}{\mathbf{B}}
\newcommand{\ID}{\mathbf{1}} 
\newcommand{\OOmega}{\mathbf{\Omega}}
\newcommand{\cH}{\mathcal{H}} 
\newcommand{\GG}{\mathbf{G}}
\newcommand{\DelBB}{\Delta\!\!\!\!\Delta} 
\newcommand{\MM}{\mathbf{M}}
\newcommand{\ZZ}{\mathbf{Z}}
\newcommand{\bxi}{\bm{\xi}}
\newcommand{\QQ}{\mathbf{Q}}
\newcommand{\GGamma}{\mathbf{\Gamma}}
\newcommand{\bfx}{\mathbf{x}}

\newcommand{\VV}{\mathbf{V}}

\newcommand{\lmax}{\lambda_{\mathrm{max}}}
\newcommand{\Pf}{ \mathrm{Pf} }
\newcommand{\Pff}[1]{ \underset{#1}{\mathrm{Pf}} }

\renewcommand{\dd}{\mathrm{d}}

\newcommand{\Z}{\mathbb{Z}}
\newcommand{\N}{\mathbb{N}}
\newcommand{\R}{\mathbb{R}}

\newcommand{\C}{\mathbb{C}}
\newcommand{\U}{\mathbb{U}}
\newcommand{\D}{\mathbb{D}}
\newcommand{\Ll}{\mathbb{L}} 
\newcommand{\Weyl}{\mathbb{W}}
\newcommand{\Quaternions}{\mathbb{H}}
\newcommand{\grG}{\mathrm{G}}

\def\Tr{{{\rm Tr}}}

\newcommand{\moy}[1]{\ensuremath{\langle #1 \rangle}}

\newcommand{\E}{\mathbb{E}}
\newcommand{\isEquivTo}[1]{\underset{#1}{\sim}} 
\renewcommand{\Pr}{\mathbb{P}} 
\newcommand{\Bevent}{\mathrm{B}}
\newcommand{\NC}{\mathrm{NC}}
\newcommand{\OU}{\mathrm{OU}}

\newcommand{\g}{\mathfrak{g}}  
\usepackage{dsfont} 
\newcommand{\1}{\mathds{1}} 
\newcommand{\Ai}{{\rm Ai}} 
\newcommand{\Ham}{\hat{ \mathcal{H}  }} 
\newcommand{\Hamm}{\hat{H}} 
\newcommand{\RT}{\mathcal{R}}
\newcommand{\ST}{\mathcal{S}}
\newcommand{\Vol}{\mathrm{Vol}}
\newcommand{\bigO}{\mathcal{O}}

\usepackage{bm}

\usepackage{empheq} 
\newcommand{\emphbe}{\begin{empheq}[box=\setlength{\fboxsep}{5pt}\fbox]{equation}} 

\renewcommand{\epsilon}{\varepsilon} 
\renewcommand{\geq}{\geqslant}

\renewcommand{\ln}{\log}

\newcommand{\ii}{\mathrm{i}}
\newcommand{\jj}{\mathrm{j}}
\newcommand{\kk}{\mathrm{k}}

\usepackage{slantsc}
\makeatletter
\patchcmd{\chaptermark}{\MakeUppercase}{\scshape\slshape}{}{}%
\patchcmd{\sectionmark}{\MakeUppercase}{\scshape\slshape}{}{}%
\patchcmd{\sectionmark}{\thesection.}{\thesection}{}{}       
\makeatother
\numberwithin{equation}{chapter}                     
\numberwithin{figure}{chapter}                       
\numberwithin{table}{chapter}                        
\renewcommand{\thechapter}{\Roman{chapter}}          
\renewcommand{\thesubsubsection}{\alph{subsubsection})} 

\usepackage{titlesec}                        
\titleformat{\chapter}[display]{\Huge\sffamily\bfseries}{\chaptertitlename~\thechapter}{1ex}{}
\titleformat{\section}[hang]{\LARGE\sffamily\bfseries}{\rlap{\thesection}}{2em}{}
\titleformat{\subsection}[hang]{\Large\sffamily\bfseries}{\rlap{\thesubsection}}{3em}{} 
\titleformat{\subsubsection}[hang]{\large\sffamily\bfseries}{\rlap{\thesubsubsection}}{2em}{} 

\usepackage{preamb/psl-cover}

\title{Stochastic and Quantum Dynamics of Repulsive Particles : \\[6pt] from Random Matrix Theory to Trapped Fermions} 

\author{Tristan Gautié}

\institute{l'École Normale Supérieure} 
\doctoralschool{Physique en \^Ile-de-France}{564}
\specialty{Physique Théorique}
\date{4 novembre 2021}

\jurymember{1}{Cécile Monthus}{CEA}{Présidente du jury} 
\jurymember{2}{Pierpaolo Vivo}{King's College}{Rapporteur}
\jurymember{3}{Malte Henkel}{Université de Lorraine}{Rapporteur}
\jurymember{4}{Reda Chhaibi}{Université Paul Sabatier}{Examinateur}
\jurymember{5}{Pierre Le Doussal}{École Normale Supérieure}{Directeur de thèse}
\jurymember{6}{Jean-Philippe Bouchaud}{CFM}{Co-directeur}
\jurymember{7}{Satya N. Majumdar}{Université Paris-Saclay}{Co-directeur}

\frabstract{ \footnotesize
\begin{spacing}{1.1}
Cette thèse de physique statistique a pour objet l’étude de trois types de systèmes présentant une interaction répulsive: valeurs propres de matrices aléatoires, marches aléatoires conditionnées à ne pas se croiser et fermions piégés. Ces modèles présentent de nombreux liens, qui peuvent être établis non seulement au niveau de leur version statique, mais aussi au niveau de leur version dynamique. Nous en proposons une analyse combinée, en exploitant les outils probabilistes des matrices aléatoires et des processus stochastiques ainsi que les outils de la mécanique quantique, dans le but de résoudre certains problèmes originaux. Outre la présentation détaillée du domaine et le compte-rendu des résultats obtenus durant le doctorat, les différents thèmes abordés dans les chapitres de la thèse permettent d’évoquer des perspectives sur des problèmes connexes. \\

Ainsi, le premier chapitre est une introduction à la théorie des matrices aléatoires; nous détaillons son évolution historique et ses nombreuses applications, et présentons en détail les concepts, constructions et résultats essentiels qui s’y rapportent. Le deuxième chapitre porte sur les marches aléatoires scalaires \emph{non-croisantes}; nous décrivons leurs liens profonds avec les processus de valeurs propres de matrices aléatoires et exposons les résultats obtenus dans le cadre de problèmes de barrière. Dans le troisième chapitre, qui aborde les processus stochastiques matriciels, nous introduisons en particulier un modèle inspiré par la récurrence aléatoire de Kesten, et mettons en lumière le lien nouveau qu’il permet d’établir entre l'ensemble de Wishart inverse et les fermions soumis au potentiel de Morse. Enfin le quatrième chapitre, centré sur le cas particulier des processus de ponts, permet de traiter conjointement les modèles scalaires et matriciels; nous y présentons une généralisation du problème de Ferrari-Spohn pour des ponts scalaires non-croisants, et, en guise d’ouverture, les connexions des ponts matriciels avec d’autres aspects des matrices aléatoires. 
\end{spacing} 
}

\enabstract{\footnotesize
\begin{spacing}{1.1}
This statistical physics thesis focuses on the study of three kinds of systems which display repulsive interactions: eigenvalues of random matrices, non-crossing random walks and trapped fermions. These systems share many links, which can be exhibited not only at the level of their static version, but also at the level of their dynamical version. We present a combined analysis of these systems, employing tools of random matrix theory and stochastic calculus as well as tools of quantum mechanics, in order to solve some original problems. Further from the detailed presentation of the field and the report of the results obtained during the PhD, the different themes exposed in the chapters of the thesis allow for perspectives on related issues.\\

As such, the first chapter is an introduction to random matrix theory; we detail its historical evolution and numerous applications, and present its essential concepts, constructions and results. The second chapter discusses non-crossing random walks; we describe the deep links they share with random matrix eigenvalue processes and showcase the results obtained in the scope of boundary problems. In the third chapter, which focuses on stochastic matrix processes, we introduce in particular a process inspired from the Kesten random recursion, and highlight the new link it allows to draw between the inverse-Wishart ensemble and fermions trapped in the Morse potential. Lastly, the fourth chapter, centred on the particular case of bridge processes, allows for a joint treatment of scalar and matrix models; therein, we develop a generalization of the Ferrari-Spohn problem for non-crossing scalar bridges and, as an opening, we exhibit the connections of matrix bridges with other aspects of random matrices.
\end{spacing} 
}

\frkeywords{ \footnotesize
Matrices aléatoires, marches aléatoires non-croisantes, fermions piégés, mouvement Brownien de Dyson, probabilités libres, systèmes désordonnés.
}

\enkeywords{ \footnotesize
Random matrices, non-crossing random walks, trapped fermions, Dyson Brownian motion, free probability, disordered systems.
}

\graphicspath{{images/}}

\begin{document}

\maketitle{}
\frontmatter
\cleardoublepage
\thispagestyle{empty}
\vspace*{5cm}
\parbox{130mm}{
\epigraph{\large Une civilisation sans la science, ce serait aussi absurde qu’un poisson sans bicyclette.}{\large Pierre Desproges \, 1939-1988}
}
\vfill
\cleardoublepage
\titlespacing*{\chapter}{0pt}{0pt}{40pt}
\begingroup
\let\cleardoublepage\relax
\chapter{Remerciements}

Aux derniers instants de rédaction de ce manuscrit qui clôt trois ans de doctorat, je souhaite adresser mes remerciements les plus sincères à toutes celles et tous ceux qui m'ont aidé à accomplir ce long travail de thèse, par leur conduite attentive, leur collaboration active ou simplement leur présence amicale.\\

Ainsi, je tiens tout d'abord à remercier ceux qui m'ont permis d'entreprendre ce voyage en physique statistique, à commencer par mon directeur de thèse, Pierre Le Doussal. J'ai beaucoup appris à ses côtés au cours des trois dernières années, entre investigations calculatoires et rédaction d'articles, des couloirs de l'Ecole Normale aux centres de physique théorique de Santa Barbara. Les longues séances de travail avec Pierre, lorsque le rythme s'accélérait en fin de projet, ont toujours été des moments d'intense activité et de belles découvertes.   \newline
Merci ensuite à mes co-directeurs, Jean-Philippe Bouchaud et Satya Majumdar, pour leur accompagnement dans ce parcours de recherche, leurs idées lumineuses et leurs conseils bienveillants. J'ai eu beaucoup de plaisir à collaborer avec eux. \newline
Je souhaite également remercier mes co-auteurs avec qui le travail a toujours été très stimulant et plaisant: Naftali Smith, Grégory Schehr, ainsi que Tony Jin et toute l’équipe. Merci aussi à Pierre Mergny pour les nombreuses discussions intéressantes que nous avons partagées.  \newline
Je remercie chaleureusement le personnel de l'ENS et de l'école doctorale, qui m'ont souvent aidé face à d'insolubles problèmes administratifs. Je pense en particulier à Laura Baron-Ledez, Jean-François Allemand, Sandrine Patacchini, Christine Chambon, Olga Hodges et Wissam Chaalal. \\

Les couloirs bigarrés de l'ENS n'auraient pas été un lieu aussi convivial sans la présence de tous ceux que j'ai eu la chance d'y cotoyer. En particulier, celui dont j'ai suivi la trace ces dernières années mérite une place spéciale dans ces remerciements: 
Alexandre Krajenbrink, mon "grand-frère" de thèse, n'a pas fini de m'impressionner par ses mille projets et ses bons tuyaux habituels. 
Les pauses-déjeuner et les bières dans le jardin ont toujours été de bons moments en compagnie de Clément Le Priol et Gabriel Gouraud, qui complètent cette belle "fratrie" de thèse, ainsi qu'avec Clément Roussel, Victor Chardès, Meriem Bensouda, Augustin Lafay, Hugo Bartolomei, Vassilis Papadopoulos et Manuel Diaz. J'ai également une pensée pour ceux que j'ai régulièrement croisés avec plaisir: Alexis Brès, Thibaud Richard, Arnaud Fanthomme, Benjamin Aubin, José Moran, Dongsheng Ge, Ludwig Hruza, Elena Bellomi, Maria Ruiz, Marko Medenjak, Stéphane d’Ascoli, Michaël Pereira... Merci enfin à toute l'équipe pour la riche aventure de Physique Pour Tous.\\

J'ai eu la chance, au cours de cette thèse, de participer à de nombreuses conférences passionnantes de Bangalore à Cargèse, dont je tiens à remercier les organisateurs. Je salue toutes les personnes formidables que j'y ai rencontrées, et particulièrement Yuan Miao, Federico Balducci, Cecilia De Fazio, Federica Montana, Sara Murciano, Alessandro Galvani, Octavio Pomponio et Federico Morelli chez les italiens; Victor Dagard, Yash Vardhan Chopra, Jyoti Sharma, Pawandeep Kaur, Mamta Yadav et Ankit Singh chez les indiens et Thibaud Maimbourg chez les corses. Merci également au professeur Takeuchi pour la visite de son laboratoire tokyoïte.\\

Ces trois années n'auraient pas eu la même saveur sans la présence constante de nombreux amis que je remercie du fond du coeur. Brice et Bruno, qui ne sont jamais bien loin pour mon plus grand bonheur; Lionel, Alice, tous les potos et nos séjours épiques à Trégastel ou Cotignac; Louis, Jean-Baptiste et tous les brillants idiots; Gustave, Robin et les squasheurs;  Andrei et les toulousains; Clara, Thibault, Benoît, Colin et Marie, Alilou... La liste est bien longue et continue encore; vous avez tous été très importants dans cette aventure, rythmée par les obsessions littéraires auvergnates, les formal dinners artisanaux et les campagnes de recrutement de l'Hôtel-Spa du Trois Bis. Mes pensées vont aussi vers Paul, qui nous a quittés.\\

Cette thèse est l'aboutissement d’un long et exaltant parcours académique. J'ai une pensée reconnaissante pour les professeurs de sciences, mais aussi de langues et d'humanités, qui m'ont souvent passionné de Charcot à Cambridge, en passant par Saint Dominique, Sainte Geneviève et Polytechnique.\\

Merci à Sophie et Christophe, mes parents, et à Joséphine, ma soeur, pour leur présence et leur soutien depuis toujours, ainsi qu'à Jean, Gisèle et Marie-Ange, sans oublier Jean Oger, mes grands-parents à qui je dédie cette thèse.\\

Enfin, je tiens à exprimer ma tendre reconnaissance à Joanne, pour tout et surtout pour avoir transformé cette année de confinements en exils flamboyants et cet été de rédaction en doux rêve breton.

\null \newpage
\clearpage 
\tableofcontents
\clearpage
\thispagestyle{empty}
\null  \newpage
\chapter{Introduction}
\vspace{1cm}

Walking down the streets of a city with a curious eye, you may glimpse the traces of a hidden universal principle. Whether glancing at parked cars along a street, perched birds on a power line or time intervals between bus arrivals at a station, you might discover systems which elements tend to be separated by a fair distance, as if repelling one another. The respective causes for this repulsion, in systems of such different natures, are necessarily unfit for comparison. However, comparing their ensuing statistics unveils a remarkable connection: in these systems, the distribution of spacings between particles follows the one of eigenvalues in random matrix theory \cite{Abul2006,Seba2009,Krbalek2000}. Further from these everyday-life examples, the universality of random matrices extends deep into physics and mathematics, making appearances in many different contexts \cite{Deift2006}. 

In recent years, the quest to exhibit such random-matrix-related universal behaviours in physical systems has driven many efforts in statistical physics. In this thesis, we focus in particular on three kinds of systems. First, eigenvalues of random matrices themselves, which exhibit repulsion as a consequence of the geometry of random matrix ensembles. Secondly, independent random walks and stochastic processes conditioned not to cross each other, which are a toy model for interface systems and for which repulsion is a consequence of their conditional definition. Lastly, fermions in a trapping potential where repulsion is a consequence of Pauli's exclusion principle. These systems share many connections which fundamentally rely on their common repulsion property, such that they can be described in the same mathematical framework, relying on determinants and Pfaffians. At the static level, these connections emerge, for some models, as similar joint distributions for the eigenvalues of random Hermitian matrices, the altitude of the non-crossing walkers at a given time, and the fermions' positions. At the dynamical level, richer connections appear between eigenvalue processes of matrices with a stochastic perturbation, non-crossing processes and time-evolving fermions.

The goal of the present thesis is to introduce the above-mentioned systems, to present their deep connections with one another and to develop the results obtained during the course of the doctoral studies for some problems surrounding this theme. In addition to this exposition, an effort is made to draw perspectives on related issues. In this purpose, the last section of each chapter is devoted to an outlook from the chapter's theme.\\[10pt]

The outline of the thesis is as follows.

The first chapter is an introduction to random matrix theory and eigenvalue processes. After brushing the historical evolution of the theory, and detailing the numerous applications it has encountered in physics, mathematics, telecommunication theory or finance, we present its main concepts and tools. As such, we define the main ensembles of random matrices and describe the crucial results concerning the global and local statistics of their eigenvalues. Herein, the link of some ensembles with quantum systems of non-interacting fermions is underlined. Finally, we give a presentation of free probability, a branch of non-commutative probability which grants powerful tools when applied to large random matrices.

The second chapter discusses non-crossing walkers, i.e. independent random walks conditioned not to cross each other. As detailed in this chapter, they share profound connections with stochastic matrix processes such as the Dyson Brownian motion, and with the eigenvalue distributions of Gaussian ensembles. The persistence properties of non-crossing walkers are treated in the context of a moving vicious boundary. Detailed results are given for this problem, in relation to a quantum fermionic system. Finally, perspectives are drawn on the connection between non-crossing random walkers and random matrices, through the imposition of a partial non-crossing condition.

After this study of non-crossing scalar processes, we turn to stochastic matrix processes in the third chapter. We introduce in particular a process inspired from the Kesten random recursion, the matrix Kesten recursion, for which we exhibit both continuous-time and discrete-time results. In the continuous setting, we highlight the new link it allows to draw between the inverse-Wishart ensemble of random matrix theory and fermions trapped in the Morse potential. In the discrete setting, we describe the results given by the tools of free probability. Further from the matrix Kesten process, we discuss processes related to it and generalizations of it. Finally, we present a Hamilton-Jacobi perspective on stochastic matrix evolutions, following recent literature.

Lastly, the fourth chapter combines the study of both scalar and matrix processes, by focusing on the particular case of bridge processes, where a final condition is fixed. After a presentation of the scalar Brownian bridge, we develop results obtained on non-crossing bridge processes in presence of a semi-circle moving boundary, which yield a generalization of the Ferrari-Spohn problem for a single walker. As an opening, we discuss the Dyson Brownian bridge process in relation to the Harish-Chandra-Itzykson-Zuber integral and its connection to the large deviations of extreme eigenvalues in some random matrix ensembles.

\clearpage  

\chapter{Publications related to this thesis}
\vspace{1cm}

\begin{enumerate}[label={(\arabic*)}]

\pubitem{publication:NonCrossingBrownianDBM}
T. Gautié, P. Le Doussal, S. N. Majumdar \& G. Schehr, {\it Non-crossing Brownian Paths and Dyson Brownian Motion Under a Moving Boundary}. \\ \doidoi{10.1007/s10955-019-02388-z}{Journal of Statistical Physics \textbf{177}(5), 752–805}, (2019). \\ \arxiv{1905.08378}.
\vspace{1.5em}

\pubitem{publication:MatrixKesten}
T. Gautié, J.-P. Bouchaud \& P. Le Doussal, {\it Matrix Kesten recursion, inverse-Wishart ensemble and fermions in a Morse potential}. \\ \doidoi{10.1088/1751-8121/abfc7f}{Journal of Physics A: Mathematical and Theoretical \textbf{54}(25) 255201}, (2021). \\ \arxiv{2101.08082}.
\vspace{1.5em} 

\pubitem{publication:ConstrainedNCFerrariSpohn}
T. Gautié \& N. R. Smith, {\it Constrained non-crossing Brownian motions, fermions and the Ferrari-Spohn distribution}. \\ \doidoi{10.1088/1742-5468/abe59c}{Journal of Statistical Mechanics: Theory and Experiment 033212}, (2021). \\ \arxiv{2011.12995}.
\vspace{1.5em}

\pubitem{publication:TransportFreeFermions}
T. Jin, T. Gautié, A. Krajenbrink, P. Ruggiero \& T. Yoshimura, {\it Interplay between transport and quantum coherences in free fermionic systems}. \\ 
 \doidoi{10.1088/1751-8121/ac20ef}{Journal of Physics A: Mathematical and Theoretical \textbf{54}(40) 404001}, (2021). \\
\arxiv{2103.13371}.

\end{enumerate}

\clearpage
\chapter{Index of notations and abbreviations}

\vfill  

\begin{tabular}{b{5cm} l}
RMT & Random Matrix Theory \\[1pt] 
GOE, GUE, GSE  & Gaussian Orthogonal, Unitary, Symplectic Ensemble \\[1pt]
LOE, LUE, LSE  & Laguerre Orthogonal, Unitary, Symplectic Ensemble \\[1pt]
COE, CUE, CSE & Circular Orthogonal, Unitary, Symplectic  Ensemble \\[1pt] 
DBM & Dyson Brownian motion \\[1pt] 
DPP & Determinantal point process \\[1pt]
TW & Tracy-Widom \\[1pt]
HCIZ & Harish-Chandra-Itzykson-Zuber \\[1pt]
FS & Ferrari-Spohn \\[1pt] 
ESD & Empirical spectral distribution \\[1pt] 
$\HH$  & Self-adjoint matrix \\[1pt]
$\UU$ & Unitary matrix \\[1pt]
$\XX$ & Generic matrix \\[1pt]
$\ID$ & Identity matrix \\[1pt]
$ \beta$ & Dyson index \\[1pt]
$\g$ & Stieltjes transform \\[1pt]
$ \mathcal{R}$  & $\mathcal{R}$-transform \\[1pt]
$ \mathcal{S}$  & $\mathcal{S}$-transform  \\[1pt]
$\Tr$ & Trace \\[1pt]
$\det$ & Determinant \\[1pt]
$\Pf$ & Pfaffian \\[20pt]

PDF & Probability distribution function \\[1pt]
JPDF & Joint probability distribution function \\[1pt]
CDF & Cumulative distribution function \\[1pt]
i.i.d. & Independent and identically distributed \\[1pt] 
FP & Fokker-Planck \\[1pt]
SDE & Stochastic differential equation \\[1pt]
$\E$ & Expectation \\[1pt]
$\Pr$  & Probability \\[1pt]
$\Pr( \ \mid \ ) $ & Conditional probability  \\[1pt]
$ P $ & Probability density \\[1pt]
$P(\  \mid \ ) $ & Conditional probability density  \\[1pt]
$\rho$ & Average density  \\[1pt]
$ S$ & Survival probability  \\[20pt]
\end{tabular} 

\newpage 
\vspace*{1.5cm}
\begin{tabular}{b{5cm} l} 
$\ket{k}$ & Single-particle quantum state \\[1pt]
$\Hamm$ & Single-particle Hamiltonian \\[1pt]
$\phi_{k}$ & Single-particle wavefunction \\[1pt]
$\epsilon_k$ & Single-particle energy level \\[1pt]
$\ket{\vec{k}}$ & Many-body quantum state \\[1pt]
$\Ham $ & Many-body Hamiltonian \\[1pt]
$\Psi_{\vec{k}}$  & Many-body wavefunction \\[1pt]
$ \epsilon_{\vec{k}}$ & Many-body energy level \\[1pt]
$G ( \ \mid \ ) $  & Quantum propagator \\[20pt]

$\ii$ & Imaginary unit \\[1pt]
$\log$ & Logarithm to base $e$ \\[1pt]
$ \Weyl_N$  & Weyl chamber for the Symmetric group $\mathrm{S}_N$ \\[1pt]
r.h.s. ,  l.h.s. & Right, left hand side \\[1pt]
$[a,b]$ ,  $]a,b[$ &  Real interval with bounds included, excluded \\[20pt]

$\Ai$ & Airy function of the first kind \\[1pt]
$\alpha_k$ & Zeros of the Airy function  \\[1pt]
$J_\alpha$ & Bessel function of order $\alpha$
\end{tabular}    
\endgroup
\titlespacing*{\chapter}{0pt}{50pt}{40pt}

\thispagestyle{empty}
\mainmatter  
\chapter{Introduction to random matrix theory and eigenvalue processes}
\chaptermark{Introduction to RMT and eigenvalue processes}
\label{chap:1}
 
The theory of random matrices has attracted immense interest in mathematics and in physics in the last hundred years, especially through the statistics of their eigenvalues. From the definition of rather mundane objects, matrices with random coefficients, random matrix theory unveils fascinating behaviours in the statistics of many different observables. Decades of scientific publications have investigated the properties of eigenvalues and other singular values, eigenvectors, condition numbers, determinants and other polynomials, integrals and linear statistics in a swarm of different matrix models. These studies require tools from an impressive collection of mathematical fields : probability and linear algebra of course, but also complex analysis, partial differential equations, quantum theory, combinatorics, group theory and even number theory.

In the first chapter of this thesis, we will give a broad and pedagogical introduction to the theory of random matrices, its tools and most important properties. We will brush the theory's history, from the first theoretical endeavours to the most recent applications, and detail the statistical properties of random eigenvalues as well as their natural extension to stochastic processes, and the more exotic aspects of free probability. This chapter presents the theoretical foundations upon which the contents of the thesis are built in the following chapters.

The first section \ref{sec:1.1} will present the historical development and fields of application of random matrix theory, along with its most significant random ensembles. The second section \ref{sec:1.2} will give a short description of the many interesting statistical properties of eigenvalue ensembles in random matrix theory, with a detailed explanation of their links with fermionic quantum systems.
Finally, section \ref{sec:1.3} will introduce the theory of free probability, a framework that proves very useful in the study of large random matrices where it generalizes the notion of independence to non-commutative random variables.
 
\section{Random matrices}
\label{sec:1.1}

Matrices are ubiquitous mathematical objects as a building block of linear algebra and natural representatives of linear applications. It is therefore not surprising that random matrices, matrices whose entries are random variables, should appear as useful tools in a large variety of settings and have been studied by experts in different fields. The history of random matrix theory (\textit{RMT}) reflects this through the variety of its contributions: mathematicians have studied it for the abstract beauty of its constructions and connection to other fields of mathematics, physicists for its natural implications in quantum theory, condensed matter physics and statistical physics, statisticians for its usefulness in multivariate statistics and signal processing, computer scientists for its applicability in numerical linear algebra and network modelling, etc.

In the first subsection, we will give a brief overview of the history and applications of RMT. In subsection \ref{subsec:1.1.2}, we will present the main constructions of the theory, the random ensembles of the Gaussian and Wishart class.

\subsection{Historical context and applications}
\label{subsec:1.1.1}

\subsubsection{History of random matrix theory}

The first hints of random matrices date back to the beginning of the XX\textsuperscript{th} century, when John Wishart introduced approaching notions in his studies of multivariate statistics in 1928 \cite{Wishart1928}. Two decades later, Eugene Wigner achieved the first feats of the theory in the context of nuclear physics \cite{Wigner1951}. At the time, experimental physicists had collected large amounts of data about the excitation spectra of heavy nuclei, and were seeking models explaining the positions of the observed resonances. A detailed explanation of the interactions between many nucleons unfortunately seemed out of reach, making it impossible to approximate the Hamiltonian of this quantum system reliably. Obtaining the energy levels, the eigenvalues of the Hamiltonian, by meticulous modelling from first principles was thus hopeless. The system of interest, the nucleus of a given heavy atom, is deterministic but highly complex, such that its configuration of energy levels is difficult to predict.

Eugene Wigner's ingenious solution was to trade complexity for randomness: he attempted to model the unknown Hamiltonian matrix by a random sample from a matrix distribution respecting the adequate symmetries of the system, i.e.~real symmetric for a time-reversal symmetric quantum system. Amazingly, even though it cannot predict the exact details of the spectrum, this ansatz effectively reproduces universal properties such as the local statistics in the energy level observations \cite{Porter1960}. Note that this constitutes a leap forward in the theory of statistical physics: in the usual statistical setting, a given system, with its associated Hamiltonian or set of dynamical rules, can occupy a number of states according to a probability distribution, such that the randomness concerns the state that the system occupies. The energy of the system is then, for example, a random observable that an observer wishes to characterize. In contrast, Wigner's proposal consists in acknowledging ignorance on the system itself, and its Hamiltonian, such that the set of accessible states itself is a random observable.

\begin{figure}[ht!]
    \centering 
    \includegraphics[width=.6 \textwidth]{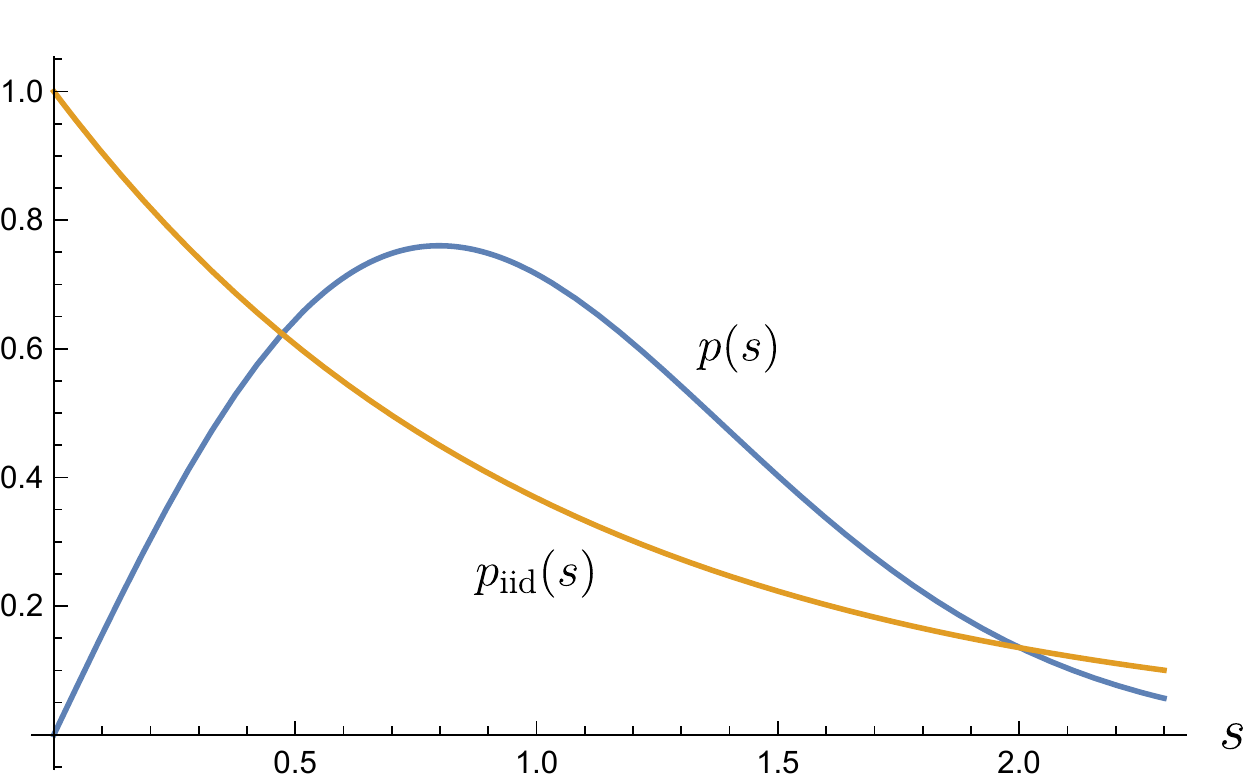} 
    \caption{Wigner's surmise: plot of the spacing distributions $p$ and $p_{\rm iid}$, with parameters fixed such that both distributions have unit mean.}
    \label{fig:surmise}
\end{figure}

The main statistical feature of Wigner's random matrix ensembles that was found to match the nuclear spectra observations is the \emph{repulsion of eigenvalues}: diagonalizing a typical sample of a random matrix ensemble produces a set of eigenvalues that have low probability of being found close to each other. This feature was famously unearthed as Wigner's surmise \cite{Wigner1957}, his conjecture for the distribution of level spacing in nuclear spectra, i.e.~the distance separating two consecutive resonances. The surmise consisted in the following probability density $p(s)$:
\be 
p(s)=a \, s  \, e^{-b s^{2}}
\ee
where $a$ and $b$ are parameters which we leave unspecified for now. This consecutive-neighbour separation distribution $p$ is indeed approximately verified for real symmetric random matrices \cite{Mehta1960} and for a class of nuclei \cite{Porter1960}. We plot this distribution in Fig \ref{fig:surmise}, along with the spacing distribution of a set of a large number of i.i.d.~random variables, constituting a Poisson point process, which can easily found to be the exponential distribution $p_{\rm iid}(s)$:
\be 
p_{\rm iid}(s) = \lambda  \,  e^{-\lambda s} \; .
\ee
It is evident in this figure that a small distance between eigenvalues of a random matrix bears much lower probability than a similar spacing in the system of independent variables. This central feature of RMT, eigenvalue repulsion, will be the focus of many of our discussions in the next sections.

Eugene Wigner also studied the global statistics of the first simple ensembles \cite{Wigner1958}. Consider for example a real symmetric $N \times N$ matrix with the entries of its upper-triangular part distributed independently from the same distribution, with variance $\frac{\sigma^2}{N}$ and finite higher moments. Wigner found that, when $N$ becomes large, the eigenvalues are systematically found on the interval $[-2\sigma,2\sigma]$, and are distributed according to the density that bears his name today: 
\be 
\text{Wigner's semicircle} \quad \quad \quad  \rho_{\rm SC} (x) = \frac{\sqrt{4\sigma^2-x^2}}{2\pi \sigma^2} \; . 
\ee

After the first successes of RMT in the description of physical reality, the study of random matrices was undertaken in full mathematical rigour by Freeman Dyson in the 1960's \cite{Dyson1962}. He developed the classification of random matrix ensembles according to their symmetry properties, the threefold way which structures the theory \cite{Dyson1962a}. He thereby formalized Wigner's ideas by constructing ensembles with the three possible symmetry classes for a quantum system's Hamiltonian. Indeed, a system's Hamiltonian should be either:
\begin{itemize}
\item complex Hermitian, if time-reversal symmetry is broken, in the presence of a magnetic field for example ;
\item real symmetric, if the system is time-reversal and spin-rotation symmetric ;
\item quaternionic self-adjoint, if the system is time-reversal symmetric with broken spin-rotation symmetry, in the presence of spin-orbit coupling for example.
\end{itemize}
This threefold way is known as the Wigner-Dyson classification. Dyson's achievement was to define, in a unified framework, random ensembles with coefficients belonging to all three number systems: real, complex and quaternionic numbers. The presentation of the main ensembles according to this hierarchy will be the focus of section \ref{subsec:1.1.2}. 

Another major contribution by Freeman Dyson was the introduction of a dynamical theory of random matrices, by his interest in the evolution of eigenvalues under stochastic perturbation of the matrix \cite{Dyson1962b}. In this effort, he constructed the Dyson Brownian motion, under which eigenvalues experience stochastic diffusion coupled with repulsion from one another, as expected from the eigenvalue repulsion presented above.  Furthermore, this model was used to build notable connections with the world of integrable systems, which was an important stimulation for subsequent research on these themes. We will present the Dyson Brownian motion in detail, including this connection to an integrable system, in relation to non-crossing walkers in chapter \ref{chap:2}.

\begin{figure}[ht!] 
    \centering
    \hspace{1 cm}
    \includegraphics[height=4cm]{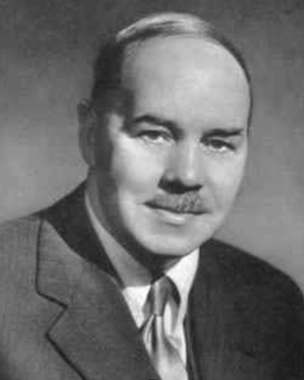} \hfill
    \includegraphics[height =4cm]{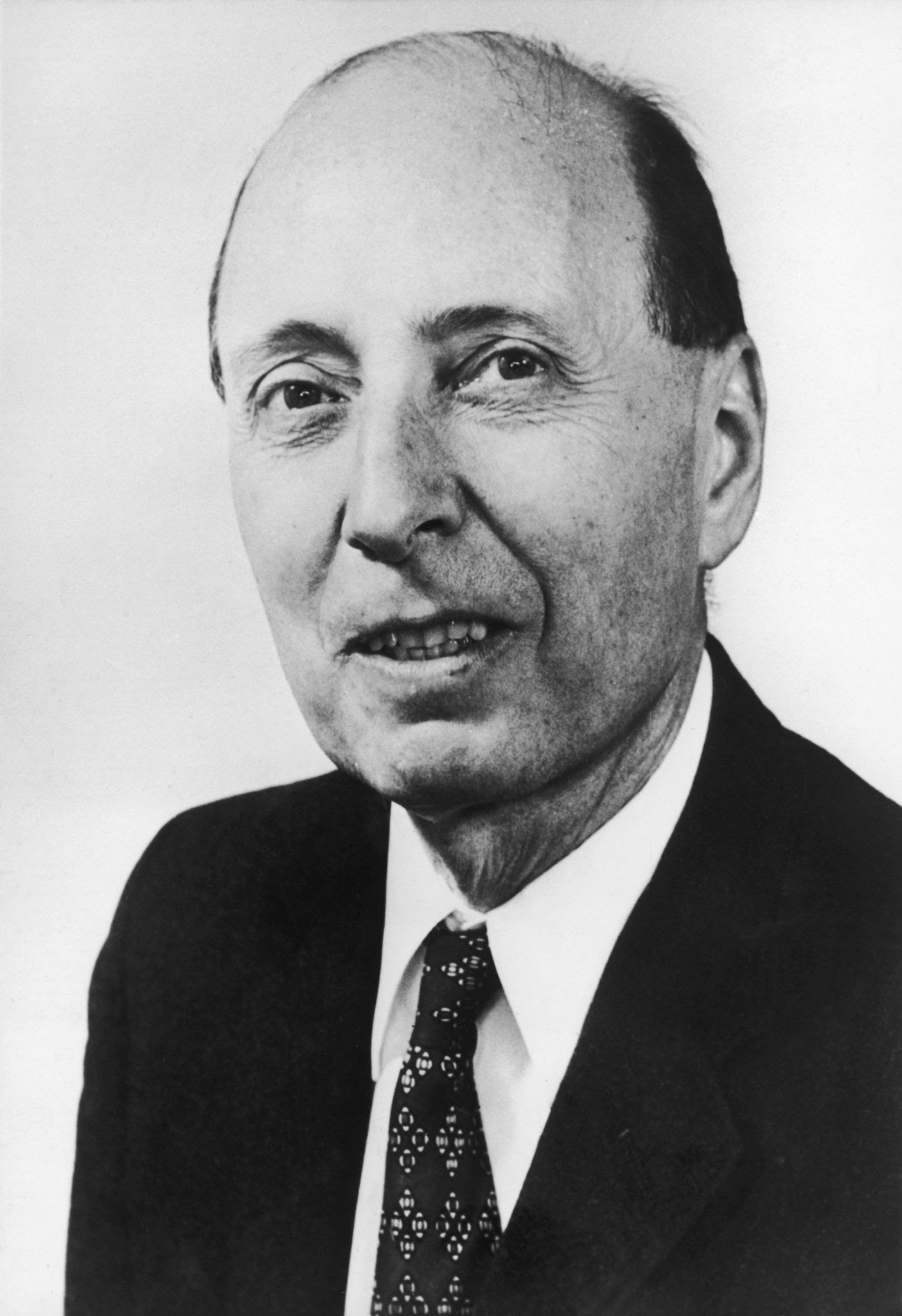} \hfill
    \includegraphics[height =4cm]{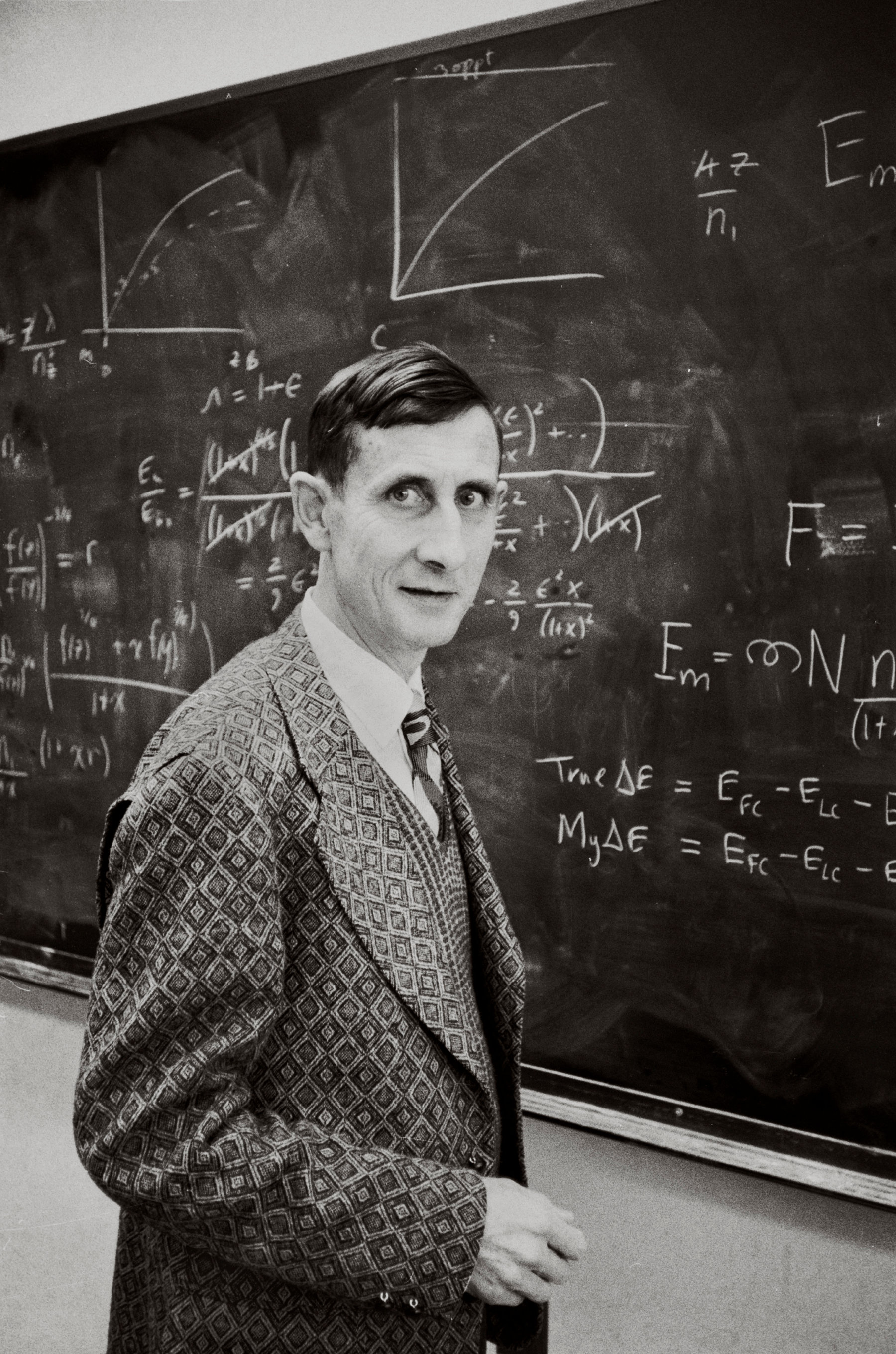}
    \hspace{1 cm}
    \caption{John Wishart, Eugene Wigner and Freeman Dyson.}
    \label{fig:portraits}
\end{figure}

After the achievements of these pioneers, whose photographs are shown in Fig.~\ref{fig:portraits} as a tribute, RMT has been developed in many research directions, both in purely mathematical studies and in applied settings. The study of eigenvalue statistics has been led to refine the knowledge of local and global properties, in the limit of large matrix size and in the finite size setting. Extremal eigenvalue statistics in particular has led to many important discoveries, along with the study of eigenvectors. The extension of results obtained initially in the Gaussian setting to more general distributions, uncovering the universality properties of RMT, is among the most notable developments. The study of matrices with a source, elliptic ensembles and many other models should also be mentioned in the history of RMT, as well as 
the onset of free probability which will be presented in section \ref{sec:1.3}. The theory has found many applications, of which we detail a few in the next paragraph.

For a more detailed introduction to the historical developments of RMT, giving due credit to many scientists that we could not mention here, we point to the review articles  \cite{Guhr1998,Forrester2003} and the books \cite{Mehta2004,Porter1965,Akemann2011}.

\subsubsection{Applications}

Apart from the nuclear physics cradle from which it emerged, RMT is deeply connected to many branches of physics, mathematics and has been successfully applied outside of these sciences in telecommunications, finance and even recently in machine learning.

Staying for now in the physics realm, the field of disordered mesoscopic systems has immensely gained from the developments of RMT. Indeed, the diffusion of waves in random media as well as the conductance of quantum particles in disordered conductors rely on the description of a random scattering matrix. The mesoscopic properties of the system, such as transport or localization of electrons through a conductor, or time delay of wave propagation, are thus guided by the random matrix statistics observed under a certain symmetry class suitable for the condensed matter system \cite{Beenakker1997}. In a related note, the extension of the notion of chaos to quantum systems has been built with regards to RMT: the repulsion feature observed in RMT is a signature of chaos in quantum systems, such as the quantum Sinai's billiard \cite{Bohigas1984}. Random matrices have been fruitfully employed in the study of fundamental interactions as well, in the context of large-$N$ limit of gauge theories \cite{Brezin1993}, quantum chromodynamics (\textit{QCD}) \cite{Verbaarschot2000}, as well as two-dimensional Liouville and Jackiw-Teitelboim (\textit{JT}) models of quantum gravity \cite{Francesco1995,Stanford2019}. Furthermore, the study of black holes reveals connections to RMT, through the Sachdev-Ye-Kitaev (\textit{SYK}) model \cite{Cotler2017}.

Classical statistical physics also has deep links with RMT. In the study of disordered spin systems, or spin glasses, the interaction between pairs of spins is modelled by a random matrix. The subtle emergent behaviour of the system thus ensues from random matrix properties \cite{Auffinger2013}. This extends to similar models of interacting agents in ecology \cite{May1972} and economy \cite{Moran2019,Moran2020,Morelli2020}. In addition, the directed polymer model, a toy model of disordered systems that has been at the core of modern studies in the field, was found to display the same law for the polymer's lowest energy as the maximal eigenvalue in some random matrix ensembles, the Tracy-Widom distribution \cite{Majumdar2007}. The connections further extend to the whole class of Kardar-Parisi-Zhang (\textit{KPZ}) growth models, that share the same universality class as the directed polymer's interface \cite{Halpin-Healy1995,Prahofer2000,Krajenbrink2019,Gueudre2016, Krajenbrink2017}. In recent studies, the Generalized Gibbs Ensembles (\textit{GGE}) of certain integrable models were related to the circular ensembles of RMT \cite{Grava2021,Spohn2021}.

In the mathematical realm, RMT connects to fields as far-reaching as combinatorics and number theory. In combinatorics for example, the Ulam problem is the study of the longest increasing subsequence (\textit{LIS}) in a random permutation of $N$ integers. For large $N$, the cardinality of the LIS fluctuates around its typical value, with a law related again to the Tracy-Widom distribution \cite{Majumdar2007,Johansson2001}. Startlingly, connections between RMT and number theory have deployed since the day Freeman Dyson and number theorist Hugh Montgomery realized, by chance, that the statistical properties of the objects each of them was studying were very similar: spacing distribution of large hermitian matrix eigenvalues on one part, and spacing distribution of the imaginary part of consecutive non-trivial zeros of Riemann's $\zeta$ function on the axis $\frac{1}{2} + i \mathbb{R}$ on the other part. This unexpected connection of RMT with a deterministic but intricate point distribution structure is reminiscent of the nuclear spectra comparisons that gave birth to the theory. After the first realization, numerical verifications and further studies have extended this connection to other $L$-functions and have aroused enthusiasm for a potential RMT track to the proof of the Riemann hypothesis, not yet successful unfortunately \cite{Keating2006,Biane2003}.

Outside of physics and pure mathematics as well, random matrices have found diverse applications. As a probabilistic theory, RMT teaches one what a noise matrix resembles. A natural application of these insights in a practical setting is the separation of the signal from the pure noise in an empirical matrix or its spectrum. In this respect, random matrix statistics are a useful tool in fields where noisy matrices appear, when one might be interested to reconstruct the original signal. This concerns of course a large variety of areas. For example, as soon as one has data samples for a collection of variables and seeks to understand their correlation properties, one is interested in the structure of the true underlying correlation matrix: approximating it by the empirical noisy correlation matrix requires RMT-inspired techniques to separate signal from noise. Interestingly, these questions have inspired theoretical modelling, which has given some rigorous answers. Practical matrix statistical studies have been developed, to name some areas of application, in the analysis of financial data \cite{Bouchaud2015} or in the biostatistics of single-cell RNA sequencing \cite{Luo2007,Landa2021}. Other original applications were mentioned in the introduction of the thesis as the repulsion properties in parked cars along a street pavement \cite{Abul2006}, perched birds along a power line \cite{Seba2009} and in the time intervals between bus arrivals in the Mexican city of Cuernavaca \cite{Krbalek2000}. In two dimensions, random matrix signatures were also found in the territorial behaviour of birds of prey \cite{Akemann2021}.

In certain fields, the use of RMT is becoming inescapable because of technological development. In the field of wireless communications, for example, where infrastructures employ multiple antennas to communicate with multiple users, information theory is pushed in the Multiple Input - Multiple Output (\textit{MIMO}) era. In the simplest linear model, the transmission is characterized by a channel matrix, which determines the level of optimal signal transmission. The modern requirements of efficient bandwidth use are an incentive to develop the theory with the tools of RMT, adapted to this regime, in order to obtain the best possible performance \cite{Tulino2004}.

Finally, recent studies suggest that RMT is the adequate theoretical framework for the study of the large-dimensional statistical problems at the heart of machine learning. Interestingly, it seems that usual machine learning techniques such as the kernel method, which relies on the construction of a kernel matrix in order to extract features from the data, are limited by the wrong low-dimensional intuitions they were built on. Indeed, studies of the model have shown that kernel methods can be improved by using \emph{non-intuitive} kernel functions, after they were suggested by RMT analysis of the model. This current direction of research provides hope that random matrices could be useful to shine light on some mysterious aspects of machine learning techniques which yet resist theoretical understanding \cite{Couillet2021}. For a broader scope on machine learning techniques, see \cite{Hastie2009,Cadene2020,Ernoult2020,Aubin2020,Roussel2021,Billaut2018}, and \cite{Dubber2020,Kirkham2018} for ethical considerations surrounding this field.

In addition to the specific references given for each of these topics, we point to \cite{Akemann2011} for an exhaustive review of the random matrix theory applications.

\subsection{Main random matrix ensembles}
 \label{subsec:1.1.2}

We turn to a technical description of random matrix theory and describe in this section the main ensembles of RMT. 

An \emph{ensemble} is defined by a set of matrices along with a probability measure supported on this set. In this thesis, we will mostly consider square matrices with real spectrum. Following the Wigner-Dyson classification presented in the historical section, this leaves three possibilities identified by the Dyson index $\beta$: real symmetric matrices ($\beta=1$), complex Hermitian matrices ($\beta =2$), or quaternionic self-adjoint matrices ($\beta=4$). 

We present the main classes of ensembles following this framework, starting with the Gaussian ensembles which are most central. We will then turn to the Wishart ensembles, before presenting briefly other important ensembles of the theory.

\subsubsection{Gaussian ensembles}

The first ensembles that were introduced in RMT are the Gaussian ensembles. As explained above, there are three Gaussian ensembles: 
\begin{itemize}
\setlength\itemsep{0em}
\item  the Gaussian Orthogonal Ensemble (\textit{GOE}), defined on the set of real symmetric square matrices ($\beta =1$), 
\item the Gaussian Unitary Ensemble (\textit{GUE}), defined on the set of complex Hermitian square matrices ($\beta =2$),
\item and the Gaussian Symplectic Ensemble (\textit{GSE}), defined on the set of quaternionic self-adjoint square matrices ($\beta =4$).
\end{itemize}

The Gaussian ensembles are central because they respect two restrictive constraints: \\ Firstly, in these ensembles, the entries of the random matrix are independent random variables, apart from the obvious symmetry constraints. An ensemble respecting this property is called a \emph{Wigner ensemble}. This condition makes for easy numerical simulations, as it is then straightforward to sample coefficients independently to obtain a valid matrix sample. Following this condition, a random matrix has no structure at the level of its entries. \\ Secondly, the probability measures of these ensembles are invariant with respect to a change of basis, meaning that a joint rotation of its set of eigenvectors leaves the probabilistic weight of a matrix unchanged. This will be formally detailed in the following paragraphs. Following this condition, a random matrix has no structure at the level of its eigenvectors. 

A standard result of RMT, the Porter-Rosenzweig theorem \cite{Porter1960}, states that the Gaussian ensembles are the only ones that respect the two conditions simultaneously. As a consequence, the Gaussian ensembles have been excellent candidates for structure-less random matrices since the birth of the theory in nuclear physics.

\paragraph{Gaussian Orthogonal Ensemble}
\subparagraph{Definition}

The Gaussian Orthogonal Ensemble is defined on the set of real symmetric $N \times N$ matrices $\HH = \HH ^T$, where $\HH ^T$ is the transpose of $\HH$. Denoting $\HH_{ij}$ the entries of $\HH$, we thus have $\HH_{ij} = \HH_{ji}$ for any $(i,j)$ pair. 

By the Wigner ensemble property, the $\frac{N(N+1)}{2}$ entries of the upper triangular part of the matrix are independent random variables. The GOE is defined by specifying the entry distributions as:
\be 
\HH_{ij} \sim \mathcal{N}(0,\sigma_{ij}^2) \quad \quad
\text{with} \quad
\sigma_{ij} = \left\{
    \begin{array}{ll}
       \sqrt{2} \sigma & \text{if} \ i=j \\
        \sigma & \text{else}
    \end{array}
\right.
\label{eq:1:DefGOE}
\ee
with $\sigma >0$ a fixed parameter. All entries are thus centered Gaussian variables, with variance $2 \sigma^2$ for the diagonal entries and variance $\sigma^2$ for the off-diagonal entries. An equivalent formulation of this definition is to introduce a $N \times N$ matrix $\XX$ with $N^2$ i.i.d.~Gaussian entries $\mathcal{N}(0,1)$ and to let
\be
\HH = \sigma \frac{\XX + \XX^T}{\sqrt{2}} \; .
\label{eq:1:DefGOEWithX}
\ee
As a consequence of this definition, the GOE is characterized by the following matrix probability distribution function (\textit{PDF}):
\be 
\mathcal{P} ( \HH) = \prod_{i} \left( \frac{1}{\sqrt{4 \pi \sigma^2}} \; e^{-\frac{\HH_{ii}^2}{4\sigma^2}} \right) \ 
\prod_{i < j} \left( \frac{1}{\sqrt{2 \pi \sigma^2}} \; e^{-\frac{\HH_{ij}^2}{2\sigma^2}} \right)
\ee 
with respect to the $\frac{N(N+1)}{2}$-dimensional Lebesgue measure:
\be 
\mathrm{d} \HH = \prod_{i \leqslant j} \mathrm{d}\HH_{ij} \; .
\ee 

In order to let the invariance properties of the ensemble appear, note that the asymmetry between diagonal and off-diagonal variances allows us to write the matrix PDF as
\emphbe  
\mathcal{P}(\HH)  
= C_1 \ e^{- \frac{1}{4 \sigma^2} \Tr \HH^2 } 
\label{eq:1:invariantdefGOE}
 \end{empheq}
where the normalization constant is
\be  
C_1= \frac{1}{2^{\frac{N}{2}} (2\sigma^2 \pi)^{\frac{N(N+1)}{4}}}  \; .
\ee
For a real symmetric matrix $\HH$, which is diagonalizable in an orthonormal basis, a rotation of the eigenbasis is implemented through conjugation by an element of the orthogonal group $\mathrm{O}(N)$ as $\HH \to \OO\HH\OO^{-1}$ where $\OO$ is an orthogonal matrix, i.e.~$\OO^T= \OO^{-1}$. It is visible in Eq.~\eqref{eq:1:invariantdefGOE} that the matrix PDF $\mathcal{P}$ is invariant under this transformation:
\be
\mathcal{P}(\OO\HH\OO^{-1}) = \mathcal{P}(\HH)  \quad \text{for any orthogonal matrix } \OO  \; .
\ee 
Since the Lebesgue measure itself is invariant as well, the probability measure of the GOE is invariant under orthogonal rotations. The ensemble takes its name from this property.

\subparagraph{JPDF of eigenvalues}

A real symmetric matrix $\HH$ is diagonalizable as $ \HH =\OO \LL \OO^{-1}$, with $\LL = \text{diag}(\lambda_1, \cdots, \lambda_N)$ the diagonal eigenvalue matrix and $\OO$ the orthogonal eigenvector matrix.

The joint probability distribution function (\textit{JPDF}) of eigenvalues of the GOE ensemble can be obtained as a marginal of the matrix distribution, integrating over the distribution of $\OO$ which, from the invariance property, is the flat Haar distribution. The Jacobian of the transformation from $\HH$ to the eigenvalue-eigenvector set $\{ \LL, \OO\}$ is shown in the next paragraph to be
\be 
J(\LL) = \abs{ \Delta (\lambda_1, \cdots, \lambda_N) }
\ee
where $\Delta$ is the Vandermonde determinant
\be 
\Delta( x_1, \cdots , x_N) =  \det\limits_{1\leqslant i,j \leqslant N}  \left(  x_j^{i-1}  \right)= 
\begin{vmatrix}
1& 1& \cdots & 1 \\
x_1 & x_2 & \cdots & x_N\\
x_1^2 & x_2^2 & \cdots  & x_N^2 \\
\cdots &  \cdots & \cdots  &  \cdots \\
x_1^{N-1} & x_2^{N-1} & \cdots & x_N^{N-1}
\end{vmatrix}
= \prod_{1\leqslant i<j \leqslant N} ( x_j - x_i) \; .
\ee 
Since this Jacobian factor is independent of the eigenvectors, integrating them out is trivial and the JPDF of eigenvalues is readily found as:
\emphbe 
P(\vec{\lambda}) =  C_1^N \   \abs{ \Delta (\vec{\lambda}) }  \ e^{-\frac{1}{4\sigma^2} \sum_{i=1}^N \lambda_i^2}
\label{eq:1:EigenvalueJPDFGOE}
\end{empheq}
where $\vec{\lambda}$ denotes the $N$-dimensional vector of eigenvalues $(\lambda_1, \cdots, \lambda_N)$, and with a normalization factor $C_1^N$ given by
\be 
C_1^N = 2 ^{-N}  (2 \sigma^2)^{-\frac{N(N+1)}{4}} \prod_{j=1}^{N}    \Gamma\left(1+\frac{j}{2} \right)^{-1}
\ee
such that  $\int_{\R^N} P(\vec{\lambda})  \, \dd \vec{\lambda} = 1$. 

\subparagraph{Jacobian proof} 
Let us prove that the Vandermonde determinant is indeed the Jacobian of the transformation $\HH = \OO \LL \OO^{T} \to \{ \LL , \OO \}$. Recalling that $\OO^T = \OO^{-1}$, differentiating $\HH$ yields:
\be
\dd \HH = \dd \OO \,  \LL   \OO^T +  \OO  \,  \dd \LL \,   \OO^T +  \OO    \LL \,  \dd  \OO^T  = \dd \OO \,   \LL   \OO^T +  \OO  \, \dd \LL \,   \OO^T  -  \OO  \LL  \OO^T   \dd \OO \,  \OO ^T
\ee
Introducing $\dd \tilde{ \HH} = \OO^T  \dd \HH \, \OO $ in order to study the variations in the eigenbasis, and noting that the final Jacobian is unchanged by substituting $\tilde{\HH}$ for $\HH$, we have
\be
\dd\tilde{ \HH} =  \dd \OOmega  \, \LL - \LL \, \dd \OOmega + \dd \LL
\ee
where  $ \dd \OOmega = \OO^T \dd \OO$ is an anti-symmetric matrix which encodes the variations of $\OO$. Note that it is natural that the variations of orthogonal matrices should appear in the form of anti-symmetric matrices, as it is the set which forms the Lie algebra of $\mathrm{O}(N)$. For the entry of index $(i,j)$, the variations read $ \dd\tilde{ \HH}_{ij} = \dd \OOmega_{ij} (\lambda_j - \lambda_i)  + \dd \lambda_i \delta_{ij} $
such that we obtain the following derivatives:
\be
\frac{ \partial \tilde{\HH}_{i j}}{\partial \lambda_{k}}=\delta_{i j} \delta_{i k}  \quad  ; \quad  \frac{\partial \tilde{\HH}_{i j}}{\partial \OOmega_{k \ell}}=\delta_{i k} \delta_{j \ell} \left(\lambda_{j}-\lambda_{i}\right) \; .
\ee 
The Jacobian matrix is defined as the matrix which displays as its coefficients the derivatives of each entry of $\tilde{\HH}$ with respect to each variable $\lambda_k$ or $\OOmega_{kl}$. By rearranging rows of this large matrix, the absolute value of its determinant, i.e.~the Jacobian of the transformation, can be seen to be $ J (\LL) = \abs{ \Delta (\vec{\lambda}) }  $.

\paragraph{Gaussian Unitary Ensemble}
\subparagraph{Definition}

The Gaussian Unitary Ensemble is defined on the set of complex Hermitian $N \times N$ matrices $\HH = \HH ^\dagger$, where $\HH ^\dagger$ is the Hermitian conjugate of $\HH$, such that $\HH_{ij} = \overline{\HH_{ji}}$ for any $(i,j)$ pair. 

By the Wigner ensemble property, the $N$ real diagonal entries and the $\frac{N(N-1)}{2}$ complex entries of the upper triangular part are independent random variables. The GUE is defined by specifying their distributions as: \vspace{-1pt}
\begin{itemize}[itemsep=.3em,topsep=4pt]
\item real centered Gaussian variables of variance $\sigma^2$, for the diagonal entries ;
\item complex centered Gaussian variables of variance $\sigma^2$, for the off-diagonal entries, meaning that their real and imaginary parts are independent centered Gaussian variables of variance $\frac{\sigma^2}{2}$.
\end{itemize}
An equivalent formulation of this definition is to introduce two independent $N \times N$ matrices $\XX$ and $\tilde{\XX}$, both with $N^2$ i.i.d.~real Gaussian entries $\mathcal{N}(0,1)$, and to let
\be
\HH = \sigma \frac{\XX + \XX^T}{2} +  \ii \sigma \frac{\tilde{\XX} - \tilde{\XX}^T}{2}  
\label{eq:1:DefGUEWithX}
\ee
where $\ii$ is the imaginary unit.

As a consequence of this definition, the GUE is characterized by the following matrix probability distribution function:
\be 
\mathcal{P} ( \HH) = \prod_{i} \left( \frac{1}{\sqrt{2 \pi \sigma^2}} \; e^{-\frac{\HH_{ii}^2}{2\sigma^2}} \right) \ 
\prod_{i < j} \left( \frac{1}{\pi \sigma^2} \; e^{-\frac{ \abs{ \HH_{ij}}^2}{\sigma^2}} \right)
\ee 
with respect to the $N^2$-dimensional Lebesgue measure $ \mathrm{d} \HH = \prod_{i } \mathrm{d}\HH_{ii}  \;  \prod\limits_{i<j} \mathrm{d} \mathrm{Re} \HH_{ij}  \; \mathrm{d} \mathrm{Im} \HH_{ij}$. In the same fashion as the GOE case, the matrix PDF can be written in the following simple way:
\emphbe 
\mathcal{P}(\HH) = C_2 \ e^{- \frac{1}{2 \sigma^2} \Tr \HH^2 } 
\label{eq:1:invariantdefGUE}
\end{empheq}
where the normalization constant is
\be 
C_2 = \frac{1}{\sigma^{N^2}(2 \pi^N)^{\frac{N}{2}}} \; .
\label{eq:1:invariantdefGUEC2Def}
\ee

As claimed, this ensemble is invariant under eigenbasis rotations. A complex Hermitian matrix $\HH$ is diagonalizable in an orthonormal basis, such that a rotation of the eigenbasis is implemented through conjugation by an element of the unitary group $\mathrm{U}(N)$ as $\HH \to \UU\HH\UU^{-1}$ where $\UU$ is a unitary matrix, i.e.~$\UU^\dagger= \UU^{-1}$. It is visible in Eq.~\eqref{eq:1:invariantdefGUE} that the matrix PDF $\mathcal{P}$ is invariant under this transformation:
\be
\mathcal{P}(\UU\HH\UU^{-1}) = \mathcal{P}(\HH)  \quad \text{for any unitary matrix } \UU  \; .
\ee 
Since the Lebesgue measure itself is invariant, the probability measure of the GUE is invariant under unitary rotations, as indicated by the ensemble's name once again.

\subparagraph{JPDF of eigenvalues}

A complex Hermitian matrix $\HH$ is diagonalizable as $ \HH =\UU \LL \UU^{-1}$, with $\LL = \text{diag}(\lambda_1, \cdots, \lambda_N)$ the real diagonal eigenvalue matrix and $\UU$ the unitary eigenvector matrix. In the complex setting, the Jacobian in the transformation from $\HH$ to the eigenvalue-eigenvector set $\{ \LL, \UU\}$ can be shown to be
\be 
J(\LL) = \abs{ \Delta (\lambda_1, \cdots, \lambda_N) }^2 \; . 
\ee
The JPDF of eigenvalues ensues:
\emphbe 
P(\vec{\lambda}) = C_2^N \  \abs{ \Delta (\vec{\lambda}) }^2  \ e^{-\frac{1}{2\sigma^2} \sum_{i=1}^N \lambda_i^2}
\label{eq:1:EigenvalueJPDFGUE}
\end{empheq}
with a normalization factor given by
\be 
C_2^N =  \sigma^{-N^2} (2 \pi)^{-\frac{N}{2}}  \prod_{j=1}^{N} \Gamma (1+ j )^{-1}   
\ee
such that  $\int_{\R^N} P(\vec{\lambda}) \, \dd \vec{\lambda} = 1$.

\paragraph{Gaussian Symplectic Ensemble}
\subparagraph{Definition}

The Gaussian Symplectic Ensemble is defined on the set of quaternionic self-adjoint $N \times N$ matrices $\HH = \HH^\star$, where $\HH^\star$ is the quaternionic adjoint of $\HH$. We recall that a quaternion $h \in \Quaternions$ can be written as 
\be 
h = x_r + \ii \, x_i + \jj \, x_j + \kk \, x_k \quad \text{with} \quad (x_r, x_i, x_j,x_k) \in \R^4
\ee
where $(1,\ii,\jj,\kk)$ is the canonical basis of $\Quaternions$, and that its conjugate is $\overline{h}= x_r - \ii \, x_i - \jj \, x_j - \kk \, x_k$ such that its squared norm is $\abs{h}^2=h\overline{h}= \sum_\alpha x_\alpha^2$. The quaternionic adjoint transformation of the matrix is the combination of matrix transposition and quaternionic conjugation of the entries. By the representation of quaternions as $2\times 2$ matrices, with a basis traditionally given by the identity and the three Pauli matrices, an equivalent definition of the GSE relies on the set of complex Hermitian self-dual $2N \times 2N$ matrices.

A quaternionic self-adjoint matrix $\HH$ is diagonalizable as $\HH = \Ss \LL \Ss^{-1}$, with $\LL$ the diagonal matrix of real eigenvalues and with $\Ss$ a symplectic matrix, i.e.~a quaternionic matrix such that $\Ss ^\star = \Ss^{-1}$. In the complex $2N \times 2N$ representation, the eigenvalues each have a multiplicity of 2, such that the spectrum is identical to the quaternionic setting.\\[1pt] 

The GSE is defined by specifying the distribution of entries of $\HH$ as:
\begin{itemize}[itemsep=.3em,topsep=4pt]
\item real centered Gaussian variables of variance $\frac{\sigma^2}{2}$ on the diagonal ;
\item quaternionic centered Gaussian variables of variance $\sigma^2$, out of the diagonal, meaning that the scalars $(x_r,x_i,x_j,x_k)$ are four independent real centered Gaussian variables of variance $\frac{\sigma^2}{4}$.
\end{itemize}
An equivalent formulation of this definition is to introduce four independent $N \times N$ matrices $\XX$ and $\tilde{\XX}^{(k)}$, each with $N^2$ i.i.d.~real Gaussian entries $\mathcal{N}(0,1)$, and to let
\be
\HH = \sigma \frac{\XX + \XX^T}{2\sqrt{2}} + \sum_{k=1}^3  \ii^{(k)} \sigma \frac{\tilde{\XX}^{(k)} - \tilde{\XX}^{(k),T}}{2\sqrt{2}}  
\label{eq:1:DefGSEWithX}
\ee
where we denote $(\ii^{(1)},\ii^{(2)},\ii^{(3)})=(\ii,\jj,\kk)$. The matrix PDF of the GSE, with respect to the $(2N^2-N)$-dimensional Lebesgue measure, is:
\emphbe 
\mathcal{P}(\HH) = C_4 \ e^{- \frac{1}{ \sigma^2} \Tr \HH^2 }
\label{eq:1:invariantdefGSE}
\end{empheq}
with a normalization constant equal to
\be 
C_4 = \frac{1}{2^{\frac{N}{2}} (\sigma^2 \pi /2 ) ^{\frac{2N^2-N}{2}} } \; .
\ee
This distribution is invariant under the action of the symplectic group $\mathrm{Sp}(N)$, giving its name to the ensemble, through conjugation by their quaternionic representatives $\Ss^\star = \Ss^{-1}$:
\be
\mathcal{P}(\Ss\HH\Ss^{-1}) = \mathcal{P}(\HH)  \quad \text{for any symplectic matrix } \Ss  \; .
\ee 

\subparagraph{JPDF of eigenvalues}

The Jacobian in the transformation from $\HH= \Ss \LL \Ss^{-1}$ to the eigenvalue-eigenvector set $\{ \LL, \Ss\}$ can be shown to be
\be 
J(\LL) = \abs{ \Delta (\lambda_1, \cdots, \lambda_N) }^4 \; . 
\ee
The JPDF of eigenvalues ensues:
\emphbe 
P(\vec{\lambda}) =  C_4^N \   \abs{ \Delta (\vec{\lambda}) }^4  \ e^{-\frac{1}{\sigma^2} \sum_{i=1}^N \lambda_i^2}
\end{empheq}
with a normalization factor such that $\int_{\R^N} P(\vec{\lambda}) \, \dd \vec{\lambda}= 1$, given by
\be 
C_4^N = \left(\frac{\sigma^2}{2}\right)^{-\frac{2N^2-N}{2}} \left(\frac{\pi}{2} \right)^{-\frac{N}{2}}   \prod_{j=1}^{N} \Gamma (1+ 2j )^{-1}  \; .
\ee

\paragraph{\texorpdfstring{$\beta$}{Beta}-ensembles and eigenvalue repulsion}

\subparagraph{A unifying framework}

Using the Dyson index $\beta = 1,2,4$ in the three cases presented above respectively, the Gaussian ensembles are unified in a common framework. Indeed, the matrix PDF of each of them is given by:
\be
\mathcal{P}(\HH) = C_\beta \ e^{-\frac{\beta}{4\sigma^2} \Tr \HH^2}
\ee
with a normalization constant
\be 
C_\beta =   \frac{\left(\frac{\beta}{2\sigma^2}\right)^{\frac{N}{2}\left(1+\frac{\beta}{2}(N-1)\right)}}{(2 \pi)^{\frac{N}{2}} \pi^{\frac{\beta}{2} \frac{N(N-1)}{2}}}  \; .
\ee
Similarly, their eigenvalue JPDF reads in all three cases:
\be
\label{eq:1:JPDFBeta}
P(\vec{\lambda}) = C_\beta^N \ \abs{\Delta(\vec{\lambda})}^\beta  \ e^{-\frac{\beta}{4 \sigma^2} \sum_{i=1}^N \lambda_j^2}
\ee
with a normalization constant which guarantees that $\int_{\R^N} P(\vec{\lambda}) \, \dd \vec{\lambda}= 1$:
\be 
C_\beta^N \ =  \frac{\left(\frac{\beta}{2\sigma^2}\right)^{\frac{N}{2}\left(1+\frac{\beta}{2}(N-1)\right)}}
{  (2 \pi )^{\frac{N}{2}}  }
\prod_{j=1}^{N} \frac{\Gamma\left(1+\frac{\beta}{2}\right)}{\Gamma\left(1+\frac{\beta}{2} j\right)}  \; .
\ee 

Within this framework, the Dyson index encodes the number of real degrees of freedom in an off-diagonal entry of the matrix: $\beta=1$ for a real variable; $\beta =2$ for complex variables with independent real and imaginary parts ; $\beta=4$ for quaternions with four independent scalar coefficients. Note that, in contrast, the diagonal entries are real in all three cases. This interpretation as a number of degrees of freedom suggests that it has an impact on the randomness of the off-diagonal coefficients' squared norm: in each case, the expectation of the squared norm is $\sigma^2$, but the fluctuations around this expected value decrease when the number $\beta$ of independent contributions grows, as illustrated by the Central Limit Theorem (\textit{CLT}). This hints that $\beta$ acts as a kind of inverse-temperature, as will be verified in section \ref{sec:1.2}.

In order to widen the scope, it is possible to define a random $N$-point process on $\R^N$ according to the JPDF of Eq.~\eqref{eq:1:JPDFBeta} for any value $\beta >0$, without restricting to the three values $\beta=1,2,4$ obtained from the random matrix construction. These general processes are called $\beta$-ensembles \cite{Forrester2011}. Constructions by Dumitriu and Edelman have interestingly shown, in return, that it is possible to find tridiagonal matrix models with a JPDF of eigenvalues matching Eq.~\eqref{eq:1:JPDFBeta} for general $\beta>0$, at the cost of losing the invariance properties \cite{Dumitriu2002}.

Note that the $\beta$-ensemble framework can be further generalized by considering a higher-degree polynomial or even a general function $V$, such that the JPDF is given by
\emphbe 
P(\vec{\lambda}) = C_{\beta,V}^N \ \abs{\Delta(\vec{\lambda})}^\beta  \ e^{- \beta \sum_{i=1}^N V(\lambda_i)} 
\label{eq:1:JPDFGeneralBetaEnsemble}
\end{empheq} 
The particular $\beta$-ensemble of Eq.~\eqref{eq:1:JPDFBeta} is called the $\beta$-Hermite ensemble, for reasons that will also become clear in section \ref{sec:1.2}. In the cases $\beta =1,2,4$ and for regular-enough functions $V$, it is also possible to define a corresponding invariant matrix ensemble as
\emphbe 
\mathcal{P}(\HH)  = C_{\beta,V}  \ e^{-\Tr \, V(\HH) }  
\end{empheq}
For details on the normalization constants of the invariant $\beta$-ensembles and the connection between the matrix PDF and the eigenvalue JPDF normalizations, see App.~\ref{app:normalization}. For details, on the $\Gamma$ function, see App.~\ref{app:specialfunctions}.
 
\subparagraph{Eigenvalue repulsion}

Having laid down the technical foundations for the pivotal Gaussian ensembles of RMT, we can now return to an important property presented in the historical section: the repulsion of random matrix eigenvalues.

This repulsion is visible in the eigenvalue JPDF of the Gaussian ensembles in Eq.~\eqref{eq:1:JPDFBeta}. The eigenvalues are distributed as Gaussian variables, with an additional term that induces a repulsive interaction. Indeed, the probabilistic weight of configurations where two eigenvalues are close-by is killed by the interaction Vandermonde term 
\be \abs{\Delta(\vec{\lambda})}^\beta= \prod\limits_{i<j} \abs{\lambda_j - \lambda_i}^\beta \ee
which is equal to zero for a configuration with a pair of equal eigenvalues. We recall that this Vandermonde term stems from the Jacobian factor of the transformation from the matrix to the eigenvalue-eigenvector decomposition, which is a geometrical measure of configuration volume stretching along this change of variables. For a choice of $\vec{\lambda}$ with two close-by eigenvalues, on the order of $\epsilon$, a small Jacobian, on the order of $\epsilon^\beta$, indicates a vanishing corresponding volume in matrix space. In other words, there are very few matrices with eigenvalues at distance $\epsilon$.

This may be counter-intuitive at first, as fixing an eigenvalue to the value of another one seems to be only fixing one real degree of freedom on the system. But in reality, more than one degree of freedom is lost. Consider for instance a real symmetric matrix $\HH$, with eigenvalues $\vec{\lambda}$ and corresponding set of eigenspaces $( E_\lambda )_{\lambda \in \vec{\lambda}}$. \\
In the generic setting where all eigenvalues are different, the $N$ eigenspaces all have dimension $1$ and are characterized by one eigenvector in $S_{N-1}=\{ \vec{x} \in \R^N \mid \norm{x}=1\}$. The number of real degrees of freedom in the choice of the orthonormal eigenbasis is $\frac{N(N-1)}{2}$: the first eigenvector $\vec{e}_1 \in S_{N-1}$ is characterized by $N-1$ angles; after this choice, $\vec{e}_2$ must be chosen orthogonal to $\vec{e}_1$, such that its choice bears $N-2$ degrees of freedom; and successively until $\vec{e}_N$ which is fixed by the choice of the $N-1$ previous vectors. The total number of degrees of freedom is thus $\frac{N(N-1)}{2}$, which is consistent with the representation of the eigenbasis as an orthogonal matrix $\OO$ and with the fact that adding the $N$ degrees of freedom of $\vec{\lambda}$ should recover the total $\frac{N(N+1)}{2}$ degrees of freedom of the real symmetric matrix $\HH$.\\
Consider now the case where two eigenvalues are equal, for example $\lambda_{N-1} = \lambda_N$. The first $N-2$ eigenvectors are chosen as above, and bear $\frac{N(N-1)}{2}-1$ degrees of freedom. The final eigenspace, corresponding to the multiplicity-two eigenvalue, is 2-dimensional. From the orthogonality constraints, it is fixed uniquely by the choice of the previous $N-2$ dimensions, such that the number of degrees of freedom in the choice of the eigenbasis is $\frac{N(N-1)}{2}-1$. As a consequence, the codimension of matrices with two equal eigenvalues in the space of real symmetric matrices is 2: one from the eigenvalue condition and one from the eigenbasis construction. This is one more than what the intuition first suggests, which explains geometrically why the Vandermonde Jacobian kills configurations with close-by eigenvalues in the GOE.\\
As suggested by the $\beta$ exponent which amplifies the effect of the Jacobian, the eigenvector-related degree of freedom loss is higher in the GUE and GSE. Indeed, a similar count as above shows that the (real) codimension of the space of matrices with two equal eigenvalues is \cite{Tao2012}:
\begin{itemize}[itemsep=0em,topsep=4pt]
\item 3 in the case of complex Hermitian matrices ($\beta =2$), with 2 from the eigenbasis construction ; 
\item 5 in the case of quaternionic self-adjoint matrices ($\beta =4$), with 4 from the eigenbasis construction.
\end{itemize}
This gives another interpretation to the exponent $\beta$: it is a measure of the geometric depletion of matrix space when restricting to configurations with close-by eigenvalues, which is stronger for greater $\beta$.

We can now give a generalization of the surmise which was conjectured for real symmetric matrices by Wigner, as presented in the historical section. Let us define the successive spacings $S_i$ and the unit-mean relative spacings $s_i$ as 
\be
S_i = \lambda_{i+1} - \lambda_i \quad  ; \quad  s_i = \frac{S_i}{\E [ S_i ] } \quad .
\ee
A more general Wigner's surmise statement is that the distribution obeyed by the relative spacings $s_i$ in samples of the Gaussian ensembles is very well approximated (and equal in the $N=2$ case) by the PDF
\be 
p_\beta(s)=a_\beta \, s^\beta  \, e^{-b_\beta s^{2}}
\label{eq:1:GeneralWignerSurmise}
\ee
with $\beta = 1,2,4$ according to the Gaussian ensemble considered. The coefficients $(a_\beta,b_\beta)$, fixed by the unit-mean and normalization conditions, are given below in Table \ref{tab:WignerSurmiseParameters} for completeness \cite{Abul-Magd1999}.
 
\begin{table}[ht!]
\centering
\begin{tabular}{||c|c|c|}
  \hline
 $ \displaystyle  \beta$  & \ $\displaystyle  a_\beta$ & \ $\displaystyle  b_\beta$   \\ \hhline{|=|=|=|}
 &&\\[-5 pt]
  1 & \quad $\displaystyle \frac{\pi}{2}$ \quad  &\quad \quad  $\displaystyle  \frac{\pi}{4}$ \quad \quad  \quad  \\[15pt]
  2 & \quad  $\displaystyle  \frac{32}{\pi^2}$ \quad & \quad \quad $\displaystyle  \frac{4}{\pi}$ \quad \quad  \quad  \\[15pt]
  4 & \quad $\displaystyle  \frac{262144}{729 \, \pi^3}$ \quad & \quad \quad $\displaystyle  \frac{64}{9  \pi}$ \quad \quad  \quad \\[15pt]
  \hline
\end{tabular}
  \caption{Coefficients $a_\beta$ and $b_\beta$ in the general Wigner's surmise distribution \myeqref{eq:1:GeneralWignerSurmise}.}
  \label{tab:WignerSurmiseParameters}
\end{table} 

\begin{figure}[ht!]
    \centering 
    \includegraphics[width=.9 \textwidth]{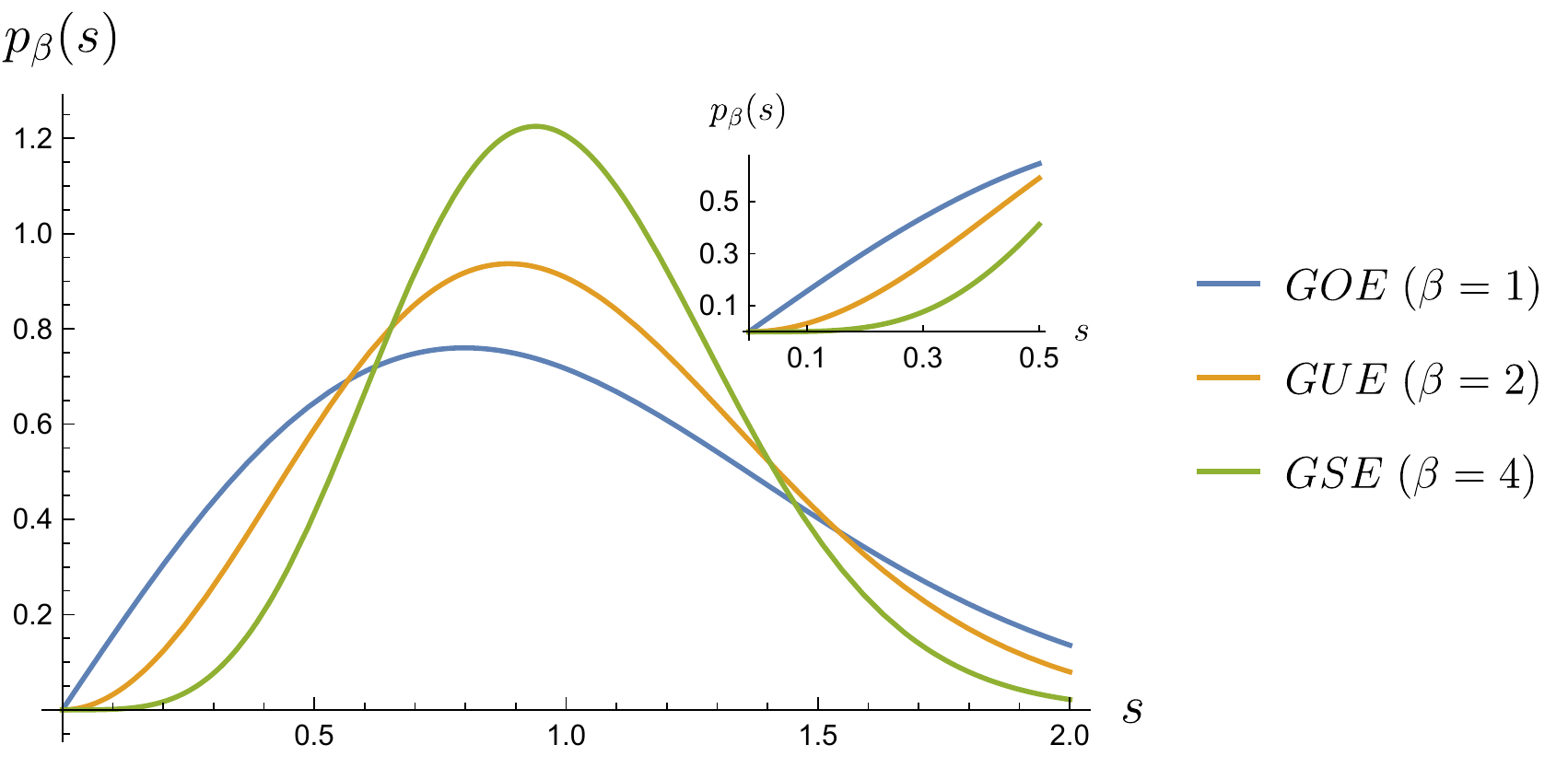} 
    \caption{Plot of the three Wigner's surmise distributions $p_\beta$ for $\beta=1,2,4$, with an inset showing a close-up on the region $s \in [0,\frac{1}{2}]$.}
    \label{fig:generalsurmise}
\end{figure}

The three PDF of the general Wigner's surmise are plotted in Fig.~\ref{fig:generalsurmise}.
In the region of small $s$, as can be seen in the inset plot, these PDF behave as $s^\beta$. As expected from the above discussion on the influence of $\beta$, this shows that the repulsion is stronger for a higher value of $\beta$, such that the probabilistic weight of low spacings is killed more effectively. On a larger scale, it appears that the width of the distribution diminishes with an increasing $\beta$, in line with the inverse-temperature interpretation mentioned above.

This closes the discussion of the generalized Wigner's surmise, which illustrates eigenvalue repulsion in the three Gaussian ensembles. Further statistical properties of the eigenvalues will be presented in section \ref{sec:1.2}.

\subsubsection{Wishart and inverse-Wishart ensembles} 

The next important ensembles that we present are the Wishart ensembles, which are of particular interest in the context of statistical applications. We will then turn to the corresponding matrix-inverse ensembles.

\paragraph{Wishart ensembles}

\subparagraph{Definition} 

According to the Wigner-Dyson classification laid out in the Gaussian case, the Wishart ensembles also come in a trio of real symmetric $(\beta=1)$, complex Hermitian $(\beta=2)$ and quaternionic self-adjoint $(\beta=4)$ invariant ensembles, such that we present them all at once for simplicity. Because of their connection to Laguerre polynomials, see section \ref{sec:1.2}, the Wishart ensembles are also  called the Laguerre Orthogonal Ensemble (\textit{LOE}), the Laguerre Unitary Ensemble (\textit{LUE}) and the Laguerre Symplectic Ensemble (\textit{LSE}) respectively.

Let $\XX$ be a rectangular $T \times N$ matrix where all entries are i.i.d.~real, complex or quaternion centered Gaussian random variables of variance $\sigma^2$. Denoting $\XX^\dagger$ the adjoint of $\XX$ in the three cases, the Wishart ensembles are composed of the $N \times N$ matrices 
\begin{empheq}[box=\setlength{\fboxsep}{8pt}\fbox]{equation}
\WW = \XX^\dagger \XX
\label{eq:1:DefWishart}
\end{empheq}
with the probability measure inherited from that of $\XX$. By construction, $\WW^\dagger =\WW$ such that $\WW$ is self-adjoint in the three cases, and has real spectrum. Furthermore, it can be noted that $\bfx^\dagger \WW \bfx \geqslant 0$ for any vector $\bfx$, such that $\WW$ is a positive semidefinite matrix, with non-negative eigenvalues.

As explained in the historical section, this ensemble was studied in theoretical statistics in the real setting by John Wishart and others, before the actual birth of RMT \cite{Wishart1928,Hsu1939,Hua1963}. Their interest in this model lies in the fact that $\WW$ has a pivotal statistical interpretation: it is related to the empirical \emph{covariance matrix} of $N$ variables, for which $T$ samples are stored in the corresponding column of $\XX$. Indeed, with this interpretation of $\XX= (X_i(t))_{ 1 \leqslant t \leqslant T,1 \leqslant i \leqslant N}$ as the data collected on $N$ centered real random variables from $T$ trials, the empirical covariance between two variables is
\footnote{Note that we have restricted the scope to centered variables. In the more general case where the variable means are unknown, the unbiased empirical covariances are given by ${\moy{X_i X_j}_\mathrm{emp.} = \frac{1}{T-1} \sum_{t=1}^T (X_i(t) - \overline{X_i})  (X_j(t)- \overline{X_j})}$ with the empirical averages $\overline{X_i}= \frac{1}{T}\sum_{t=1}^T X_i(t) $, such that the structure of Eq.~\eqref{eq:1:DefWishart} remains after centering the data.} :
\be 
\moy{X_i X_j}_\mathrm{emp.} = \frac{1}{T} \sum_{t=1}^T X_i(t) X_j(t) = \frac{1}{T} \WW_{ij} \; .
\ee
When the different trials $t \in [1 \ldotp \ldotp T]$ are independent from each other, $\frac{1}{T} \WW$ is an unbiased estimator of the real covariance matrix $(\moy{X_i X_j})_{1 \leqslant i,j \leqslant N}$, meaning that $\E \left[ \frac{1}{T} \WW_{ij} \right] = \moy{X_i X_j}$. Because of this property, the empirical covariance matrix is a useful tool in many situations to gain information about the structure of the data. In financial applications for example, the variables $X_i(t)$ can represent stock price variations observed at $T$ instants in time. Finding hints about the correlations between price changes, i.e.~reconstructing the correlation matrix, can be very useful to optimize a portfolio or to build hedging strategies \cite{Bouchaud2003,Walter2013,Bun2017,Potters2020,Bun2016a}.

In the zero-signal case where the random variables $(X_i)_{1\leqslant i \leqslant N}$ are i.i.d.~and Gaussian, the empirical covariance matrix is a sample of the Wishart ensemble. Understanding the properties of this model can then help build statistical tools for practical applications. Furthermore, the statistical technique of \emph{principal component analysis} (\textit{PCA}) gives a strong incentive to specifically study the eigenvalues of this model. This widely used technique aims at compressing the data to retain only its most significant features: it consists in finding the high-variance linear combinations of the variables $X_i$, which can be shown to be encoded in the eigenvectors corresponding to the highest eigenvalues of the empirical covariance matrix. In the ideal i.i.d.~Gaussian case, such properties are governed by the statistics of eigenvalues in the Wishart ensemble, which can thus give deep insights on more general applications of the PCA technique.  For more details about PCA, we point to the introductory note \cite{Shlens2014}.

\subparagraph{Matrix PDF} We present now the matrix PDF of the Wishart ensembles, with respect to the adequate Lebesgue measures. Omitting a normalization constant, the matrix distribution of $\XX$, as defined above Eq.~\eqref{eq:1:DefWishart}, is 
\be 
\mathcal{P}(\XX) \propto e^{- \frac{\beta}{2\sigma^2} \Tr \XX^\dagger \XX } \; .
\ee
The Jacobian of the transformation $\XX \to \WW =  \XX^\dagger \XX$ can be shown to be equal to $\det(\WW)^{ \frac{\beta}{2} (T-N+1)-1}$ such that the matrix PDF of the Wishart ensembles is:
\emphbe
\mathcal{P}(\WW) =  K_{\beta}  \ 
\det (\WW)^{ \frac{\beta}{2} (T-N+1)-1}  
\ e^{-  \frac{\beta}{2\sigma^2} \Tr \WW} 
\label{eq:1:MatPDFWishart}
\end{empheq}
where the normalization constant $K_{\beta}  $ is 
\be
K_{\beta} =
\left(\frac{ \beta}{2\sigma^2}  \right) ^{\frac{\beta}{2} N T}    \frac{\pi^{- \frac{\beta}{2} \frac{N(N-1)}{2}  }}{    \prod_{j=1}^N \Gamma \left( \frac{\beta}{2}( T - N +j) \right)}  \; .
\label{eq:1:MatNormalizationConstantWishart}
\ee

Note in the matrix PDF of Eq.~\eqref{eq:1:MatPDFWishart} that the Wishart ensembles are invariant under orthogonal, unitary and symplectic conjugation respectively, in the same way as the Gaussian ensembles.

\subparagraph{JPDF of eigenvalues} 
From the invariance properties, the eigenvalue JPDF is derived from the matrix PDF as in the Gaussian case and yields:
\emphbe
P(\vec{\lambda} ) =
K_{  \beta}^N  \ \abs{\Delta(\vec{\lambda})}^{\beta}  \ \prod_{i=1}^{N} \lambda_{i}^{\frac{\beta}{2}(T-N+1)-1}  \
e^{- \frac{\beta}{2\sigma^2} \sum_{i=1}^N \lambda_{i}} 
\label{eq:1:JPDFEigenvaluesWishartEnsemble}
\end{empheq}
where the normalization constant $K_{ \beta}^N  $, which ensures $\int_{{\R^+}^N} P(\vec{\lambda}) \, \dd \vec{\lambda}= 1$, is:
\be
K_{\beta}^N  =\left(\frac{ \beta}{2\sigma^2}  \right) ^{\frac{\beta}{2} N T} \ \prod_{j=1}^{N} \frac{\Gamma\left(1+\frac{\beta}{2}\right)}{\Gamma\left(1+\frac{\beta}{2} j\right) \Gamma\left(\frac{\beta}{2} ( T-N+j)\right)}
\; .
\label{eq:1:EigNormalizationConstantWishart}
\ee

The eigenvalues of the Wishart ensembles are distributed as Gamma random variables 
\be 
X \sim \Gamma( k , \theta ) \quad \quad  
P_X(x) =  \frac{1}{\Gamma(k) \theta^{k}} \  x^{k-1} \ \mathrm{e}^{-\frac{x}{\theta}} \ \1_{\R^+}(x) \quad \quad   
\text{with} \quad 
 \left\{
    \begin{array}{ll}
        k =  \frac{\beta}{2}(T-N+1) \\
        \theta = \frac{2 \sigma^2}{\beta}
    \end{array}
\right.
\ee
with an additional Vandermonde term that induces repulsion among the eigenvalues. This process is called the Laguerre $\beta$-ensemble, and is generalizable to arbitrary $\beta >0$ and to non-integers $N$ and $T$, as long as $T> N-1$ which guarantees normalizability. This $\beta$ ensemble is the particular case of Eq.~\eqref{eq:1:JPDFGeneralBetaEnsemble} with 
\be 
V(\lambda)=  \frac{1}{2 \sigma^2 } \lambda - \left( \frac{1}{2}(T-N+1) - \frac{1}{\beta} \right) \log \lambda \; .
\label{eq:1:VWishart}
\ee 
As in the Gaussian case, tridiagonal matrix models respecting the Wishart eigenvalue JPDF have been constructed for non-standard values of $\beta$ \cite{Dumitriu2002}.

\paragraph{Inverse-Wishart ensembles}
 
\subparagraph{Definition} 
The inverse-Wishart ensembles are the three ensembles defined by the matrices
\be 
\InverseW = \WW^{-1}
\ee
where $\WW$ belongs to the corresponding Wishart ensemble. Taking into account the Jacobian of the $\WW \to \WW^{-1}$ transform, equal to $\det (\WW) ^{-\beta(N-1)-2}$, we obtain the inverse-Wishart matrix probability density function as
\emphbe
\mathcal{P}\left( \InverseW \right) = K_{ \beta}   \
\operatorname{det}(\InverseW)^{ -\frac{\beta}{2} (T + N -1 ) -1 } \
 e^{-  \frac{\beta}{2 \sigma^2} \Tr \InverseW^{-1}}
 \label{eq:1:MatPDFInverseWishart}
\end{empheq}
with the normalization constant given in Eq.~\eqref{eq:1:MatNormalizationConstantWishart}.
 
\subparagraph{JPDF of eigenvalues} 
The corresponding joint eigenvalue distribution is
\emphbe
P(\vec{\lambda} ) =K_{\beta}^N  \ \abs{\Delta(\vec{\lambda})}^{\beta}  \ \prod_{i=1}^{N} \lambda_{i}^{-\frac{\beta}{2} (T + N -1 ) -1 } \
e^{- \frac{\beta}{2\sigma^2} \sum_{i=1}^N \frac{1}{\lambda_{i}}  } 
 \label{eq:1:EigPDFInverseWishart}
\end{empheq}
with the normalization constant given in Eq.~\eqref{eq:1:EigNormalizationConstantWishart}, such that it is the particular case of Eq.~\eqref{eq:1:JPDFGeneralBetaEnsemble} with 
\be 
V(\lambda)=  \frac{1}{2 \sigma^2 } \frac{1}{\lambda} + \left( \frac{1}{2}(T+N-1) + \frac{1}{\beta} \right) \log \lambda \; .
\ee 

\subsubsection{Other ensembles}

In this last subsection, we briefly mention other interesting ensembles of RMT.

\paragraph{Jacobi ensembles} 

The Gaussian and Wishart ensembles are natural matrix generalizations of the Gaussian and Gamma laws of real random variables, as seen above. Interestingly, a similar generalization exists for the Beta distribution: the Jacobi ensembles. The Beta distribution is the law of a real random variable $X \in [0,1]$ sampled from two Gamma random variables $Y_1$ and $Y_2$ as
\begin{equation}
X = \frac{Y_1}{Y_1 + Y_2}= \frac{1}{1 + Y_1^{-1} Y_2}  \quad \text{with} \quad Y_{1,2} \sim \Gamma(c_{1,2},1)
\end{equation}
and it has the PDF
\begin{equation}
P_X(x)=\frac{\Gamma\left(c_{1}+c_{2}\right)}{\Gamma\left(c_{1}\right) \Gamma\left(c_{2}\right)} x^{c_{1}-1}(1-x)^{c_{2}-1} 
\end{equation} 
which is normalized per the Beta function, see App.~\ref{app:specialfunctions}.

In a parallel construction, the Jacobi ensembles consist of $N \times N$ self-adjoint matrices $\JJ$ with eigenvalues in $[0,1]$ sampled from two Wishart matrices $ \BB_1$ and $\BB_2$ as:
\begin{equation}
\JJ = ( \ID + \BB_1^{1/2}  \BB_2^{-1}  \BB_1^{1/2} ) ^{-1}
\end{equation}
where $\BB_{1,2} = \XX_{1,2}^\dagger \XX_{1,2}$ with $\XX_{1,2}$ a rectangular matrix of size $T_{1,2} \times N$ with all entries i.i.d. Gaussian; and where $ \ID$ is the identity matrix. With $Z_\beta$ and $Z_\beta^N$ fixed by normalization, we give the matrix PDF of this ensemble:
\begin{equation}
\mathcal{P}( \JJ) =Z_\beta \  \det \JJ ^{\beta (T_1-N+1) / 2-1}   \det( \mathbf{1}-\JJ)^{\beta(T_2-N+1) / 2 -1}  
\end{equation}
and the corresponding eigenvalue JPDF:
\begin{equation}
P(\vec{\lambda}) = Z_\beta^N  \  \abs{\Delta(\vec{\lambda)}}^\beta  \ \prod_{i=1}^N \lambda_i^{\beta (T_1-N+1) / 2-1} \   \prod_{i=1}^N (1-\lambda_i)^{\beta(T_2-N+1) / 2 -1} 
\end{equation} 
such that the Jacobi $\beta$-ensemble is the particular case of Eq.~\eqref{eq:1:JPDFGeneralBetaEnsemble} with 
\be 
V(\lambda)=  - A_1  \log \lambda  - A_2 \log ( 1- \lambda) \quad \text{;} \ A_{1,2}={\frac{1}{2}(T_{1,2}-N+1)  -\frac{1}{\beta}}   \; .
\label{eq:VJacobi}
\ee  
See \cite{Potters2020,Forrester2010} for the value of the Jacobi ensemble normalization constants $Z_\beta$ and $Z_\beta^N$.

We summarize in Table \ref{tab:MainEnsembles} the four $\beta$-ensembles presented in the previous pages, with the support of the matrix spectrum in these ensembles and their eigenvalue JPDF.

\begin{table}[ht!]
\centering
\begin{tabular}{||c|c|c|}
  \hline
 Ensemble  & \ Support & \ Eigenvalue JPDF : $ P(\vec{\lambda}) \propto$  \\ \hhline{|=|=|=|}
 && \\[0pt]
  Gaussian & $\R$  &  \quad  $ \abs{\Delta(\vec{\lambda})}^\beta  \ \prod_{i=1}^N \ e^{-\frac{\beta}{4 \sigma^2} \lambda_i^2} $  \quad \quad  \\[20pt]
  Wishart &  $\R^+$ &   $  \abs{\Delta(\vec{\lambda})}^\beta  \ \prod_{i=1}^{N} \ \lambda_{i}^{\frac{\beta(T-N+1)}{2}-1}  
e^{- \frac{\beta}{2\sigma^2} \lambda_{i}}  $  \\[20pt]
  inverse-Wishart &  $\R^+$ &   $   \abs{\Delta(\vec{\lambda})}^\beta  \ \prod_{i=1}^{N} \ \lambda_{i}^{-\frac{\beta(T+N-1)}{2}-1}  
e^{- \frac{\beta}{2\sigma^2 \lambda_{i}} }  $  \\[20pt]
  Jacobi &  $[0,1]$ & $  \abs{\Delta(\vec{\lambda})}^\beta  \ \prod_{i=1}^N \  \lambda_i^{\frac{\beta}{2} (T_1-N+1) -1}  (1-\lambda_i)^{ \frac{\beta}{2} (T_2-N+1)  -1}  $ \\[15pt]  
  \hline
\end{tabular} 
\caption{Main random matrix ensembles}
\label{tab:MainEnsembles}
\end{table}

Finally, another ensemble with an interesting matrix construction is the Cauchy ensemble obtained with $V(\lambda) = (\frac{N-1}{2}+\frac{1}{\beta})  \log(1+\lambda^2)$ \cite{Burda2002}. Note that the general JPDF Eq.~\eqref{eq:1:JPDFGeneralBetaEnsemble} allows one to generalize any real distribution, but that this does not necessarily come with a matrix interpretation.

\paragraph{Ensembles with complex spectrum} 
Relaxing the restriction to ensembles of self-adjoint matrices with real spectrum, it is possible to consider more general ensembles with complex-valued eigenvalues.

\subparagraph{Circular ensembles} 
The ensembles of orthogonal matrices in $\mathrm{O}(N)$, unitary matrices in $\mathrm{U}(N)$ and symplectic matrices in $\mathrm{Sp}(N)$, distributed according to the uniform Haar measure of the corresponding group, are called the circular ensembles. More precisely, they are the Circular Orthogonal Ensemble (\textit{COE}), the Circular Unitary Ensemble (\textit{CUE}) and the Circular Symplectic Ensemble (\textit{CSE}) respectively.

The eigenvalues of these matrices lie on $\U= \{ z \in \C  \mid \abs{z}=1  \}$, the unit circle of the complex plane, such that the set of eigenvalues can be denoted as $( e^{i \theta_k})_{1\leqslant k\leqslant N}$ with $\theta \in [0,2\pi]^N$. With the Dyson index $\beta$ distinguishing the three cases as usual, the JPDF of the real $\theta$ variables is:
\be 
P( \vec{\theta}) = Q_\beta^N \prod_{i<j  }\left|e^{\ii \theta_{i}}-e^{\ii \theta_{j}}\right|^{\beta}
\ee
with $Q_\beta^N$ fixed by normalization. 

\subparagraph{Ginibre ensembles}
By relaxing the symmetry constraints of the three Gaussian ensembles, one obtains ensembles of real, complex and quaternionic matrices where the $N^2$ entries are i.i.d.~centered Gaussian random variables of variance $\sigma^2$: this defines the Ginibre ensembles \cite{Ginibre1965}, denoted $GinO \! E,GinU\! E$ and $GinS\! E$. We will not detail here the JPDF of the complex eigenvalues, however note the interesting property that, after a rescaling of the matrix by $\frac{1}{\sigma \sqrt{N}}$, the eigenvalues of a Ginibre matrix tend, in the large-$N$ limit, to spread uniformly on  $\D = \{ z \in \C  \mid \abs{z} \leqslant 1  \}$, the unit disc of the complex plane \cite{Bordenave2012}. Note however that the situation is subtle at large but finite $N$: in the $GinO \! E$, a positive fraction of the complex eigenvalues is then found on the real axis \cite{Baik2020}.

\subparagraph{Elliptic ensembles}
In the Gaussian ensembles, two symmetrical entries of the matrix are conjugate (or equal in the real case): $\HH_{ji}= \overline{\HH_{ij}}$. In the Ginibre ensembles, they are completely independent. The elliptic ensembles are an interpolation of these two extreme cases by considering real, complex and quaternionic matrices where entries are centered $\sigma^2$-variance Gaussian random variables verifying $\E \Big[ \HH_{ji} \overline{\HH_{ji}} \Big] = \tau \sigma^2$, with $\tau \in [0,1]$.

For $\tau =1$, the covariance of the Gaussian random variables $(\HH_{ij},\HH_{ji})$ saturates their shared variance such that $\HH_{ji}= \overline{\HH_{ij}}$ and the Gaussian ensembles are recovered. For $\tau =0$, these Gaussian random variables are independent such that the Ginibre ensembles are recovered. 

For intermediate values of $\tau$, the properties of the eigenvalues interpolate between those of the real spectrum of the Gaussian ensembles and those of the complex Ginibre spectrum. In particular after a rescaling by $\frac{1}{\sqrt{N}}$, the eigenvalues spread, in the large-$N$ limit, on an ellipse in the complex plane, with an eccentricity parameter expressed in terms of $\tau$ \cite{Khoruzhenko2009,Potters2020}.

What is more, the range of $\tau$ can be further extended to $[-1,1]$. Indeed for ${\tau = -1}$, $\HH_{ji} = - \overline{\HH_{ij}}$ such that $\HH$ is anti-symmetric, anti-Hermitian, anti self-adjoint for $\beta = 1,2,4$ respectively, and the spectrum is then supported on the imaginary axis of the complex plane. For $-1 < \tau < 0$, the properties of the eigenvalues of the elliptic ensembles interpolate between those of this imaginary spectrum and those of the complex Ginibre spectrum. \\[5pt]

Besides this brief overview, many other interesting ensembles and matrix models have been defined and studied. For complementary elements to the presentation of random matrix theory given in this section, we point to the books 
\cite{Mehta2004,Potters2020,Akemann2011,Forrester2010,Livan2018,Tao2012} and the theses \cite{Nadal2011,Grabsch2018,Marino2015}. In the next section, we detail further statistical properties of the eigenvalues in random matrix ensembles.

\section{Eigenvalue statistics and fermion systems}
\label{sec:1.2}

As discussed and illustrated in the previous section, eigenvalues of random matrices interact repulsively, such that configurations with small spacings occur infrequently. In addition to the behaviour of nearest-neighbour spacings, we will see in this section that it is possible to investigate many other statistical properties of the eigenvalues in $\beta$-ensembles.

In the first subsection, we will discuss the typical densities observed as the matrix size becomes large, investigating the global statistics of eigenvalues. In subsection \ref{subsec:1.2.2}, we will turn to the local statistics and detail connections to fermion systems and determinantal point processes. In the last subsection \ref{subsec:1.2.3}, we will discuss the behaviour of the maximal eigenvalue, i.e.~the extreme value statistics of the system.

\subsection{Global statistics: typical densities}
\label{subsec:1.2.1}

Consider the general $\beta$-ensemble defined in Eq.~\eqref{eq:1:JPDFGeneralBetaEnsemble}, in which $N$ eigenvalues are confined by a potential $V$ while repelling each other. How do these two influences balance out in the system when $N$ becomes large ? What does a typical configuration resemble ? We address these questions for the main ensembles, by first giving a physical interpretation to the system.

\subsubsection{Coulomb gas interpretation}

The JPDF of eigenvalues in a $\beta$-ensemble given by Eq.~\eqref{eq:1:JPDFGeneralBetaEnsemble} can be rewritten directly in the following way:
\be 
P(\vec{\lambda}) = \frac{1}{Z_{\beta,V}} \  e^{- \beta \, \mathcal{H}(\vec{\lambda})}  
\label{eq:1:CoulombGas}
\ee 
where
\be 
\mathcal{H}(\vec{\lambda}) =  \sum_{i=1}^N V(\lambda_i) -  \sum_{i<j}  \log \abs{\lambda_j - \lambda_i} \; .
\label{eq:1:HamiltonianCoulombGas}
\ee

In this form, the probabilistic system directly gains a statistical physics interpretation: the eigenvalues $\vec{\lambda}$ are the positions of particles on a line, governed by the classical Hamiltonian $\mathcal{H}$, such that they are placed in an external potential $V$ and repel each other logarithmically. Since the logarithm is the Green's function of the Laplacian operator in two dimensions, this interaction is the repulsion of charged particles in two-dimensional space and this system is called a \emph{Coulomb gas}.

In this statistical physics perspective, the Dyson index is equal to the inverse of the system's temperature $\beta =\frac{1}{T}$, as hinted by observations in the previous section. Finally, the normalization constant appears now as the inverse of the partition function of the system $Z_{\beta,V}$.

The question of global statistics becomes the following: in the external potential $V$, how does a plasma of same-charge particles arrange itself in space in order to reach equilibrium ? More precisely, when the number of particles become large, an interesting observable is the average density of particles defined as
\be 
\rho(\lambda) = \frac{1}{N} \sum_{i=1}^N \E \left[ \delta(\lambda - \lambda_i) \right] \; .
\label{eq:1:defDensity}
\ee
This is the marginal distribution for one eigenvalue, such that it is normalized and can also be written
\be 
\rho(\lambda) = \int_{\R^{N-1}} P(\lambda, \lambda_2, \cdots, \lambda_N) \ \prod_{i=2}^N \dd \lambda_i \; .
\ee 

In the large-$N$ limit, the average density of $\beta$-ensembles, which is also the equilibrium configuration of the Coulomb gas, converges to a certain typical density which depends on the external potential. We will present this typical density in the Gaussian case where the potential is $V(\lambda) = \frac{\lambda^2}{4\sigma^2}$ and in the Wishart case where the potential is $V(\lambda) = A \lambda + B \log \lambda$ with constants given in Eq.~\eqref{eq:1:VWishart}.

\subsubsection{Wigner's semicircle}

As mentioned in the historical section, Eugene Wigner was the first to study the Gaussian ensembles and their global statistics. He exhibited the typical density in which the eigenvalues of these ensembles arrange at large $N$, which now bears his name: Wigner's semicircle \cite{Wigner1958}. In this section, we discuss the global statistics of the Gaussian ensembles and subsequently fix the harmonic potential
\be 
V(\lambda) = \frac{\lambda^2}{4\sigma^2} \; .
\ee

\paragraph{Density convergence}

Because of the repulsion, the typical lengthscale $\ell$ on which the eigenvalues spread out must grow with $N$. To find this rate of expansion, notice that the two terms in Eq.~\eqref{eq:1:HamiltonianCoulombGas} must scale identically with $N$ in the typical configuration, in order to respect the equilibrium between confinement in the quadratic potential $V(\lambda) = \frac{\lambda^2}{4\sigma^2}$ and logarithmic repulsion. The potential term scales as $N \ell ^2$, while the interaction term scales as $N^2$, such that the typical support of the density must grow as $\ell = \sqrt{N}$. The typical density is observed after a rescaling by this factor; in the following we denote $\mu$ the variable in rescaled space. 

\begin{figure}[ht!]
    \centering 
    \includegraphics[width=.6 \textwidth]{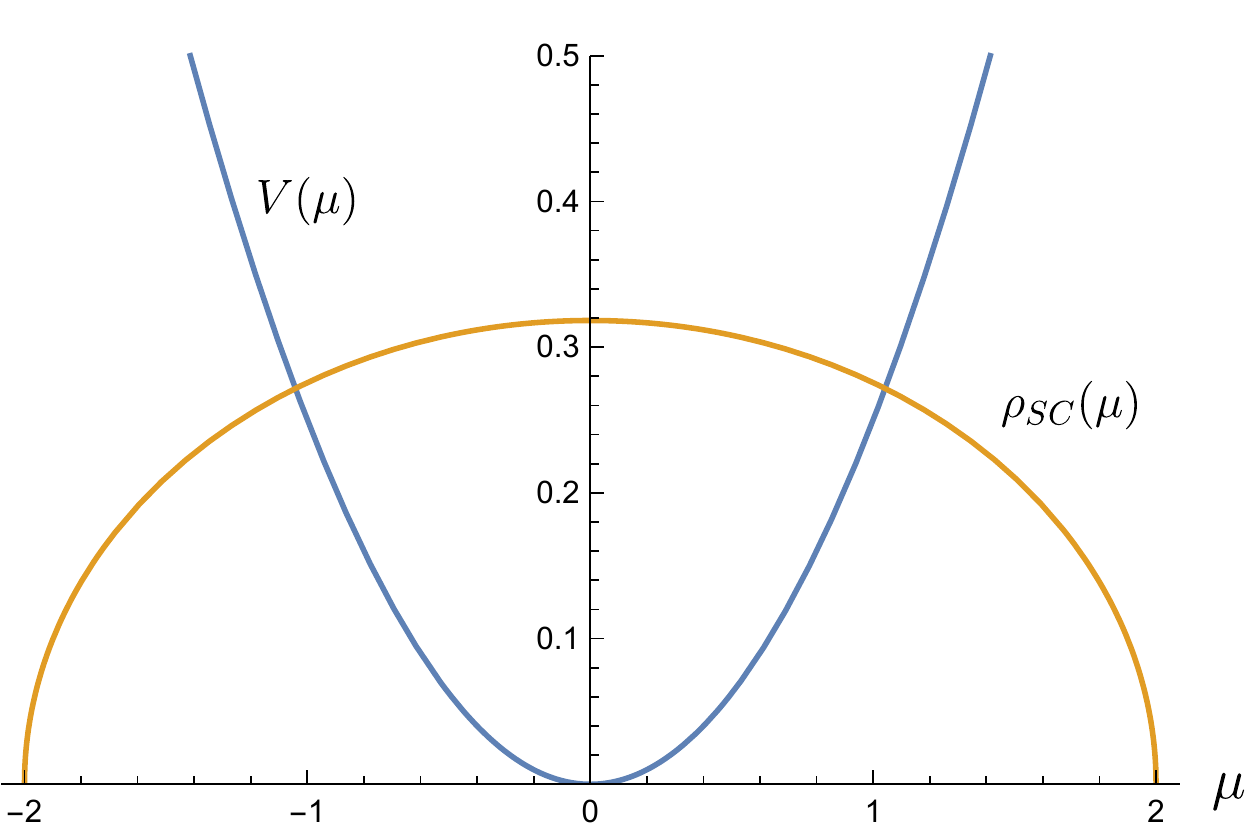} 
    \caption{Wigner's semicircle and quadratic potential of the Gaussian ensembles, for $\sigma=1$.}
    \label{fig:WignerSemicircle}
\end{figure}

Let $\HH$ be a $N \times N$ matrix from one of the three Gaussian ensembles and let $\rho_\mathrm{\sqrt{N}}$ be the average density of the rescaled matrix $\MM=\frac{\HH}{\sqrt{N}}$. Then:
\be 
\rho_\mathrm{\sqrt{N}} (\mu) = \frac{1}{N} \sum_{i=1}^N \E \left[ \delta \left(\mu - \frac{\lambda_i}{\sqrt{N}}\right) \right]     \quad \quad \xrightarrow[N \to \infty]{}  \quad \quad \rho_{\rm SC} (\mu) 
\label{eq:1:WignerSemicircleConvergence}
\ee
in the weak sense of distribution convergence, where $\rho_{\rm SC}$ is Wigner's semicircle distribution
\begin{empheq}[box=\setlength{\fboxsep}{8pt}\fbox]{equation}
\rho_{\rm SC} (\mu) = \frac{\sqrt{4\sigma^2-\mu^2}}{2\pi \sigma^2}
\label{eq:1:WignerSemicircleDefinition}
\end{empheq}
This distribution is supported on $[-2\sigma, 2\sigma]$ and plotted in Fig.~\ref{fig:WignerSemicircle} with the quadratic potential of the Gaussian ensembles.

Strikingly, this convergence result on the density of $\MM = \frac{\HH}{\sqrt{N}}$ extends to the empirical density: if one replaces in Eq.~\eqref{eq:1:WignerSemicircleConvergence} the average density of the rescaled matrix by its empirical spectral distribution (\textit{ESD}) $\rho_{\rm emp}$ defined as
\be 
\rho_{\rm emp}(\mu)= \frac{1}{N} \sum_{i=1}^N  \delta(\mu - \frac{\lambda_i}{\sqrt{N}}) \; ,
\ee
then the convergence to Wigner's semicircle is true in the almost-sure sense. This shows a general feature of random matrices: the eigenvalues of a large matrix self-average to the equilibrium configuration.

In the next paragraphs, we outline a few trails of proof, in order to show how the eigenvalues of the Gaussian ensembles arrange themselves in the semicircle distribution, first by a matrix analysis, then by using the Coulomb gas interpretation presented in the previous section.

\paragraph{Resolvent derivation}

The first trail we take to the semicircle is by a direct analysis of the matrix structure, restricting to the real case for simplicity. A useful tool is the resolvent matrix $\GG_{\MM}(z)$ obtained from the rescaled GOE matrix $\MM$ and depending on a complex parameter $z$. Its normalized trace $\g(z)$ is called the \emph{Stieltjes transform}, or also simply \emph{resolvent}. Denoting the eigenvalues of $\MM$ as $\mu_i = \frac{\lambda_i}{\sqrt{N}}$, both objects are defined as:
\be
\GG_{\MM}(z)= \left(z \ID - \MM \right)^{-1}
\label{eq:1:DefResolvent}
\ee
and  
\be 
 \mathfrak{g}(z) = \frac{1}{N} \Tr \, \GG_{\MM}(z) = \frac{1}{N} \sum_{i=1}^N \frac{1}{z- \mu_i} = \int_{-\infty}^\infty \frac{\rho_{\rm emp}(\mu)}{z- \mu} d \mu \; .
 \label{eq:1:DefStieltjes}
 \ee
We see that the Stieltjes transform $\g$ is defined directly from the empirical density $\rho_{\rm emp}$. Conversely, it is also possible to extract $\rho_{\rm emp}$ from the knowledge of $\mathfrak{g}$ close to the real axis, from the Sokhotski–Plemelj formula \cite{Plemelj1908,Muskhelishvili2008}:
\be
\rho_{\rm emp}(\mu) = \frac{1}{\pi}  \lim\limits_{\ \ \eta \to 0^+} \operatorname{Im} \g ( \mu- \ii \eta) \; .
\label{eq:1:SokhotskiPlemeljFormula}
\ee

A trail to the semicircle is then to first obtain the large-$N$ limit of the Stieltjes transform of $\MM$. Applying Schur's complement formula 
\footnote{
Let $\AAA$ an invertible matrix, divided in four blocks:
$\AAA = ( ( \AAA_{11}, \AAA_{12}) , (\AAA_{21}, \AAA_{22}))$ 
and
$\AAA^{-1} = \GG = ( ( \GG_{11}, \GG_{12}) , (\GG_{21}, \GG_{22}))$.
The upper-left block of $\mathbf{G}$ is given by Schur's complement formula \cite{Potters2020} as:
$
\mathbf{G}_{11}^{-1}=\AAA_{11}-\AAA_{12}\left(\AAA_{22}\right)^{-1} \AAA_{21}
$.
}
to $\AAA = z\mathbf{1} -\MM=\GG_{\MM}(z)^{-1}$ with $\AAA_{11}$ its top-left element and $\AAA_{22}$ the $(N-1)\times (N-1)$ submatrix of $\AAA$ with the first row and column removed gives directly:
\be
\label{eq:1:SchurGM}
\frac{1}{{\GG_{\MM}(z)}_{11}} = \AAA_{11}- \sum_{k,l=2}^N \AAA_{1k}\left(\AAA_{22}\right)^{-1}_{kl} \AAA_{l1}
\ee
Taking the expectation on the r.h.s.~over the variables of the first line and column $\AAA_{1k} = \AAA_{k1} = - \MM_{1k}$, we obtain: $\E_{\left\{\MM_{1k}\right\}}\left[\AAA_{1 k}\left(\AAA_{22}\right)_{kl}^{-1} \AAA_{l 1 }\right]=\frac{\sigma^2}{N}\left(\AAA_{22}\right)_{kk}^{-1} \delta_{kl}$ such that the sum converges to the Stieltjes transform of the $(N-1) \times( N-1)$ submatrix of $\MM$ with first line and column removed, which is proportional to $\mathfrak{g}$ in the large-$N$ limit:
\be
\E_{\left\{\MM_{1 i}\right\}}\left[\sum_{k, l=2}^{N} \AAA_{1 k}\left(\AAA_{22}\right)_{k l}^{-1} \AAA_{l 1}\right]
=\frac{\sigma^2}{N} \Tr \left[\left(\AAA_{22}\right)^{-1}  \right]
\ \ \xrightarrow[N \to \infty]{}  \ \ \sigma^2 \mathfrak{g}(z)
\ee

With ${\mathbf{G}_{\MM}(z)}_{11}$ having the same distribution as all other diagonal entries by rotational invariance, we have $\E\left[{\mathbf{G}_{\MM}(z)}_{11}\right]=\frac{1}{N} \E\left[\Tr \, \mathbf{G}_{{\MM}}(z)\right]\rightarrow \mathfrak{g}(z)$. In the large-$N$ limit where fluctuations are small, the expectation of \eqref{eq:1:SchurGM} then gives the following self-consistent equation for the Stieltjes transform $\mathfrak{g}$:
\be 
\frac{1}{\mathfrak{g}(z)} = z - \sigma^2 \mathfrak{g}(z)
\ee
The Stieltjes transform of a large GOE matrix is then the complex function:
\be
\g(z)=\frac{z-  \sqrt{z^2-4\sigma^2 }}{2\sigma^2}
\label{eq:1:SelfConsistentG}
\ee
where the branch is chosen such that $ \g(z) \simeq \frac{1}{z}$ as $z \to \infty$.
From this solution, we obtain by the Sokhotski–Plemelj formula \myeqref{eq:1:SokhotskiPlemeljFormula} the semicircle density $\rho_\mathrm{SC}$ of Eq. \eqref{eq:1:WignerSemicircleDefinition}.  

\paragraph{Coulomb gas derivations}

We briefly brush two other trails to the semicircle, from the perspective of the Coulomb gas physics. Interestingly, these methods apply directly for all three Gaussian ensembles, since $\beta$ appears only as the inverse-temperature of the Coulomb gas.

\subparagraph{Equilibrium configuration} 

One possibility is to investigate the equilibrium configuration of the Coulomb gas at fixed $N$, in order to deduce the equilibrium density at large $N$. The equilibrium configuration is found by differentiating the Hamiltonian $\cH$ of \myeqref{eq:1:HamiltonianCoulombGas} with respect to the particle positions:
\be
\frac{\partial \mathcal{H}}{\partial \lambda_{i}}=0 \quad \quad \text{such that } \quad \quad  
 \lambda_{i} = 2\sigma^2 \sum_{j \neq i} \frac{1}{\lambda_{i}-\lambda_{j}} \; .
 \label{eq:1:EquilibriumConfiguration}
 \ee
This can be used, as in \cite{Potters2020}, to constrain the (rescaled) Stieltjes transform $\g$ defined in \myeqref{eq:1:DefStieltjes} such that
\be
\label{eq:1:FiniteNselfconsistent}
z\mathfrak{g}(z) -1  = \sigma^2 \mathfrak{g}(z) ^2+ \sigma^2 \frac{\mathfrak{g}'(z)}{N} \; .
\ee 
In the large-$N$ limit, we obtain the same equation as \myeqref{eq:1:SelfConsistentG}, which leads again to the semicircle distribution. 
 
\subparagraph{Saddle-point optimization}
Instead of studying the minimum-energy configuration of the Coulomb gas, another approach is to consider directly the $N \to \infty$ limit and introduce an arbitrary density $\omega$ corresponding to a certain distribution of the rescaled eigenvalues $\mu_i$. The Hamiltonian $\cH$ can be used to assign an exponential functional weight to this density $\omega$, with an additional sub-dominant entropy term $ \int \omega \log \omega$ which accounts for the likelihood of this density with respect to the underlying discrete configurations, as:
\be 
\begin{aligned} P(\omega)= Z^{-1} \exp \bigg(& - \beta N^{2}\left[\int_{\R}  \dd x   \,  \omega(x) \frac{x^2}{4 \sigma^2}- \frac{1}{2} \int_{\R^2} \dd x  \,  \dd  y  \, \omega(x) \omega(y) \log \abs{x-y} \right] 
 \\ &  -N \int_{\R}  \dd x \, \omega(x) \log \omega(x)\bigg) \end{aligned}
\ee 
The goal is then to find the equilibrium density $\omega^*$, which maximizes this functional weight. Applying a saddle-point method, one obtains unsurprisingly:
\be
\omega^*(\mu) = \rho_\mathrm{SC}(\mu)= \frac{\sqrt{4\sigma^2 - \mu^2}}{2 \pi \sigma^2} \; .
\ee
See \cite{Nadal2011} for more details on these Coulomb gas techniques.

\subsubsection{\texorpdfstring{Mar\v{c}enko}{Marcenko}-Pastur distribution}

\paragraph{Wishart density convergence}

The similar study of the global arrangement of eigenvalues in the Wishart ensembles yields the Mar\v{c}enko-Pastur distribution as the typical eigenvalue density \cite{Marcenko1967}. More precisely, this convergence holds in the large-size limit, when the dimensions of the intermediate $T \times N$ matrix $\XX$ in the definition of a Wishart matrix $\WW = \XX^\dagger \XX$, see \myeqref{eq:1:DefWishart}, both tend to $+\infty$ with fixed ratio
\be
q =\frac{N}{T} \; .
\ee
The normalizability conditions of the Wishart ensembles impose $0 \leqslant q \leqslant 1$, otherwise the rank of the matrix implies that a macroscopic fraction of eigenvalues are equal to zero. Recall that the potential corresponding to the Wishart ensembles is 
\be 
V(\lambda)= A \lambda -B \log \lambda 
\ee 
with constants given in Eq.~\eqref{eq:1:VWishart}.

Let us denote $\ell$ the lengthscale over which $N$ eigenvalues spread under the coupled actions of potential confinement and pairwise repulsion. Notice that the interaction term in  Eq.~\eqref{eq:1:HamiltonianCoulombGas} is of order $N^2$, as in the Gaussian case, while the confinement term is now of order $N \ell$. The typical support of the density must then grow as $\ell = N$, and the typical density is observed after an order-$N$ rescaling. Specifically, we will rescale the matrix by a factor $T$ in the following.

Let $\rho_T$ be the average density of the rescaled matrix $\frac{\WW}{T}$. In the limit of large $N$ and $T$ with $q=\frac{N}{T}$ fixed, we have
\be 
\rho_T(\mu) = \frac{1}{N} \sum_{i=1}^N \E \left[ \delta \left(\mu - \frac{\lambda_i}{T}\right) \right]     \quad  \xrightarrow[N,T \to + \infty]{q=\frac{N}{T}}   \quad \rho_{\rm MP} (\mu) 
\label{eq:1:MarcenkoPasturConvergence}
\ee
where $\rho_{\rm MP}$ is the Mar\v{c}enko-Pastur distribution of parameter $q$
\emphbe
\rho_{\rm MP} (\mu) = \frac{\sqrt{4  q\sigma^2 \, \mu -(\mu+\sigma^2(q-1))^{2}}}{2 q \pi \sigma^2 \,  \mu}
\label{eq:1:MarcenkoPasturDefinition}
\end{empheq}
which is supported on
\be 
\left[ \mu_-^{MP} , \mu_+^{MP} \right]  \ = \ \left[ \, \sigma^2(1-\sqrt{q})^2 \, ,\, \sigma^2 (1+\sqrt{q})^2 \, \right]   \; .
\label{eq:1:BoundsMarcenkoPastur}
\ee
\begin{figure}[ht!]
    \centering 
    \includegraphics[width=.9 \textwidth]{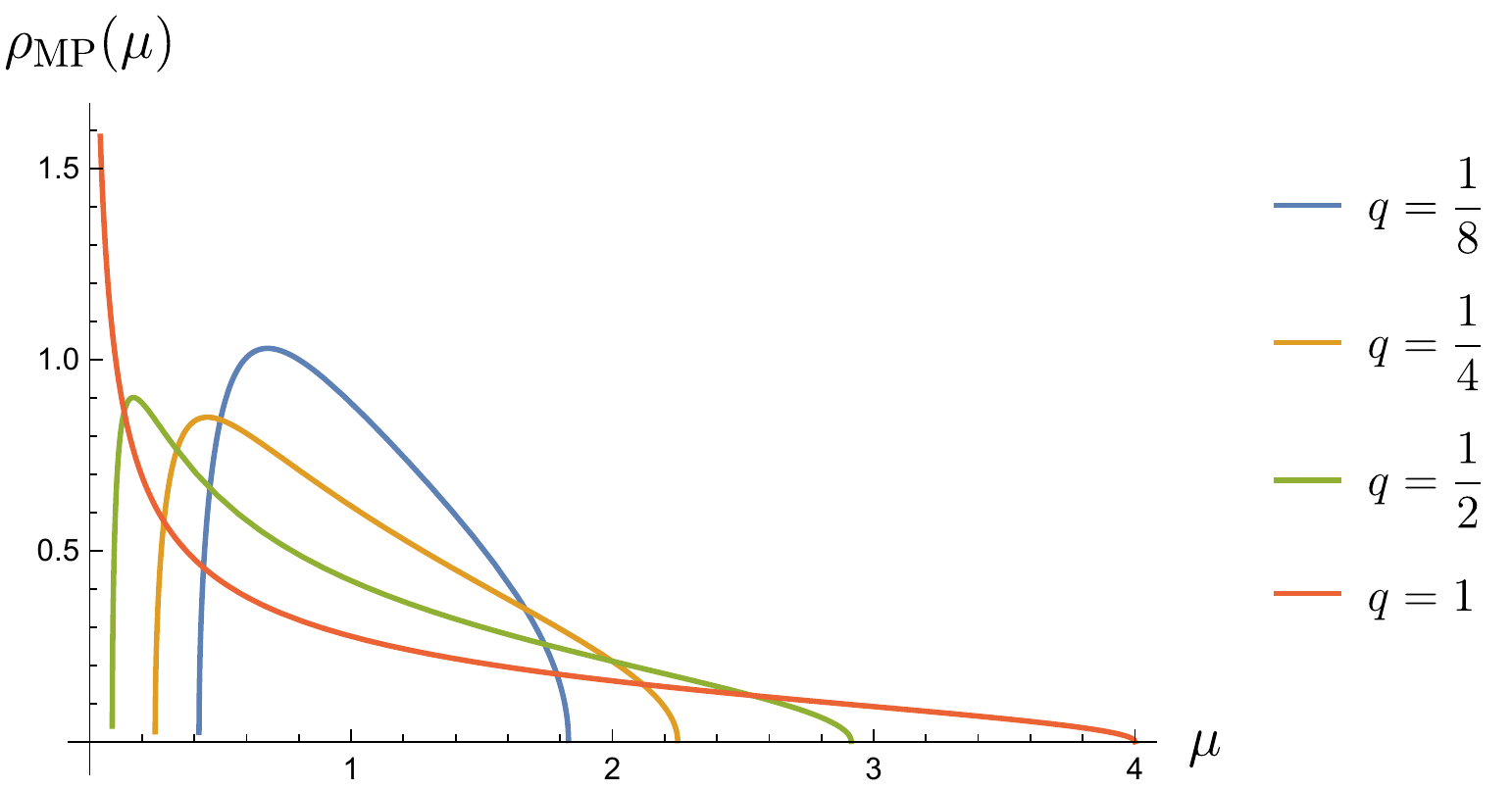} 
    \caption{Plots of the Mar\v{c}enko-Pastur distribution, with parameters $\sigma=1$ and $q \in \{\frac{1}{8},\frac{1}{4},\frac{1}{2},1\}$.}
    \label{fig:MarcenkoPastur}
\end{figure} 

Note that one can also re-write this conveniently as
\be 
\rho_{\rm MP} (\mu) = \frac{\sqrt{(\mu_+^{MP} - \mu)(\mu - \mu_-^{MP})}}{2 q \pi \sigma^2 \,  \mu} \; .
\label{eq:1:MarcenkoPasturRewritten}
\ee
As in the Gaussian case, this convergence also holds almost-surely for the empirical density of the rescaled Wishart matrix. 

The Mar\v{c}enko-Pastur distribution is plotted in Fig.~\ref{fig:MarcenkoPastur} for several values of $0 \leqslant q \leqslant 1$. For the critical case $q=1$, the density is supported on $[0,4 \sigma^2]$ and diverges at zero, as any $q>1$ would lead to a macroscopic fraction of eigenvalues being strictly equal to zero. Furthermore, for $q=1$, the Wishart density has an interesting connection with the Gaussian setting: changing variables to $\nu = \sqrt{\mu}$, the transformed Mar\v{c}enko-Pastur distribution is 
\be 
\rho_{\rm QC} (\nu)  =   \frac{\sqrt{4 \sigma^2 - \nu^2}}{2 \pi \sigma^2}
\quad \quad
\nu \in [0, 2 \sigma ] \; .
\label{eq:1:QuarterCircleDistribution}
\ee
This is Wigner's semicircle \eqref{eq:1:WignerSemicircleDefinition} restricted to $\R^+$, such that it is called the \emph{quarter-circle} distribution.

\paragraph{Inverse-Wishart density convergence}

For the inverse-Wishart ensemble, the limit density is directly obtained by a change of variables from the previous results. Firstly, in the large $(N,T)$ limit with $q =\frac{N}{T}$ fixed, the inverse-Wishart matrix $\InverseW = \WW^{-1}$ should be scaled-up by a factor $T$, from the asymptotics of the Wishart ensemble. We will also introduce a factor $(1-q)$ in order to obtain a unit-mean distribution.

Let $\rho_{(1-q)T}$ and $\rho^{\rm emp}_{(1-q)T}$ be the average and empirical spectral distributions of the rescaled matrix $ (1-q)T \times \InverseW$. By the previous results, both these distributions converge, in the large $(N,T)$ limit with $q =\frac{N}{T}$ fixed, to the inverse Mar\v{c}enko-Pastur distribution
\emphbe
\rho_{\mathrm{IMP}}(\mu)= \frac{(1-q) \sqrt{\left(\mu_{+}^{IMP} -\mu \right)\left(\mu-\mu_{-}^{IMP}\right)}}{2 q \pi \sigma^2 \, \mu^{2}}
\label{eq:1:IMPDistribution}
\end{empheq}
supported on 
\be 
[\,  \mu_{-}^{IMP} ,\mu_{+}^{IMP}  \, ]= \left[ \
\frac{1-q}{\mu_+^{MP}} \ , \  \frac{1-q}{\mu_-^{MP}}  \
\right] \; .
\ee
\begin{figure}[ht!]
    \centering 
    \includegraphics[width=.9 \textwidth]{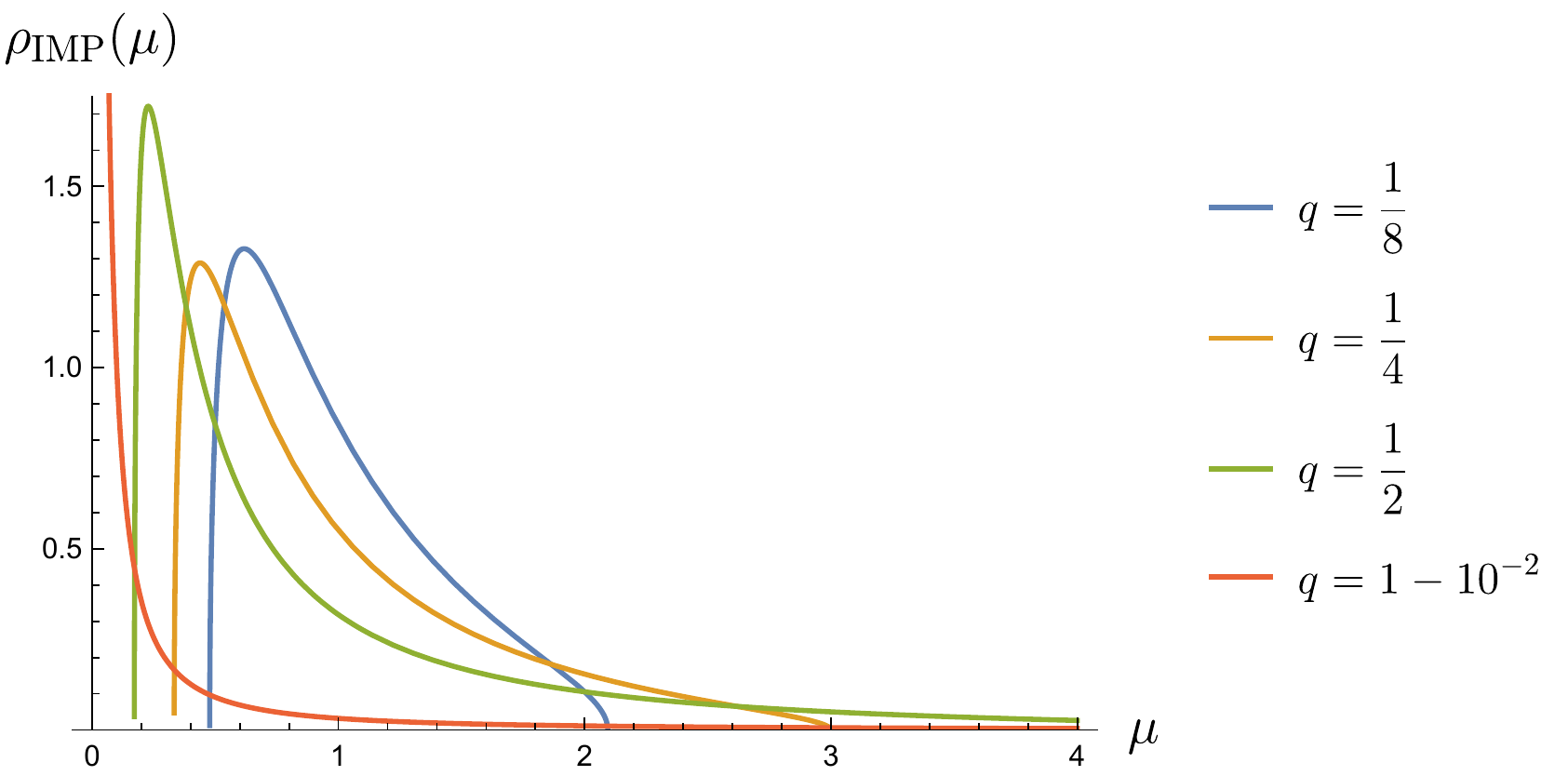} 
    \caption{Plots of the Inverse Mar\v{c}enko-Pastur distribution, with parameters $\sigma=1$ and $q \in \{\frac{1}{8},\frac{1}{4},\frac{1}{2},1- 10^{-2}\}$.}
    \label{fig:InverseMarcenkoPastur}
\end{figure} 

Introducing 
\be 
\kappa = \frac{1-q}{2q}
\label{eq:1:DefKappaParameterIMP}
\ee
such that $q =\frac{1}{2 \kappa +1}$, the inverse Mar\v{c}enko-Pastur distribution can be re-written
\be 
\label{eq:1:IMPDistributionRewritten}
\rho_{\mathrm{IMP}}(\mu)= \kappa \frac{ \sqrt{\left(\mu_{+}^{IMP} -\mu \right)\left(\mu-\mu_{-}^{IMP}\right)}}{ \pi \sigma^2 \, \mu^{2}}
\ee
with  $
 \mu_{\pm}^{IMP} =\frac{1}{\kappa \sigma^2} \left( \kappa+1 \pm \sqrt{2 \kappa+1} \right) $.

The inverse-Mar\v{c}enko-Pastur distribution is plotted in Fig.~\ref{fig:InverseMarcenkoPastur} for several values of $0 \leqslant q < 1$. Close to $q=1$, the divergence of the Mar\v{c}enko-Pastur distribution at zero induces a diverging mean for the inverse eigenvalues, such that the unit-mean inverse-Mar\v{c}enko-Pastur distribution $\rho_{\mathrm{IMP}}$ is not defined in the critical case $q=1$.

\subsection{Local statistics: determinantal processes and fermion systems}
\label{subsec:1.2.2}

In the previous section, we have detailed the global statistics that are observed at large $N$ in the distribution of eigenvalues, on the full scale of the spectrum. At this global scale, we have described the typical density of eigenvalues. The previous study does not however shed light on the system's behaviour at a scale where individual eigenvalues reappear, by zooming in on a small region of the typical density. This issue is that of the local statistics of the eigenvalues, which we turn to in this section. 

The study is simpler in the case of unitary ensembles ($\beta=2$) because the determinantal structure allows the use of orthogonal polynomials, as we explain in the first paragraph. This structure is also the key element in the important bridges that exist between eigenvalue ensembles and fermion systems, which will be a central element of the next chapters of this thesis. Finally, we explain how a different structure, relying on Pfaffian calculus, appears in the orthogonal and symplectic cases ($\beta=1,4$).

\subsubsection{Eigenvalues of unitary ensembles (\texorpdfstring{$\beta =2$}{Beta=2})}

In this section, we restrict to the unitary ensembles characterized by the Dyson index $\beta=2$. In order to study the local statistics, we will need to investigate the properties hidden in the JPDF \myeqref{eq:1:JPDFGeneralBetaEnsemble}. The special case $\beta=2$ allows to transform this JPDF advantageously in order to introduce orthogonal polynomials and exhibit a determinantal process.

\paragraph{Orthogonal polynomials}

\subparagraph{JPDF transformation}
For a general potential $V$, we consider the corresponding unitary ensemble with eigenvalue JPDF
\be 
P(\vec{\lambda}) = C_{2,V}^N \ \abs{\Delta(\vec{\lambda})}^2  \ e^{-2 \sum_{i=1}^N V(\lambda_i)} \;.
\label{eq:1:JPDFBeta2}
\ee 

An interesting remark is that for any family $(R_k)_{1\leqslant k \leqslant N}$ of monic polynomials such that $\deg R_k = k-1$, the Vandermonde determinant is equal, by standard operations on rows, to
\be
\Delta(\vec{\lambda}) =  \det\limits_{1\leqslant i,j \leqslant N}  \left(  \lambda_i^{j-1}  \right) = \det\limits_{1\leqslant i,j \leqslant N}  \left(  R_j(\lambda_i)  \right)  \; .
\ee
Recall that a polynomial is called monic when the coefficient of the highest power is one. Inserting arbitrary positive constants $(h_k)_{1\leqslant k \leqslant N}$ for future needs, this becomes:
\be
\Delta(\vec{\lambda}) = \prod_{k=1}^N \sqrt{h_k} \ \det\limits_{1\leqslant i,j \leqslant N}  \left( \frac{1}{\sqrt{h_j}} R_j(\lambda_i)  \right)
\ee
The squared Vandermonde term is then 
\be 
\abs{\Delta(\vec{\lambda})}^2 =\prod_{k=1}^N h_k \ \left[  \det\limits_{1\leqslant i,j \leqslant N}  \left(  \frac{1}{\sqrt{h_j}} R_j(\lambda_i)  \right)  \right]^2 =  \prod_{k=1}^N h_k \   \left(   \det \mathbf{R} \right)^2
\label{eq:1:VandermondeSquaredOrthPol}
\ee 
where $\mathbf{R}$ is the matrix defined by $\mathbf{R}_{ij}=   \frac{1}{\sqrt{h_j}} R_j(\lambda_i) $. Since $\left(   \det \mathbf{R} \right)^2 = \det \mathbf{R}  . \det \mathbf{R}=  \det \mathbf{R}  . \det \mathbf{R}^T = \det   \mathbf{R}  \mathbf{R}^T$, we obtain
\be 
\abs{\Delta(\vec{\lambda})}^2 =\prod_{k=1}^N h_k \   \det   \mathbf{R}  \mathbf{R}^T =   \prod_{k=1}^N h_k  \ \det_{1\leqslant i,j \leqslant N}  \left(   \sum_{k=1}^N  \frac{1}{h_k}  R_k( \lambda_i) R_k(\lambda_j)  \right)  
\ee
where each eigenvalue appears in one row and one column of the matrix $\mathbf{R}\mathbf{R}^T$. The eigenvalue JPDF of \myeqref{eq:1:JPDFBeta2} can thus be written in a new form as:
\be 
P(\vec{\lambda}) = C_{2,V}^N \prod_{k=1}^N h_k \   \det_{1\leqslant i,j \leqslant N}  \left(   \sum_{k=1}^N \frac{1}{h_k}   R_k( \lambda_i) R_k(\lambda_j)  \right)   \ e^{-2 \sum_{i=1}^N V(\lambda_i)}
\ee
or in the compact form
\emphbe 
\label{eq:1:TransformedJPDFPol}
P(\vec{\lambda}) = C_{2,V}^N  \prod_{k=1}^N h_k \   \det_{1\leqslant i,j \leqslant N}  \left(  e^{- V(\lambda_i) - V(\lambda_j) } \sum_{k=1}^N \frac{1}{h_k}   R_k( \lambda_i)R_k(\lambda_j)  \right)
\end{empheq}

An interesting choice for the family of polynomials, which is arbitrary as of now, is the family of orthogonal polynomials with respect to the weight function $\omega(\lambda) = e^{-2V(\lambda)}$. 

\subparagraph{Definition}
Let us consider a general non-negative weight function $\omega$, which we assume either to have sufficiently fast decay at $\pm \infty$, or to vanish outside of a bounded support. There exists a unique set of monic polynomials $(R_k)_{1\leqslant k \leqslant N}$ which respects the orthogonality conditions
\be 
\int_\R \omega(x)  R_i(x) R_j(x) \dd x = 0  \quad \text{if} \ i \neq j 
\ee
and are such that $\deg R_k = k-1$ for all $k$. This set can be constructed sequentially through the Gram-Schmidt orthogonalisation procedure, which gives a direct proof of the previous statement \cite{Doman2015}. From this set of polynomials are defined the following quantities
\be 
h_k = \int_\R \omega(x)  R_k(x)^2 \dd x  \  > 0 \; .
\ee 

\subparagraph{Application} We suppose now that the polynomials $R_k$ in \myeqref{eq:1:TransformedJPDFPol} are the monic orthogonal polynomials with respect to $\omega(\lambda) = e^{-2V(\lambda)}$, and that the constants $h_k$ are such that
\be 
\int_\R  e^{-2V(\lambda)}  \  R_i(\lambda) R_j(\lambda) \ \dd \lambda =  h_i \, \delta_{i,j} \; .
\ee
This assumption gives a helpful tool to manipulate the JPDF, and in particular to apply successive integrations. In this regard, let us further introduce a useful algebraic tool called the \emph{continuous Cauchy-Binet} (or Andréief's) \emph{formula} \cite{Andreief1883,Forrester2018}:
\be 
\int_{\R^{N}} \det_{1 \leqslant i,j \leqslant N}  \left[f_{i}\left(x_{j}\right)\right] \det_{1 \leqslant i,j \leqslant N} \left[g_{i}\left(x_{j}\right)\right] \prod_{i=1}^{N} 
\dd \mu\left(x_{i}\right)=
N ! \det_{1 \leqslant i,j \leqslant N}  \left[\int_{\R} f_{i}(x) g_{j}(x) \mathrm{d} \mu(x)\right]
\label{eq:1:ContinuousCauchyBinet}
\ee
where $(f_i,g_i)_i$ are two sets of functions and $\dd \mu$ is a measure supported on $\R$.

As an example of a computation made possible by the orthogonality constraint on the polynomial family, let us study the following integral
\be 
I_N = \int_{\R^N}   \prod_{k=1}^N \dd \lambda_k \ \abs{\Delta(\vec{\lambda})}^2  \ e^{-2 \sum_{i=1}^N V(\lambda_i)}
\ee
which determines the normalization constant. Rewriting the Vandermonde term, we have by a direct application of the continuous Cauchy-Binet formula: 
\bea
I_N &=&  \int_{\R^N}   \prod_{k=1}^N (h_k \dd  \lambda_k ) \  
\left(  \det_{1 \leqslant i,j \leqslant N}  \left[
\frac{R_i(\lambda_j)}{\sqrt{h_i}}
\right]  \right)^2
\ e^{-2 \sum_{i=1}^N V(\lambda_i)} \\
&=& N! \,  \prod_{k=1}^N  h_k   \det_{1 \leqslant i,j \leqslant N}  \left[\int_{\R} \mathrm{d} \lambda \frac{R_i(\lambda)R_j(\lambda)}{\sqrt{h_i h_j}} e^{-2V(\lambda)}  \right] \\
&=& N! \,  \prod_{k=1}^N  h_k  \det_{1 \leqslant i,j \leqslant N}  \delta_{i,j}  \\
I_N &=& N! \,  \prod_{k=1}^N  h_k 
\eea 
As a consequence of this computation, we deduce the normalization constant
\be 
 C_{2,V}^N = \left( N! \, \prod_{k=1}^N h_k \right)^{-1}
\ee
such that, with this choice of orthogonal polynomials and related constants $h_k$, \myeqref{eq:1:TransformedJPDFPol} can be written in a compact form where the prefactor becomes very simple: 
\emphbe  
P(\vec{\lambda}) = \frac{1}{N!} \   \det_{1\leqslant i,j \leqslant N}  \left(  e^{- V(\lambda_i) - V(\lambda_j) } \sum_{k=1}^N \frac{1}{h_k}   R_k( \lambda_i)R_k(\lambda_j)  \right) 
\label{eq:1:TransformedJPDFOrthPol} 
\end{empheq}

In a more general setting, let us consider a function $f$ and compute the average of the product $\prod_k (1+f(\lambda_k))$. This general computation will allow to obtain interesting quantities such as the $m$-particle marginal PDF for all $0 \leqslant m \leqslant N$. Up to a constant, $\E \left[ \prod_k \left(1+f(\lambda_k) \right) \right]$ is given by the integral:
\be
I_f = \int_{\R^N} \prod_{k=1}^N  \dd \lambda_k  \ \abs{\Delta(\vec{\lambda})}^2  \ e^{-2 \sum_{i=1}^N V(\lambda_i)} \ \prod_{k=1}^N (1+f(\lambda_k)) \; .
\label{eq:1:GeneralIntegralf}
\ee 
Fixing $f=0$ recovers directly the previous computation. In the general case, a similar computation as above yields:
\be 
I_f  = N!  \, (  \prod_{k=1}^N  h_k )  \,    \det_{1\leqslant i,j \leqslant N} \left(      \delta_{i,j} +         
 \int_{\R}   \dd \lambda   \frac{R_{i}(\lambda) R_{j} (\lambda) }{\sqrt{h_{i} h_{j}}} e^{-2 V(\lambda)} f(\lambda)  
 \right) \; .
\ee 
Since for an arbitrary $N \times N$ matrix $\mathbf{C}$, we have the determinant expansion
\be 
\operatorname{det}(\ID+\mathbf{C})=1 + \sum_{m=1}^{N} \frac{1}{m !} \sum_{i_{1}, \ldots, i_{m}=1}^{N} \det_{1 \leqslant r, s \leqslant m}\left(   \mathbf{C}_{i_{r} i_{s}}   \right)
\ee 
we obtain:
\be
I_f = N!   \, ( \prod_{k=1}^N  h_k )   \, \left(   
1 + \sum_{m=1}^{N} \frac{1}{m !} \sum_{i_{1}, \ldots, i_{m}=1}^{N} \det_{1 \leqslant r, s \leqslant m}\left(
\psi_{i_r,i_s}
\right)
 \right)
\ee
where $ \psi_{i,j} =  \int_{\R}   \dd \lambda   \frac{R_{i}(\lambda) R_{j} (\lambda) }{\sqrt{h_{i} h_{j}}} e^{-2 V(\lambda)} f(\lambda)  $. By a reverse application of the continuous Cauchy-Binet formula \eqref{eq:1:ContinuousCauchyBinet} for the $m\times m$ determinant:
\bea
\det_{1 \leqslant r, s \leqslant m}\left(
\psi_{i_r,i_s}
\right) 
&=& \frac{1}{m!} \int_{\R^m}  
\left( \det_{1 \leqslant r, s \leqslant m} 
 \frac{R_{i_r}(\lambda_s) }{\sqrt{h_{i_r}}} 
\right)^2
\prod_{k=1}^m e^{-2 V(\lambda_k)} f(\lambda_k) \dd \lambda_k \\
&=&  \frac{1}{m!} \int_{\R^m}  
 \det_{1 \leqslant r, s \leqslant m} 
\left( \sum_{k=1}^m
 \frac{R_{i_k}(\lambda_r) R_{i_k}(\lambda_s)  }{h_{i_k}} 
\right)
\prod_{k=1}^m e^{-2 V(\lambda_k)} f(\lambda_k) \dd \lambda_k \; . \nonumber
\eea
We have by symmetry
\bea
 \frac{1}{m!}  \sum_{i_{1}, \ldots, i_{m}=1}^{N}
 \det_{1 \leqslant r, s \leqslant m} 
\left( \sum_{k=1}^m
 \frac{R_{i_k}(\lambda_r) R_{i_k}(\lambda_s)  }{h_{i_k}} 
\right)
&=& 
 \sum_{i_{1}, \ldots, i_{m}=1}^{N}
  \det_{1 \leqslant r, s \leqslant m} 
\left(
 \frac{R_{i_r}(\lambda_r) R_{i_r}(\lambda_s)  }{h_{i_r}} 
\right) \nonumber \\
&=& 
 \det_{1 \leqslant r, s \leqslant m} 
\left( \sum_{i=1}^N
 \frac{R_{i}(\lambda_r) R_{i}(\lambda_s)  }{h_{i}} 
\right)
\eea
such that 
\be 
 \sum_{i_{1}, \ldots, i_{m}=1}^{N} \det_{1 \leqslant r, s \leqslant m}\left( \psi_{i_r,i_s} \right) =   \int_{\R^m}    
 \det_{1 \leqslant r, s \leqslant m}\left(\sum_{i=1}^N 
  \frac{ R_{i}(\lambda_r) R_{i} (\lambda_s) }{h_i}
 \right)
   \prod_{k=1}^m \dd \lambda_k e^{-2 V(\lambda_k)} f(\lambda_k)  
 \ee
and we finally obtain $I_f$ for a general function $f$ as
\be
\label{eq:1:SolutionIntegralGeneralf}
I_f = N! (  \prod_{k=1}^N  h_k ) \left(   
1 + \sum_{m=1}^{N} \frac{1}{m !} 
\int_{\R^m}    
 \det_{1 \leqslant r, s \leqslant m}\left(\sum_{i=1}^N 
  \frac{ R_{i}(\lambda_r) R_{i} (\lambda_s) }{h_i}
 \right)
   \prod_{k=1}^m \dd \lambda_k e^{-2 V(\lambda_k)} f(\lambda_k) 
 \right) \; .
 \ee
These computations are very general for now, but can be applied as follows in order to obtain the $m$-particle marginal PDF $
 P(\lambda_1, \cdots, \lambda_m) = 
 \int_{\R^{N-m}} P(\vec{\lambda}) \prod_{k=m+1}^N \dd \lambda_k$.  Indeed, functionally differentiating $I_f$ in \myeqref{eq:1:GeneralIntegralf} $m$ times with respect to $f$, at points $(\lambda_1, \cdots, \lambda_m)$, and then setting $f=0$ yields:
\be 
I_m(\lambda_1, \cdots, \lambda_m) =  \frac{N!}{(N-m)!}  \int_{\R^{N-m}} \prod_{k=m+1}^N  \dd \lambda_k  \ \abs{\Delta(\vec{\lambda})}^2  \ e^{-2 \sum_{i=1}^N V(\lambda_i)}  
\ee 
which is related to the $m$-particle marginal density through
\be
 P(\lambda_1, \cdots, \lambda_m) =  
  C_{2,V}^N  \frac{(N-m)!}{N!} I_m(\lambda_1, \cdots, \lambda_m) \; .
\ee
Applying the same operation on \myeqref{eq:1:SolutionIntegralGeneralf}, we obtain directly:
\be 
I_m(\lambda_1, \cdots, \lambda_m) =N! (  \prod_{k=1}^N  h_k )
 \det_{1 \leqslant r, s \leqslant m}\left(\sum_{i=1}^N 
  \frac{ R_{i}(\lambda_r) R_{i} (\lambda_s) }{h_i}
 \right) e^{-2 \sum\limits_{k=1}^m V(\lambda_k)}  
\ee 
and we are finally able to conclude that the $m$-particle marginal density is given by
\emphbe
P(\lambda_1, \cdots, \lambda_m) =  \frac{(N-m)!}{N!} \   \det_{1\leqslant i,j \leqslant m}  \left(  e^{- V(\lambda_i) - V(\lambda_j) } \sum_{k=1}^N \frac{1}{h_k}   R_k( \lambda_i)R_k(\lambda_j)  \right) 
\label{eq:1:MarginalDensityOrthPol}
\end{empheq}
Thanks to the introduction of orthogonal polynomials, we have obtained in this formula the marginal densities for all number of particles $m \leqslant N$ in a very elegant form as a $m\times m$ determinant, which is a generalization of the case $m=N$ given as a $N\times N$ determinant in the full JPDF of \myeqref{eq:1:TransformedJPDFOrthPol}. It is truly noteworthy that the marginal densities should keep the same structure after successive integrations, and we insist that the key element in this result are the orthogonal conditions imposed on the polynomials $R_k$. 

In the next section, we will introduce the mathematical framework which defines such a structure: the theory of determinantal point processes. However, we give before that some details on the classical orthogonal polynomials, obtained with respect to usual weight functions.

\subparagraph{Hermite, Laguerre and Jacobi} 

Some families of orthogonal polynomials which prove useful in RMT are presented below. 

\begin{figure}[ht!]
    \centering 
    \includegraphics[width=.9 \textwidth]{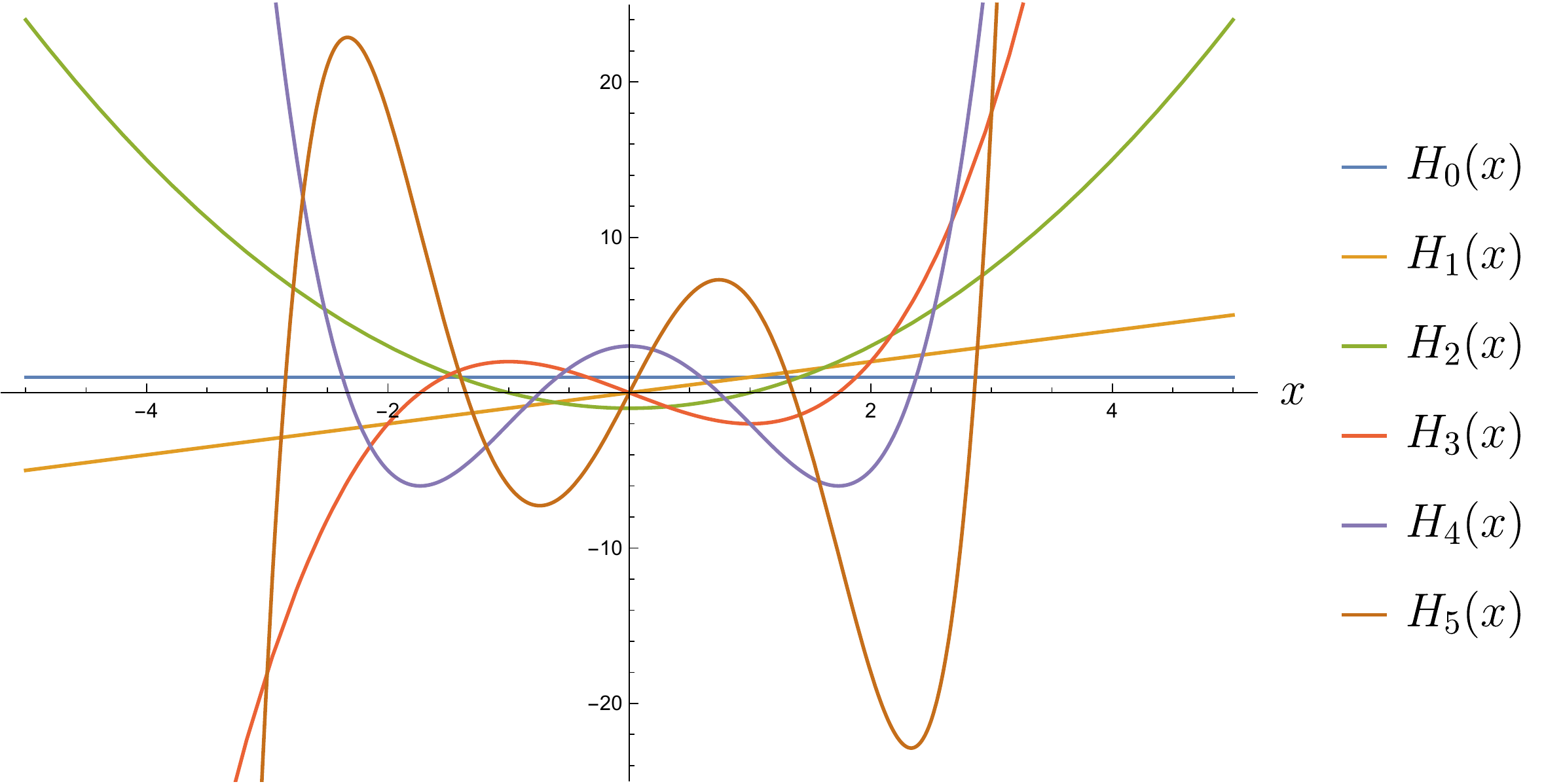} 
    \caption{Plot of the first six Hermite polynomials $H_k$.}
    \label{fig:Hermite}
\end{figure}

The Hermite polynomials $(H_k)_{k\geqslant 0}$ are the monic orthogonal polynomials with respect to the Gaussian weight function supported on $\R$:
\be 
\omega(x) = e^{-\frac{x^2}{2}}  \; .
\label{eq:1:HermiteGaussianWeight}
\ee
They obey $ \int_{\R} H_{i}(x) H_{j}(x) \mathrm{e}^{ - \frac{x^2}{2}} \dd x= i ! \sqrt{2 \pi} \,  \delta_{i, j}$. Note that $\omega(x) = e^{-2 V(x)}$ with $V(x) = \frac{x^2}{4}$ such that the Hermite polynomials are useful in the Gaussian ensembles, with a possible rescaling if a factor $\sigma^2$ is introduced as in the previous sections. This explains their alternate name of $\beta$-Hermite ensembles. The first six Hermite polynomials $(H_k)_{0 \leqslant k \leqslant 5}$ are plotted in Fig.~\ref{fig:Hermite}.

The Laguerre polynomials $(L^{(\alpha)}_k)_{k\geqslant 0}$ are the orthogonal polynomials with respect to the weight function supported on $\R^+$:
\be 
\omega(x) = x^\alpha e^{-x}  \; .
\ee
They are usually defined such that the coefficient of their highest-degree monomial is $\frac{(-1)^k}{k!}$ and they are thus not monic, in contrast with the assumption of the previous pages. From the form of the weight function, the Laguerre polynomials are directly useful in the Wishart ensembles, which are also called $\beta$-Laguerre ensembles as a consequence.

Finally, we also mention the Jacobi polynomials $(P_k^{(\alpha,\beta)})_{k\geqslant 0}$ which are the orthogonal polynomials with respect to the weight function supported on $[-1,1]$:
\be
\omega(x)= (1-x)^{\alpha}(1+x)^{\beta} \; .
\ee
The variable $x$ can be translated and rescaled to make the support match with the interval $[0,1]$, such that the polynomials $P_k^{(\alpha,\beta)}$ appear in the study of the Jacobi ensembles, which directly take their name from this polynomial family.

We point to the book \cite{Doman2015} for more details on the classical families of orthogonal polynomials and their properties.

\paragraph{Determinantal point processes} 

The special structure exhibited in \myeqref{eq:1:TransformedJPDFOrthPol} and \myeqref{eq:1:MarginalDensityOrthPol} for the JPDF and marginal densities of the eigenvalue process in the unitary ensembles is characteristic of a determinantal point process (\textit{DPP}). We give the definition and main properties of these objects and show how this framework can be applied to study the local statistics of the system.

\subparagraph{Definition and properties}

A \emph{point process} is a probability measure on all point configurations $\{x_1, x_2, \cdots\} = \mathcal{X}$ on the real line, and the special case where the cardinality of the set of points $\mathrm{card}(\mathcal{X})$ is almost-surely equal to $N$ is called an $N$-point process. It is then characterized by its JPDF $P(x_1, \cdots, x_N)$ which is invariant under permutations.

A \emph{determinantal point process} is an $N$-point process with JPDF and marginal densities given, for every $0 \leqslant m \leqslant N$, by
\be 
P( x_1 , \cdots, x_m )  = \frac{(N-m)!}{N!} \det_{1 \leqslant i,j \leqslant m} \left( K(x_i,x_j) \right) 
\label{eq:1:DefDPPMarginal}
\ee
where the function $K$ is called the correlation kernel of the DPP. $K$ is crucial in this definition as it contains all the information about the system, such that all quantities of interest can be expressed in term of it. Note that the kernel is not unique as the conjugation with any positive function $h$ gives an equivalent kernel
\be 
\tilde{K}(x,y) = h(x) K(x,y) h(y)^{-1}
\ee
which is easily seen to describe the same DPP. Finally, it should be noted that this kernel $K$ naturally defines an integral operator on $\Ll^2(\R)$ acting as $K_f(t)= \int_\R K(t,s) f(s) \dd s$.

This definition of the DPP can be extended to general point processes, where the number of points is not necessarily fixed with probability 1, by using correlation functions instead of marginal densities such that the prefactor, with the particular integer $N$, disappears in \myeqref{eq:1:DefDPPMarginal}. We will however only consider $N$-point processes in the following.

We now state a useful theorem \cite{Soshnikov2000}, which gives the conditions for a kernel function to be suitable for the definition of a corresponding determinantal point process. For a kernel $K$ respecting the following three conditions
\begin{itemize}[itemsep=.3em,topsep=10pt]
\item trace-class property : $\operatorname{Tr} K=\int_{\mathbb{R}} K(x, x) \dd x= N <+\infty$ ; 
\item positivity : $\det\limits_{1 \leqslant i, j \leqslant N}  K\left(x_{i}, x_{j}\right)  \geqslant 0$ for every $\vec{x} \in \R^N$ ;
\item reproducing property :  $ K(x, y)=\int_{\mathbb{R}} K(x, s) K(s, y) \mathrm{d} s$ for every $x,y \in \R$ ;
\end{itemize} 
then  $ P( \vec{x} ) = \frac{1}{N!}\det\limits_{1 \leqslant i, j \leqslant N}  K\left(x_{i}, x_{j}\right)$ is a probability measure on $\R^N$, invariant under permutations, which defines a DPP with kernel $K$. Note that the reproducing property implies that the spectrum of $K$ is bounded above by one. Interestingly, replacing reproducibility by spectrum-boundedness in the conditions preserves the result, see \cite{Soshnikov2000}. As a consequence, the last two conditions can be replaced by the mere requirement that all eigenvalues of $K$ belong to $[0,1]$.

Let us review some properties of DPP. An interesting element on a point process is the statistics of the number of particles $\ell_B =\mathrm{card}(\mathcal{X} \cap B) $ found in a given compact subset $B$, say for simplicity an interval of $\R$. This is usually called the full counting statistics (\textit{FCS}) in the physics literature. The key observable in the study of the random variable $\ell_B$ is the generating function of its probability distribution:
\be 
F_B(z) = \E \left[    z^{\ell_B}  \right] \; .
\ee
Taylor expanding this expression close to $z=1$ gives:
\be
F_B(z) = 1 + \sum_{k=1}^\infty  (z-1)^k \ \E   \binom{\ell_B}{k} 
\ee
where $\E  \binom{\ell_B}{k}$  is the expected number of $k$-element groups which can be formed, with the constraint that they all belong to $B$. It is thus equal to the number of possible $k$-element groups multiplied by the probability that the chosen elements all belong to $B$:
\be
 \E  \binom{\ell_B}{k} = \binom{N}{k}  \int_{B^k} \prod_{i=1}^k \dd x_i   P(x_1, \cdots, x_k) = \frac{1}{k!}  \int_{B^k} \prod_{i=1}^k \dd x_i \det_{1 \leqslant i,j \leqslant k } K(x_i,x_j) 
\ee
such that:
\be 
F_B(z) =  1 + \sum_{k=1}^\infty  \frac{(z-1)^k}{k!} \   \int_{B^k} \prod_{i=1}^k \dd x_i \det_{1 \leqslant i,j \leqslant k } K(x_i,x_j)  \; .
\ee
This quantity is called the \textit{Fredholm determinant} of the trace-class integral operator $K$, with prefactor $(z-1)$, restricted to the subset $B$ \cite{Forrester2010} and it is denoted:
\be 
F_B(z) = \det \bigg(\ID + \left. \left(z-1\right) K\right|_{B}\bigg) 
\ee
where $\ID$ is the identity operator. This generating function encodes the full probability distribution of $\ell_B$. As a special case, the probability that no particle is found in $B$ is called the \emph{gap probability} and can be expressed directly as
\be 
\mathbb{P}(\ell_B = 0)= F_B( 0) = \det \bigg(\ID - \left. K \right|_{B}\bigg)  \; .
\label{eq:1:GapProbabilityDPP}
\ee

For more details about DPP and proofs of some statements made above, we point to the pedagogical presentation \cite{Girotti2015} and to the more complete articles \cite{Hough2006,Johansson2006,Soshnikov2000,Borodin2016}. DPP have found a large number of applications in physics and combinatorics, but also in computer science and machine learning where their repulsive feature proves useful to ensure the diversity of data samples picked from a dataset. This diverse sampling can be used, for example, to ensure high-quality search results by avoiding redundant answers that would be close to each other in a representative space \cite{Kulesza2012}.

\subparagraph{Application}  

This presentation has shown that the eigenvalue process in a unitary ensemble of RMT ($\beta=2$) with potential $V$ is indeed a determinantal point process, from the structure of the JPDF and subsequent marginal densities in \myeqref{eq:1:MarginalDensityOrthPol}, with a kernel given by:
\emphbe 
K(x,y) =  e^{- V(x) - V(y) } \sum_{k=1}^N \frac{1}{h_k}   R_k( x ) R_k( y )  
\label{eq:1:KernelEigenvalueProcessOrthPol}
\end{empheq}
where $(R_k)_{k \geqslant 0}$ is the orthogonal monic polynomial family with respect to $\omega(\cdot)= e^{-2V(\cdot)}$ and with corresponding $h_k=\int_\R R_k(x)^2 e^{-2 V(x)} \dd x$.

Assuming that we had not already derived the decisive set of marginal densities in \myeqref{eq:1:MarginalDensityOrthPol}, we can verify that this kernel indeed verifies the three conditions given by the suitability theorem of the previous section:
\begin{itemize}[itemsep=.3em,topsep=10pt]
\item the trace of the operator $K$ is
\be \Tr K=\int_{\mathbb{R}} K(x, x) \dd x= 
\int_\R e^{-2 V(x)  } \sum_{k=1}^N \frac{1}{h_k}   R_k( x )^2  \dd x=
 \sum_{k=1}^N \frac{h_k}{h_k} = N <+\infty
\ee
such that the trace-class condition is verified  ; 
\item the $N\times N$ determinant of the kernel on an arbitrary vector $\vec{x} \in \R^N$ is : 
\be 
\det\limits_{1 \leqslant i, j \leqslant N}  K\left(x_{i}, x_{j}\right) = (\det \mathbf{R})^2 e^{-2 \sum_{i=1}^N V(x_i)}  \geqslant 0 
\ee  
where $\mathbf{R}$ is the matrix introduced in \myeqref{eq:1:VandermondeSquaredOrthPol}, such that the positivity condition is verified ; 
\item and finally for arbitrary $x,y \in \R$:
\bea
\int_{\mathbb{R}} K(x, s) K(s, y) \mathrm{d} s &=&  
\sum\limits_{1\leqslant k,l \leqslant N} \frac{e^{-V(x) - V(y)} R_k(x) R_l(y) }{h_k h_l} \int_\R R_k(s) R_l(s) e^{-2 V(s) } \dd s \nonumber \\
&=& 
\sum\limits_{1\leqslant k,l \leqslant N} \frac{e^{-V(x) - V(y)} R_k(x) R_l(y) }{h_k h_l} h_k \delta_{k,l}
= K(x, y)
\eea
such that the reproducing property is verified.
\end{itemize} 

Any statistical property of eigenvalues that one might be interested in can be expressed in terms of this kernel $K$. As announced at the beginning of this section, we are interested in particular in the local statistics of the system. However, let us first make a small detour back to the eigenvalue density, studied in the previous section \ref{subsec:1.2.1}. The determinantal structure gives a direct expression for the average density $\rho(\lambda) $ defined in \myeqref{eq:1:defDensity}, which coincides with the 1-particle marginal, as:
\be 
\rho(\lambda)   = \frac{1}{N}  K(\lambda, \lambda )=   \frac{e^{- 2 V(\lambda)  }}{N}  \sum_{k=1}^N \frac{R_k( \lambda )  ^2}{h_k}  =  
\frac{R_N(\lambda)R_{N+1}'(\lambda) - R_N'(\lambda) R_{N+1}(\lambda) }{N h_N \, e^{ 2 V(\lambda)  } }  
\label{eq:1:DensityOrthPol}
\ee
where the last equation is obtained by the Christoffel-Darboux formula, which applies to all families of orthogonal polynomials \cite{Doman2015}. The prime indicates derivation with respect to the polynomial's argument.

The convergence of the eigenvalue density of the GUE towards Wigner's semicircle stated in \myeqref{eq:1:WignerSemicircleConvergence} can thus be reformulated as the convergence of 
\be 
\rho_{\sqrt{N}} (\lambda)  =  \sqrt{N} \rho( \sqrt{N} \lambda) = 
\frac{H_N(  \sqrt{N} \lambda)H_{N+1}'( \sqrt{N} \lambda) - H_N'( \sqrt{N} \lambda) H_{N+1}( \sqrt{N} \lambda) }{ \sqrt{N} h_N \, e^{ N \lambda^2 /2   } }
\ee
towards $\rho_{\mathrm{SC}}(\lambda)$ when $N$ grows to infinity, with $V(\lambda) = \frac{\lambda^2}{4}$ and $H_k$ the Hermite polynomials, where we have fixed $\sigma=1$ here. As a consequence, this is now a simple exercise of Hermite polynomial asymptotics. A similar statement can be made in the LUE with the Laguerre polynomials and the Mar\v{c}enko-Pastur distribution about the convergence of \myeqref{eq:1:MarcenkoPasturConvergence}.

Finally, we turn to the precise matter of the local statistics of the unitary ensembles, equipped with the powerful determinantal tools. 

\subparagraph{Universal local statistics}

The issue of the local statistics consists in determining the behaviour of the system at a small scale compared to the lengthscale given by the support of the typical density. In the unitary ensembles, the statistics of the eigenvalue process are characterized by the kernel $K$. As a consequence, investigating the properties on a small scale of the spectrum requires to determine what the kernel becomes when we zoom in to this microscopic scale. The asymptotic kernel will then bear all the information on the rescaled system, opening the road to the local statistics. 

Strikingly, only a few asymptotic kernels are found as the limit of the rescaling procedure of $K$, whatever the choice of the potential $V$. Which class the result falls in depends on which region of the spectrum we investigate. There are two choices: either we zoom in on some place in the interior of the typical spectrum, the \emph{bulk}, or we zoom in on a boundary of the typical spectrum, the \emph{edge}, where there are far fewer eigenvalues and the behaviour is different. In the latter option, two cases arise: either the potential has a finite slope at the edge such that the edge is \emph{soft}, or the potential rises abruptly and imposes a hard constraint on the system, such that the edge is \emph{hard}. We present the asymptotic kernel in these three cases.

\begin{itemize}
\item 
In the bulk, the limit kernel obtained in the microscopic scale, where the mean distance between eigenvalues is 1, is the sine kernel \cite{Dyson1962}
\be
K^{\mathrm{sine}}(x, y)=\frac{\sin \pi(x-y)}{\pi(x-y)} \; .
\label{eq:1:DefSineKernel}
\ee 
As an illustration in the GUE with $\sigma=1$, this means that the following convergence is observed, when choosing a point $x_0 \in ]-2,2[$ in the bulk of $\rho_{\mathrm{SC}}$:
\be 
\frac{1}{ \sqrt{N}  \rho_{\mathrm{SC}}(x_0) }  K\left(  \sqrt{N} x_0 +  \frac{x}{ \sqrt{N} \rho_{\mathrm{SC}}(x_0)}  ,  \sqrt{N} x_0 +  \frac{y}{ \sqrt{N} \rho_{\mathrm{SC}}(x_0)} \right)   \quad  \xrightarrow[N \to \infty] \quad
K^{\operatorname{sine}}(x, y)
\ee
as we focus around $\sqrt{N} x_0$ on the scale of the typical distance between two eigenvalues, which is $ \frac{1}{\sqrt{N} \rho_{\mathrm{SC}}(x_0) }$. Indeed, we can see that the number of particles found in an interval of order $\sqrt{N}$ around $x_0$ is of order $N \rho_\mathrm{SC}(x_0)$.

Note that the sine kernel is invariant under translation $(x,y) \to (x+ d, y+d)$ as expected from the bulk behaviour.

\item At a soft edge, the limit kernel in the microscopic scale is the Airy kernel \cite{Tracy1994}
\be 
K^{\mathrm{Airy}}(x, y)=\frac{\Ai(x) \Ai^{\prime}(y)-\Ai^{\prime}(x) \Ai(y)}{x-y}
\label{eq:1:AiryKernel}
\ee 
where $\Ai$ is the Airy function, see App.~\ref{app:specialfunctions}. In the GUE with $\sigma=1$, the following convergence is observed at the soft edge $2\sqrt{N}$:
\be 
\frac{1}{ N^{\frac{1}{6}} }  K\left(   2 \sqrt{N} +  \frac{x}{ N^{\frac{1}{6}} }  , 2 \sqrt{N} +  \frac{y}{N^{\frac{1}{6}}}  \right)   \quad  \xrightarrow[N \to \infty] \quad
K^{\mathrm{Airy}}(x, y)
\ee
with a point density of the order of $N^{\frac{1}{6}}$ around the edge. This scaling can be obtained from the square-root behaviour of $\rho_{\mathrm{SC}}$ close to the edge, such that the number of particles found in $[(2-t)\sqrt{N},2 \sqrt{N}]$ for small $t >0$ is of order $N \int_0^t \sqrt{u} \dd u= N t^{3/2}$. One should impose $t=N^{-2/3}$ to observe a number of particles of order 1, implying a lengthscale of order $N^{-1/6}$. As expected, the edge density $N^{1/6}$ is thus much smaller than the bulk density of order $\sqrt{N}$. For a general potential $V$, this lengthscale is related to the slope of the potential at the edge.

\item At a hard edge, the limit kernel in the microscopic scale is the Bessel kernel \cite{Tracy1994a}
\be
\label{eq:1:BesselKernel}
K^{\mathrm{Bessel}}_\alpha(x, y)
=
\frac{\sqrt{y}  J_{\alpha}^{\prime}(\sqrt{y} )  J_{\alpha}(\sqrt{x}) -\sqrt{x} J_{\alpha}^{\prime}(\sqrt{x}) J_{\alpha}(\sqrt{y})}{2(x-y)}
\ee
where $J_\alpha$ is the Bessel function of order $\alpha$, see App.~\ref{app:specialfunctions}. Common examples of hard edge situations are the lower edge in the Laguerre-Wishart unitary ensemble which is supported on $\R^+$, or the two edges of the Jacobi unitary ensemble supported on $[0,1]$. 

For example in the LUE with $\sigma=1$, the hard-edge appears when the left edge is equal to zero such that the boundary of the support imposes a hard constraint on the spectrum. As seen in \myeqref{eq:1:BoundsMarcenkoPastur} and illustrated in Fig. \ref{fig:MarcenkoPastur}, this happens for $q=1$, such that $T = N + \alpha$, with $\alpha$ a constant. The large-$N$ kernel convergence can then be stated as
\be 
\frac{1}{ 4N }  K\left(   \frac{x}{ 4N}  ,  \frac{y}{4N}  \right)   \quad  \xrightarrow[N \to \infty] \quad
K^{\mathrm{Bessel}}_\alpha(x, y)
\label{eq:1:LimitBesselKernelHardEdgeLUE}
\ee
with the natural order-$N$ scaling coinciding with the typical density scaling, see \myeqref{eq:1:MarcenkoPasturConvergence}.

Note that the Airy and Bessel kernels are not invariant under translation $(x,y) \to (x+ d, y+d)$ as expected from the edge behaviour.
\end{itemize}

These kernels are universal in the sense that they characterize the microscopic limit of the process regardless of the particular shape of the potential $V$. For details on the extension of universality in local statistics of RMT, we point to \cite{Kuijlaars2007}.

Other kernels do arise in special situations, of which we give an example. In the above, we have assumed implicitly that the support of the spectrum is an interval, as suggested by the Gaussian and Wishart ensembles; if we however allow potentials such that the support is a reunion of intervals, a different behaviour appears when two disconnected intervals reunite. The asymptotic kernel at the cusp point, the point in the bulk where the density vanishes, is the Pearcey kernel \cite{Brezin1998}.

\subsubsection{Connections with fermion systems}
\label{subsubsec:1.2.2.b}

In the previous section, we have detailed the determinantal structure of the unitary ensembles and the resulting local statistics. Before turning to the different structure arising in the orthogonal and symplectic ensembles, let us make a digression and note in this section the important connections of the eigenvalues in the unitary ensembles with quantum systems of fermions.

\paragraph{The physics of cold fermions}

Consider a system of $N$ spin-less fermions in one-dimensional space, trapped by a potential $V_\mathrm{f}$ and denote by $(\hat{x}_i, \hat{p}_i)_{1 \leqslant i \leqslant N}$ their positions and impulsions. We will restrict to the study of non-interacting fermions, for which the quantum Hamiltonian of the system is
\be
\Ham =   \sum_{i=1}^N \Hamm_i = \sum_{i=1}^N H(\hat{x}_i, \hat{p}_i) \quad \quad    \text{where} \quad   H(\hat{x}_i, \hat{p}_i)  = \frac{\hat{p}^2}{2m}  + V_\mathrm{f} (\hat{x}) \; .
\label{eq:1:DefNbodyHamiltonian}
\ee 

The one-particle Hamiltonian $\Hamm$ is characterized by the set of its eigenstates $(\ket{k})_{k \geqslant 0}$ defined by
\be 
\Hamm \ket{k} = \epsilon_k \ket{k}
\ee
where $\epsilon_k$ is the corresponding eigenvalue or \emph{energy}. In position space, the corresponding wavefunctions $\phi_k(x) = \bra{x}\ket{k}$ are eigenfunctions of the position space Hamiltonian, such that
\be \left(-\frac{\hbar^2}{2m} \frac{\partial^2}{\partial x^2} + V_\mathrm{f}(x) \right) \phi_k(x) = \epsilon_k \phi_k(x) \; .
\label{eq:1:EigenProblemWavefunction}
\ee
In the following, we will simplify the conventions by fixing $m = \hbar =1$. The wavefunctions form an orthonormal basis of the quantum system's Hilbert space as
\be 
\int_\R \dd x \, \phi_k(x) \,  \overline{ \phi_l(x) } = \delta_{k,l}
\ee
where the Hermitian inner product is used. From quantum principles, the position of a fermion occupying the state $\ket{k}$ is a random variable distributed according to the PDF
\be 
P(x) = \abs{ \phi_k(x) }^2 \; .
\label{eq:1:PDFFromWavefunction}
\ee

For $N$ fermions in the $N$-particle eigenstate $\ket{\vec{k}}=\ket{k_1, \cdots , k_N}$, the energy of the system is directly obtained from \myeqref{eq:1:DefNbodyHamiltonian} as $E_{\vec{k}} = \sum_{i=1}^N \epsilon_{k_i}$ and the $N$-particle wavefunction is the Slater determinant
\be 
\Psi_{\vec{k}} ( \vec{x}) = \frac{1}{\sqrt{N!}} \det_{1 \leqslant i,j \leqslant N} \phi_{k_i}(x_j) \; .
\label{eq:1:SlaterDeterminantManyBodyWavefunction}
\ee 
The $N$-particle position JPDF is given by its squared modulus as
\be 
P(\vec{x}) = \frac{1}{N!} \abs{\det_{1 \leqslant i,j \leqslant N} \phi_{k_i}(x_j)}^2 \; .
\ee
The Pauli exclusion principle states that no two fermions should occupy the same state, i.e. $k_i \neq k_j$ for all $(i,j)$, and is enforced by the Slater determinant which associates a trivial wavefunction to a forbidden state. The admissible set of eigenstates is thus 
\be 
 \Omega_N = \{ \vec{k} \in \mathbb{N}^N \mid k_1 < k_2 < \cdots < k_N    \}
 \label{eq:1:DefOmegaNManyBodyEigenstates}
 \ee
and the ground-state of the system is $\vec{0}= \ket{0, \cdots , N-1}$, such that the position JPDF is given at zero temperature by:
\be 
P(\vec{x}) = \frac{1}{N!} \abs{\det_{1 \leqslant i,j \leqslant N} \phi_{i-1}(x_j)}^2  = \frac{1}{N!} \det_{1 \leqslant i,j \leqslant N}  K_\mathrm{f}(x_i, x_j) 
\ee
with the kernel
\be 
K_\mathrm{f}(x_i, x_j)  = \sum_{k=0}^{N-1} \phi_k(x) \overline{\phi_k(y)} \; .
\label{eq:1:KernelFermions}
\ee 
The position statistics of zero temperature fermions on the line is then a DPP with kernel $K_\mathrm{f}$, which can be shown to respect the three conditions of trace-class, positivity and reproducibility. It is now obvious that the statistics of fermions should be related to RMT, as we detail in the next section. 

As a side note, we mention that the determinantal structure is lost for a system of $N$ fermions at positive temperature, where all states $\ket{\vec{k}}$ appear with probability $e^{-\beta E_{\vec{k}}}$ as per the canonical ensemble description, with $\beta$ the inverse-temperature. However, when describing the system in the grand canonical ensemble, where the number of particles is free to fluctuate, the determinantal structure reappears with a kernel depending on all single-particle states with a prefactor given by the Fermi-Dirac statistics. 

For more details on the theory of quantum mechanics in general and fermion systems in particular, see \cite{Basdevant2002,Giorgini2008,Lacroix-A-Chez-Toine2019,Dean2016}. For reviews of experimental realizations in the field of cold fermion systems, see \cite{Ketterle2008,Nascimbene2010,Bloch2008}. In particular, the recent development of quantum microscopes for Fermi gases gives a practical motivation for the theoretical investigations into the spatial correlations of these systems \cite{Anderson2016}. In a wider scope, cold boson systems are also a subject of intense experimental activity \cite{Bakkali-Hassani2021,Saint-Jalm2019}.

\paragraph{Links with random matrix ensembles}

The statistics of $N$ zero-temperature non-interacting fermions on the line is equivalent to that of the $N$ eigenvalues of a random unitary matrix ensemble as soon as the kernels in Eqs.~\eqref{eq:1:KernelEigenvalueProcessOrthPol} and \eqref{eq:1:KernelFermions} are equal.

On the random matrix side, we recall that the kernel $K$ depends on the potential $V$ and ensuing orthogonal polynomials $R_k$; while on the fermions side, the kernel $K_f$ depends on the potential $V_f$ and ensuing wavefunctions $\phi_k(x)$ obtained as solutions of \myeqref{eq:1:EigenProblemWavefunction}. We present in the following three cases of such a correspondence.

\begin{figure}[ht!]
    \centering 
    \includegraphics[width=.9 \textwidth]{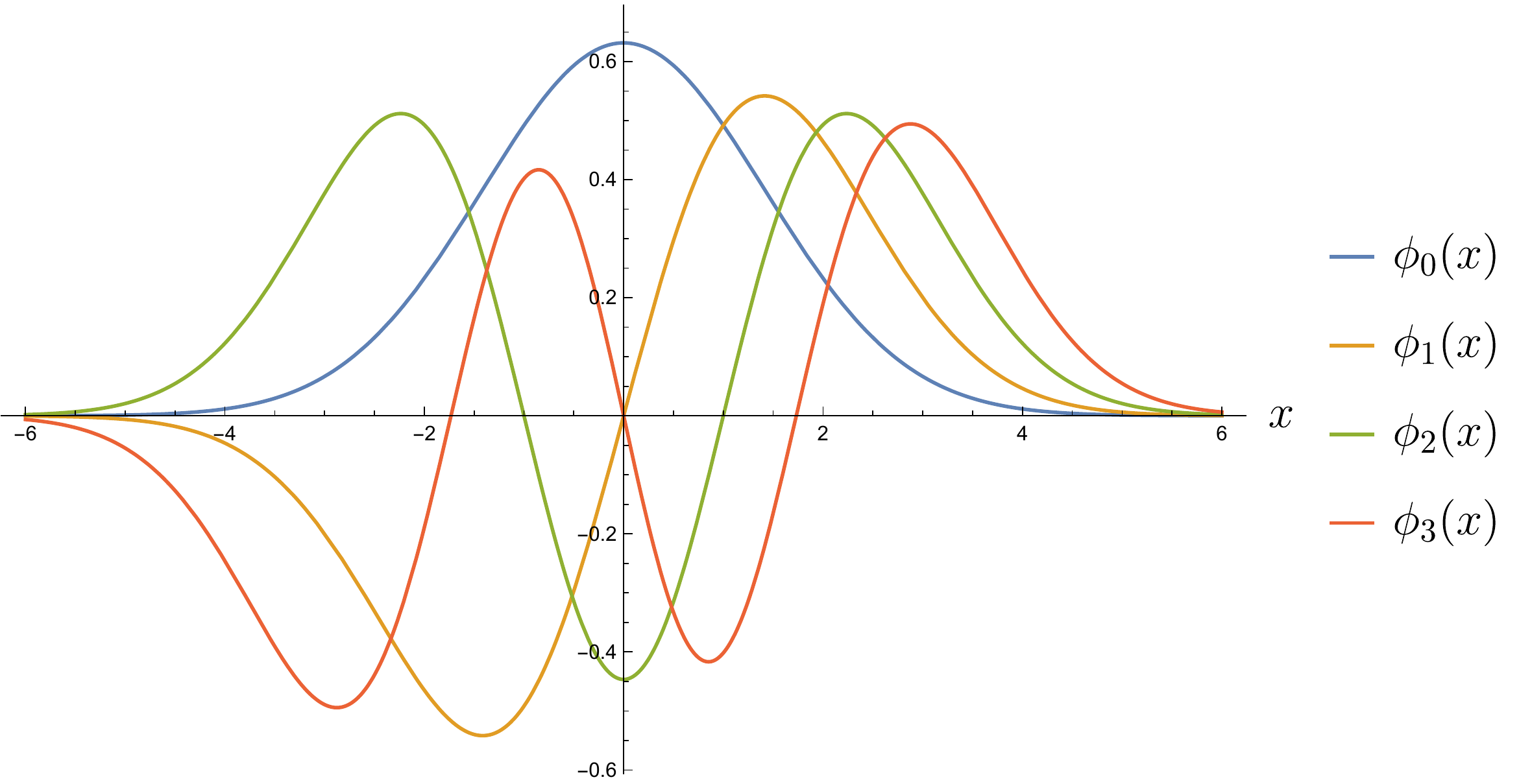} 
    \caption{Plot of the wavefunctions corresponding to the four lowest-lying quantum states for a particle in a harmonic trap.}
    \label{fig:HarmonicWavefunctions}
\end{figure}

\subparagraph{Harmonic potential and the GUE}
We first consider the case of fermions trapped in the harmonic potential $V_\mathrm{f}(x)=\frac{1}{2} \omega^{2} x^{2}$. In this case, the single-particles wavefunctions are given in terms of the Hermite polynomials by the real functions
\be 
\phi_k(x) = \sqrt{\frac{\sqrt{\omega/\pi}}{k!}} \   H_k( \sqrt{2\omega} x)  \ e^{-\omega x^2/2} 
\ee
with associated energy $\epsilon_k = (k+\frac{1}{2}) \omega$. Fixing $\omega=\frac{1}{2\sigma^2}$ allows to recover the GUE kernel, as we illustrate with $\sigma=1$ in which case
\be
\phi_k(x) = \sqrt{\frac{1}{k!\sqrt{2\pi}}} \   H_k(   x)  \ e^{- x^2/4} =  \frac{H_k(x)}{\sqrt{h_k}} e^{-\frac{x^2}{4}}
\label{eq:1:HarmonicWavefunctions}
\ee
such that the kernels Eqs.~\eqref{eq:1:KernelEigenvalueProcessOrthPol} and \eqref{eq:1:KernelFermions} are indeed equal for the GUE potential $V(\lambda) = \frac{\lambda^2}{4}$ and the fermion potential $V_\mathrm{f}(x) = \frac{x^2}{8}$. The similarity between these two harmonic potentials is however not a general fact, and we see in the next examples that the fermionic potential can be very different from the corresponding RMT potential. The first four harmonic wavefunctions $(\phi_k)_{0\leqslant k \leqslant 3}$ are plotted in Fig.~\ref{fig:HarmonicWavefunctions}. 

\subparagraph{Potential \texorpdfstring{$V_\mathrm{f}(x) = Ax^2 +B/x^2$}{Vf(x)=Ax2+B/x2} and the LUE} The second correspondence we discuss concerns fermions restricted to $x\geqslant 0$, in the potential
\be 
V_\mathrm{f}(x) = \frac{b^{2}}{2} x^{2}+\frac{\alpha(\alpha-1)}{2 x^{2}}
\ee
where we follow \cite{Dean2019} for the convenient parametrization with $b >0, \alpha >1$. The wavefunctions $\phi_k(x)$ can be expressed in terms of the Laguerre polynomials as
\be 
\phi_{k}(x)=c_{k} \ L_{k}^{\left(\alpha-\frac{1}{2}\right)}\left(b x^{2}\right)  x^{\alpha}  e^{-\frac{b}{2} x^{2}}
\ee 
with corresponding energy $\epsilon_k =b\left(2 k+\alpha+\frac{1}{2}\right)$. Changing variables to $y_{i}=x_{i}^{2}$, the kernel for the process of the $y$ variables can be shown to match that of the LUE, where the eigenvalues are restricted to be in $\R^+$ with however a potential $V$ which is very different of $V_\mathrm{f}$, see \myeqref{eq:1:VWishart}.

\subparagraph{Hard box potential and the JUE} Finally, fermions confined in a hard-box, such that $V_\mathrm{f}(x) =0$ on $[-1,1]$ and $V_\mathrm{f}(x) = \infty$ if $x \notin [-1,1]$, can be related to the eigenvalues of the JUE. Indeed, changing variables to $u_i =\frac{1}{2} ( 1+ \sin(\frac{\pi x_i}{2}))\in [0,1]$, the process of the $u$ variables has the same statistics as the eigenvalues of the Jacobi Ensemble with potential $V$ given in \myeqref{eq:VJacobi} where the two Jacobi parameters are fixed as $A_1=A_2=\frac{1}{4}$. Furthermore, the JUE with arbitrary parameters can similarly be obtained as a variable change of a fermion system with potential $V_\mathrm{f}(x) = \frac{A}{\sin(x/2)^2} + \frac{B}{\cos(x/2)^2}$.

For a review of these connections between RMT and fermions on the real line, and details on the results mentioned above, we point to \cite{Dean2019,Forrester2010}. 

As presented above, there are a small number of known exact mappings between fermion systems and RMT ensembles. In the course of the research projects related to this thesis and presented in publication \pubref{publication:MatrixKesten}, we were led to exhibit a previously unknown connection between the inverse-Wishart ensemble and fermions in the renowned Morse potential $V_\mathrm{f}(x) \propto (1- e^{-ax})^2$, as will be explained in chapter \ref{chap:3}. 

\subparagraph{Higher dimensions}
Complex spectra of non-Hermitian random matrices can also be mapped to certain two-dimensional systems of spinless fermions in rotation \cite{Lacroix-A-Chez-Toine2019,Ginibre1965}, to one-dimensional fermions with spin \cite{Chui1975} and to two-dimensional one-component plasma (\textit{2DOCP}) \cite{Alastuey1981,Forrester1998}.

Fermions can also be studied in higher dimensions, in contrast with eigenvalues of RMT. Investigations in higher-dimensional fermion systems show that they exhibit similar kind of universal local statistics as seen in the $d=1$ case presented in the previous section for the equivalent RMT systems \cite{Dean2015,Lacroix-A-Chez-Toine2018}. 
In this general $d>1$ setting, investigations have also concerned the full counting statistics and gap probabilities of the non-interacting fermions \cite{Gouraud2021,Smith2021}.

\subsubsection{Eigenvalues of orthogonal and symplectic ensembles (\texorpdfstring{$\beta = 1,4$}{Beta=1,4})}
 
The previous sections have dealt in length with the unitary ensembles ($\beta=2$), where the squared Vandermonde determinant induces the powerful determinantal structure of the eigenvalue JPDF, and allows a mapping to fermion systems. In this section, we turn to the remaining orthogonal ($\beta=1$) and symplectic ($\beta=4$) ensembles, and present briefly how a parallel structure intervenes.

\paragraph{Pfaffian point processes} 

This structure requires the definition of the Pfaffian of a $2N \times 2N$ anti-symmetric matrix $\AAA$, which is a polynomial of degree $N$ in the matrix entries such that
\be 
( \Pf \AAA)^2 = \det \AAA \; .
\ee  
It is formally defined as a sum on the $(2N)!$ permutations of $[1\ldotp \ldotp 2N]$, denoted $(i)$ with sign $\sigma(i)$, as
\be 
 \Pf \AAA = \frac{1}{2^N N !} \sum_{(i)}  \sigma(i) \prod_{k=1}^{N} \AAA_{i_{2k-1}, i_{2k}} \; . 
 \label{eq:1:DefPfaffian}
 \ee
 
In parallel with the construction of the determinantal point process, we define a Pfaffian point process (\textit{PPP}) as the probability measure on the configurations of $N$ points on the real line, given by the JPDF and marginal densities given for $0 \leqslant m \leqslant N$ by
\be 
P(x_1, \cdots , x_m) = \frac{(N-m)!}{N!} \ \Pff{1 \leqslant i,j \leqslant N} K(x_i , x_j)
 \label{eq:1:DefPPP}
\ee
where the kernel K is now a $2 \times 2$ matrix
\be 
K(x, y)=\left(\begin{array}{ll}K_{11}(x, y) & K_{12}(x, y) \\ K_{21}(x, y) & K_{22}(x, y)\end{array}\right)  \; .
\ee 
For the $2N \times 2N$ matrix entering the Pfaffian to be anti-symmetric, the kernel is constrained to verify $K_{12}(x,y) = - K_{21}(y,x)$, with anti-symmetric diagonal entries $K_{ii}(x,y) = - K_{ii}(y,x)$, $i=1,2$.

A PPP can be viewed as a generalization of a DPP. Indeed, in the particular case where the diagonal parts of the kernel vanish, such that $K_{11} = K_{22} = 0$,  the Pfaffian appearing in \myeqref{eq:1:DefPPP} reads
\be
 \Pff{1 \leqslant i,j \leqslant N} K(x_i , x_j) = \det_{1 \leqslant i,j \leqslant N} K_{12}(x_i , x_j) 
\ee
and the point process is in fact a DPP with kernel $K_{12}$.

\paragraph{Orthogonal and symplectic ensembles} 
 
The eigenvalue processes of orthogonal and symplectic ensembles, i.e. $\beta$-ensembles with $\beta=1$ and $4$, can be shown to be Pfaffian point processes. The corresponding $2 \times 2$ matrix anti-symmetric kernel can be constructed from the family of \emph{skew-orthogonal polynomials} associated with the potential $V$ and corresponding weight function $\omega(\cdot) = e^{- \beta V(\cdot)}$.

We will not detail the construction in this thesis, but instead point to \cite{Soshnikov2003} for the exact mapping of the eigenvalue JPDF in a Pfaffian form, to \cite{Ferrari2003} for the application to the GOE and GSE particular cases and to \cite{Mehta2004,Forrester2010} for extensive details on the skew-symmetric constructions.

Remarkably, the universal local statistics observed in the unitary case extend to the PPP structure of the orthogonal and symplectic cases. On the microscopic scale, the bulk, soft edge and hard edge statistics are then respectively described by the sine, Airy and Bessel $2\times 2$ matrix kernels \cite{Kuijlaars2007}, which are different in the $\beta =1$ and $\beta =4$ cases, and are of course different from the scalar kernels presented in the $\beta=2$ case.

\subsection{Extreme value statistics}
\label{subsec:1.2.3}

In this section, we present another aspect of random matrix eigenvalues: the statistics of the extreme values of the spectrum. These properties have raised much interest in RMT, and many efforts have been made until recent years to refine the knowledge on this aspect of the eigenvalue statistics. 

The interest in the properties of the maximal eigenvalue of a random matrix spectrum can be explained by the fact that they are non-standard with respect to usual independent probability theory. Indeed, for a set of i.i.d.~random variables, the distribution of the maximal value follows either the Gumbel, Fréchet or Weibull law according to the characteristics of the initial distribution's tail \cite{Majumdar2020}. The maximum of a random matrix spectrum, where the eigenvalues form a set of strongly-correlated variables, does not fit in this classification.  

\subsubsection{Tracy-Widom distribution}

Craig Tracy and Harold Widom were the first to study the statistics of the maximal eigenvalue of the main RMT ensembles \cite{Tracy1994,Tracy1996}. They exhibited the distribution of the maximal eigenvalue $\lmax$ of Gaussian ensembles around its typical value, which is now called the Tracy-Widom (\textit{TW}) distribution. This distribution depends on the index $\beta$ and describes the fluctuations of $\lmax$ on the microscopic scale around the typical edge of the spectrum. The cumulative distribution function (\textit{CDF}) of the TW-$\beta$ law is usually denoted $\mathcal{F}_{\beta}$.

Let us state the precise result for the Gaussian $\beta$ ensemble with $\sigma=1$. We saw in section \ref{subsec:1.2.2} that the eigenvalue density around the edge $2\sqrt{N}$ is of order $N^{1/6}$, such that the typical lengthscale is then $N^{-1/6}$. The maximal eigenvalue then follows the TW distribution, in the large-$N$ limit, after a corresponding rescaling as
\be 
\frac{ \lmax - 2 \sqrt{ N}}{N^{- \frac{1}{6} }} \  \sim \  \chi 
\ee 
where $\chi$ is a TW-distributed random variable, such that $\mathbb{P}[\chi \leqslant x]=\mathcal{F}_{\beta}(x)$. 

The probability $\mathbb{P}[\chi \leqslant x]$ is the probability that the rescaled process has no particles above the position $x$. In the case $\beta =2$, the local Airy kernel DPP structure at the soft edge allows to write this gap probability on $[x,+\infty[$ in the Fredholm determinant form of \myeqref{eq:1:GapProbabilityDPP} such that:
\be
\mathcal{F}_{2}(x) = \mathbb{P}[\chi \leqslant x]= \det \bigg(\ID - \left. K^{\mathrm{Airy}} \right|_{[x,+\infty[}\bigg)  \; .
\label{eq:1:TracyWidom2}
\ee
In the cases $\beta=1$ and $4$, the PPP structure implies an expression of $\mathcal{F}_{1,4}$ as Fredholm Pfaffians, which are an extension of Pfaffians to skew-symmetric operators similarly as the Fredholm version of the determinant. The TW distribution has been further extended to general $\beta>0$ in the study of the stochastic Airy operator \cite{Ramirez2011}.

\begin{figure}[ht!]
    \centering 
    \includegraphics[width=.8 \textwidth]{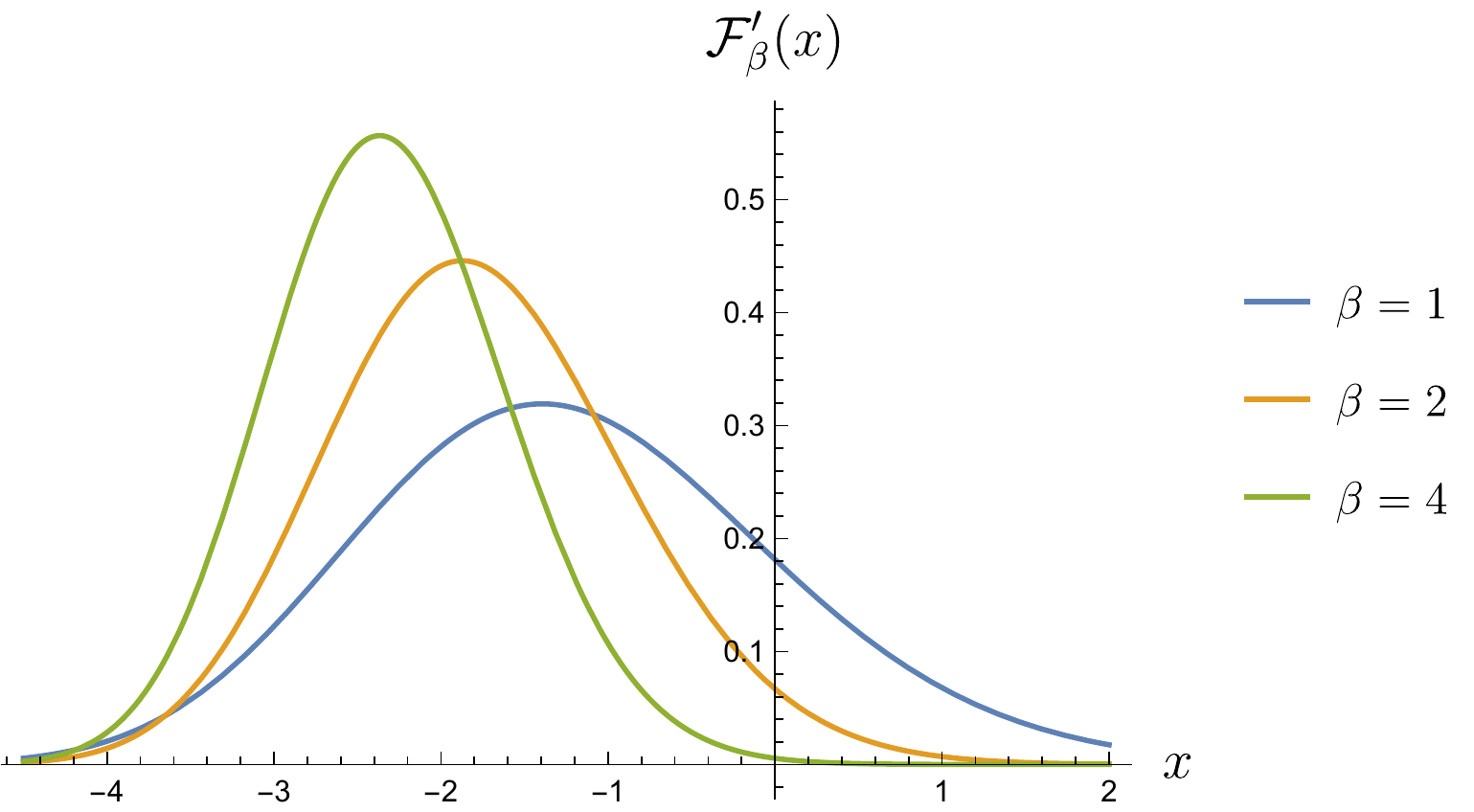} 
    \caption{The Tracy-Widom PDF $\mathcal{F}_\beta^\prime(x)$ for $\beta=1,2,4$.}
    \label{fig:TracyWidom}
\end{figure}

The PDF $\mathcal{F}_\beta^\prime(x)$ of the TW distribution is plotted for the three values of $\beta$ in Fig.~\ref{fig:TracyWidom}, where their asymmetry is apparent. Indeed, the two tails of the distribution are not equivalent and obey the following asymptotics:
\be 
\mathcal{F}_{\beta}^{\prime}(x) \simeq  
\left\{\begin{array}{ll} 
e^{ -\frac{\beta}{24}\abs{x}^{3} }  & \text{ as }  x \rightarrow-\infty \; , \\[5pt]
 e^{ -\frac{2 \beta}{3} x^{\frac{3}{2}} } & \text{ as } x \rightarrow+\infty \; . 
 \end{array}\right.
\ee
The left tail vanishes much faster than the right tail, as expected from the fact that a large maximum value only requires to have an unexpected position for one eigenvalue, while a low maximum value requires to push a large number of eigenvalues to the left. For more technical precisions about the TW distribution, and in particular its connections to the Painlevé II nonlinear differential equation, see \cite{Mehta2004,Nadal2011,Krajenbrink2021a}.

The universality of the Airy kernel on the local scale at a soft edge implies the universality of the Tracy-Widom distribution which therefore appears in many random matrix ensembles. Furthermore, this distribution appears outside of RMT in a wide range of problems such as the directed polymer model and other growth processes \cite{Prahofer2000} or the combinatorics of longest increasing subsequences \cite{Majumdar2007}. It has also been observed in experiments on nematic liquid crystal behaviour \cite{Takeuchi2011} and coupled lasers \cite{Fridman2012}. An explanation for the universal emergence of the Tracy-Widom distribution in many different contexts has been suggested through the mechanism of a third-order phase transition \cite{Majumdar2014}.

\subsubsection{Large deviations}

Beyond the typical fluctuations of the maximal eigenvalue around the edge of the spectrum, the rare events obey a large deviation principle \cite{Touchette2009}. These rare events can be either an extremely high or an extremely low value of $\lmax$, such that the distance of $\lmax$ with the expected edge is of the order of the global scale of the spectrum. 

For example, in the Gaussian $\beta$ ensembles, the large deviation tails behave asymptotically in the large-$N$ limit as
\be 
\partial_{x} \ \mathbb{P} \left[\lambda_{\max } \leqslant  \sqrt{N}x  \right] \simeq 
\left\{\begin{array}{ll}
\exp \left(-\beta N^{2} \, \Phi_{-}
(x)\right) &  \text{for } x <2  \\[10pt]
\exp \left(-\beta N \, \Phi_{+}
(x)\right) &  \text{for } x >2 
\end{array} \right.
\ee 
where the large deviations functions $ \Phi_{-}$ and $\Phi_{+}$ were obtained for general values of $\beta$ in \cite{Dean2006,Dean2008,Majumdar2009}. Once again, the asymmetry between the left and right tail is visible in the different exponent appearing in the prefactor. In physical terms, pulling a single particle of the Coulomb gas to a high value costs an energy of order $N$, while pushing the whole gas of particles such that all the particles stay to the left of a low barrier costs an energy of order $N^2$. The difference between these two tails has similarities with the Gross-Witten-Wadia phase transition of lattice QCD \cite{Majumdar2014}. 

The large deviations of the extreme eigenvalues of some RMT ensembles are further discussed in relation to the HCIZ integral in chapter \ref{chap:4}. On this topic of recent activity and for studies of other ensembles, one may also consult \cite{Majumdar2014,Borot2011,Nadal2011}.

\section{Free probability for large matrices}
\label{sec:1.3}
 
In this last section of the introduction chapter, we give a presentation of the theory of free probability. This young branch of probability theory was founded at the beginning of the 1980's by Dan-Virgil Voiculescu who studied Von Neumann operator algebras, and was in particular led to investigate properties of random objects in these non-commutative algebras. He thereupon defined the property of freeness for non-commutative random variables \cite{Voiculescu1985} and found a major field of application of this notion in the study of large random matrices \cite{Voiculescu1991,Voiculescu1995}. The constructions and methods of free probability are now an essential part of the standard toolbox of random matrix theory. 

In the context of random matrices, a first idea of free probability can be given as follows. In general, the knowledge of the eigenvalues of two matrices is not sufficient to deduce the spectrum of the matrix sum, unless the matrices commute and are then diagonalizable in the same basis. Free probability introduces a condition under which the spectrum of the sum can be obtained from the two initial spectra. This freeness condition must then imply that eigenvectors are immaterial, and we will see that the condition is indeed respected for large matrices whose eigenbases are randomly rotated with respect to each other.

In addition to its ties with random matrices, free probability is connected to the beautiful combinatorics of non-crossing partitions and to many other fields such as representation theory of large groups, quantum groups, statistical inference and quantum information theory \cite{Mingo2010}. 

In the first subsection, we define the notion of freeness and give an overview of the combinatorial structure of the theory.
In subsection \ref{subsec:1.3.2}, we detail the properties of the sum of free random matrices and the relevant tool of this issue, the $\RT$-transform.
In subsection \ref{subsec:1.3.3}, we detail the properties of the product of free random matrices and the relevant tool of this issue, the $\ST$-transform.
For more comprehensive presentations of free probability, we refer the reader to \cite{Potters2020,Speicher2003,Speicher1997,Novak2012,Tulino2004,Mingo2010,Nica2006,Bun2017}.


\subsection{Definition and properties of freeness}
\label{subsec:1.3.1}

We define the notion of freeness for random matrices, which applies in the large-size limit. We will see that this notion leads naturally to the definition of \emph{free cumulants}, in an elegant connection to the combinatorics of non-crossing partitions.

\subsubsection{Asymptotic freeness}

The random variables that we ought to consider are sequences of Hermitian $N \times N$ matrices $\MM_N$ which converge to a well-defined infinite-size random matrix $\MM$. Note that for the limit to be well-defined, the matrix sequence must be normalized such the asymptotic spectrum is compactly supported. The limit $\MM$ is then characterized by its eigenvalue density $\rho_{\rm emp} = \rho$, which is equal to the average density by the self-averaging property of large matrices. In full rigour, the limit of the sequence belongs to a non-commutative operator algebra, to which the free probability results can be formally applied, but we will informally consider that we are handling a large random matrix $\MM$ in the following.

 A useful observable is then the expected renormalized trace as:
\be 
\phi ( \MM )  = \lim_{N \to \infty} \frac{1}{N} \E \left[ \Tr \, \MM_N \right] =  \int_\R \lambda \, \rho(\lambda) \,  \dd \lambda \; .
\ee 
Note that $\phi$ is linear and verifies $\phi ( \ID ) = 1$. This allows to define the $k$\textsuperscript{th} asymptotic moment as 
\be
m_k= \phi (\MM^k ) = \int_\R \lambda^k \rho(\lambda) \dd \lambda \; .
\ee
With these conventions in place, we can define the notion of \emph{asymptotic freeness}.

\paragraph{Definition}
The random matrices $\AAA$ and $\BB$ are asymptotically free if for all $\ell$ and for all polynomials $p_i$ and $q_i$ ($1 \leqslant i \leqslant \ell$) such that $ \phi ( p_i ( \AAA )) = \phi ( q_i ( \BB )) = 0 $, we have:
\emphbe 
\phi \bigg( p_1 (\AAA )  q_1 (\mathbf{B } ) \cdots   p_\ell (\AAA )   q_\ell (\mathbf{B } ) \bigg)= 0 
\label{eq:1:DefFreenessPhiEqualsZero}
\end{empheq}
Of course this definition extends to several matrices, such that $\AAA_1, \cdots, \AAA_m$ are asymptotically free if $
\phi \bigg( p_1 ( \AAA_{j_1} ) p_2 ( \AAA_{j_2} ) \cdots  p_\ell ( \AAA_{j_\ell} ) \bigg) = 0 $
whenever $\phi( p_i ( \AAA_{j_i} ) ) = 0$ for all $i$, and all consecutive indices are different: $j_i \neq j_{i+1}$. Note that this is a statement on the limit matrices, hence the asymptotic nature of the notion in the RMT setting.

The construction is intended to be a non-commutative version of independence. Indeed, if we consider $m$ distinct asymptotically free matrices $(\AAA_{i})_{1 \leqslant i \leqslant m}$, the freeness property implies for example that for any polynomials  $(p_{i})_{1 \leqslant i \leqslant m}$:
\be 
\phi \bigg( p_1 (\AAA_1) \cdots p_m (\AAA_m) \bigg) = 
\phi \big( p_1 (\AAA_1) \big) \cdots \phi \big( p_m (\AAA_m) \big) 
\ee
which is reminiscent of the expectation factorization of independent scalar random variables $(X_i)_{1 \leqslant i \leqslant m} $:
\be
\E \big[ p_1 (X_1)  \cdots p_m( X_m)  \big] = \E \big[ p_1 (X_1)  \big] \cdots \E \big[ p_m (X_m)   \big]  \; .
\ee 
However in the commutative case, if two variables are the same $X_i = X_j$, they can be regrouped before the factorization as:
\be 
\E \bigg[ X  Y X  Y\bigg] = \E \big[ X^2  \big]  \  \E \big[ Y^2 \big]  \; .
\ee
For free matrices $\AAA$ and $\BB$ on the other hand, non-commutativity changes things drastically when a matrix appears several times in the product. For example, by introducing centered polynomials such as $\AAA - \phi(\AAA)$ and by successive applications of \myeqref{eq:1:DefFreenessPhiEqualsZero}, a few lines of algebra yield
\be
\phi \bigg ( \AAA   \BB  \AAA   \BB  \bigg) 
=
\phi(\AAA)^2 \phi (\BB^2) + 
\phi(\BB)^2 \phi (\AAA^2)  -
\phi(\AAA)^2 \phi(\BB)^2  \; .
\ee
We note that $\phi ( \AAA \BB \AAA \BB )$, in addition to featuring three terms when the usual scalar version only gives one, does not even display the term $\phi (\AAA^2) \phi (\BB^2) $ that would be expected from the scalar setting. As a non-commutative version of independence, freeness has many differences from the usual version, which we will detail after giving a few results on the conditions for random matrices to be asymptotically free.

\paragraph{Sufficient conditions}
An important question is the following: when do two large random matrices constitute pair of free variables, in the asymptotic sense ? In the following lines, we give interesting results on this issue.
\begin{itemize}
\item Any random matrix and the identity $\ID$ are asymptotically free.
\item If two matrices are asymptotically free and they commute, then one of them is a multiple of $\ID$. Such a scalar matrix is also called \emph{constant}, from the point of view of the eigenvalue distribution.
\item A random matrix which is free with itself is constant.
\item $X_1 \ID$ and $X_2 \ID$, where $X_1$ and $X_2$ are two independent scalar random variables with non-zero variance, are not asymptotically free.
\item Independent standard Gaussian matrices are asymptotically free.
\end{itemize}  
Finally, the most important result for the application of free probability to RMT is the following theorem:
\begin{itquote}
If $\AAA$ and $\BB$ are random matrices (with compactly supported asymptotic ESD) and if $\mathbf{U}$ is a (Haar-distributed) random unitary matrix, then $\UU \AAA \UU^\dagger$ and $\BB$ are asymptotically free. 
\end{itquote} 

This result shows the importance of free probability theory in random matrix theory: any two matrices that are randomly rotated with respect to each other are asymptotically free. In the large size limit, they can be treated as free random variables. For more precise statements and proofs of results cited above, we point to \cite{Mingo2010,Tulino2004,Potters2020}.

\subsubsection{Non-crossing partitions and free cumulants}

As defined above, freeness is a rule that allows the computation of mixed moments of $\AAA$ and $\BB$ at all orders, given the moments of $\AAA$ and the moments of $\BB$. The application of this rule quickly becomes cumbersome and it is useful to introduce a tool called the \emph{free cumulant}. Its construction relies on the combinatorics of non-crossing partitions.

\paragraph{Non-crossing partitions}

A partition $\omega= \{ V_1, \cdots, V_k\}$ of the set $\mathbb{S}_n=\{1, \cdots, n\}$, where the blocks $V_i$ are non-empty subsets of $\mathbb{S}_n$ with no shared element such that $\cup_{i=1}^k V_i = \mathbb{S}_n$, is called \emph{non-crossing} if there are no four elements $ 1 \leqslant  p_1 < q_1 < p_2 < q_2 \leqslant n$ such that:
\be 
 p_1, p_2 \in V_i \quad , \quad q_1, q_2 \in V_j  \quad \quad  \text{ with } i \neq j .
 \ee
A partition where this occurs is said to be crossing. In the following, we denote $\mathcal{P}_n$ as the set of all partitions of the set $\mathbb{S}_n$, and $\mathcal{P}^{\mathrm{NC}}_n$ as the subset of $\mathcal{P}_n$ consisting of the non-crossing partitions.

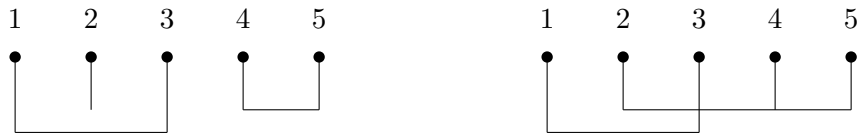
\begin{figure}[ht!]
\centering
\vspace{.5cm}
\begin{tikzpicture}
\draw (0,0) -- (0,1); 
\draw (1,.3) -- (1,1); 
\draw (2,0) -- (2,1);
\draw (3,.3) -- (3,1);
\draw (4,.3) -- (4,1);
\draw (7,0) -- (7,1);   
\draw (8,.3) -- (8,1);
\draw (9,0) -- (9,1); 
\draw (10,.3) -- (10,1); 
\draw (11,.3) -- (11,1); 
\draw (0,0) -- (2,0); 
\draw (3,.3) -- (4,.3); 
\draw (7,0) -- (9,0);
\draw (8,.3) -- (11,.3); 
\node at (0,1)[circle,fill,inner sep=1.5pt]{};
\node at (1,1)[circle,fill,inner sep=1.5pt]{};
\node at (2,1)[circle,fill,inner sep=1.5pt]{};
\node at (3,1)[circle,fill,inner sep=1.5pt]{};
\node at (4,1)[circle,fill,inner sep=1.5pt]{};
\node at (7,1)[circle,fill,inner sep=1.5pt]{};
\node at (8,1)[circle,fill,inner sep=1.5pt]{};
\node at (9,1)[circle,fill,inner sep=1.5pt]{};
\node at (10,1)[circle,fill,inner sep=1.5pt]{};
\node at (11,1)[circle,fill,inner sep=1.5pt]{};
\node[draw=none] at (0,1.5) {1};
\node[draw=none] at (1,1.5) {2};
\node[draw=none] at (2,1.5) {3};
\node[draw=none] at (3,1.5) {4};
\node[draw=none] at (4,1.5) {5};
\node[draw=none] at (7,1.5) {1};
\node[draw=none] at (8,1.5) {2};
\node[draw=none] at (9,1.5) {3};
\node[draw=none] at (10,1.5) {4};
\node[draw=none] at (11,1.5) {5};
\end{tikzpicture}
\vspace{.5cm}
\caption{Two partitions of $\mathbb{S}_5=\{1,2,3,4,5\}$.}
\label{fig:Twopartitions}
\end{figure}

As an illustration, the non-crossing partition $\{\{1,3\},\{2\},\{4,5\}\}$ and the crossing partition $\{\{1,3\},\{2,4,5\}\}$ of $\mathbb{S}_5$ are represented in Fig. \ref{fig:Twopartitions}, where elements in the same block are connected by a line, such that a partition is crossing if and only if two lines intersect.

\begin{figure}[ht!]
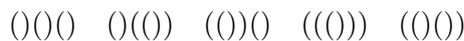

\be
()()() \quad ()(()) \quad (()) () \quad ((())) \quad (()())
\nonumber
\ee
\caption{Five orderings of three pairs of parentheses.} 
\label{fig:Parentheses}
\end{figure}

The number of non-crossing partitions of $\mathbb{S}_n$ is the $n$-th Catalan number $C_n$:
 \be 
C_n = \mathrm{card}\left(  \mathcal{P}^{\mathrm{NC}}_n  \right) =   \frac{1}{n+1} \binom{2n}{n}  \; .
 \ee
Interestingly, $C_n$ is also the number of non-crossing pair partitions of $\mathbb{S}_{2n}$, i.e. elements of $\mathcal{P}^{\mathrm{NC}}_{2n}$ consisting of $n$ pairings of elements. Denoting $\mathcal{P}^{\mathrm{NC}(2)}_{2n}$ their ensemble, $C_n =  \mathrm{card}\left( \mathcal{P}^{\mathrm{NC}(2)}_{2n} \right)$. Elements of $\mathcal{P}^{\mathrm{NC}(2)}_{2n}$ can be seen as Dyck words of length $2n$, i.e. words that can be written using pairs of left and right parentheses without closing an unopened parenthesis. For example $C_3= 5$ as seen in Fig. \ref{fig:Parentheses}, because there are 5 ways to order 3 pairs of parentheses.

%

\paragraph{Standard cumulants}

Before introducing free cumulants, we briefly recall the standard construction of cumulants $k_n$ of a scalar random variable $X$, in relation with its moments $M_n = \E[X^n]$. The cumulants are defined as the coefficients of the following series:
\begin{equation}
\ln \E \left[ e^{tX} \right] = \sum_{n \geqslant 1 } k_n \frac{t^n}{n!}  
\end{equation}
and they are related to the moments through
\be 
M_n = \sum_{ \omega \in \mathcal{P}_n} \prod_{i =1}^k k_{\ell_i}= \sum_{ \omega \in \mathcal{P}_n}  k_{\ell_1} \cdots k_{\ell_k}  \; .
\label{eq:1:StandardMomentsfromCumulants}
\ee
where we denote $\omega =\{V_1, \cdots, V_k\} \in \mathcal{P}_n$ a partition and $\ell_i = \mathrm{card}( V_i )$ the cardinalities of its blocks.

As their name suggests, the main interest of these cumulants is that they behave well under the addition of two independent random variables. For any two independent random variables $X$ and $Y$ and for all integers $n$, with $k_n(\cdot)$ the $n$\textsuperscript{th} cumulant function :
\be 
k_n( X + Y)  = k_n ( X ) + k_n(  Y)  
\ee
which is easily proven as:
\be 
\sum_{n \geqslant 1 } k_n(X+Y) \frac{t^n}{n!} = \ln \E \left[ e^{t(X+Y ) } \right] = \ln \left( \E \left[ e^{t(X ) } \right] \E \left[ e^{t(Y ) } \right] \right) = \sum_{n \geqslant 1 } \left( k_n(X) + k_n(Y) \right)  \frac{t^n}{n!}  \;.
\ee

\paragraph{Free cumulants} 

In the non-commutative realm also, it is possible to define free cumulants which behave well under the addition of free random matrices. For a random matrix $\MM$ with a given sequence of moments $m_n = \phi( \mathbf{M}^n)$, the sequence of free cumulants $\kappa_n$ is defined recursively such that
\be  
m_n = \sum_{ \omega \in \mathcal{P}^{\mathrm{NC}}_n} \prod_{i =1}^k \kappa_{\ell_i}= \sum_{ \omega \in \mathcal{P}^{\mathrm{NC}}_n}  \kappa_{\ell_1} \cdots \kappa_{\ell_k} 
\label{eq:1:FreeMomentsfromCumulants}
\ee 
where the sum is taken on non-crossing partitions only, in contrast with the standard case. Denoting $\kappa_n(\mathbf{M})$ the $n$\textsuperscript{th} cumulant of a random matrix $\mathbf{M}$, it is a theorem that for any two free random matrices $\AAA$ and $\BB$ and for all integers $n$:
\begin{equation}
\kappa_n ( \AAA + \BB)  = \kappa_n ( \AAA ) + \kappa_n (  \BB)   \; .
\label{eq:1:AdditivityFreeCumulants}
\end{equation} 

This property translates the information carried by the property of freeness, i.e. that alternating and centered mixed moments vanish, in a very simple form with free cumulants. A proof of this theorem requires the additional definition of mixed cumulants $\kappa_n( \AAA , \BB,  \mathbf{C} ,...)$, which can be shown to vanish for free matrices. The quantity $\kappa_n( \AAA + \BB)$ can then be written as a sum on all possible mixed cumulants of $\AAA$ and $\BB$, where only $ \kappa_n ( \AAA )$ and $ \kappa_n ( \BB )$ survive. 

For a proof of \myeqref{eq:1:AdditivityFreeCumulants}, showing how the structure of non-crossing partitions used in the definition of free cumulants provides a useful tool to simplify the analysis of free variables addition, we point to \cite{Speicher2003,Nica2006,Speicher1997,Potters2020}.

In Table \ref{tab:Cumulants}, we give the first expressions of moments in terms of cumulants both in the standard and free cases, obtained from \myeqref{eq:1:StandardMomentsfromCumulants} and \myeqref{eq:1:FreeMomentsfromCumulants}. We note that the first three terms are identical, as all partitions of $\mathbb{S}_n$ are non-crossing for $n \leqslant 3$. For $n=4$, the difference is the removal of the unique crossing partition $\{ \{ 1, 3\}\{2,4\} \}$ such that the coefficient of $\kappa_2^2$ is 2 instead of 3.

\begin{table}[ht!]
\centering
\vspace{.5cm}
\begin{tabular}{||p{0.3 cm}|p{6 cm}|p{6 cm}|}
  \hline
 $n$  & \hfil $M_n$ &  \hfil  $m_n$  \\ \hhline{|=|=|=|}
 && \\[-8pt]
1 & $k_1 $ &  $ \kappa_1$  \\[10pt]
2 & $ k_2 + k_1^2$ & $ \kappa_2 + \kappa_1^2$  \\[10pt]
3 & $ k_3 + 3k_2 k_1 + k_1^3 $  & $\kappa_3 + 3 \kappa_2 \kappa_1 + \kappa_1^3 $ \\[10pt]
4 & $ k_4 +  4 k_3 k_1 + 3 k_2^2 + 6 k_2 k_1^2 + k_1^4 $ & $\kappa_4 +  4 \kappa_3 \kappa_1 + 2 \kappa_2^2 + 6  \kappa_2  \kappa_1^2 +  \kappa_1^4 $ \\[10pt]  
5 & $k_{5}+5 k_{4} k_{1}+10 k_{3} k_{2}+10 k_{3} k_{1}^{2}+15 k_{2}^{2} k_{1}+10 k_{2} k_{1}^{3}+k_{1}^{5}
$ & $\kappa_{5}+5 \kappa_{4} \kappa_{1}+5 \kappa_{3} \kappa_{2}+10 \kappa_{3} \kappa_{1}^{2}+10 \kappa_{2}^{2} \kappa_{1}+10 \kappa_{2} \kappa_{1}^{3}+\kappa_{1}^{5}$\\[20pt] 
  \hline
\end{tabular} 
\caption{Five first moments expressed in terms of cumulants, in standard and free cases.} 
\label{tab:Cumulants}
\end{table}

\subsection{Sum of free random matrices}
\label{subsec:1.3.2}

In the previous section, we have seen that freeness provides a free-cumulant addition formula \myeqref{eq:1:AdditivityFreeCumulants}. We now turn to a practical use of this property, to solve the problem mentioned at the beginning of the section: from the knowledge of the spectra of two free random matrices $\AAA$ and $\BB$, how can we characterize the spectrum of $\AAA + \BB$ ? This problem of free addition of random variables was first studied in \cite{Voiculescu1986} and the tools introduced therein are now commonly used in RMT.

\subsubsection{The \texorpdfstring{$\RT$}{R}-transform} 

The interesting property on free cumulants will be transformed on an addition property for the generating function of these free cumulants, the $\RT$-transform, defined in the following. 

Recall the definition of the Stieltjes transform $\g$ of our large matrix $\MM$ in terms of the density $\rho$ in \myeqref{eq:1:DefStieltjes}, and note that it admits a large-$z$ series expansion where the coefficients are the moments $m_n$:
\be
 \mathfrak{g}(z)  =  \frac{1}{z} +\sum\limits_{n \geqslant 1} \frac{m_n }{z^{n+1} } \; .
\ee
Denoting $[a,b]$ the support of the density $\rho$ on the real line, $\g$ is defined on $ \mathbb{C}\setminus[a,b]$. Its functional inverse is called the $\mathcal{B}$ (or Blue) transform:
\be
\mathcal{B}(\mathfrak{g}(z)) = z \quad ,  \quad \forall z \in \mathbb{C}\setminus[a,b]
\ee
and the $\RT$ (or Red) transform is defined from it as:
\be
\RT(\omega ) = \mathcal{B}(\omega) - \frac{1}{\omega} \; .
\label{eq:1:DefRTransform}
\ee
Both functions are defined on $\C^*$.

It can be shown that the $\RT$-transform is precisely the free cumulant generating function, such that it admits the following Taylor expansion as $\omega \to 0$:
\be
\RT(\omega) = \sum\limits_{n \geqslant 1}  \kappa_n \, \omega^{n-1}
\ee
where the coefficients are the free cumulants of $\MM$. This is proven by analyzing the functional relation of $\RT$ with the generating function of the moments $\g (\frac{1}{z})$, in regards to the combinatorial definition \myeqref{eq:1:FreeMomentsfromCumulants}, see
\cite{Nica2006,Speicher1997,Potters2020}.

%

The addition property of free cumulants directly yields:
\begin{empheq}[box=\setlength{\fboxsep}{8pt}\fbox]{equation}
\RT_{\AAA+\BB}(\omega)  =\RT_{\AAA }(\omega)  + \RT_{ \BB}(\omega)  
\label{eq:1:FreeAdditionRTransform}
\end{empheq}
for two random matrices $\AAA$ and $\BB$. This relation characterizes exactly the asymptotic spectrum of the sum $\AAA + \BB$ from the knowledge of the spectra of $\AAA $ and $\BB $: the knowledge of $\RT_{\AAA+\BB}$ indeed allows to recover the Stieltjes transform, which in turn characterizes the eigenvalue distribution exactly.

\subsubsection{Wigner's semicircle and the free central limit theorem}

We give in the following paragraph an example of $\RT$ transform for the large-size limit of the Gaussian ensembles, characterized by Wigner's semicircle eigenvalue density, as presented in section \ref{subsec:1.2.1}.

\paragraph{The free cumulants of Wigner's semicircle}
For $\MM$ a rescaled Gaussian matrix (with either $\beta=1,2$ or $4$) of large size, the eigenvalue density is characterized by Wigner's semicircle density $\rho_\mathrm{SC}$ supported on $[-2\sigma, 2\sigma]$, see \myeqref{eq:1:WignerSemicircleDefinition}. The Stieltjes transform of this distribution can be computed as \myeqref{eq:1:SelfConsistentG}, which is inverted as $\mathcal{B}(\omega) = \sigma^2 \omega + \frac{1}{\omega}$ such that the $\RT$-transform is linear in $\omega$:
\be
\RT(\omega) = \sigma^2 \omega \; .
\ee
The free cumulants of the semicircle distribution are directly read as: 
\be
\left\{
    \begin{array}{ll}
        \kappa_1 = m_1 = 0  \\
       \kappa_2 = \sigma^2 \\
       \kappa_n =0 \quad  \text{for } n \geqslant 3 \; .
    \end{array}
\right.
\ee
The first cumulant is zero, as every odd moment of $\MM$, because the distribution is centered. The semicircle distribution is further characterized by the fact that all cumulants of order larger than 2 vanish. This can be understood from the study of the moments, which can be computed in terms of the Catalan numbers as:
\be
m_n  = \int_\R \lambda^n \rho_\mathrm{SC}(\lambda) \dd \lambda =
\left\{
    \begin{array}{lll}
        0   & \text{for $n$ odd;}  \\
       \sigma^n \ C_{n/2} & \text{for $n$ even.}
    \end{array}
\right.
\ee
With the interpretation of Catalan numbers as the number of non-crossing pair partitions, we have:
\be
m_n = \sigma^n \ C_{n/2} = \sum_{\mathcal{P}^{\mathrm{NC}(2)}_{n}} (\sigma^2)^{n/2}
= \sum_{\mathcal{P}^{\mathrm{NC}(2)}_{n}} \kappa_2^{n/2}
\ee
such that only one kind of term is non-zero in the sum relating $m_n$ to free cumulants \eqref{eq:1:FreeMomentsfromCumulants}, corresponding to the non-crossing partitions where all blocks have two elements. All other partitions have vanishing contributions such that $\kappa_n = 0$ for all $n \geqslant 3$. 

These vanishing free cumulants give rise to a linear $\RT$-transform which has an interesting consequence. If $\AAA$ and $\BB$ have semicircle distributions with variance $\sigma^2$ and they are free, then their sum verifies $ \RT_{\AAA + \BB } (\omega) = \RT_{\AAA }(\omega) +\RT_{\BB}  ( \omega) = 2 \sigma^2 \omega $
such that $\AAA + \BB $ has semicircle distribution of variance $ 2 \sigma^2$. Up to a rescaling, the semicircle distribution is stable under free addition.

The semicircle distribution thus plays, in the free world, the role of the Gaussian distribution which is stable under the addition of independent scalar variables. In the same way, the Gaussian distribution is characterized by vanishing standard cumulants $k_n=0$ for $n\geqslant 3$, as a corollary of Wick's theorem. This plays a crucial role in the CLT, which can thus be extended to the non-commutative case \cite{Pata1996}, along with the law of large numbers \cite{Bercovici1996}.

The correspondence between the properties of the Gaussian distribution in standard theory and those of the semicircle distribution in the free setting is an example of the \emph{Bercovici-Pata bijection}, which associates to any standard distribution of finite cumulants $(k_n)_{n\geqslant 1}$ a free random variable with the same sequence of free cumulants $\kappa_n = k_n$ \cite{Bercovici1995}.

\paragraph{Free Law of Large Numbers}
The sum of $L$ free identically distributed random matrices $(\MM_k)_{1 \leqslant k \leqslant L}$, of mean $\phi(\MM_k)=m$ and finite other moments, normalized by $L$ converges to a constant with the same mean:
\be 
\frac{1}{L} \sum_{k=1}^L \MM_k   \quad 
\xrightarrow[L \to \infty]{} \quad
m \mathbf{I}
\ee
as proven by noticing that the additive free cumulants converge to
\be 
\kappa_n(  \sum_{k=1}^L \frac{\MM_k}{L} ) = \frac{L}{L^n} \kappa_n   \longrightarrow
\left\{
    \begin{array}{ll}
        \kappa_1 = m & \mbox{if } n=1 \\
        0  & \mbox{if }  n \geq 2
    \end{array}
\right.
\ee
such that $\frac{1}{L} \sum_{k=1}^L \MM_k  $ converges to a variable with all free cumulants equal to zero except the first one, which is a constant matrix $ m \mathbf{I}$.

\paragraph{Free Central Limit Theorem}
The sum of $L$ free identically distributed centered random matrices $(\MM_k)_{1 \leqslant k \leqslant L}$, of variance $\phi(\MM_k^2)=\kappa_2$ and finite other moments, normalized by $\sqrt{L}$ converges to a semicircle-distributed random matrix $\MM_\infty$ with the same variance $\kappa_2$:
\be
\frac{1}{\sqrt{L}} \sum_{k=1}^L \MM_k  
\quad
\xrightarrow[L \to \infty]{} 
\quad 
\MM_\infty
\ \sim \text{ Semicircle}(\kappa_2)
\ee
as proven by noticing that the additive free cumulants converge to
\be
\kappa_n(  \sum_{k=1}^L \frac{\MM_k}{\sqrt{L}} ) = \frac{L}{L^{n/2}} \kappa_n   \longrightarrow
\left\{
    \begin{array}{ll}
        0 & \mbox{if } n=1 \\
        \kappa_2   & \mbox{if } n=2 \\
        0  & \mbox{if }  n \geq 3
    \end{array}
\right.
\ee
such that $\frac{1}{\sqrt{L}} \sum_{k=1}^L \MM_k  $ converges to a variable with second cumulant $\kappa_2$ and all others equal to zero, which follows a semicircle distribution with parameter $\kappa_2$.
 
It should be noted that the condition in these two theorems is that all variables $\MM_k$ are collectively free, not just pairwise free. The same warning is in order in the commutative setting.

\subsection{Product of free random matrices}
\label{subsec:1.3.3}

As we have seen in the previous section, it is possible to retrieve the spectrum of the free sum $\AAA + \BB$ from the spectra of $\AAA$ and $\BB$. We now turn to the same question for the free product: it is possible to retrieve the spectrum of $\AAA . \BB$ from the spectra of the free random matrices $\AAA$ and $\BB$ ? In the context of self-adjoint random matrices, the free product must be defined as $\sqrt{\AAA} \; \BB \, \sqrt{\AAA}$, in order to ensure that the ensuing spectrum is real. As for the free addition, the free product was first studied by Voiculescu \cite{Voiculescu1987}.

We introduce additional entire functions in order to study the spectrum of the free product. The $\mathcal{T}$-transform is defined from the Stieltjes transform as:
\be
\mathcal{T}(z) = z \, \g(z) - 1 =\int_\R \frac{u \, \rho(u)}{z-u} \mathrm{d} u   = \sum\limits_{n \geqslant 1}  \frac{m_n }{z^{n} } \; .
\ee 
Denoting $\mathcal{T}^{-1}$ the functional inverse of the $\mathcal{T}$-transform, the $\ST$-transform is defined for $\omega \in \C$ as:
\be
\ST(\omega) = \frac{\omega +1 }{\omega \mathcal{T}^{-1}(\omega) } \; .
\ee 
The $\ST$-transform is formally related to the $\RT$-transform as:
\be
\label{eq:1:RelationSR}
\ST(\omega) = \frac{1}{\RT( \omega \ST(\omega) )  }\quad \quad \quad
\RT(\omega) = \frac{1}{\ST( \omega \RT(\omega) ) }
\ee 
and, from the series expansion of the $\mathcal{T}$-transform, it can be shown that the following expansion, in terms of the moments and free cumulants of the distribution, holds for small $\omega$:
\bea
\ST (\omega) &=&  \frac{1}{m_{1}}+\frac{\omega}{m_{1}^{3}}\left(m_{1}^{2}-m_{2}\right)+\frac{\omega^{2}}{m_{1}^{5}}\left(2 m_{2}^{2}-m_{2} m_{1}^{2}-m_{3} m_{1}\right)+\mathcal{O}\left(\omega^{3}\right)  \\
&=& \frac{1}{\kappa_{1}}-\frac{\kappa_{2}}{\kappa_{1}^{3}} \omega+\frac{2 \kappa_{2}^{2}-\kappa_{1} \kappa_{3}}{\kappa_{1}^{5}} \omega^{2}+\mathcal{O}\left(\omega^{3}\right) \; .
\eea
This shows that the $\ST$-transform of a centered distribution, with $m_1=\kappa_1=0$, is ill-defined. It can however be defined by introducing a translation to break the symmetry, see Table \ref{tab:RMTtransforms} for the semicircle distribution.

The interest of this transform appears in the following theorem. For two free self-adjoint matrices $\AAA$ and $\BB$, the $\ST$-transform of the self-adjoint product $\sqrt{\AAA} \; \BB \, \sqrt{\AAA}$ verifies the following factorization property:
\begin{empheq}[box=\setlength{\fboxsep}{8pt}\fbox]{equation}
\label{eq:1:FreeMultiplicationSTransform}
\ST_{\sqrt{\AAA} \BB \sqrt{\AAA}} (\omega) = \ST_\AAA(\omega) \times  \ST_\BB(\omega) 
\end{empheq}

For proofs of this theorem and presentations of the combinatorial structures related to the free product, we point to \cite{Potters2020,Speicher1997,Nica2006}. Note that the scalar equivalent of the $\ST$-transform is the Mellin transform $ \mathcal{M}_X(s) = \E[ X^{s-1}]$ of a random variable $X$, which is multiplicative for the product of independent variables.  \\

To conclude this introduction to free probability, we give in Table \ref{tab:RMTtransforms} the Stieltjes, $\RT$ and $\ST$ transforms for the main RMT ensembles' asymptotic distributions \cite{Bun2017}, i.e. the semicircle, Mar\v{c}enko-Pastur and inverse Mar\v{c}enko-Pastur distributions defined respectively in Eqs.~\eqref{eq:1:WignerSemicircleDefinition}, \eqref{eq:1:MarcenkoPasturDefinition} and \eqref{eq:1:IMPDistributionRewritten}. For the sake of well-posedness, the $\ST$-transform of the semicircle distribution is in fact that of the $\mu$-translated distribution, with $\mu>0$. For $\mu <0$, the branch should be chosen with minus sign in front of the square-root to ensure regular behaviour as $\omega \to 0$.

\vspace{1cm}
  
\begin{table}[ht!] 
\hspace*{-1cm}
\begin{tabular}{| c | c | c  | c  | }
  \hline
Distribution &    $\g(z)$     &   $\RT(\omega)$ &  \hfil  $\ST(\omega)$  \\ \hhline{|=|=|=|=|}
 &&& \\[15pt] 
Semicircle &  $\displaystyle \frac{z-\sqrt{z^2-4\sigma^2 }}{2\sigma^2}$ &  $\displaystyle \sigma^2 \omega $ & $\displaystyle  \frac{-\mu + \sqrt{\mu^2 + 4 \sigma^2 \omega} }{ 2 \sigma^2 \omega}  $ \\[35pt]
 \shortstack{Mar\v{c}enko \\ Pastur }  &  $\displaystyle \frac{z+\sigma^2(q-1)-\sqrt{  z - \mu_-^{MP} } \sqrt{ z-\mu_+^{MP} }}{2 q \sigma^2 \, z} $  & $\displaystyle   \frac{1}{\frac{1}{\sigma^2} - q  \omega}$ & $\displaystyle \frac{1}{\sigma^2} \frac{1}{1+q \omega}$  \\[35pt]
 \shortstack{Inverse \\  Mar\v{c}enko \\ Pastur } & $\displaystyle \frac{z(\kappa+1)-\frac{\kappa}{\sigma^2}-\kappa \sqrt{z- \mu_-^{IMP} } \sqrt{z-\mu_+^{IMP} } }{z^{2}}$ & $ \displaystyle  \frac{\kappa-\sqrt{\kappa\left(\kappa-\frac{2}{\sigma^2} \omega\right)}}{\omega} $  & $ \displaystyle \sigma^2 \left(  1 - \frac{\omega}{2 \kappa} \right) $ \\[35pt] 
  \hline
\end{tabular} 
\caption{Stieltjes, $\RT$ and $\ST$ transforms of the main asymptotic distributions.}
\label{tab:RMTtransforms}
\end{table}   

\vspace{1cm}
This ends our presentation of random matrix theory, eigenvalue statistics on the global and local scales, and free probability tools related to the study of large random matrices and their asymptotic spectrum.

\thispagestyle{empty}
\chapter{Non-crossing walkers and vicious boundary problems}
\chaptermark{Non-crossing walkers and boundary problems}
\label{chap:2}
  
As presented in chapter \ref{chap:1}, the theory of random matrices has spawned from its applications in nuclear physics, where eigenvalue repulsion matched heavy nuclei energy level statistics, and has developed through the study of properties implied by this fundamental repulsive feature. This gives a direct motivation to study other systems displaying repulsive interactions and investigate the links they may share with random matrices. An example of such a system is the simple exclusion process (\textit{SEP}), exactly solvable in its symmetric and asymmetric versions \cite{Schutz2015,Golinelli2006}, where particles are free to perform a random walk on $\Z$ under the restriction that no two particles may occupy the same site. Remarkably, the current fluctuations of the totally asymmetric SEP is related to the Tracy-Widom distribution \cite{Johansson2000}. Another example of spatial exclusion is given by random domino tilings of the Aztec diamond domain, where the spatial configurations of dominos are highly constrained and where determinantal processes appear as in RMT \cite{Johansson2002}. Finally, other systems which display repulsion are models of interfaces, where a portion of space is divided into different regions separated by interfaces which should then remain distinct. An example of these is the multi-layer polynuclear growth model (\textit{PNG}) which describes multiple layer growth on a substrate in a super-saturated vapor, and is linked to Pfaffian processes and random matrices \cite{Prahofer2000,Ferrari2004}.

In this chapter, we turn to the simplest model of random one-dimensional interfaces: \emph{non-crossing walkers}. In this system of $N$ Brownian motions on the line, repulsion is provided by conditioning the paths not to cross each other. The resulting conditioned process gives a simple model for horizontal interfaces splitting the plane in $N+1$ layers, and it is also naturally a model of repulsive stochastic particles in one-dimension. As such, it is intimately related to the Dyson Brownian motion (\textit{DBM}), the diffusion of eigenvalues obtained in the dynamic extension of RMT to stochastic processes, where eigenvalues diffuse randomly while repelling each other. These two systems are presented in detail in this chapter.  

The focus of publication \pubref{publication:NonCrossingBrownianDBM} was to study the properties of these systems under a moving vicious boundary. Considering that the system is "killed" if a particle touches the boundary, the main properties of interest concern the decay of the survival probability $S(t)$, the probability that the system is not killed before time $t$. As will be illustrated in this chapter, the most interesting case is that of the critical square-root boundary $g(t) = W \sqrt{t}$, which grows at the same speed as the stochastic expansion of the walkers. It was found in this case that the survival probability decays as $S(t) \sim t^{-\beta (N,W)}$, where the exponent $\beta(N,W)$ was found to be the ground-state energy of a particular system of non-interacting fermions and studied in a variety of different scaling limits. Furthermore, the joint distribution of particles under the boundary at large time was also studied.

The first section will present the dynamic extension of RMT which gives rise to stochastic evolution of eigenvalues and derive the DBM stochastic differential equation as well as the connections of the model to fermion systems. We will then present the system of non-crossing walkers in the context of random interfaces, and detail its links with the DBM.
The second section \ref{sec:2.2} will present the results of publication \pubref{publication:NonCrossingBrownianDBM} and the resolution of the square-root boundary problem for non-crossing walkers and the DBM, after detailing the simpler one-particle case. Finally, section \ref{sec:2.3} will give an additional perspective into the connections between non-crossing walkers and RMT by constructing an interpolating $G \beta E$ measure through a partial non-crossing condition.

\section[Non-crossing walkers and dynamical random matrix theory]{Non-crossing walkers and the dynamical extension of random matrix theory} 
\sectionmark{Non-crossing walkers and dynamical RMT}
\label{sec:2.1}

In this section, we complete the presentation of RMT of chapter \ref{chap:1} by the discussion of its dynamical extension. Following the first steps made by Freeman Dyson \cite{Dyson1962b,Dyson1965}, the theory of random matrices can indeed be extended to stochastic processes where a matrix evolves dynamically in time. In this context, the most important construction is the Dyson Brownian motion, which we present thoroughly here, before presenting the system of non-crossing walkers and the links shared by these two systems of diffusing particles.

In the first subsection, we define the DBM and present its dynamical connections to fermions in the Calogero-Moser Hamiltonian. In subsection \ref{subsec:2.1.2}, we define the system of non-crossing walkers and detail its links with the DBM.

\subsection{The Dyson Brownian motion}
\label{subsec:2.1.1}

\subsubsection{Definition and properties}

Let us recall, as a preamble, the definition of the one-dimensional Brownian motion $(B_t)_{t \geqslant 0}$ as the unique stochastic process on $\R$ which verifies the following properties:
\begin{itemize}[itemsep=.3em,topsep=4pt] 
\item $B_0 = 0$ ; 
\item $B_t$ is almost-surely continuous ;
\item all increments are independent ;
\item and an increment $B_{t+T} - B_t$ is distributed as $\mathcal{N}(0,T)$.
\end{itemize}
The Brownian motion has become a central construction in mathematics and physics after its insightful introduction by Albert Einstein in the context of small particle suspensions in liquids \cite{Einstein1905,Renn2005}. The myriad of its applications \cite{Duplantier2005} include statistical physics and studies on non-equilibrium dynamics \cite{Kubo1986,Majumdar2006,Ferretti2021,Krajenbrink2019a}, biological physics and soft matter science \cite{Frey2005,Ronteix2021,Gerber2019}, solid state fracture analysis \cite{Bouchaud1993,Finel2018}, interstellar medium research and astrophysics \cite{Regaldo-SaintBlancard2020} and the surge in financial modelling of the second part of the last century \cite{Walter2013}. For a presentation of stochastic calculus, discussions of the Brownian motion and of its detailed properties, we point to \cite{Applebaum2009,Varadhan2007,Mansuy2008,Revuz1995}.

The Brownian motion is a diffusion process on the line which is distributed as a standard Gaussian random variable at time 1, i.e. $B_1 \sim \mathcal{N}(0,1)$. The Dyson Brownian motion is the stochastic process obtained by promoting all independent real Gaussian variables appearing in the Gaussian ensembles ($\beta=1,2$ or $4$) to standard Brownian motions. It is thus a first step in a dynamical extension of RMT, where random matrices are promoted to stochastic processes \cite{Erdos2017}.

\paragraph{Definition}
Let us define the Dyson Brownian motion $\HH_t^\beta$ in the three usual cases, and denote it the DBM-$\beta$. We introduce $\XX_t$ a real matrix with $N^2$ independent standard Brownian motions as its entries, and $\tilde{\XX}_t^{(k)}$ independent versions of $\XX_t$. With these notations, the DBM-$1$ is the real symmetric matrix process:
\emphbe
\HH_t^1 = \frac{\XX_t + \XX_t^T}{\sqrt{2}} \; ;
\label{eq:2:DefDBMGOE}
\end{empheq}
the DBM-2 is the complex Hermitian matrix process:
\emphbe
\HH_t^2 =  \frac{\XX_t + \XX_t^T}{2} +  \ii \, \frac{\tilde{\XX}_t - \tilde{\XX}_t^T}{2} \; ;
\label{eq:2:DefDBMGUE}
\end{empheq}
and the DBM-4 is the following self-adjoint quaternionic matrix process, denoting the canonical basis of quaternions $(1,i^{(1)},i^{(2)},i^{(3)})$:
\emphbe
\HH_t^4 =   \frac{\XX_t + \XX_t^T}{2\sqrt{2}} +  \sum_{k=1}^3  \ii^{{(k)}}  \frac{\tilde{\XX}^{(k)}_t - \tilde{\XX}^{(k),T}_t}{2 \sqrt{2}} \; .
\label{eq:2:DefDBMGSE}
\end{empheq}

In all three cases, the construction ensures that $\HH_1^\beta$, the evaluation of the process at time $t=1$, is distributed as a $G \beta E$ matrix with parameter $\sigma=1$, as can be seen from Eqs.~\eqref{eq:1:DefGOEWithX}, \eqref{eq:1:DefGUEWithX} and \eqref{eq:1:DefGSEWithX}. For general time $t$, $\HH_t^\beta$ is distributed as a $G \beta E$ matrix with parameter $\sigma= \sqrt{t}$. 

The DBM is formally a Brownian motion in the space of self-adjoint matrices: it has continuous paths and independent increments distributed according to the corresponding Gaussian ensemble, with variance given by the time increment. As in the scalar case where Donsker’s invariance principle implies the convergence of a random walk with i.i.d.~centered finite-variance steps to the Brownian motion, the DBM can be interpreted as the continuous limit of a large sum of independent infinitesimal matrix increments \cite{Potters2020}. In a more abstract scope, Brownian motions on general Lie groups have been widely studied in mathematics \cite{Ito1950,Chhaibi2013}.


\paragraph{Langevin equation for the eigenvalues}

By extension, the Dyson Brownian motion also refers to the stochastic process of the $N$ real eigenvalues $(\lambda_{i,t})_{1 \leqslant i \leqslant N}$ of $\HH^\beta_t$. In this section, we derive the stochastic differential equation (\textit{SDE}), or Langevin equation, obeyed by this $\R^N$-valued process, for $\beta = 1, 2$ or $4$.

\subparagraph{Perturbation theory}
Let us consider a fixed time $t$. From the addition properties of the Brownian motion, the evolution of the DBM between $t$ and infinitesimally larger time $t+\dd t$ is distributed as
\be 
\DelBB_{\dd t} =  \HH_{t+\dd t}^\beta - \HH_{t}^\beta  \ \  \sim \ \  \sqrt{\dd t}  \  \tilde\HH_{1}^\beta 
\label{eq:2:PerturbationTheoryBeginning}
\ee
with $\tilde{\HH}$ an independent version of the DBM. The infinitesimal matrix $\DelBB_{\dd t} $ is thus distributed as $\sqrt{\dd t}$ times a unit-variance $G\beta E$ matrix $\DelBB$, and we write $\DelBB_{\dd t} = \sqrt{\dd t} \, \DelBB$.

In order to deduce the eigenvalue SDE, we are interested in the infinitesimal evolution of the eigenvalues of $\HH^\beta_t$ under the addition of this infinitesimal matrix increment  $\DelBB_{\dd t}$. This amounts to develop a perturbation theory for eigenvalues, which is well-know in the context of quantum mechanics where it is used to determine the evolution of energy levels under slight modification of the Hamiltonian \cite{Basdevant2002}. In reference, we develop the argument with notations inspired from this setting.

Let us denote $( \ket{v_{i,t}} )_{1 \leqslant i \leqslant N}$ the set of orthonormal eigenvectors of the matrix $\HH_t^\beta$, corresponding respectively to eigenvalue $\lambda_{i,t}$. The perturbed eigenvalues and eigenvectors can be written in an expansion on the infinitesimal factor $\epsilon = \sqrt{\dd t}$:
\be 
\lambda_{i,t+\dd t} = \lambda_{i,t} +  \epsilon \lambda_i^{(1)} + \epsilon^2  \lambda_i^{(2)} \cdots 
\ee 
and
\be 
\ket{v_{i,t+\dd t}} = \ket{v_{i,t}} +  \epsilon \ket{v_{i}}^{(1)} + \epsilon^2  \ket{v_{i}}^{(2)} \cdots  
\ee 
with the orthonormality constraint $ \bra{v_{i,t+\dd t}} \ket{v_{j,t+\dd t}}  = \delta_{i,j}$. The expansion coefficients $\lambda_i^{(k)}$ and $\ket{v_{i}}^{(k)}$ are found recursively on $k$ by enforcing the eigenvalue equation
\be
\HH_{t+\dd t}^\beta \ket{v_{i,t+\dd t}} =   \left( \HH_{t}^\beta + \epsilon \, \DelBB  \right) \ket{v_{i,t+\dd t}} = \lambda_{i,t+\dd t} \ket{v_{i,t+\dd t}} 
\ee 
such that the perturbed eigenvalues and eigenvectors are given at first orders by \cite{Basdevant2002}:
\be 
\lambda_{i,t+\dd t} = \lambda_{i,t} +  \epsilon \bra{v_{i,t}} \DelBB \ket{v_{i,t}} + \epsilon^2 \sum_{\substack{1 \leqslant j \leqslant N \\j\neq i}} \frac{\abs{ \bra{v_{j,t}} \DelBB \ket{v_{i,t}}   }^2}{\lambda_{i,t}-\lambda_{j,t}} + \bigO(\epsilon^3)
\label{eq:2:EigenvaluePerturbationTheory}
\ee
and 
\be 
\ket{v_{i,t+\dd t}} = \ket{v_{i,t}} +  \epsilon   \sum_{\substack{1 \leqslant j \leqslant N \\j\neq i}} \frac{  \bra{v_{j,t}} \DelBB \ket{v_{i,t}}  }{\lambda_{i,t} - \lambda_{j,t}}  \ket{v_{j,t}} + \bigO(\epsilon^2) \; .
\ee

From the distribution of $\DelBB$ as a unit-variance rotation-invariant Gaussian matrix, we have on one hand that $\bra{v_{i,t}} \DelBB \ket{v_{i,t}}\sim \mathcal{N}(0, \frac{2}{\beta})$ and on the other that $\abs{ \bra{v_{i,t}} \DelBB \ket{v_{j,t}} }^2 -1$ has mean zero, such that \myeqref{eq:2:EigenvaluePerturbationTheory} yields, in the $\epsilon \to 0$ limit:
\be 
\lambda_{i,t+\dd t} \simeq \lambda_{i,t} +  \epsilon \sqrt{\frac{2}{\beta}} X_i + \epsilon^2 \sum_{\substack{1 \leqslant j \leqslant N \\j\neq i}} \frac{1}{\lambda_{i,t}-\lambda_{j,t}}  
\ee
where $X_i \sim \mathcal{N}(0,1)$. Introducing $N$ independent standard Brownian motions $(B_{i,t})_{1 \leqslant i \leqslant N}$ and replacing $\epsilon = \sqrt{dt}$, this yields the $N$ stochastic differential equations for the eigenvalues of the Dyson Brownian motion as:
\emphbe 
\dd \lambda_{i,t} =  \sqrt{\frac{2}{\beta}} \dd B_{i,t} + \sum_{\substack{1 \leqslant j \leqslant N \\j\neq i}} \frac{1}{\lambda_{i,t}-\lambda_{j,t}}  \dd t   
\label{eq:2:DBMEigenvalueSDE}
\end{empheq}
For details on the convergence of the perturbation result to this SDE, in particular for the rigorous treatment of the $(\abs{ \DelBB_{ij}}^2 -1)$ term, see \cite{Tao2012}. This SDE is also often found in the literature in the time-rescaled form, with $t \to \frac{\beta}{2} t$, as
\be 
\dd \lambda_{i,t} =    \dd B_{i,t} +  \frac{\beta}{2} \sum_{j\neq i} \frac{1}{\lambda_{i,t}-\lambda_{j,t}}  \dd t \; .
\label{eq:2:DBMEigenvalueSDETimeRescaled}
\ee

\subparagraph{Itô's formula}

Another derivation of the SDE \eqref{eq:2:DBMEigenvalueSDE} can be obtained by a formal application of Itô's formula \cite{Applebaum2009}. Indeed, the DBM matrix processes in Eqs.~\eqref{eq:2:DefDBMGOE}, \eqref{eq:2:DefDBMGUE} and \eqref{eq:2:DefDBMGSE} are defined from the joint process of the $\beta \, N^2$ independent Brownian motions in the $\XX$ matrices. The eigenvalues $\lambda_{i,t}$ are a function of the entries of $\HH_t^\beta$, and thus of these independent Brownian motions, such that Itô's formula can be applied to obtain their stochastic evolution, which recovers \eqref{eq:2:DBMEigenvalueSDE}. We point to \cite{Potters2020} for the detailed computation in the $\beta=1$ case.

\subparagraph{Properties}

The evolution equation of eigenvalues given by \myeqref{eq:2:DBMEigenvalueSDE} combines two ingredients. On the one hand, the eigenvalues diffuse according to independent Brownian motions, following the stochastic diffusion of the matrix entries. On the other hand, they repel each other through the drift term, as expected from the statistics of eigenvalues in static RMT presented in the first chapter. 

Note that this drift term is similar to the one of the Bessel process $(R_t)_{t \geqslant 0}$ which describes the norm of the $D$-dimensional Brownian motion $R_t = \norm{\vec{B}_t}_D$ and obeys the following SDE \cite{Lawler2019}
\be 
\dd R_t = \dd B_t + \frac{D-1}{2} \frac{1}{R_t} \dd t \; .
\label{eq:2:BesselSDE}
\ee
By the recurrence properties of the Brownian motion, $R_t$ is repelled away from the origin and stays almost-surely strictly positive as soon as $D \geqslant 2$. 

The process $(\lambda_{i,t})_{t \geqslant 0}$ can be seen as a generalization on $\R^N$ of the Bessel process, restricted to stay in the Weyl chamber $\Weyl_N$ defined as
\be 
\Weyl_N= \{ \vec{\lambda} \in \R^N  \mid \lambda_1 \geqslant \lambda_2 \geqslant \cdots \geqslant \lambda_N \}  \; .
\label{eq:2:DefWeylChamber}
\ee 
We also introduce the notation $\Weyl_N^*$ for the interior of the Weyl chamber, where the inequalities are strict. For both the DBM and the Bessel processes, the repulsive drift term is the inverse-distance to the forbidden boundary. From the properties of the Bessel process, the boundary is almost-surely never touched when the prefactor of this drift term is at least $\frac{1}{2}$, such that the prefactor $\frac{\beta}{2}$ in the time-rescaled DBM SDE \eqref{eq:2:DBMEigenvalueSDETimeRescaled} ensures that eigenvalues of the DBM never meet each other in the three cases $\beta=1,2,4$ \cite{Tao2012,Grabiner1999}. 

For a more practical point of view, consider a situation where two diffusing DBM eigenvalues $\lambda_i > \lambda_j$ get very close to each other at some time $t$. The difference of their stochastic evolutions \eqref{eq:2:DBMEigenvalueSDETimeRescaled} yields the SDE for the difference $\lambda_i - \lambda_j$ as
$
\dd(\lambda_i - \lambda_j) \simeq \sqrt{2} \dd B_t + \frac{\beta}{\lambda_i - \lambda_j} \dd t
$
or after a time-rescaling $t \to t/2$:
\be 
\dd(\lambda_i - \lambda_j) \simeq  \dd B_t + \frac{\beta}{2} \frac{1}{\lambda_i - \lambda_j} \dd t \; .
\label{eq:2:DBMSDECloseToACollision}
\ee
After this time-rescaling, the difference of two close-by DBM eigenvalues is distributed as the $(\beta+1)$-dimensional Bessel process, such that indeed two eigenvalues of the DBM almost-surely never collide as soon as $\beta \geqslant1$. For more details on the DBM as a multivariate extension of the Bessel process, see \cite{Katori2016} where connections are also presented between the Schramm-Loewner Evolution (\textit{SLE}) \cite{Cardy2005,Bauer2006,Baverez2020} and the Bessel flow.

The dynamical extension of RMT given by the Dyson Brownian motion can also be seen, at the eigenvalue level, as a dynamical extension of the Coulomb gas framework of \myeqref{eq:1:CoulombGas}, as particles experience stochastic diffusion in addition to repulsion and generate configuration samples of the Coulomb gas. We recover in \myeqref{eq:2:DBMEigenvalueSDE} the inverse-temperature interpretation of $\beta$ since the coefficient of $\dd B_t$ is expected, in the Coulomb gas with stochastic diffusion, to be proportional to the square-root temperature by the Stokes-Einstein law \cite{Cruickshank1924}.

\paragraph{Stieltjes transform evolution}

As seen in the first chapter, the Stieltjes transform $\g$ defined in \myeqref{eq:1:DefStieltjes} is a useful tool to characterize the spectrum of a random matrix, especially in the large-$N$ setting, in parallel with other transforms defined from it as seen in section \ref{sec:1.3}. In this paragraph, we derive the evolution of $\g$ under the DBM matrix diffusion. 

More precisely, we define the time-dependent Stieltjes transform of the rescaled DBM process $\tilde{\HH}_t^\beta = \frac{1}{\sqrt{N}} \HH_t^\beta $ as
\be 
 \mathfrak{g}(z,t) = \frac{1}{N} \sum_{i=1}^N \frac{1}{z- \lambda_{i,t}/\sqrt{N}}  
\ee
where the $\sqrt{N}$ scaling factor is the one that appeared naturally in the Stieltjes transform study of the Gaussian ensembles in section \ref{subsec:1.2.1}. 

\subparagraph{Itô's formula}
The Stieltjes transform $ \mathfrak{g}(z,t) $ is a function of the time-evolving eigenvalues $\lambda_{i,t}$ and verifies:
\be 
\frac{\partial \g}{\partial \lambda_{i,t}} = \frac{1}{N\sqrt{N}} \frac{1}{(z- \lambda_{i,t}/\sqrt{N})^2} 
\ee
and
\be 
\frac{\partial^2 \g}{\partial \lambda_{i,t}^2} = \frac{2}{N^2} \frac{1}{(z- \lambda_{i,t}/\sqrt{N})^3} \; . 
\ee
A direct application of Itô's formula then gives the evolution of $\g(z,t)$ as:
\bea
\dd \g(z,t) &=& \sum_{i=1}^N  \frac{\partial \g}{\partial \lambda_{i,t}} \dd \lambda_{i,t} + \frac{1}{2} \sum_{i=1}^N \frac{\partial^2 \g}{\partial \lambda_{i,t}^2}  \frac{2}{\beta} \dd t \\
 &=& \frac{1}{N}\sqrt{\frac{2}{\beta N}} \sum_i \frac{dB_{i}}{(z-\frac{\lambda_i}{\sqrt{N}})^2}  + \frac{\dd t}{N\sqrt{N}} \sum_{i\neq j} \frac{1}{(\lambda_i - \lambda_j) (z-\frac{\lambda_i}{\sqrt{N}})^2} + \frac{2\dd t}{\beta N^2 } \sum_{i} \frac{1}{(z-\frac{\lambda_i}{\sqrt{N}})^3} \nonumber
\eea
where we have dropped time indices for simplicity. A usual computation is then to rewrite the second term as:
\bea
 \frac{\dd t}{N\sqrt{N}} \sum_{i\neq j} \frac{1}{(\lambda_i - \lambda_j) (z-\frac{\lambda_i}{\sqrt{N}})^2}  &=& \frac{1}{2} \frac{\dd t}{N\sqrt{N}} \sum_{i\neq j} \bigg[ \frac{1}{(\lambda_i - \lambda_j) (z-\frac{\lambda_i}{\sqrt{N}})^2} + \frac{1}{(\lambda_j - \lambda_i) (z-\frac{\lambda_j}{\sqrt{N}})^2} \bigg] \nonumber \\
&=&  \frac{\dd t}{2N^2} \sum_{i\neq j}  \frac{  2z -  \frac{\lambda_i + \lambda_j}{\sqrt{N}} }{(z-\frac{\lambda_i}{\sqrt{N}})^2(z-\frac{\lambda_j}{\sqrt{N}})^2} \\
&=& \frac{\dd t}{N^2} \sum_{i \neq j} \frac{1}{(z-\frac{\lambda_i}{\sqrt{N}})^2(z-\frac{\lambda_j}{\sqrt{N}})} \\
&=&  \frac{\dd t}{N^2} \sum_{i,j}  \frac{1}{(z-\frac{\lambda_i}{\sqrt{N}})^2(z-\frac{\lambda_j}{\sqrt{N}})} - 
\frac{\dd t}{N^2} \sum_{i} \frac{1}{(z-\frac{\lambda_i}{\sqrt{N}})^3} \;. 
\eea
The first term in this sum can be recognized to be equal to $-\g \frac{\partial \g}{\partial z} \dd t$ and the second one to $\frac{-1}{2N}\frac{\partial^2 \g}{\partial z^2}$, such that the SDE satisfied by the Stieltjes transform combines as:
\be 
\dd \g(z,t) = \frac{1}{N}\sqrt{\frac{2}{\beta N}} \sum_i \frac{dB_{i}}{(z-\frac{\lambda_i}{\sqrt{N}})^2} - \g \frac{\partial \g}{\partial z}  \dd t + (\frac{2}{\beta}-1)\frac{\dd t}{2N}  \frac{\partial^2 \g}{\partial z^2} \; .
\label{eq:2:SDEStieltjesFiniteN}
\ee

At large $N$, the self-averaging Stieltjes transform evolves according to the differential equation obtained by taking the expectation of this SDE. Dropping the sub-dominant $ \frac{\partial^2 \g}{\partial z^2}$ term, the partial differential equation (\textit{PDE}) obeyed by the large-$N$ Stieltjes transform is then the inviscid complex Burger's equation \cite{Polyanin2003,Menon2017}:
\emphbe 
\frac{\partial \g(z,t)}{\partial t} = - \g(z,t) \frac{\partial \g(z,t)}{\partial z} 
\label{eq:2:BurgersEquationDBM}
\end{empheq}
This PDE is renowned in its real form, because it occurs as a conservation equation in many fields such as fluid mechanics, nonlinear acoustics, gas dynamics and traffic flow \cite{Bonkile2018,Menon2012}. It can exhibit shocks in this inviscid form \cite{Sinai1992}, if it is not dampened by the dissipative diffusion term $ \frac{\partial^2 \g}{\partial z^2}$ which was dropped in the large-$N$ limit step of our computation. See \cite{Said2021,Said2021a} for recent results on the dispersive Burgers equation. 

The appearance of this equation in random matrix theory was first obtained in \cite{Blaizot2010}, suggesting that shocks in hydrodynamical conservation equations can explain some properties of random matrices. This link was further investigated in \cite{Allez2012,Blaizot2015,Krajenbrink2021,Grela2021}. 

\subparagraph{Solution by the method of characteristics}

This equation can be solved by the method of complex characteristics, which involves an ingenious change of variables and allows to reduce the non-linear PDE to an ordinary differential equation. We point to \cite{Menon2017} for the specificities of the complex version of this technique. Denoting $\g_0(z) = \g(z,0)$ the initial condition of the PDE, the solution is found as the solution of \cite{Polyanin2003,Menon2017,Potters2020}:
\be 
\g(z,t) = \g_0 \bigg(z- t \, \g(z,t) \bigg)  \; .
\label{eq:2:SolutionBurgersCharacteristics}
\ee 

With the initial value of our Dyson Brownian motion being the zero matrix, the initial condition is $\g_0(t) = \frac{1}{z}$ such that the solution for $\g$ is found as
\be 
\g(z,t) = \frac{z - \sqrt{z^2-4t}}{2t} 
\label{eq:2:SolutionSemiCircle}
\ee
where the branch is chosen such that the boundary condition $\g(z,t)  \underset{\abs{z} \to \infty}{\scalebox{1.5}[1]{$\sim$}} \frac{1}{z}$ is verified. As expected from the fixed-time distribution of the DBM as a Gaussian matrix, this large-$N$ result is the Stieltjes transform associated to Wigner's semicircle in \myeqref{eq:1:SelfConsistentG}, with $\sigma = \sqrt{t}$ .

\subsubsection{From Fokker-Planck to Schrödinger}
\label{subsubsec:2.1.1.b}

As a stochastic process on $\R^N$, the DBM describes the diffusion of $N$ repulsive particles on the line. An interesting property of such a stochastic system, as shown for a single-particle system in the next paragraph, is that it can be mapped to a quantum system. In the second paragraph, we will apply this general framework to the Dyson Brownian motion and detail its well-known mapping with the Calogero-Moser system. This property of the DBM is thus an opportunity for us to describe in this section the dynamical mapping that can hold between the Fokker-Planck equation of a stochastic system to the Schrödinger equation of a quantum system.

\paragraph{Single particle}
Let us consider a single-particle stochastic system in a potential $V$, such that the position of the particle $x_t$ evolves according to the SDE:
\be 
\dd x_t  = \sqrt{2 D} \, \dd B_t -  V'(x_t)  \, \dd t \; .
\label{eq:2:SDEMappingQuantum}
\ee
Here the coefficient of the Brownian term is expressed in terms of a constant diffusion coefficient $D$.

\subparagraph{Evolution mapping}

The associated Fokker-Planck (\textit{FP}) equation driving the evolution of the position PDF $P(x,t)$  \cite{Risken1996,Applebaum2009,Varadhan2007} is:
\be 
\frac{ \partial P(x,t)}{\partial t} =  \frac{ \partial^2 }{\partial x^2} \bigg( D P(x,t)  \bigg)  + \frac{\partial}{\partial x}\bigg(V'(x) P(x,t) \bigg) = - \hat{H}_{FP} \,  P(x,t) 
\ee 
where this equation defines the Fokker-Planck operator $\hat{H}_{FP}$. This evolution equation characterizes in particular the steady-state solution $P_\infty$, observed for large times, as the solution of 
\be 
0=  \frac{ \partial }{\partial x} \bigg( D \frac{\partial P_\infty(x)}{\partial x}   + V'(x) P_\infty(x) \bigg) \; .
\ee
It can be shown in this setting that the solution has zero flux such that
$
0 = D \frac{\partial P_\infty(x)}{\partial x}   + V'(x) P_\infty(x)  
$
and the solution is found as
\be 
P_\infty(x) =   A \ e^{-\frac{V(x)}{D}} 
\label{eq:2:SingleParticleSteadyState}
\ee
where $A$ is a normalization constant. Introducing the auxiliary function
\be
\label{eq:2:IntroducingPhi}
\phi(x,t) = \frac{P(x,t)}{\sqrt{P_\infty(x)}} \; ,
\ee
the evolution equation of $\phi$ is given by the following PDE
\be 
\frac{ \partial \phi(x,t)}{\partial t} =  D \frac{\partial^2  \phi(x,t)}{\partial x^2} -   \left(\frac{{V'(x)}^2}{4D} -  \frac{V''(x)}{2}  \right)  \phi(x,t)  \; .
\ee
By a change of perspective, the initial diffusion dynamics are now characterized by the following Schrödinger equation, in imaginary time, describing the evolution of a quantum wavefunction $\phi$ as
\be
\frac{\partial \phi}{\partial t} = -  \hat{H} \phi  
\label{eq:2:SchrodingerMapping}
\ee
with the Hamiltonian given by
\be 
\hat{H} = - D \frac{\partial^2}{\partial x^2}+  V_\mathrm{f}(x)  \quad \text{where} \quad V_\mathrm{f}(x) =\left(  \frac{{V'(x)}^2}{4D} - \frac{V''(x)}{2} \right)  \; .
\label{eq:2:HamiltonianQuantumMapping}
\ee 
Formally, the previous computations show that the Fokker-Planck operator and the quantum Hamiltonian are related as:
\emphbe
\hat{H}_{FP} = {P_\infty}^{\frac{1}{2}} \ \hat{H} \  {P_\infty}^{-\frac{1}{2}} 
\label{eq:2:OperatorFormalMapping}
\end{empheq}
Notice the formal similarity of this mapping with Doob's $h$-transform of probability theory \cite{Doob1949}.

\subparagraph{Propagator mapping}

The mapping of evolution operators can be used to express the stochastic transition probabilities in terms of the quantum propagator
\be 
G(x,t \mid x_0 , 0) = \bra{x} e^{-t \hat{H}} \ket{x_0}
\label{eq:2:DefQuantumPropagator}
\ee
which evolves according to $\partial_t G = - \hat{H}G$ and verifies the initial condition $G(x,0 \mid x_0 , 0)  = \delta(x-x_0)$. Indeed, let us introduce in parallel the stochastic propagator defined as the probability density of the particle's position, conditioned on a known initial position $x_0$ at time $0$, and denoted $P(x, t \mid x_0 , 0 )$. This stochastic propagator evolves according to $\partial_t P= - \hat{H}_{FP} P$ and also verifies $P(x, 0 \mid x_0 , 0 ) = \delta(x-x_0)$. By the connection, given in \myeqref{eq:2:IntroducingPhi}, between these two solutions of their respective evolution equation, the propagators are proportional up to a factor $\sqrt{P_\infty(x)}$ with a constant fixed by the initial condition such that:
\be 
P(x, t \mid x_0 , 0 ) = \sqrt{\frac{P_\infty(x)}{P_\infty(x_0)}} \ G(x,t \mid x_0 , 0)  \; .
\label{eq:2:PropagatorMappingSingleParticle}
\ee 
This relation can also be seen formally by injecting the operator relation \myeqref{eq:2:OperatorFormalMapping} into the propagator definition \myeqref{eq:2:DefQuantumPropagator}.

As a consequence, all properties of the single-particle stochastic system defined by SDE \eqref{eq:2:SDEMappingQuantum} can be obtained from solving the quantum problem with Hamiltonian $\hat{H}$ in \myeqref{eq:2:HamiltonianQuantumMapping}. This fact will prove useful in the rest of this chapter, as this mapping was a key element in publication \pubref{publication:NonCrossingBrownianDBM}.
More generally, these computations hint that stochastic and quantum systems share deep links. For a presentation of these links far beyond this simple mapping and with a focus on discrete-space Markov chains, see \cite{Baez2012,Guerra1983}.

\paragraph{Calogero-Moser: application to the Dyson Brownian motion}

The link between the diffusion of a stochastic particle in a potential $V$ and the quantum mechanics of a single-particle quantum system in a related potential $V_\mathrm{f}$ can be extended to many-body systems as the Dyson Brownian motion, which is the focus of the current section. In this paragraph, we apply the mapping presented above to this $N$-particle system.

Let us first note that the trapping potential, which is a key element in the mapping as it ensures the existence of a steady-state, is not present in the DBM SDE \myeqref{eq:2:DBMEigenvalueSDE}. As a consequence, the particles expand indefinitely, as illustrated in the large-$N$ limit in \myeqref{eq:2:SolutionSemiCircle}. Adding a potential $V$ can be done by modifying the set of SDEs to
\be 
\dd \lambda_{i,t} =  \sqrt{\frac{2}{\beta}} \dd B_{i,t} - V'(\lambda_{i,t}) \dd t  +   \sum_{\substack{1 \leqslant j \leqslant N \\j\neq i}} \frac{1}{\lambda_{i,t}-\lambda_{j,t}}  \dd t  \; .
\label{eq:2:DefDBMWithPotential}
\ee
The steady-state distribution of this stochastic $N$-particle system, which can be derived similarly as in the single-particle case, is the probability measure of the $\beta$-ensemble with potential $V$ defined in \myeqref{eq:1:JPDFGeneralBetaEnsemble}:
\be 
P_\infty(\vec{\lambda}) = C_{\beta,V}^N \ \abs{\Delta(\vec{\lambda})}^\beta  \ e^{- \beta \sum_{i=1}^N V(\lambda_i)} \; .
\ee 
As an illustration, the one-particle case, where the Vandermonde term disappears, matches with the steady state solution of \myeqref{eq:2:SingleParticleSteadyState}, after injecting $D \to \frac{1}{\beta}$.

\subparagraph{Evolution mapping}

The multivariate Fokker-Planck equation associated with this SDE sequence is
\be 
\frac{\partial P(\vec{\lambda},t)}{\partial t} = \sum_{i=1}^N \frac{\partial }{\partial \lambda_i}
\left( 
 \frac{1}{\beta} \frac{\partial P(\vec{\lambda},t)}{\partial \lambda_i} - \mathcal{F}_i (\vec{\lambda}) \, P(\vec{\lambda},t) 
\right)
= - \Ham_{FP}  \, P(\vec{\lambda},t)
\ee
where $\Ham_{FP} $ is the associated Fokker-Planck operator, and where the force acting on the $i$\textsuperscript{th} particle is
\be 
\mathcal{F}_i(\vec{\lambda}) =   \sum_{\substack{1 \leqslant j \leqslant N \\j\neq i}} \frac{1}{\lambda_{i}-\lambda_{j}}  - V'(\lambda_i) \; .
\ee 
As in the previous paragraph, we introduce
\be 
\phi(\vec{\lambda},t) = \frac{P(\vec{\lambda},t)}{ \sqrt{P_\infty(\vec{\lambda})}} \; .
\ee 
Note that the derivative of the square-root steady-state distribution with respect to an eigenvalue is
\be 
\partial_{\lambda_i} \sqrt{P_\infty(\vec{\lambda})} = \frac{\beta}{2} \sqrt{P_\infty}\,  \mathcal{F}_i
\ee
such that the evolution of $\phi$ is obtained as
\be 
\frac{\partial \phi(\vec{\lambda},t) }{\partial t} = \sum_{i=1}^N \left(   \frac{1}{\beta} \frac{\partial^2\phi(\vec{\lambda},t)  }{\partial \lambda_i^2} 
- \left( \frac{\beta}{4} \mathcal{F}_i^2 + \frac{1}{2} \frac{\partial \mathcal{F}_i}{\partial \lambda_i}  \right)
\phi(\vec{\lambda},t) 
\right) 
\label{eq:2:EvolutionAuxiliaryPhiDBM}
\ee 
where we omit the dependence on $\vec{\lambda}$ in the forces $\mathcal{F}_i(\vec{\lambda})$. We see that we recover a Schrödinger-like evolution, in imaginary time, where the potential felt by a particle, i.e. the prefactor of $\phi$, can be expanded as:
\be
\begin{aligned}
\frac{\beta}{4} \mathcal{F}_i^2 + \frac{1}{2} \frac{\partial \mathcal{F}_i}{\partial \lambda_i}  = \frac{\beta}{4} \sum_{\substack{j \neq i \\k\neq i}} \frac{1}{(\lambda_i - \lambda_j)(\lambda_i- \lambda_k) } - \frac{1}{2} \sum_{j \neq i} \frac{1}{(\lambda_i - \lambda_j)^2}  \\ 
 - \frac{\beta V'(\lambda_i)}{2} \sum_{j \neq i} \frac{1}{\lambda_i - \lambda_j} + \frac{\beta}{4} V'(\lambda_i)^2 - \frac{1}{2}V''(\lambda_i)
\end{aligned}
\ee
where the first two terms are proportional since $\sum_{i \neq j \neq k} \frac{1}{(\lambda_i - \lambda_j)(\lambda_i- \lambda_k) }  =0$ by symmetry, such that:
\be
\frac{\beta}{4} \mathcal{F}_i^2 + \frac{1}{2} \frac{\partial \mathcal{F}_i}{\partial \lambda_i}  = 
\frac{\beta - 2}{4} \sum_{j \neq i} \frac{1}{(\lambda_i - \lambda_j)^2}  - \frac{\beta V'(\lambda_i)}{2} \sum_{j \neq i} \frac{1}{\lambda_i - \lambda_j} + \frac{\beta}{4} V'(\lambda_i)^2 - \frac{1}{2}V''(\lambda_i) \; .
\label{eq:2:QuantumPotentialDBMMappingGeneralV}
\ee
This potential exhibits a one-particle contribution $ \frac{\beta}{4} V'(\lambda_i)^2 - \frac{1}{2}V''(\lambda_i)$ reminiscent of \myeqref{eq:2:HamiltonianQuantumMapping} and two interaction terms, one of which depends on the RMT potential $V$. 

Until this point, we have kept full generality in this potential $V$. In order to get rid of the potential-dependent interaction term, we now make the usual assumption \cite{Potters2020,Forrester2010} that we are considering the quadratic potential $V(\lambda) = \frac{\lambda^2}{4}$, such that the large-time distribution of the particles is that of the $G \beta E$ with $\sigma=1$. In this case indeed, the evolution of $\phi$ is the Schrödinger equation
\be 
\frac{\partial \phi(\vec{\lambda},t) }{\partial t} = - \Ham \ \phi(\vec{\lambda},t)
\ee
with the $N$-particle Hamiltonian $\Ham$ given by
\be 
 \Ham = - \frac{1}{\beta} \sum_{i=1}^N \frac{\partial^2}{\partial \lambda_i^2}    + V_\mathrm{f}(\vec{\lambda}) 
 \label{eq:2:DefQuantumHamDBM}
\ee
where the $N$-particle potential $V_\mathrm{f}$ is given by
\bea
V_\mathrm{f}(\vec{\lambda})  &=& \sum_{i=1}^N 
\left(
\frac{\beta}{4} \mathcal{F}_i^2 + \frac{1}{2} \frac{\partial \mathcal{F}_i}{\partial \lambda_i} \right)  \\
&=& 
\frac{\beta }{16} \sum_{i=1}^N \lambda_i^2  
+ \frac{\beta -2 }{4} \sum_{\substack{1 \leqslant i,j \leqslant N \\i\neq j}} \frac{1}{(\lambda_i - \lambda_j)^2}
- \frac{N}{4} - \frac{\beta N(N-1)}{16}
\; .
\label{eq:2:PotentialCalogeroMoser}
\eea
Under this mapping, the DBM has been connected to an $N$-particle quantum system with quadratic confinement for each particle as $\frac{\beta}{16} \lambda_i^2$, and inverse-square interaction between each pair of particles as $\frac{\beta-2}{4} \frac{1}{(\lambda_i-\lambda_j)^2}$. This is the \emph{Calogero-Moser} Hamiltonian, first introduced by Francesco Calogero in 1971 \cite{Calogero1971}, which is integrable both classically and quantum-mechanicallly \cite{Polychronakos2006,Corrigan2002}. The special inverse-square interaction term indeed endows the system with an integrable structure that makes it fully-solvable, as in the case of the Sutherland-type $1/\sinh^2$ term and, on the circle, the $1/\sin^2$ term. More generally, the name \emph{Calogero-Sutherland} is given to a large class of quantum many-body systems related to root systems of classical reflection groups \cite{Olshanetsky1983}. For a presentation of integrable systems, we point to \cite{Sutherland2004,Babelon2003,Shabat2010,Maillet2020,Lafay2021}.

The formal operator mapping, unveiled above, between the DBM FP operator $\Ham_{FP}$ and the Calogero-Moser Hamiltonian $\Ham$ reads
\emphbe
\Ham_{FP} = {P_\infty}
^{\frac{1}{2}} \ \Ham \  {P_\infty}
^{-\frac{1}{2}} 
\label{eq:2:OperatorFormalMappingDBM}
\end{empheq}

\subparagraph{Boundary conditions and fermions}
In addition to this evolution-operator connection, the boundary conditions of the system are fixed by the fact that the DBM particles have probability zero of crossing, as explained earlier from the Bessel process. This implies that, as soon as $\lambda_i= \lambda_j$ for $i \neq j$, $P(\vec{\lambda},t) =0 \implies \phi(\vec{\lambda},t)=0$. Assuming that the eigenvalues are ordered in the initial condition, the stochastic system is then constrained inside the Weyl chamber $\Weyl_N$ defined in \myeqref{eq:2:DefWeylChamber}. Since the wavefunction $\phi(\vec{\lambda},t)$ is zero on the boundary of $\Weyl_N$, the system can be naturally specified to have fermionic symmetry, such that the mapping yields a system of $N$ fermions in the Hamiltonian $\Ham$. As a consequence, the mapping relates the large-time position statistics of $N$ fermions in $\Ham$ to the $G \beta E$ eigenvalue JPDF $P_\infty(\vec{\lambda})$ as
\be 
\abs{ \phi( \vec{\lambda}, t \to \infty ) }^2 = \left( \frac{P( \vec{\lambda}, t \to \infty ) }{\sqrt{P_\infty(\vec{\lambda})}}  \right)^2  = P_\infty(\vec{\lambda})  \; .
\ee
In this imaginary-time Schrödinger setting, the large-time wavefunction $\phi( \vec{\lambda}, t \to \infty )$ is the ground-state of the system, as contribution from all other states are killed exponentially fast in time.

In the case $\beta =2$ where the interaction term of $V_\mathrm{f}(\vec{\lambda})$ vanishes, we recover the connection laid out in section \subsubref{subsubsec:1.2.2.b}{I.2.2} between eigenvalues of the GUE and zero-temperature non-interacting fermions in the potential $V_\mathrm{f}(x) = \frac{x^2}{8}$.

In the cases $\beta =1,4$ where the interaction term remains, the above results grant a connection between the statistics of the GOE and GSE eigenvalues and those of zero-temperature inverse-square interacting fermions in a quadratic potential.

\subparagraph{Ornstein-Uhlenbeck matrix diffusion}

As an aside, note that in the special case of the Harmonic RMT potential $V(\lambda) = \frac{\lambda^2}{4}$ chosen here to simplify the quantum potential, the SDEs \eqref{eq:2:DefDBMWithPotential} that were introduced artificially at the beginning of the section can in fact be obtained elegantly from the matrix setting. Indeed, this is done simply by replacing the independent Brownian motions in the $\XX$ matrices of the DBM definition in Eqs.~(\ref{eq:2:DefDBMGOE}-\ref{eq:2:DefDBMGUE}-\ref{eq:2:DefDBMGSE}) by independent identically distributed Ornstein-Uhlenbeck (\textit{OU}) processes \cite{Applebaum2009}, i.e. processes obeying the following SDE
\be 
\dd x_t  =  \dd B_t -  \frac{x_t}{2}  \dd t \; .
\ee 
In this case, the linear drift implies that the perturbative computations go through, under the replacement $\DelBB \to \DelBB - \frac{\epsilon}{2} \HH_t^\beta$ in \myeqref{eq:2:EigenvaluePerturbationTheory}, and indeed yield \myeqref{eq:2:DefDBMWithPotential}. Besides, we note that this Ornstein-Uhlenbeck version was the original model introduced by Freeman Dyson \cite{Dyson1962b,Dyson1965}, with the aim of dynamically generating samples of the Gaussian ensembles. For a generalization of interacting Ornstein-Uhlenbeck walkers with coloured Gaussian noise, entering the realm of active matter, see \cite{Martin2021}.

As suggested by the links of Wishart and Jacobi ensembles with fermion systems mentioned in section \subsubref{subsubsec:1.2.2.b}{I.2.2}, stochastic matrix models also exist for these models and grant a connection to a Calogero-Moser Hamiltonian with adequate potential. For the Wishart ensemble for example, the stochastic extension is obtained by upgrading all independent Gaussian entries of the matrix $\XX$ in \myeqref{eq:1:DefWishart} to independent Brownian motions. A similar parameter-dependent ensemble construction exists for circular ensembles as well \cite{Forrester2010}. 

However this cannot be fully generalized. From the constraints imposed by \myeqref{eq:2:QuantumPotentialDBMMappingGeneralV}, a non-interacting system of fermions for $\beta=2$ can only be achieved for certain RMT potentials $V$.

\subsection{Non-crossing walkers}  
\label{subsec:2.1.2}

In the previous subsection, we have completed our introduction to random matrix theory by the presentation of its dynamical extension to stochastic matrix processes such as the Dyson Brownian motion. As laid out above, this extension naturally brings the repulsive feature of eigenvalues in RMT into $N$-particle stochastic systems. This is a direct motivation to study in parallel another model of repulsive diffusive particles, which we present in this section.

\begin{figure}[ht!]
    \centering 
    \includegraphics[width=.85 \textwidth]{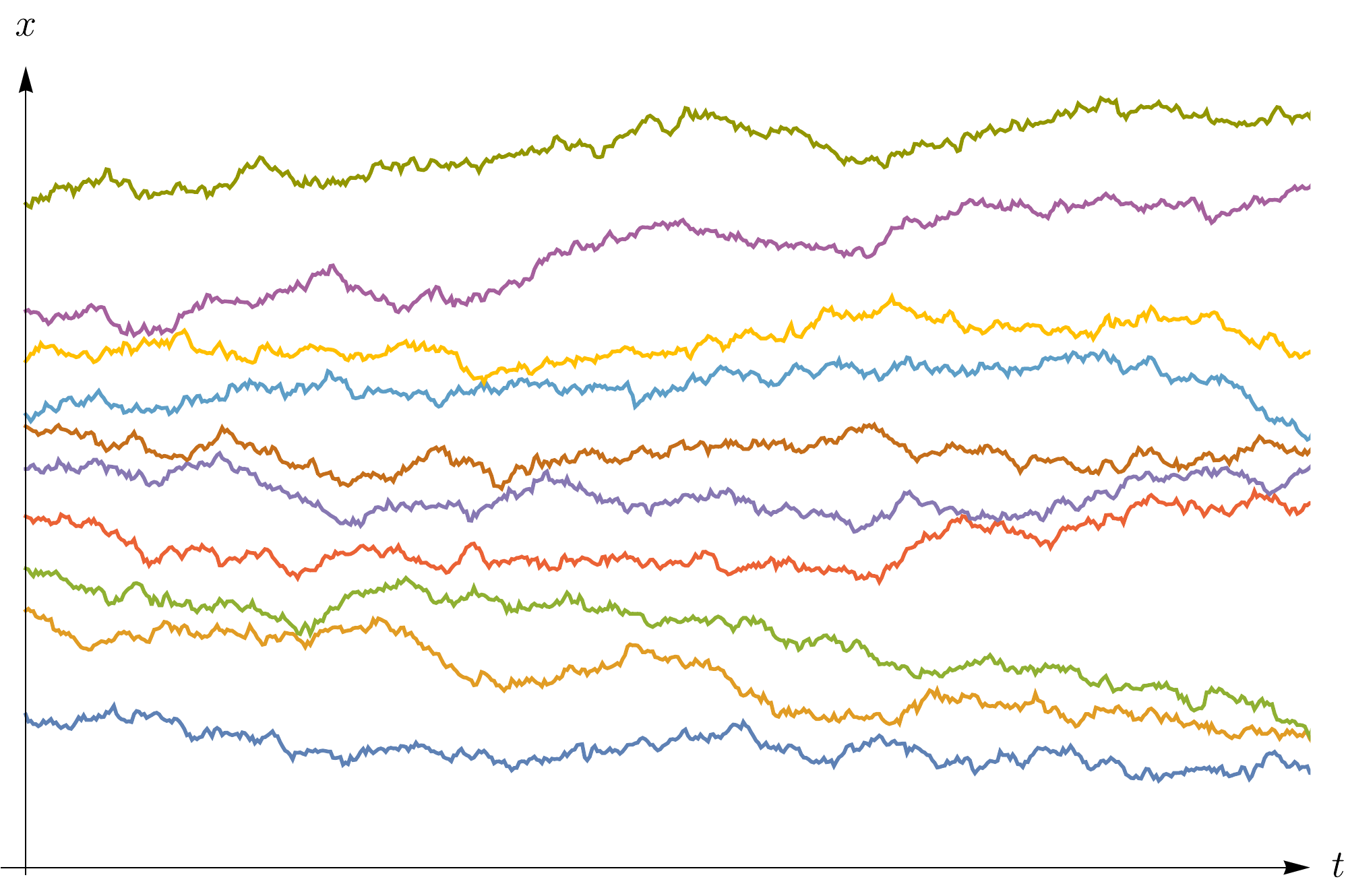} 
    \caption{A configuration of $N=10$ non-crossing paths, for random walkers diffusing in one-dimensional space $(x)$ over time $(t)$.}
    \label{fig:NonCrossingConfiguration}
\end{figure}

\emph{Non-crossing walkers} are a system of $N$ standard one-dimensional random walks that are conditioned not to cross each other, as illustrated in Fig. \ref{fig:NonCrossingConfiguration}. The non-crossing condition acts as an interaction between otherwise independent random processes. We will mostly consider in the following that space is continuous and that the processes are standard Brownian motions. The model of non-crossing (or non-intersecting, or non-colliding) walkers is also called (surviving) \emph{vicious walkers}: considering a reaction-diffusion system of $N$ independent random walkers who kill each other upon crossing, such that they are dangerously vicious indeed, the full-survival configurations are exactly those of non-crossing walkers, and bear the same probabilistic weight when conditioning on the survival of the full vicious system.

This system has been widely studied because it is a toy model for an interacting system, which can be physically relevant in the modelling of random interfaces, and where first-passage properties play a direct role in the behaviour of the system \cite{Redner2007,Bray2013,Nadal2011}. As we will further detail in this subsection, the non-crossing condition gives birth to a determinantal probabilistic structure which allows to solve it completely and grants connections with random matrix theory and the Dyson Brownian motion.

\subsubsection{A model of random interfaces}

The first model of one-dimensional non-crossing polymers was introduced in 1968 by Pierre-Gilles de Gennes, as a toy model for lipid-water interfaces \cite{DeGennes1968}. He studied a collection of thin, long, flexible chains in two dimensions. The chains being in thermal equilibrium under a strong unidirectional stretching force, each one can be modelled by a one-dimensional Brownian motion. In order to introduce a steric constraint, he excluded configurations where different chains intersect each other, such that his model is exactly a system of non-crossing walkers. In the philosophy of the stochastic-to-quantum mapping presented in the previous subsection, he solved the system by a path integral method for a fermionic system. 
This theoretical work was a first important step in the theory of lipid-water lamellar interfaces where lines of lipids separate thick sheets of water, although physically relevant elements such as rigidity and long-range interactions are absent from the toy model.

\begin{figure}[ht!]
    \centering 
    \includegraphics[width=.7 \textwidth]{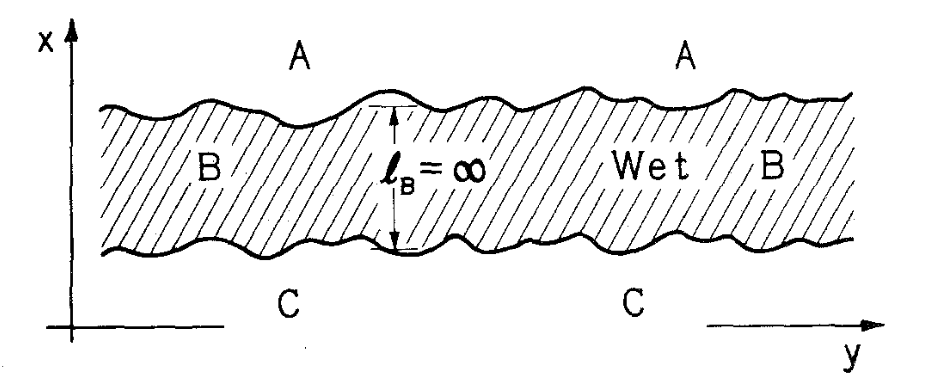} 
    \caption{Full wetting of two 2D liquid phases $A$ and $C$ by an intermediate phase $B$, leading to two non-intersecting linear interfaces \cite{Fisher1984}.}
    \label{fig:FisherWetting}
\end{figure}

\begin{figure}[ht!]
    \centering 
    \includegraphics[width=.6 \textwidth]{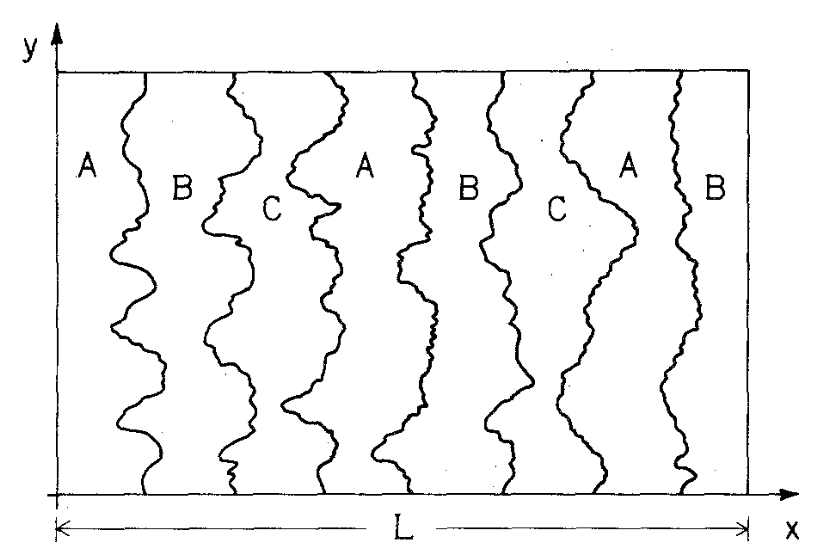} 
    \caption{Multiple linear interfaces between three commensurate phases $A,B,C$ of adsorbed atoms on a two-dimensional lattice substrate  \cite{Fisher1984}.}
    \label{fig:FisherCommensurate}
\end{figure}

Two decades later, non-crossing walkers were studied in a large variety of settings by Michael Fisher, who coined the name of "vicious" walkers . He studied 
many properties of the walkers in presence or not of an absorbing wall, and applied these results to physical models of random interfaces. For example in a two-dimensional multiphase liquid, the full wetting of two phases by an intermediate one leads to two non-intersecting linear interfaces as seen in Fig. \ref{fig:FisherWetting}, which can be modelled efficiently by non-crossing walkers. A multiple-non-crossing-line problem arises in the interfaces between commensurate phases of adsorbed atoms on an underlying two-dimensional substrate lattice, as seen in Fig. \ref{fig:FisherCommensurate}. Further, the applications proposed by Fisher extend to domain walls in magnetic systems such as the Ising model, the denaturation transition of a biopolymer molecule such as Deoxyribonucleic Acid (\textit{DNA}) and dislocation topological defects in discrete lattices.

Another type of interfaces which can be modelled, at first order, by non-crossing walkers is that of terrace width distributions obtained after a cut of a discrete 3D lattice. When the cut surface is chosen along a misoriented surface, i.e. with a small angle with respect to a symmetry surface of the crystal, a vicinal crystal is obtained, which displays a sequence of terraces oriented in the high-symmetry direction and separated by steps typically one atomic layer high. The edges of the terraces then form a sequence of non-crossing lines, which can be modelled by non-crossing walkers. More precise investigations have shown that these lines behave as Calogero-Sutherland models with a $\frac{\beta-2}{4}$ interaction prefactor as in \myeqref{eq:2:PotentialCalogeroMoser}, with a general $\beta$ not restricted to the RMT classes \cite{Einstein2003,Richards2005}.

As illustrated in the previous lines, many physical problems of random interfaces can be naturally described by non-crossing walkers. This is of course a first theoretical step, as this model leaves aside any specificity of the system apart from the non-crossing property. The main advantage of this model is that its simple definition comes with a structure that allows the computation of many probabilistic observables, as we will see in the next section.

Before proceeding, let us note that another track of research on the physics of elastic interfaces concerns avalanches that occur in a random environment when a force is gradually applied to the interface \cite{LePriol2020a,LePriol2021,LeDoussal2013}. For a presentation of the experimental, numerical and theoretical aspects of this field, see \cite{LePriol2020}.

\subsubsection{Non-crossing statistics}
\label{subsubsec:II.1.2.b}

\paragraph{Determinantal structure}

In order to derive analytical results on the statistics of non-crossing Brownian motions, a pivotal tool is the determinantal structure that appears naturally in the joint probabilistic weight of non-crossing independent paths. This was first realized by Samuel Karlin and James McGregor \cite{Karlin1959} such that the corresponding result is now called the Karlin-McGregor theorem.

\subparagraph{Karlin-McGregor} To state this theorem, let us consider a continuous Markov process on the line, with transition probability density from $x$ to $y$ in time $t$ given by $P(y, t \mid x,0)$. This is the PDF corresponding to the event where, starting from a known initial position $x$, the process is found at $y$ after a time interval of $t$. Note that the standard Brownian motion respects the above conditions, with $P(y, t \mid x,0)$ being the PDF of a variance-$t$ Gaussian random variable as
\be 
P(y, t \mid x,0)= \frac{1}{\sqrt{2\pi t}} \, e^{  - \frac{(y-x)^2}{2t}  } \; .
\label{eq:2:PropagatorBM}
\ee
Let us further introduce $(\vec{x}, \vec{y}) \in (\Weyl_N^*)^2$ respectively composed of $N$ ordered distinct positions on the real line. 

\begin{figure}[ht!]
    \centering 
    \includegraphics[width=.8 \textwidth]{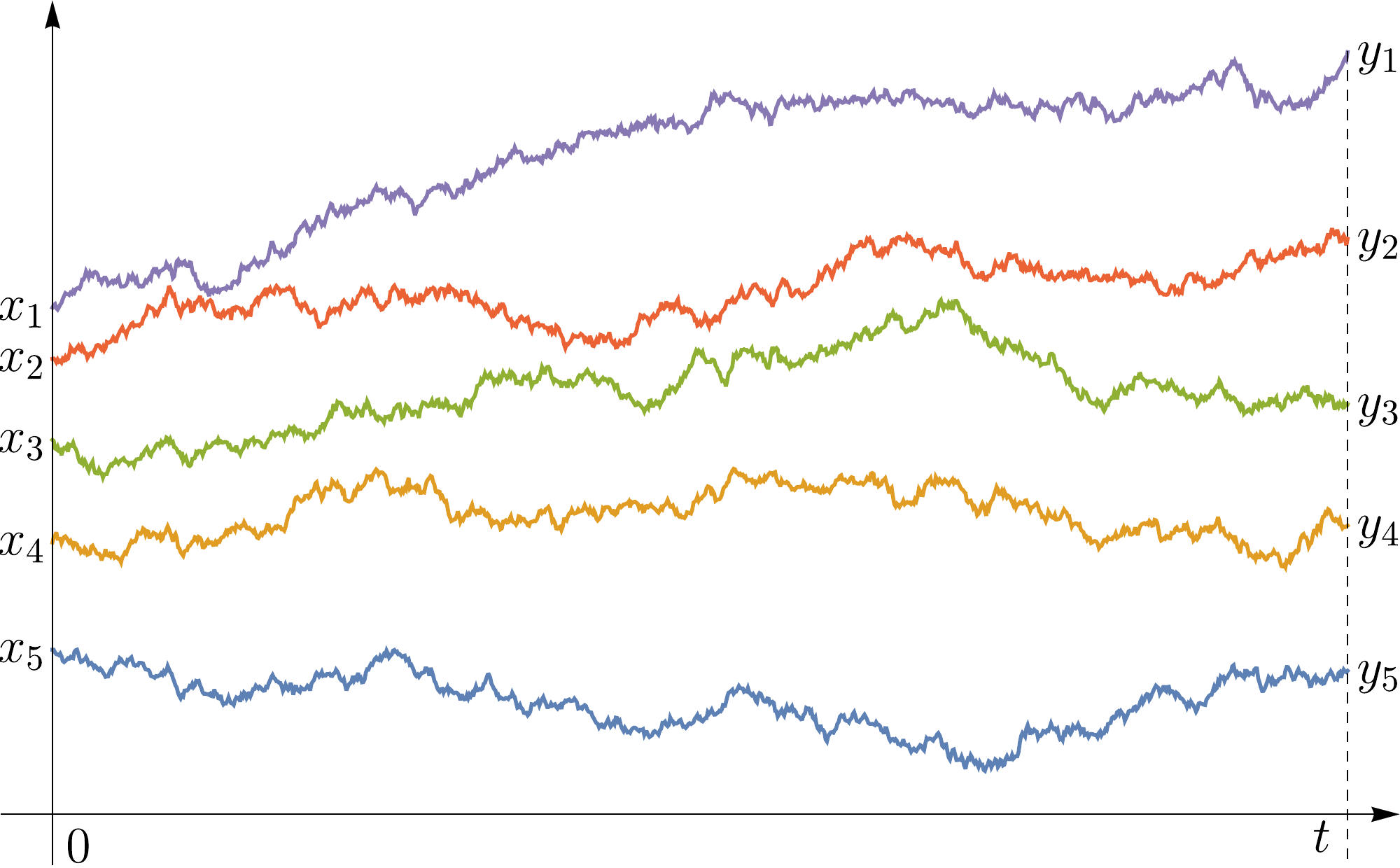} 
    \caption{Non-intersecting paths started from initial positions $\vec{x}$ and ending at positions $\vec{y}$ at time $t$.}
    \label{fig:KarlinMcGregorIllustration}
\end{figure}

The Karlin-McGregor theorem \cite{Karlin1959,Karlin1988,Bohm1997} states that for $N$ independent processes distributed as the one introduced above, the joint probability density for the $N$ particles to go, \emph{without crossing each other}, from $\vec{x}$ to $\vec{y}$ in time $t$ is
\emphbe 
P^{NC}(\vec{y},t \mid \vec{x}, 0 ) = \det_{1 \leqslant i,j \leqslant N} P(y_i , t  \mid x_j , 0) 
\label{eq:2:KarlinMcGregorTheorem}
\end{empheq}
See Fig.~\ref{fig:KarlinMcGregorIllustration} for an illustration of the walkers started from initial positions $\vec{x}$ and ending at positions $\vec{y}$ at time $t$, evolving along a non-crossing configuration.

This result is elegantly proven as we now describe, following \cite{Karlin1959}. Let us denote the $N$ processes $(B_{i,\tau})_{\substack{1\leqslant i \leqslant N \\ 0 \leqslant \tau \leqslant t}}$ and assume that they verify $B_{i,0}= x_i$. For a permutation $(k)$ of $[1\ldotp \ldotp N]$, let us denote $A_{(k)}$ the event 
\be 
A_{(k)} = \{ B_{i,t}= y_{k_i} \, , \,  \forall i \in [1\ldotp \ldotp N]  \}
\ee
and  $A^{NC}_{(k)}$ the event where the non-crossing condition between $0$ and $t$ is added to the previous definition
\be 
A^{NC}_{(k)} = A_{(k)} \  \bigcap  \ \{ B_{i,\tau} \neq B_{j,\tau} \, , \, \forall i \neq j , \forall \tau \in [0,t]   \} \; .
\ee
The proof consists in showing the following equality:
\be 
\sum_{(k)} \sigma(k) \, P \left( A_{(k)} \right) = \sum_{(k)} \sigma(k) \, P \left( A^{NC}_{(k)} \right) \; . 
\label{eq:2:KarlinMcGregorCrucialEquality}
\ee 
where $P$ designates the probability density of these singular events, which is defined with respect to the $N$-dimensional Lebesgue measure.

Assuming that this equality holds, the only admissible non-crossing contribution on the r.h.s~is the one where $(k)$ is the identity permutation $(\mathrm{id})$. For a continuous-path process on the line, the particles must indeed have the same order at initial and final conditions in order to be non-crossing, and the r.h.s~of \myeqref{eq:2:KarlinMcGregorCrucialEquality} is then simply equal to $P \left( A^{NC}_{(\mathrm{id})} \right) = P^{NC}(\vec{y},t \mid \vec{x},0) $. As a consequence of independence, we then have the Karlin-McGregor formula:
\be 
 P^{NC}(\vec{y},t \mid \vec{x}, 0) = \sum_{(k)} \sigma(k) \, \prod_{i=1}^N P(y_{k_i}, t \mid x_i , 0) 
=  \det_{1 \leqslant i,j \leqslant N} P(y_j , t \mid x_i , 0 ) \; .
\ee 

Finally, let us explain why \myeqref{eq:2:KarlinMcGregorCrucialEquality} holds. In order to do this, let us fix $(k)$ and choose a specific crossing configuration of $A_{(k)}$. We denote $\tau_C$ the first crossing time of the configuration, such that $\tau_C$ is the smallest time where two particles $i$ and $j$ are such that $B_{i,\tau_C}=  B_{j,\tau_C}$. 

Denoting $(k')$ the same permutation with simply $i$ and $j$ inverted such that $k'_i=k_j$ and $k'_j=k_i$, a configuration of $A_{(k')}$ is naturally defined from our initial configuration as being exactly the same for $\tau < \tau_C$ and the same up to $i \leftrightarrow j$ inversion for $\tau > \tau_C$. From i.i.d.~and Markov properties, the two configurations constructed above have the same probabilistic weight. 

\begin{figure}[ht!]
    \centering 
    \includegraphics[width=.45 \textwidth]{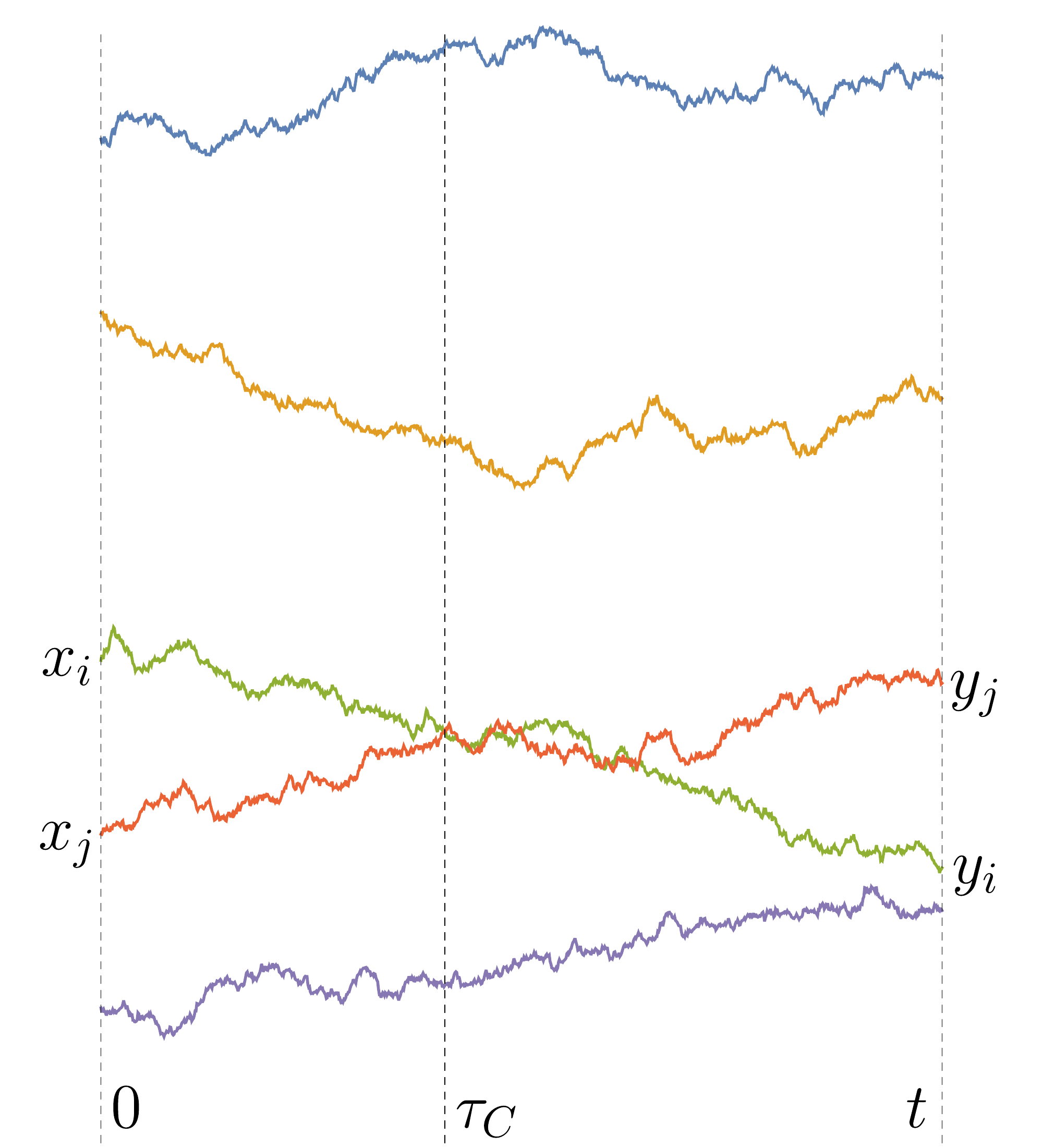} 
    \hfill
    \vbox to 180pt {\vfil  \hbox to 1cm{ $\boldsymbol{\Longleftrightarrow}$ } \vfil }
    \hfill
    \includegraphics[width=.45 \textwidth]{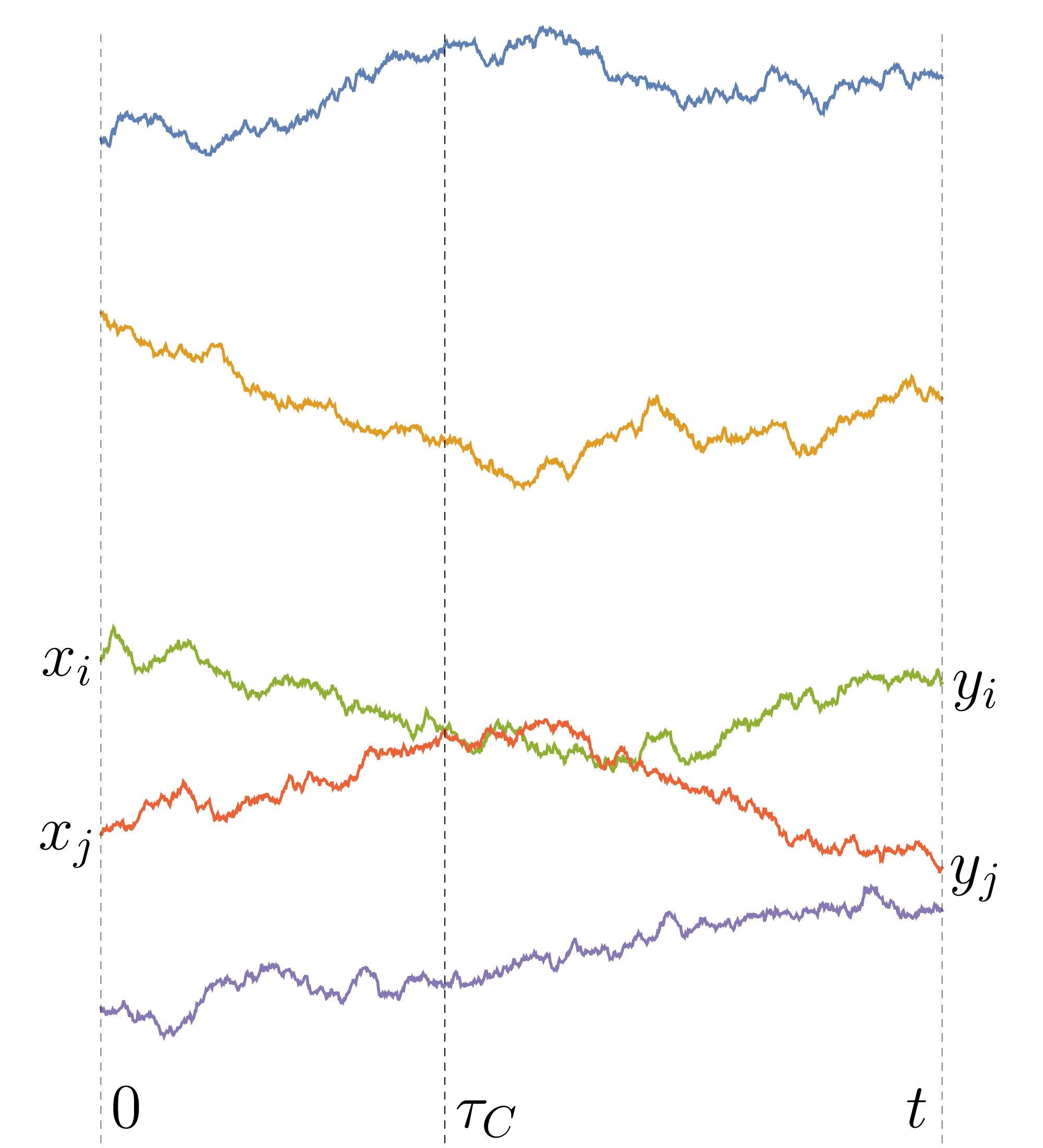} 
    \caption{Schematics of the crossing configuration involution employed in the proof of the Karlin-McGregor theorem. Under this transformation, the first two crossing particles $i$ and $j$ are inverted after their first crossing time $\tau_C$. The configuration on the right is crossing as well, since the positions of the particles coincide at $\tau_C$.}
    \label{fig:InvolutionSchema}
\end{figure}

The configuration transformation described above, and illustrated in Fig.~\ref{fig:InvolutionSchema}, is easily seen to be an involution. As a consequence, any configuration of $A_{(k)}$ has a unique counterpart, which belongs to $A_{(k')}$ where $(k')$ is such that $\sigma(k')=- \sigma(k)$. This ensures by symmetry that all crossing configurations vanish from the l.h.s~of \myeqref{eq:2:KarlinMcGregorCrucialEquality} such that the equality holds.
 
The general Karlin-McGregor theorem \myeqref{eq:2:KarlinMcGregorCrucialEquality} holds in a higher-dimensional metric space, where the r.h.s.~cannot be reduced to the identity permutation as in the one-dimensional case. Another special case is obtained when placing the process on the circle, such that the r.h.s.~is restricted to cyclic permutations.

\subparagraph{Lindström-Gessel-Viennot}

In the setting of discrete combinatorics, the determinantal structure underpinning the enumeration of non-crossing paths on a general lattice was obtained by Bert Lindström \cite{Lindstrom1973} and developed by Ira Gessel and Xavier Gérard Viennot \cite{Gessel1985,Gessel1989}. In this context, the counterpart to the Karlin-McGregor theorem is called the Lindström-Gessel-Viennot theorem. 
For a process on a finite metric space, the combinatorial and probabilistic results are exactly equivalent. The continuous-space Karlin-McGregor result can thus be seen as an infinitesimal limit of the Lindström-Gessel-Viennot result.

The discrete result is most useful in applications to combinatorics of semi-standard Young tableaux \cite{Guttmann1998,Krattenthaler2000}, as well as in random tilings and dimer models \cite{Johansson2001,Johansson2002}.

\paragraph{Connections with random matrix theory}

The determinantal structure laid out above will allow us to present the well-known links between non-crossing walkers and RMT: we will first detail the static links with the Gaussian ensembles before turning to the dynamical links with the DBM.

\subparagraph{Links with Gaussian ensembles}

The links between non-crossing walkers and Gaussian ensembles, first obtained from the mapping of discrete random walks to random permutations \cite{Baik1999,Forrester2001,Johansson2001,Johansson2002}, rest on the determinantal probabilistic structure granted by the Karlin-McGregor theorem. 

Let us consider $N$ independent standard Brownian motions $(B_{i,t})$. The following results hold:
\begin{itquote}
\begin{itemize}
\item Conditioning the walkers to be non-crossing on $]0,t]$, the joint position distribution at time $t$ is given by the GOE eigenvalue JPDF with parameter $\sigma = \sqrt{t/2}$. 
\item Conditioning the walkers to be non-crossing on $]0,+\infty[$, the joint position distribution at time $t$ is given by the GUE eigenvalue JPDF with parameter $\sigma = \sqrt{t}$.
\end{itemize}
\end{itquote}
Note that, being standard Brownian motions, the processes are all started from $0$, which is singular with respect to the non-crossing condition. However, the distributions described above can be shown to make sense in the conditional sense, as will be clear in the following.

Let us prove the GOE statement first for a given time $t$, by choosing the initial positions as $B_{i,0}= x_i = \epsilon \, t \, z_i$ with fixed ordered $\vec{z} \in \Weyl_N$. This regularization allows us to use the Karlin-McGregor formula before taking the $\epsilon \to 0$ limit. Let us introduce a notation for the event that particles are observed to be non-crossing on a certain interval as
\be 
\NC_{[a,b]}  = \{  B_{i,\tau } \neq B_{j, \tau} \ , \  \ \forall \tau \in [a,b]  , \forall i \neq j  \} \; .
\label{eq:2:DefNonCrossingEvent}
\ee
With this notation, the JPDF of positions at time $t$, knowing the initial condition $\vec{x}$ and conditioned on the non-crossing property on $]0,t]$, is: 
\be 
P\left(\vec{y},t \mid \vec{x}, 0, \NC_{]0,t]} \right) = \frac{ \det\limits_{1 \leqslant i,j \leqslant N} P(x_i , y_j ,t) }{ \int_{\R^N} \ \dd \vec{z}  \  \det\limits_{1 \leqslant i,j \leqslant N} P(x_i , z_j ,t) } \; .
\label{eq:2:NonCrossingJPDFGOE}
\ee
The numerator of this expression can be expressed in our Brownian setting as
\be
\det\limits_{1 \leqslant i,j \leqslant N} P(x_i , y_j ,t) = \det\limits_{1 \leqslant i,j \leqslant N} \left( \frac{1}{\sqrt{2\pi t}} \, e^{  - \frac{(y_j-x_i)^2}{2t}  } \right) = \frac{1}{(2 \pi t)^{\frac{N}{2}} }  e^{-\sum_{i=1}^N \frac{x_i^2 + y_i^2}{2t}  } \det\limits_{1 \leqslant i,j \leqslant N}  e^{\frac{y_i x_j}{t}}  
\ee
where the determinant is $ \det\limits_{1 \leqslant i,j \leqslant N}  e^{\frac{y_i x_j}{t}}  = \det\limits_{1 \leqslant i,j \leqslant N}  e^{ \epsilon  y_i  z_j }   $.  As proven in App.~\ref{app:VandermondeLimit}, the following equivalent holds in the $\epsilon \to 0$ limit:
\be 
 \det\limits_{1 \leqslant i,j \leqslant N}  e^{ \epsilon y_i z_j }  \ \simeq \ \epsilon^{\frac{N(N-1)}{2}}  \frac{ \Delta ( \vec{z} )}{\prod_{i=1}^{N-1} i! } \  \Delta ( \vec{y} )  \; .
\ee
Removing all constants not depending on the positions $\vec{y}$ and removing $\epsilon^{\frac{N(N-1)}{2}}$ factors appearing both on the numerator and denominator of \myeqref{eq:2:NonCrossingJPDFGOE}, the conditional JPDF is then, in the small-$\epsilon$ limit, proportional to:
\emphbe 
P\left(\vec{y},t \mid \vec{0}, 0, \NC_{]0,t]}\right) \  \propto \  \abs{   \Delta ( \vec{y} )  }  \ e^{- \frac{1}{2t} \sum_{i=1}^N  y_i^2  }  
\label{eq:2:ProofNonCrossingGOE}
\end{empheq}
In this equation, an absolute value has been harmlessly added as the $\vec{y}$ positions are necessarily ordered from the non-crossing condition and the ordered initialization. As claimed, this is the eigenvalue JPDF of the GOE with $\sigma= \sqrt{t/2}$, see \myeqref{eq:1:EigenvalueJPDFGOE}. The fact that it converges to a well-defined JPDF further proves well-posedness of the singular conditionment. 

For the proof of the GUE case, we regularize the initial condition in the same fashion as above. Consider that in order to condition the walkers to remain non-crossing on $]0, +\infty[$, we can consider first that this condition is verified until large time $T$ where we can further assume that the walkers come back to their initial positions $x_i = \epsilon  t z_i$. This artificial assumption is harmless, as the $T\to \infty$ limit will dissipate any influence of this constraint on the JPDF at fixed time $t$, and we leave the corresponding term $P\left( \vec{x},T \mid \vec{x}, 0,  \NC_{[0,T]}\right)$ away in the next lines. The conditioned JPDF can then be decomposed, by the Markov property, into two parts:
\be 
P\left(\vec{y},t  \mid \{\vec{x}, 0\}, \{\vec{x},T\} , \NC_{[0,T]}\right) \propto  P\left(\vec{y},t \mid \vec{x}, 0, \NC_{[0,t]}\right)  P\left(\vec{x},T \mid \vec{y}, t, \NC_{[t,T]} \right)  \;. 
\ee
The first term is directly given by the previous GOE computations leading to \myeqref{eq:2:ProofNonCrossingGOE}. The second term can be reduced to the same form by time-reversibility of the Brownian motion as:
\be 
P\left(\vec{0},T \mid \vec{y}, t, \NC_{[t,T]} \right) = P\left(\vec{y},T-t \mid \vec{0}, 0, \NC_{[0,T-t]} \right)   \propto \Delta(\vec{y}) \ e^{-\frac{1}{2(T-t)} \sum\limits_{i=1}^N y_i^2  }  \  \xrightarrow[T \to \infty]{} \   \Delta(\vec{y}) \; .
\label{eq:2:VandermondeFromInfiniteConditioning}
\ee
In the $(\epsilon \to 0,T\to \infty)$ limit, the distribution at time $t$ is then
\emphbe 
P(\vec{y},t \mid \vec{0}, 0, \NC_{]0,\infty[}) \   \propto   \  \abs{   \Delta ( \vec{y} )  }^2  \ e^{- \frac{1}{2t} \sum_{i=1}^N  y_i^2  }
\end{empheq}
which is, as claimed, the JPDF of the GUE with $\sigma= \sqrt{t}$, see \myeqref{eq:1:EigenvalueJPDFGUE}.

\subparagraph{Links with the Dyson Brownian motion}
 
We turn to the dynamical links of non-crossing walkers with the DBM, where we are interested in the law of the process instead of its mere fixed-time distributions. In this scope, a central result is as follows \cite{Rambeau2011,Potters2020,Grabiner1999}~: 

\begin{itquote}
The law of $N$ Brownian motions conditioned to never intersect is equal to that of the DBM-2.
\end{itquote}

When conditioning to avoid crossings on $]0, + \infty [$, we have seen in the previous paragraph that fixed-time distributions of non-crossing walkers started from $\vec{0}$ are equal to that of the DBM with $\beta=2$. The result cited above states that the two processes are in fact identical.  

In order to prove this result, let us first relate the propagator of the $\beta=2$ Dyson Brownian motion \be P_{DBM}( \vec{y},t_2 \mid \vec{x}, t_1)
\ee with the one of vicious Brownian motions 
\be P(\vec{y},t_2,\NC_{[t_1,t_2]} \mid \vec{x}, t_1) \; .
\ee
Note that this propagator is the probability that the Brownian motions go from positions $\vec{x}$ to $\vec{y}$ in time $t_2 - t_1$ \emph{and} do not cross in the meantime. This can be directly expressed by the Karlin-McGregor formula as a determinant of the one-particle propagators as in \myeqref{eq:2:KarlinMcGregorTheorem}. The Fokker-Planck equations respected by these one-particle propagators $\partial_t P = \frac{1}{2} \partial_y^2 P$ imply directly that
\be 
\frac{\partial P(\vec{y},t_2,\NC_{[t_1,t_2]} \mid \vec{x}, t_1) }{\partial t_2} 
= \frac{1}{2} \sum_{i=1}^N\frac{\partial^2 P(\vec{y},t_2,\NC_{[t_1,t_2]} \mid \vec{x}, t_1) }{\partial y_i^2} \; .
\label{eq:DiffusionEquationViciousWalkers}
\ee
In other words, vicious walkers respect the diffusion equation of independent Brownian motions, as they should, with null boundary condition at the boundary of the Weyl chamber.

Recalling the quantum mapping of the DBM in section \subsubref{subsubsec:2.1.1.b}{II.1.1}, we see that this equation \eqref{eq:DiffusionEquationViciousWalkers} is the same as the evolution of the auxiliary function in \myeqref{eq:2:EvolutionAuxiliaryPhiDBM}, in the case of the DBM-2, with $\beta=2$ and $V=0$ where only the diffusion term remains. This correspondence allows us to express the vicious walker propagator as:
\be 
P(\vec{y},t_2,\NC_{[t_1,t_2]} \mid \vec{x}, t_1) =  \frac{\Delta(\vec{x})}{\Delta(\vec{y})}  \ P_{DBM}( \vec{y},t_2 \mid \vec{x}, t_1) 
\label{eq:2:PropagatorRelationViciousWalkerDBM2}
\ee 
where the numerator $\Delta(\vec{x})$ is needed to ensure that the initial conditions $P(\vec{y},t,\NC_{[t,t]} \mid \vec{x}, t)  = P_{DBM}( \vec{y},t \mid \vec{x}, t) = \prod_{i=1}^N \delta(y_i - x_i)$ are verified.

Finally, we relate this to our pursued process of Brownian motions conditioned to never intersect. The propagator of this process is 
\be P(\vec{y},t_2, \mid \vec{x}, t_1,\NC_{[0,\infty[}) \; . 
\ee
This joint probability distribution function can be transformed by modifying the conditions and splitting the probabilities from Markov property as
\bea
P(\vec{y},t_2, \mid \vec{x}, t_1,\NC_{[0,\infty[}) 
&=&
\frac{P(\vec{y},t_2, \NC_{[t_1,t_2] }, \NC_{[t_2,\infty[ } \mid \vec{x}, t_1) }{
P(\NC_{[t_1,\infty[} \mid \vec{x}, t_1)
}\\
&=& 
\frac{ P( \NC_{[t_2,\infty[ }\mid \vec{y}, t_2)}{
P(\NC_{[t_1,\infty[} \mid \vec{x}, t_1)
}
P(\vec{y},t_2, \NC_{[t_1,t_2]} \mid \vec{x}, t_1 ) 
\eea
We have transformed the propagator into the vicious walkers' propagator with a prefactor. This prefactor presents ill-defined probabilities, but note that one could regularize with a large time $T \to \infty$ such that the limit of the fraction is well-defined. By summing on all possible configurations at the regularizing endtime $T$ and through manipulations on the conditions, this limit can be obtained from \myeqref{eq:2:VandermondeFromInfiniteConditioning} as
\be 
\frac{ P( \NC_{[t_2,\infty[ } \mid \vec{y}, t_2)}{
P(\NC_{[t_1,\infty[} \mid \vec{x}, t_1)
}
=\frac{\Delta(\vec{y})}{\Delta(\vec{x})}
\ee
such that
\be
P(\vec{y},t_2, \mid \vec{x}, t_1,\NC_{[0,\infty[}) =   \frac{\Delta(\vec{y})}{\Delta(\vec{x})}  \
P(\vec{y},t_2,\NC_{[t_1,t_2]} \mid \vec{x}, t_1)  \; . 
\label{eq:2:NeverCrossingInTermsofIntermediatePropagator}
\ee 
We can finally conclude that walkers conditioned to never intersect for $t \in \R^+$ and the DBM-2 are identical processes since their propagators are equal:
\emphbe 
P(\vec{y},t_2, \mid \vec{x}, t_1,\NC_{[0,\infty[})  = P_{DBM}( \vec{y},t_2 \mid \vec{x}, t_1)
\end{empheq}

\subparagraph{Other links}

The analytical connections of non-crossing walkers extend beyond the fundamental examples given above. In the presence of a repulsive substrate, they are connected to the Wishart ensembles of RMT \cite{Nadal2009}. Connections with the Gaussian and Wishart ensembles have also been unveiled from the extremal properties of non-crossing walkers \cite{Nguyen2017,Nguyen2017a,Kobayashi2008}. On a larger scope, the reunion probabilities of non-crossing walkers have been mapped to the partition function of two-dimensional Yang-Mills theory on the sphere \cite{Forrester2011a}. Finally, the applications of the model, further from the physical modelling of random interfaces, extend to computer science \cite{Bonichon2003}.

\paragraph{Survival probability decay and first-passage}

As detailed above, non-crossing Brownian motions have deep links with random matrix theory. However, many questions of interest go beyond these situations where walkers are assumed to have necessarily survived. For example, a problem that has been widely studied \cite{Fisher1984,Huse1984,Essam1995} is that of the \emph{decay of the survival probability}.

The survival probability, at time $t$, of vicious walkers started at a given initial position $\vec{x}$, is obtained by summing the non-crossing probability on all possible endpoints as
\be 
\label{eq:2:SurvivalProbabilityDefinition}
S(t \mid \vec{x}, 0) =  \int_{\R^N} \ \dd \vec{z}  \ P( \vec{z} ,t , \NC_{[0,t]}  \mid \vec{x} , 0) \; .
\ee
The quantity $S(t \mid \vec{x}, 0)$ depends on the details of the initial condition $\vec{x}$ but its large-time behaviour is universal. In this large-time limit, $S(t)$ decays algebraically, i.e.~as a power-law, according to
{\begin{empheq}[box=\setlength{\fboxsep}{8pt}\fbox]{equation} 
S(t) \ \simeq \ t^{-\frac{N(N-1)}{4}} 
\label{eq:2:ViciousWalkersSurvivalProbDecay}
\end{empheq}
where the exponent is known as the Fisher exponent \cite{Fisher1984,Huse1984,Guttmann1998}. We will prove this statement in a larger scope in the next section. Indeed, it will be a by-product of more general computations related to the boundary problems.

Many peripheral results have been obtained on this topic. Survival probability computations have been extended to vicious walkers in a potential \cite{Bray2004}, and vicious walkers divided in families such that family members are harmless with each other \cite{Cardy2003}. Other interesting aspects include the distribution of positions in certain situations such as the flat-to-flat boundary conditions \cite{Grela2021} or with a vicious boundary, see next section, and extremal statistics of vicious walkers \cite{Schehr2008,Schehr2012,Schehr2013,Rambeau2011,Rambeau2010,Borodin2009}, see \cite{Majumdar2020} for a review.

The survival probability for non-crossing walkers
is a typical \emph{first-passage property} of the multiple Brownian motion system. Indeed, it relies on a variable reaching a specific value for the first time, here the distance between two walkers reaching zero. The first-passage properties of a stochastic system are also called \emph{persistence} properties, as they describe the statistics of remaining away from the given bound. For example, vicious walkers persist in surviving as long as they remain away from each other, which happens with vanishing probability $S(t)$.  

First-passage problems have been widely studied, see \cite{Aurzada2015} for a mathematical review, and have found many applications in engineering and in physics, where persistence properties have attracted a lot of interest in the context of non-equilibrium statistical mechanics of spatially extended systems, both theoretically and experimentally \cite{Bray2013}. We point to \cite{Redner2007} for a pedagogical introduction to first-passage probabilities and related processes.

In many systems, such as stochastic diffusion in finite or semi-infinite domains, on the line or in higher dimensions, with spherical symmetry or in wedges \cite{Redner2007}, first-passage probabilities decay algebraically as does the survival probability of vicious walkers in \myeqref{eq:2:ViciousWalkersSurvivalProbDecay}. As a consequence many physical models where first-passage plays a role, from coarsening dynamics to fluctuating interfaces or polymer chains, present an algebraically-decaying persistence probability, in accordance with experiments \cite{Bray2013}. The exponent of the algebraic decay, or \emph{persistence exponent}, thus bears high theoretical and experimental interest, and will be the focus of next section in the setting of boundary problems.

\section{Vicious boundary}
\label{sec:2.2}

In this section, we turn to the study of vicious boundary problems for one-dimensional walkers, where a deterministic time-dependent moving boundary kills any walker that reaches it. Such a vicious boundary can be seen in a pictorial way as a \textit{moving cliff} \cite{Redner2007}, off which an unlucky walker will fall if he steps too far. For a unique Brownian motion, this is a well-known problem in probability theory \cite{Doob1949,Breiman1966,Salminen1988,Aurzada2013} and a lynchpin of first-passage processes \cite{Redner2007,Krapivsky1996,Majumdar1999}. Another broadly-studied problem is the cage model with \emph{two} absorbing time-dependent boundaries \cite{Novikov1981,Bray2007,Krapivsky1996,Turban1992}. In statistics, this version is furthermore related to the Kolmogorov-Smirnov (\textit{KS}) distribution test \cite{Kolmogorov1933,Smirnov1936}: quantifying the survival probability in an expanding cage can be used to improve the sensitivity of the KS test in the tails \cite{Chicheportiche2012}.  

The Brownian motion diffusing below an absorbing moving boundary displays an algebraically decaying survival probability for a large range of boundary shapes. As presented in the first subsection \ref{subsec:2.2.1}, the behaviour of the system depends on the speed of the moving boundary $g(t)$ compared with the standard deviation of the Brownian motion growing as $\sqrt{t}$. In particular, it will be shown that the critical case is obtained for a square-root boundary $g(t) = W \sqrt{t}$. 

In publication \pubref{publication:NonCrossingBrownianDBM}, I studied an extension of this classical problem to non-crossing walkers and investigated the survival probability exponent decay of vicious walkers under a square-root vicious boundary $g(t) = W \sqrt{t}$, as presented in subsection \ref{subsec:2.2.2}. From the mapping presented in the previous section, these results can be applied to the Dyson Brownian motion as developed in subsection \ref{subsec:2.2.3}.

\subsection{Single Brownian motion under a moving boundary}
\label{subsec:2.2.1}

We consider in this subsection a single Brownian motion $(B_t)_{t \geqslant 0}$, started from position $x$, under a deterministic moving boundary $g(t)$. We present the properties of the survival probability
\be 
S (t \mid x , 0) = \Pr \left( B_\tau < g( \tau ) \ , \  \forall \tau \in [0,t]  \mid B_0 = x \right) 
\ee
and in particular the exponent of its algebraic decay, in different cases.

\subsubsection{Fixed boundary: method of images}

In the case of a fixed boundary $g(t) = g$, the symmetry of the problem grants us with an elegant solution through the \emph{method of images}. 

\paragraph{Electrostatic image}
This method is well-known in the context of electrostatics \cite{Jackson1999}. Consider a point charge $q$ facing a perfect planar conductor with fixed electrostatic potential $\Phi = 0$. What is the potential field in the half-space domain, from this charge distribution and boundary condition ? 

The method of images consists in switching to an equivalent problem, where the planar conductor is replaced by a mere point charge $-q$, located symmetrically from the initial charge with respect to the conductor's boundary, see Fig.~\ref{fig:JacksonMethodImages}. This problem is equivalent at the level of the potential since it respects the same Poisson (or Maxwell-Gauss) equation in the half-space domain, and since the boundary condition $\Phi =0$ still stands by symmetry. The electrostatic problem is greatly simplified by this change of point of view: the equivalent formulation allows to conclude directly that, in the half-space, the potential is given by $\Phi = \Phi_q  + \Phi_{-q}$ where $ (\Phi_q , \Phi_{-q} )$ are the potential fields generated by the two charges.

\begin{figure}[ht!]
    \centering 
    \includegraphics[width=.6 \textwidth]{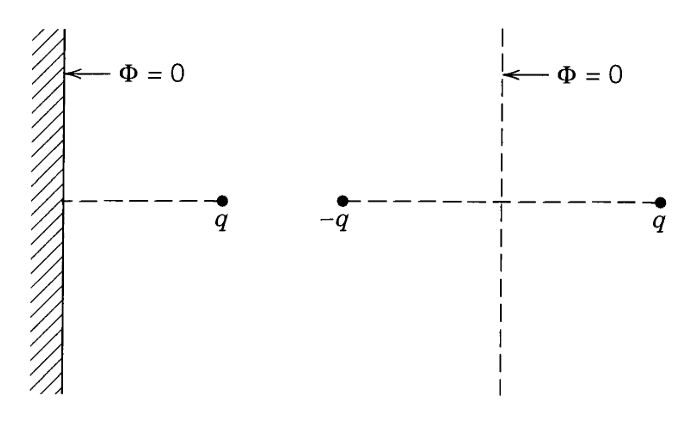} 
    \caption{Electrostatic equivalence of the problem with a charge facing a perfect conductor (left) and the image-problem with two opposite charges (right) \cite{Jackson1999}.}
    \label{fig:JacksonMethodImages}
\end{figure}

\paragraph{Stochastic image}
As in the electrostatic situation, symmetry can be used in the first-passage problem of a Brownian motion under a fixed boundary, in a stochastic method of images which relies on a similar equivalence. Let us introduce the following notation for the event where the Brownian motion remains below the boundary:
\be 
\Bevent_{[a,b]}  = \{ B_\tau < g(\tau)  \ , \  \forall \tau \in [a,b] \}   
\label{eq:2:DefBoundaryNonCrossingEvent}
\ee
where the boundary is fixed in this section as $g(\tau) = g$. The PDF for the walker, diffusing under the vicious boundary, to be found at position $y$ at time $t$ after starting from initial position $x$, is
\be  
P \left( y  , t ,  \Bevent_{[0,t]}  \mid  x , 0 \right)
\ee 
and verifies the diffusion evolution in the free half-space $]-\infty, g[$
\be 
\frac{\partial \, P \left( y  , t ,  \Bevent_{[0,t]}  \mid  x , 0 \right)  }{\partial t }
=
\frac{1}{2}
\frac{\partial^2 \, P \left( y  , t ,  \Bevent_{[0,t]}  \mid  x , 0 \right)  }{\partial y^2 }
\ee
with boundary condition
\be 
P \left( g  , t ,  \Bevent_{[0,t]}  \mid  x , 0 \right) = 0 \; .
\ee

The method of images consists in removing the boundary, and considering two standard Brownian motions started at the symmetrical positions $x$ and $2g-x$, with propagator given in \myeqref{eq:2:PropagatorBM}. The difference of their propagators $P_\mathrm{diff} (y,t)= P(y , t \mid x , 0)  - P(y , t \mid 2g-x , 0)$ verifies by linearity
\be 
\frac{\partial \ P_\mathrm{diff} (y,t)  }{\partial t }
=
\frac{1}{2}
\frac{\partial^2 \ P_\mathrm{diff} (y,t)  }{\partial y^2 }
\ee
and  $ P_\mathrm{diff} (g,t) = 0$ by symmetry. It is thus the solution of the boundary problem: 
\emphbe
P \left( y  , t ,  \Bevent_{[0,t]}  \mid  x , 0 \right) 
=
\frac{1}{\sqrt{2 \pi t }} \left(  e^{- \frac{(y-x)^2}{2t}} - e^{- \frac{(y+x-2g)^2}{2t}} \right) 
\label{eq:2:ImageMethodPropagatorSolution}
\end{empheq}

This solution can be explained intuitively through the reflection symmetry of the Brownian motion's law. Consider an arbitrary path starting from initial position $x$ at time $0$ and ending at a given position $y$ below the barrier at time $t$. If the path crosses the boundary and $\tau_C$ is the first crossing time, a path can be constructed between $2g-x$ and $x$, by reflecting $(B_\tau)_{0 \leqslant \tau \leqslant \tau_C}$ with respect to the $g$ axis. Every crossing configuration from $x$ to $y$ then has an analogue configuration from $2g-x$ to $y$, of same probabilistic weight by reflection symmetry  \cite{Applebaum2009,Varadhan2007,Mansuy2008}. The non-crossing propagator is then given by the difference \eqref{eq:2:ImageMethodPropagatorSolution}. The construction is illustrated in Fig.~\ref{fig:StochasticImages}.

\begin{figure}[ht!]
    \centering 
    \includegraphics[width=.7 \textwidth]{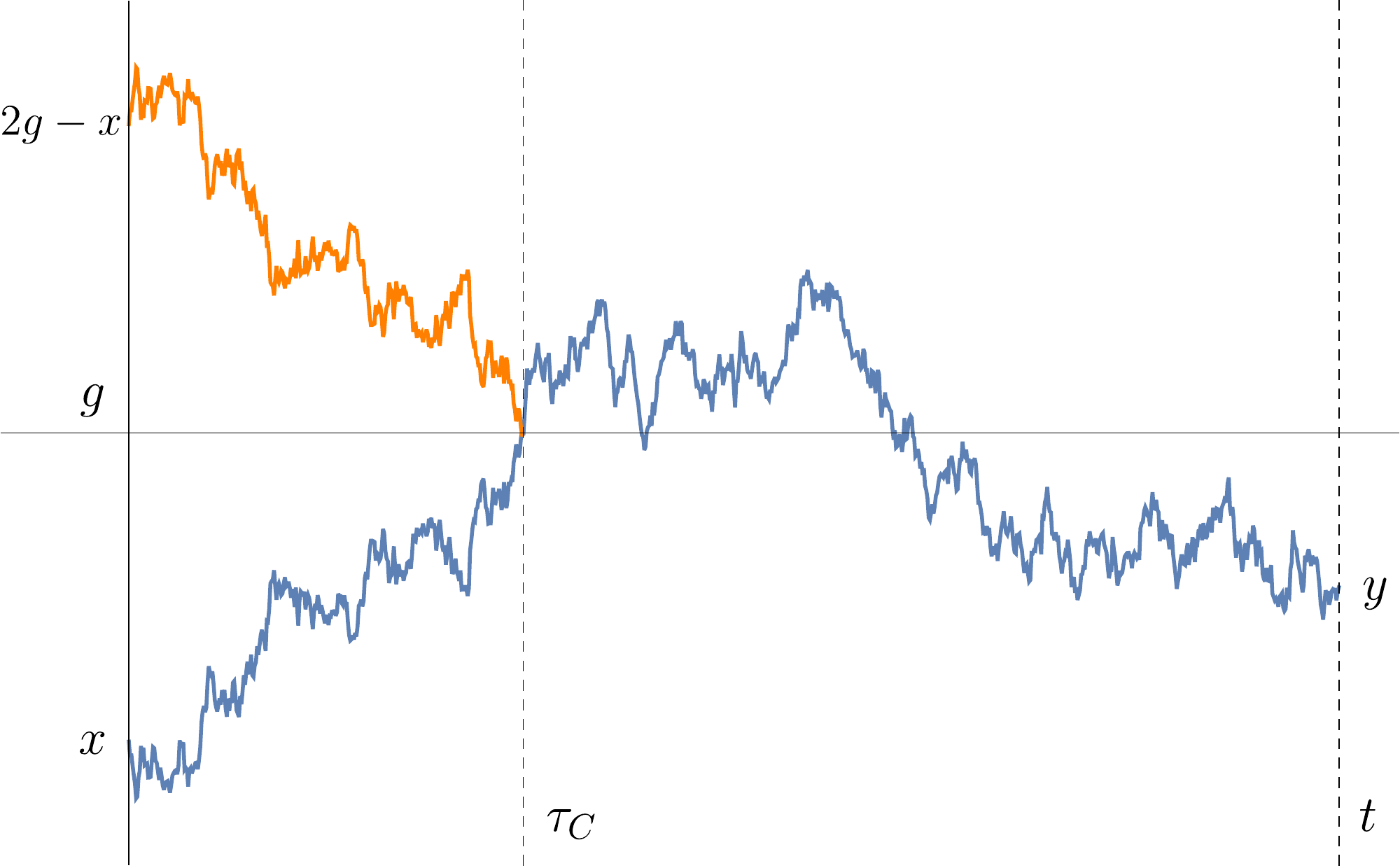} 
    \caption{Crossing configuration from $x$ to $y$ (blue) and equivalent configuration from $(2g-x)$ to $y$ (orange before crossing time $\tau_C$ and blue afterwards). Both configurations bear the same probabilistic weight by reflection symmetry of the Brownian motion.}
    \label{fig:StochasticImages}
\end{figure}

The Brownian motion considered here is driftless, but the method of images presented above can in fact be extended to the case of a boundary moving linearly in time, as to encompass the drifting Brownian motion with fixed boundary as well \cite{Redner2007}.

\paragraph{Survival probability decay}

The survival probability can be obtained directly from the analytic solution presented above, as 
\be 
S(t \mid x, 0) = \int_{-\infty}^g P \left( y  , t ,  \Bevent_{[0,t]}  \mid  x , 0 \right) = \erf \left( \frac{g -x}{\sqrt{2  t}} \right)
\ee
where $\erf$ is the \emph{error function} of the Gaussian distribution, defined by
\be 
\erf (z) = \frac{2}{\sqrt{\pi}}  \int_{0}^z e^{-u^2} \dd u  \; .
\label{eq:2:DefErrorFunction}
\ee 
From the asymptotics $\erf(\epsilon) \simeq \frac{2}{\sqrt{\pi}} \epsilon$ for $\epsilon \to 0$ \cite{Abramowitz1964}, we have at large time:
\emphbe
S(t \mid x, 0)  \simeq (g-x) \sqrt{\frac{2 }{\pi t}}  \ \propto  \ t^{- \frac{1}{2}}
\label{eq:2:LargeTimeSurvivalProbaFixedBoundary}
\end{empheq} 
We have recovered here the well-known power-law decay with exponent $\frac{1}{2}$ of the survival probability for a standard Brownian under a fixed absorbing boundary. See \cite{Redner2007} for an in-depth presentation of the previous computations and other analogies between first-passage properties and electrostatics. 

Note that the decay exponent found here is identical to the one of two vicious walkers in \myeqref{eq:2:ViciousWalkersSurvivalProbDecay} with $N=2$. Indeed, the difference of two independent Brownian motions is equal in law to $\sqrt{2}$ times a Brownian motion. The system of two vicious walkers is then equivalent to a single Brownian motion above a fixed boundary at $0$.
 
\subsubsection{Slow boundary: universal exponent}

Let us now consider a moving boundary, such that $g(t)$ recovers its time-dependence. Let us further restrict it to having slow enough growth at infinity compared to the standard deviation of the Brownian motion, which grows as $\sqrt{t}$. Then, the decay exponent of the survival probability is the same as the above result for a fixed barrier. The exponent $\frac{1}{2}$ is thus universal for slow boundaries:
\be 
S_{\mathrm{slow}}(t) \ \propto  \ t^{-1/2} \; .
\ee

More precisely, Uchiyama showed in \cite{Uchiyama1980} that this exponent holds for $g$ continuous and either concave or convex, respecting the following condition
\be
\int_{1}^{\infty}  \abs{g(\tau)} \, \tau^{-3 / 2} \, \dd \tau<\infty \; .
\ee
This covers in particular all power-law barriers slower than $\sqrt{t}$:
\be
g(t) = W t^\alpha  \quad \quad \text{with} \quad \alpha < \frac{1}{2} \; ,
\ee
such that the boundary motion is asymptotically irrelevant in these cases, and the large-time behaviour is identical to the fixed-boundary case.

Although the exact solution for an arbitrary slow boundary cannot be derived by the image method presented above, the large-time behaviour can be heuristically obtained by a quasi-static (or adiabatic) approximation. Since the boundary is slow, the PDF is expected to be linear on a region of scale $\sqrt{t}$ close to the boundary. Further away, the particle is expected to diffuse independently of the boundary. Matching these two regions yields the square-root decay of the total surviving probability, in the slow-boundary case \cite{Redner2007}.

\subsubsection{Fast boundary}

For a fast boundary, compared with the $\sqrt{t}$ expansion rate, the behaviour is also very simple.

For a fast boundary with positive velocity, receding away from the walker, the survival probability does not decay but instead converges to a non-zero limit, as
\be 
S_{\mathrm{fast}}^>(t) \to S_\infty  > 0  
\ee
where $S_\infty$ can be calculated explicitly in some cases \cite{Novikov1981}. 

For a fast boundary with negative velocity, approaching the walker as $g(t)=-At^\alpha$ with $\alpha>\frac{1}{2}$, the survival probability vanishes faster than any power law as \cite{Redner2007}
\be 
S_{\mathrm{fast}}^<(t) \ \propto \  e^{ - A \, t^{2\alpha -1} } \; .
\ee

\subsubsection{Critical square-root boundary}

As seen in the previous paragraphs, the algebraic decay of the survival probability is very simple if the boundary is either slow or fast. The interesting case, at the level of the decay exponent, thus concerns the critical square-root boundary
\be 
g(t ) \ = \ W \sqrt{t} \; .
\label{eq:2:SquareRootBoundary}
\ee
In this case, the survival probability is characterized by a non-universal decay exponent $\beta(W)$ which depends on the boundary prefactor $W$, such that \cite{Redner2007,Breiman1966,Krapivsky1996,Turban1992,Uchiyama1980}
\emphbe 
S_{W} (t) \ \propto \ t^{- \beta(W) } 
\label{eq:2:DecayExponentOneWalkerSquareRoot}
\end{empheq}
The survival exponent $\beta(W)$ can be computed for a general value of $W$ as
the smallest root of the following equation
\be
D_{2\beta(W)} (-W) = 0    
\label{eq:ConditionExponentOneWalkerParabolic}
\ee
where $D_\nu(x)$ is the parabolic cylinder function of index $\nu$, see App.~\ref{app:specialfunctions}. This exponent has the following asymptotics \cite{Abramowitz1964}
\be
\beta(W) \longrightarrow  
 \left\{
    \begin{array}{ll}
        \displaystyle \frac{1}{2} & \mbox{as } W \to 0 \; ;  \\[8pt]
          \displaystyle  0 & \mbox{as } W \to + \infty \;  ; \\[5pt]
            \displaystyle  + \infty & \mbox{as } W \to - \infty  \; .
    \end{array}
\right.
\ee
We see that these asymptotics match the behaviour of the slow and fast boundaries described above.

\subsection{Non-crossing walkers under a square-root boundary}
\label{subsec:2.2.2}

In the previous sections, we have described the decay of the survival probability of vicious walkers, see \myeqref{eq:2:ViciousWalkersSurvivalProbDecay}, and the decay of the survival probability of a single Brownian motion under a moving boundary, see \myeqref{eq:2:DecayExponentOneWalkerSquareRoot}. It is thus natural to extend and reunite these problems, and study the properties of $N$ non-crossing walkers under a square-root boundary, shown in the single-particle case to be the critical setting of interest. This was done in publication \pubref{publication:NonCrossingBrownianDBM}, the results of which we present here. This study gives exact results on a system of interacting particles in the presence of a time-dependent external potential, which is a central generic problem in non-equilibrium statistical physics, both for classical and quantum systems.

In the following, the moving boundary is $g(t) = W \sqrt{t}$ as in \myeqref{eq:2:SquareRootBoundary}. The $N$ walkers are independent Brownian motions $(B_{i,t})_{\substack{1 \leqslant i \leqslant N \\ 0 \leqslant t }}$. The main object of interest will be the survival probability between times $t_0$ and $t$, defined as
\be 
S (t \mid \vec{x} , t_0) = \Pr \left( \NC_{[t_0,t]}  \ , \  \Bevent_{[t_0,t]}  \mid \vec{B}_{t_0} = \vec{x} \right) 
\label{eq:2:DefSurvivalProbabilityBrowniansSquareroot}
\ee
where $\NC_{[t_0,t]}$ is the event that the $N$ walkers do not touch each other as in \myeqref{eq:2:DefNonCrossingEvent}, $\Bevent_{[t_0,t]} $ is the event that none of the walkers touches the boundary $g$ as in \myeqref{eq:2:DefBoundaryNonCrossingEvent}, and where the initial condition is fixed at starting time $t_0$ by the $N$ positions $\vec{x}$. A configuration of non-crossing walkers under a square-root boundary is illustrated in Fig. \ref{fig:NonCrossingConfigurationSquareRootBoundary}.
 
 \begin{figure}[ht!]
    \centering 
    \includegraphics[width=.8 \textwidth]{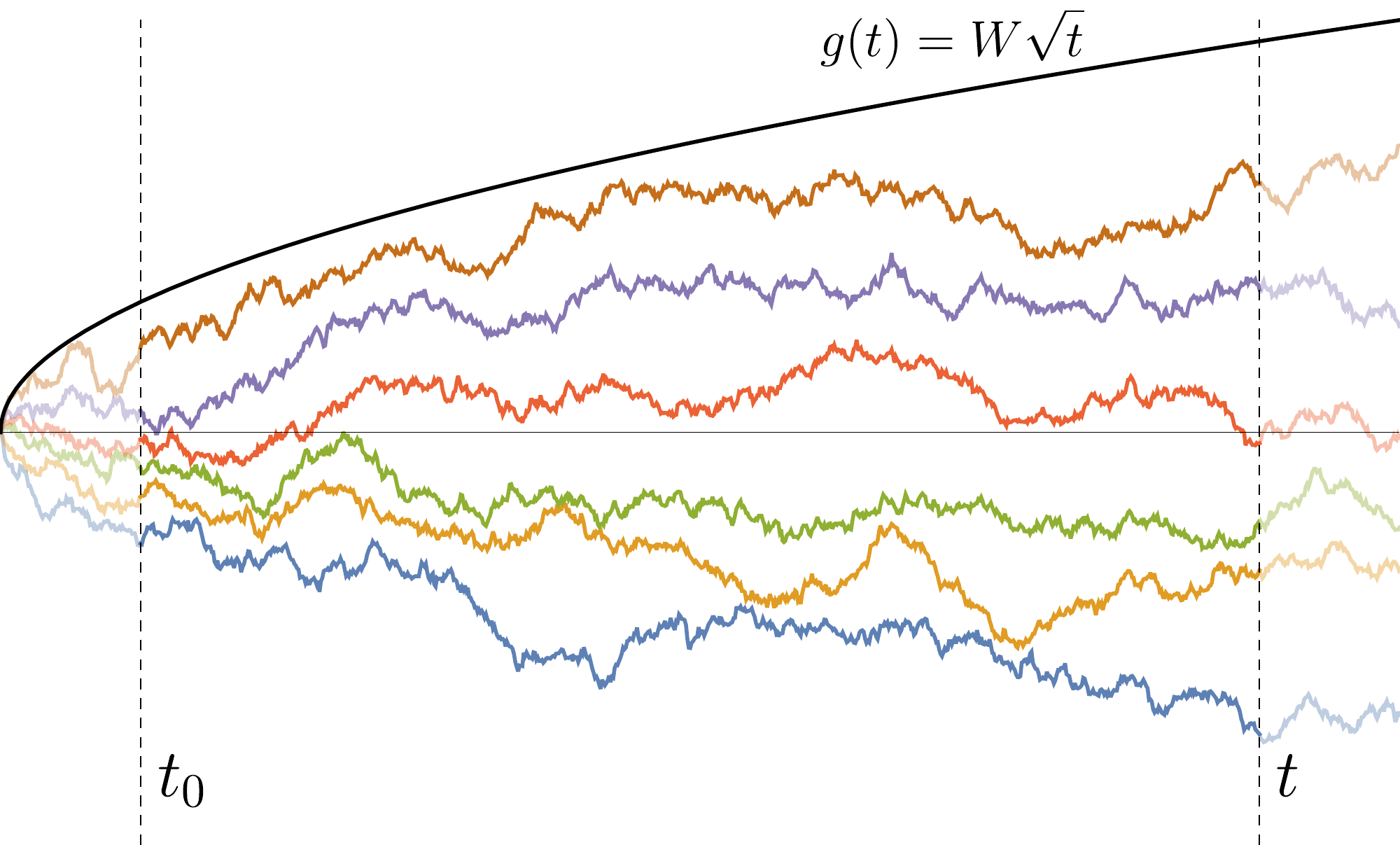} 
    \caption{Configuration of $N=6$ Brownian motions which remain non-crossing among each other and with the square-root boundary, between times $t_0$ and $t$.}
    \label{fig:NonCrossingConfigurationSquareRootBoundary}
\end{figure}

The starting time $t_0>0$ introduced above is useful to regularize the survival probability: for $t_0 \to 0$, the probability that $N$ Brownian motions started from the same point do not cross vanishes. The same can be said about their crossing of the boundary as $g(0)=0$. As in the single-particle case, the choice of initial positions $\vec{x}$ is not significant in the large-time limit. 

As can be expected from the previous sections, the main result presented in the following is that the survival probability of this system behaves for large time as
\emphbe 
S(t \mid \vec{x} , t_0)  \ \propto \ t^{- \beta(N,W)}
\end{empheq}
where the exponent $\beta(N,W)$, which depends on the prefactor of the square-root boundary and on the number of walkers, is the ground-state energy of a quantum system as shown below. 

\subsubsection{Mapping to fermions in a quadratic potential}

The stochastic system described above can be mapped to a quantum system of fermions in a quadratic potential, after a first mapping to Ornstein-Uhlenbeck processes called the Lamperti mapping.

\paragraph{Lamperti mapping}

The \emph{Lamperti} mapping states that a standard Brownian motion $B_t$ and an Ornstein-Uhlenbeck process $M_T$ with restoring coefficient $\frac{1}{2}$ are related by the following correspondence
\be 
M_{\ln(t)} \ \isEquivTo{\mathrm{law}} \ \frac{B_t}{\sqrt{t}} \; .
\ee
Indeed, applying Itô's formula on the process
\be 
M_T = e^{-\frac{T}{2}} B_{e^T}
\label{eq:2:LampertiMappingMT}
\ee
yields
\be 
\dd M_T =  - \frac{e^{-\frac{T}{2}}}{2}  B_{e^T} \dd T +  e^{-\frac{T}{2}} \dd B_{e^T}  = - \frac{M_T}{2} \dd T + \dd B_T
\label{eq:2:OrnsteinUhlenbeckSDE}
\ee
such that $(M_T)_{0 \leqslant T}$ is an OU process with restoring coefficient $\frac{1}{2}$. The mapping is named in reference to the fact that it is a special case of Lamperti's transformations for scale-invariant self-similar stochastic processes applied to the Brownian Motion \cite{Lamperti1972,Borgnat2005}. 
This mapping is also sometimes called Doob's transform \cite{Salminen1988}, since it is given by the application of Doob's transformation theorem \cite{Doob1949} to the OU process. 

The interest of this transformation is that it maps our system of non-crossing Brownian motions under a square-root barrier $g(t) = W \sqrt{t}$ to a system of non-crossing Ornstein-Uhlenbeck processes under a \emph{fixed} barrier at $W$. Indeed, the processes $M_{i,T}$, obtained as in \myeqref{eq:2:LampertiMappingMT} from the Brownian motions $B_{i,t}$, remain below a fixed barrier at $W$ if and only if the $B_{i,t}$ remain below $g(t)$. This trick was used by Breiman in \cite{Breiman1966} to study a similar problem for a single particle, and in later works in the physics literature \cite{Majumdar1999,Bray2013,Majumdar1996,Bray2004}. Furthermore, the processes $M_{i,T}$ are non-crossing if and only if the initial processes $B_{i,t}$ are. The survival probabilities of both systems are thus equal.

More precisely, let us relate the propagators of the two systems. The propagator of the vicious walkers killed by the boundary is 
\be 
P\left( \vec{y}, t , \NC_{[t_0,t]} , \Bevent_{[t_0,t]} \mid \vec{x}, t_0 \right)
\ee
which, we recall, is the probability density for $N$ walkers to go from $\vec{x}$ to $\vec{y}$ \emph{and} survive. 
We denote
\be 
P_{\OU}\left( \vec{Y}, T ,  \NC_{[T_0,T]} , \Bevent_{[T_0,T]} \mid \vec{X}, T_0 \right)
\label{eq:2:DefPropagatorViciousOUWalkers}
\ee
the propagator of $N$ vicious OU processes independently obeying SDE \eqref{eq:2:OrnsteinUhlenbeckSDE}, where the notations are similar as above with the boundary now fixed at $W$. By the Lamperti mapping, the propagators are related as
\be 
P\left( \vec{y}, t , \NC_{[t_0,t]} , \Bevent_{[t_0,t]} \mid \vec{x}, t_0 \right)  \; \dd \vec{y}  
= P_{\OU}\left( \vec{Y}, T ,  \NC_{[T_0,T]} , \Bevent_{[T_0,T]} \mid \vec{X}, T_0 \right)  \;   \dd \vec{Y} 
\label{eq:2:BrownianOUPropagatorMapping}
\ee
where 
\be 
\begin{array}{cc}
\displaystyle t= e^T  & \displaystyle  t_0 = e^{T_0} \\[5pt]
\displaystyle  \vec{Y} = \frac{\vec{y}}{\sqrt{t}} &  \displaystyle  \vec{X} = \frac{\vec{x}}{\sqrt{t_0}} \; . \\
\end{array}
\label{eq:2:VariableRelations}
\ee

\paragraph{Quantum mapping}

In the spirit of section \subsubref{subsubsec:2.1.1.b}{II.1.1}, the stochastic system of $N$ vicious OU processes can be mapped to a many-body quantum system of non-interacting fermions.

\subparagraph{Single-particle mapping}

The OU process obeying SDE \eqref{eq:2:OrnsteinUhlenbeckSDE}, i.e. \myeqref{eq:2:SDEMappingQuantum} with $V(x) = \frac{x^2}{4}$ and $D = \frac{1}{2}$, is mapped by \myeqref{eq:2:OperatorFormalMapping} to a quantum system with Hamiltonian
\be
\hat{H} = - \frac{1}{2} \frac{\partial^2}{\partial X^2}+ \frac{1}{8}X^2 - \frac{1}{4}  \; .
\label{eq:2:HamiltonianQuadraticMapping}
\ee
The wavefunctions $\phi_k(X)$ of this quadratic potential are given in \myeqref{eq:1:HarmonicWavefunctions}. Taking into account the $-\frac{1}{4}$ translation, the corresponding energy levels are $\epsilon_k= \frac{k}{2}$.

\subparagraph{Hard wall} An additional element of our system of vicious OU walkers is the presence of the wall, which imposes that the position PDF vanishes at position $W$. In a quantum system, an equivalent boundary condition is imposed by an infinite potential in the forbidden region. For a single particle, the presence of the boundary thus simply changes the above quantum Hamiltonian to
\be
\hat{H}^W =
 \left\{
    \begin{array}{ll}
        \displaystyle  - \frac{1}{2} \frac{\partial^2}{\partial X^2}+ \frac{1}{8}X^2 - \frac{1}{4}    & \mbox{for } X  < W \; ,  \\[8 pt]
        \displaystyle   + \infty    & \mbox{for } X \geqslant W \;  . \\
    \end{array}
\right.
\label{eq:2:HamiltonianQuadraticHardWall}
\ee
Let us denote $(\phi_k^W,\epsilon_k^W)_{k \geqslant 0}$ the normalized wavefunctions and corresponding energy levels of this Hamiltonian, such that the dependence on the position of the hard wall is made explicit.

\subparagraph{Multiple walkers} Finally, as illustrated by the DBM in section  \subsubref{subsubsec:2.1.1.b}{II.1.1}, the non-crossing property of a vicious stochastic system directly leads, in the stochastic to quantum mapping, to the fermionic symmetry of a many-body quantum system. As a consequence, the vicious OU processes under an absorbing boundary at $W$ are mapped to a system of $N$ non-interacting fermions in the Hamiltonian 
\be
 \Ham = \sum_{i=1}^N \hat{H}^W_i 
 \label{eq:2:HamiltonianNoninteractingFermionsWithWallatW}
\ee
where $\hat{H}^W_i $ is obtained by replacing $X \to X_i$ in \myeqref{eq:2:HamiltonianQuadraticHardWall}.

The $N$-particle mapping can be proven at propagator level as we detail now. The propagator of the vicious OU walkers below the boundary, defined in \myeqref{eq:2:DefPropagatorViciousOUWalkers}, is equal by the Karlin-McGregor theorem to the determinant of the one-particle OU propagators under the boundary as
\be 
P_{\OU}\left( \vec{Y}, T ,  \NC_{[T_0,T]} , \Bevent_{[T_0,T]} \mid \vec{X}, T_0 \right) =
\det_{1 \leqslant i, j \leqslant N} 
P_{\OU}\left( Y_j, T ,  \Bevent_{[T_0,T]} \mid X_i, T_0 \right)  \; .
\ee
The single-particle propagator mapping of \myeqref{eq:2:PropagatorMappingSingleParticle} yields
\be 
P_{\OU}\left( \vec{Y}, T ,  \NC_{[T_0,T]} , \Bevent_{[T_0,T]} \mid \vec{X}, T_0 \right) 
= e^{-\frac{1}{4} \sum\limits_{i=1}^{N} (Y_i^2 - X_i^2)   } \det_{1 \leqslant i,j \leqslant N}  G ( Y_i , T | X _ { j } , T _ { 0 }  )  
\ee 
where $G$ is the quantum propagator of the quantum system with Hamiltonian $\hat{H}^W$, defined by
\be
G(Y,T \mid X, T_0)  = \bra{Y} e^{-(T-T_0) \hat{H}^W} \ket{X}  =  \sum \limits_{k \geqslant 0} \phi_k^W(Y) {\phi_k^W}^*(X) e^{-\epsilon_k^W (T-T_0)}  \;.
\ee
By the discrete Cauchy-Binet formula, the determinant is given by
\be
 \det_{1 \leqslant i,j \leqslant N}  G ( Y_i , T | X _ { j } , T _ { 0 }  )  
 = 
 \frac{1}{N!}  \sum_{k_1, \cdots, k_N \geqslant 0}   \det_{1 \leqslant i,j \leqslant N}  \phi_{k_i}^W(Y_j)
 \det_{1 \leqslant i,j \leqslant N}  {\phi_{k_i}^W}^*(X_j)  e^{- (T-T_0) \sum\limits_{\ell = 1}^N \epsilon_{k_\ell}^W  }
\ee
where the r.h.s. is related to the quantum propagator $G_N ( \vec{Y} , T | \vec{X} , T _ { 0 }  ) $ of $N$ non-interacting fermions with Hamiltonian $\Ham$ defined in \myeqref{eq:2:HamiltonianNoninteractingFermionsWithWallatW}. Indeed, the $N$-particle wavefunction $\Psi_{\vec{k}}^W(\vec{X})$ are given by a Slater determinant, see \myeqref{eq:1:SlaterDeterminantManyBodyWavefunction}, and the corresponding energy level is $\epsilon_{\vec{k}}^W= \sum\limits_{\ell =1}^N \epsilon_{k_\ell}^W$. With a symmetry factor $N!$ from the order constraint in the admissible set of eigenstates $\Omega_N$ defined in \myeqref{eq:1:DefOmegaNManyBodyEigenstates}, we have:
\be
\frac{1}{N!} \det_{1 \leqslant i,j \leqslant N}  G ( Y_i , T | X _ { j } , T _ { 0 }  )  
 = 
 \sum_{\vec{k} \in \Omega_N}  \Psi_{\vec{k}}^W(\vec{Y}) \
{\Psi_{\vec{k}}^W}^*(\vec{X}) \ e^{- (T-T_0) \ \epsilon_{\vec{k}}^W  } 
=
G_N ( \vec{Y} , T \mid \vec{X} , T _ { 0 }  )   \; .
\label{eq:2:QuantumPropagatorRelation}
 \ee
The $N$-particle stochastic and quantum propagators are finally related as
\be
P_{\OU}\left( \vec{Y}, T ,  \NC_{[T_0,T]} , \Bevent_{[T_0,T]} \mid \vec{X}, T_0 \right)  =
N! \ e^{-\frac{1}{4} \sum\limits_{i=1}^{N} (Y_i^2 - X_i^2)   }
G_N ( \vec{Y} , T \mid \vec{X} , T _ { 0 }  )   \; .
\label{eq:2:OUQuantumPropagatorMapping}
\ee 
 
\paragraph{Summary} 
In summary, the two mappings \myeqref{eq:2:BrownianOUPropagatorMapping} and \myeqref{eq:2:OUQuantumPropagatorMapping} have transformed our initial problem of vicious Brownian motions under a square-root boundary to a quantum system of non-interacting fermions in a quadratic potential with a hard wall. The relation between the propagators of these two systems is
\emphbe 
P\left( \vec{y}, t , \NC_{[t_0,t]} , \Bevent_{[t_0,t]} \mid \vec{x}, t_0 \right) \ \dd \vec{y}
= 
N! \ \exp \left( \sum\limits_{i=1}^{N} \frac{X_i^2 - Y_i^2}{4}   \right)
G_N ( \vec{Y} , T \mid \vec{X} , T _ { 0 }  )  \ \dd \vec{Y}
\label{eq:2:SummaryStochasticQuantumPropagators}
\end{empheq}
where the variables are related as in \myeqref{eq:2:VariableRelations}.
 
\subsubsection{Survival probability decay}

The change of point of view operated in the previous section allows us to characterize the behaviour of the survival probability at large time $t$.

\paragraph{Decay exponent and ground-state energy}

The survival probability of the system defined in \myeqref{eq:2:DefSurvivalProbabilityBrowniansSquareroot} can be obtained from the quantum propagator by summing on all possible end configurations as
\bea
S (t \mid \vec{x} , t_0) 
&=&
\int_{\Weyl_N^W} \dd \vec{y} \
P\left( \vec{y}, t , \NC_{[t_0,t]} , \Bevent_{[t_0,t]} \mid \vec{x}, t_0 \right)  \\
&=& N! \  e^{\sum_{i=1}^N \frac{X_i^2}{4}}  \int_{\Weyl_N^W} \dd \vec{Y} \  e^{- \sum_{i=1}^N \frac{Y_i^2}{4}} \
G_N ( \vec{Y} , T \mid \vec{X} , T _ { 0 }  )
\eea
where we denote
\be 
\Weyl_N^W = \{ \vec{\lambda} \in \R^N  \mid W \geqslant \lambda_1 \geqslant \lambda_2 \geqslant \cdots \geqslant \lambda_N \}   \; .
\ee

\subparagraph{Large-time limit}

As can be seen in the series expansion of $G_N$ in \myeqref{eq:2:QuantumPropagatorRelation}, the dominating contribution at large $t= e^T$ corresponds to the ground-state $\vec{0}= \ket{0, \cdots , N-1}$, which has lowest energy level, such that
\be 
G_N ( \vec{Y} , T \mid \vec{X} , T _ { 0 }  )  \simeq 
\Psi_{\vec{0}}^W(\vec{Y}) \
{\Psi_{\vec{0}}^W}^*(\vec{X}) \ e^{- (T-T_0) \ \epsilon_{\vec{0}}^W  }  \; .
\label{eq:2:LargeTimePropagatorApproximation}
\ee
This approximation is accurate for $T - T_0 \gg \Delta^{-1}$, where $\Delta=\epsilon_N^W - \epsilon_{N-1}^W$ is the gap between the many-body ground state and the first excited state $\ket{0,1,\dots,N-2,N}$.

Replacing the $T$ and $X$ variables by the $t$ and $x$ variables of the original problem, the large-time survival probability decays algebraically as
\begin{empheq}[box=\setlength{\fboxsep}{8pt}\fbox]{equation} 
\begin{split}
S (t \mid \vec{x} , t_0) \ \simeq  \ A(\vec{x} ,t_0) \ t^{ - \beta(N,W) } \\[5pt]
\text{where} \quad 
\beta(N,W)= \epsilon_{\vec{0}}^W  = \sum_{k=0}^{N-1} \epsilon_{k}^W  
\end{split}
\label{eq:2:ResultLargeTimeSurvivalProbabilityDecay}
\end{empheq}
and with
\be 
A(\vec{x} ,t_0)  =
N! \  e^{\sum_{i=1}^N \frac{x_i^2}{4 t_0}}  \ {\Psi_{\vec{0}}^W}^*(\frac{\vec{x}}{\sqrt{t_0}}) \  t_0^{ \epsilon_{\vec{0}}^W  }  \   \int_{\Weyl_N^W} \dd \vec{Y} \ e^{- \sum_{i=1}^N \frac{Y_i^2}{4}} \
\Psi_{\vec{0}}^W(\vec{Y}) \; .
\label{eq:2:PrefactorSurvivalProbability}
\ee 
As announced earlier, the survival probability of our problem decays as a power-law, with a decay exponent characterized as the ground-state energy of a system of non-interacting fermions with Hamiltonian $\Ham$ of \myeqref{eq:2:HamiltonianNoninteractingFermionsWithWallatW}.

\subparagraph{Simple cases}

The exponent $\beta(N,W)$ can be obtained straightforwardly in a few special cases.

\begin{itemize}
\item In the limit $W \to + \infty$:\\
$\hat{H}^W$ tends to $\hat{H}$, the simple harmonic oscillator hamiltonian without a wall, which has energy levels $\epsilon_k^\infty = \epsilon_k = \frac{k}{2}$, such that
\be
\lim\limits_{W\to +\infty} \beta(N,W) \quad = \quad \sum_{k=0}^{N-1} \frac{k}{2} \quad = \quad \frac{N(N-1)}{4} \; .
\label{eq:2:DecayExponentWithoutWall}
\ee
This proves the Fisher exponent announced in \myeqref{eq:2:ViciousWalkersSurvivalProbDecay}, which characterizes the decay of the survival probability for vicious walkers without a boundary.
\item If $W= 0$:\\
With respect to the previous case, the hard wall imposes a cancellation of the wavefunction at $W=0$, such that the wavefunctions are simply the odd wave-functions of the harmonic oscillator. The ground-state energy is then obtained by populating the first $N$ odd levels of the free harmonic oscillator:
\be 
\beta(N,W=0) \quad = \quad \sum_{k=0}^{N-1} \frac{2k+1}{2} \quad = \quad \frac{N^2}{2} \;.
\label{eq:2:DecayExponentViciousWalkersFixedWall}
\ee
This exponent characterizes the survival probability's decay for vicious walkers under a fixed boundary, recovering the results of \cite{Katori2003,Krattenthaler2000}.
\item In the limit $W \to -\infty$:\\
The wall is very far to the left of the center at a potential energy 
$V(W)= \frac{W^2}{8}$. Expanding $X= W + \tilde X$ for small $\tilde X<0$ the potential energy is
$V(X)= \frac{W^2}{8} + \frac{W}{4} \tilde X + \bigO(W^0)$. The energy levels are therefore equal to those of a particle in a linear potential with a slope $\frac{W}{4}$ and confined on the negative axis. These single-particle energy levels are given by
\be 
\epsilon_k^W = \frac{W^2}{8}  - \frac{1}{2^{5/3}} \alpha_{k+1} \abs{W}^{2/3} 
\ee 
where $\alpha_m$ is the $m$\textsuperscript{th} zero of the Airy function, see App.~\ref{app:specialfunctions}. Hence we find the following asymptotics for $\beta(N,W)$ for fixed $N$ and $W \to -\infty$
\be 
\beta(N,W) \simeq \frac{N W^2}{8} -  A_N \abs{W}^{2/3} \quad , \quad A_N = \frac{1}{2^{5/3}} \sum_{m=1}^N \alpha_m \;.
\ee 
This limit is treated at length in chapter \ref{chap:4}, in the context of non-crossing bridges. We simply remark that the eigenfunctions are Airy functions centered at $X_k =  \alpha_{k+1} (2/W)^{1/3}$ which is much smaller than $\abs{W}$ for fixed $k$ and $W \to - \infty$. As a consequence, one can indeed neglect the term $\propto X^2 =\bigO(1)$ in the potential and the linear potential approximation becomes exact.
\end{itemize}

\subparagraph{General $W$}

For a general value of the wall's position $W$, the single-particle energy levels $\epsilon_k^W$ can be expressed as follows. 
The spectrum of the single particle Hamiltonian $\hat{H}_W$ is obtained by solving
the eigenvalue equation with a Dirichlet boundary condition at $X=W$
\be   
\hat{H}_W \phi = \epsilon \phi   \quad \iff \quad \left\{
    \begin{array}{ll}
        \frac{\partial^2 \phi}{\partial X^2}  + (  2 \epsilon + \frac{1}{2} - \frac{1}{4} X^2   ) \phi = 0  \quad  \text{for $X<W$}   \\[8pt]
        \phi (W) = 0  \quad , \quad  \phi (-\infty) = 0 \;.
    \end{array}
\right.
\ee
The solution of the differential equation, which vanishes as $X \to - \infty $, is the parabolic cylinder function
\be  
\phi (X) = C \, D_{2 \epsilon}(-X) 
\ee
with $C$ a normalization constant. The boundary condition at $W$ quantizes 
the eigenvalues $\epsilon=\epsilon_k^W$ according to the equation 
\be 
D_{2 \epsilon} (-W) = 0 \;.
\label{eq:2:EigenvalueEquationVanishingBoundaryCondition}
\ee
The energy levels $( \epsilon_k^W)_{ k \geqslant 0 } $ are thus given by the ordered sequence of solutions of this equation. The corresponding wavefunctions are
\be
\phi_k^W( X) = C_{k} \, D_{2 \epsilon_k^W}(-X) \;.
\label{eq:2:GeneralWSingleParticleWavefunction}
\ee
As an illustration, $D_{2 \epsilon} (-W)$ is plotted as a function of $\epsilon$ in Fig. \ref{fig:ParabolicCylinder}, for a given value of $W=6$. The first 11 zeros can be seen, and the amplitude of the oscillations quickly increases. We note that the first seven levels are very close to what would be expected for a harmonic oscillator $(0, \frac{1}{2}, 1, \frac{3}{2}, 2, \frac{5}{2}, 3 )$, but that the following levels start deviating from the $\frac{k}{2}$ levels, as the effect of the hard wall becomes more important for large $\epsilon$.

 \begin{figure}[ht!]
    \centering 
    \includegraphics[width=.65 \textwidth]{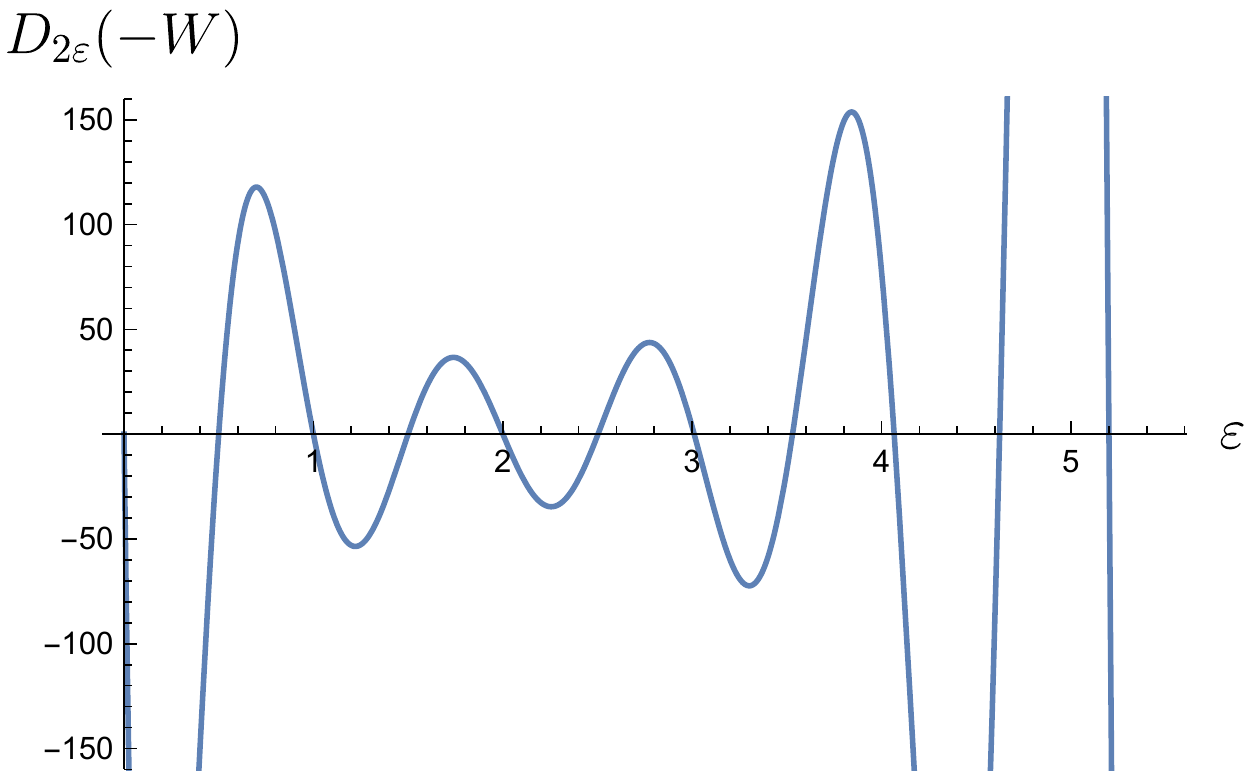} 
    \caption{Plot of the function $D_{2 \epsilon} (-W)$ versus $\epsilon$ for $W=6$. Note the first cancellation close to $\epsilon = 0$.}
    \label{fig:ParabolicCylinder}
\end{figure}

For general $W$, the decay exponent $\beta(N,W)$ is given in \myeqref{eq:2:ResultLargeTimeSurvivalProbabilityDecay} as the sum of the first $N$ solutions of \myeqref{eq:2:EigenvalueEquationVanishingBoundaryCondition}. We recover from this result the exponent $\beta(W)$ for the special case $N=1$, as announced in \myeqref{eq:ConditionExponentOneWalkerParabolic}.

The decay exponent $\beta(N,W)$ can be profitably studied in some limiting cases, further from the simple situations described in the previous section. In the following, we describe firstly the perturbative analysis that can be developed for $W \sim 0$, and secondly the semi-classical scaling obtained for large $N$ with $W = \bigO(\sqrt{N})$.

\paragraph{Perturbative expansion for the slow boundary} 

For a slow boundary $\abs{W} \ll 1$, i.e. when the hard wall is close to the minimum of the harmonic trap, we can calculate the energy levels $\epsilon_k^W$ and the exponent $\beta(N,W)$ perturbatively. 

When $W=0$ precisely, the spectrum is composed of the odd eigenstates of the free harmonic oscillator with wavefunctions $\phi_{2k+1}(x)$ and energies $\epsilon_{k}^0 =  \epsilon_{2 k+1} = \frac{(2k+1)}{2}$, as explained in \myeqref{eq:2:DecayExponentViciousWalkersFixedWall}.

For $\abs{W} \ll 1$, let us change variables to $Y= X - W \in ]- \infty , 0]$, such that the hard wall is at the origin in the new frame. In the $Y$-coordinate the Hamiltonian becomes 
\bea  
\hat{H}' = -\frac{1}{2}\frac{\partial^2}{\partial Y^2} + \frac{1}{8} (Y+W)^2 - \frac{1}{4}  
= \hat{H}^{W=0}   + W \Delta\hat{H} + \frac{1}{8}W^2  \;,
\eea
with $ \Delta\hat{H} = \frac{1}{4} Y$. The energy levels of the perturbed hamiltonian $\hat{H}'$ can be obtained by perturbation theory \cite{Basdevant2002} as explained in \myeqref{eq:2:EigenvaluePerturbationTheory}. The energy level $\epsilon_k^W$ is thus given, up to second order in $W$, by:
\be 
\epsilon_{k}^W - \epsilon_{k}^{W=0} = \Delta E _{k}
= \Delta E _{k}^{(1)}  \ W+  \Delta E _{k}^{(2)}  \ W^2 + \frac{1}{8}W^2 + \mathcal{O}(W^3)
\ee
with 
\be 
    \begin{cases}
       \displaystyle    \Delta E _{k}^{(1)} = \bra{2k+1} \Delta\hat{H} \ket{2k+1} \\
       \displaystyle  \Delta E _{k}^{(2)} = \sum\limits_{k' \geqslant 0, k'\neq k} \frac{\abs{ \bra{2k'+1}\Delta\hat{H} \ket{2k+1} }^2 }{\epsilon_{2k+1} -\epsilon_{2k'+1} }  
    \end{cases}
\ee
where $\left(\ket{2k+1}\right)_{k \geqslant 0}$ denote the odd levels of the free harmonic oscillator, however normalized to unity over the half-space $\R^-$. Computing the matrix elements on the half-space, see \pubref{publication:NonCrossingBrownianDBM}, one obtains explicitly
\be
\begin{cases}
\Delta E_{k}^{(1)} =  -\frac{1}{\sqrt{2\pi}2^{2k}} \frac{(2k+1)!}{k!^2} \\[5pt]
\Delta E_{k}^{(2)} = \frac{(2k+1)!}{\pi \, 2^{2k+1}\, (k!)^2}\sum\limits_{k'\geq 0, k' \neq k} \frac{1}{k-k'} \frac{(2k'+1)!}{2^{2k'} (k'!)^2} \frac{1}{(4(k-k')^2 - 1)^2} 
\end{cases}
\ee
One can show that the intricate expression for $\Delta E_{k}^{(2)}$ has a simple limit
\be 
\lim_{k \to \infty} \Delta E_{k}^{(2)} = \frac{1}{\pi^2} - \frac{1}{8} = -0.0236788 \ldots 
\ee

The decay exponent $\beta(N,W)=\sum\limits_{k=0}^{N-1} \epsilon_{k}^W$ is then obtained, in the small-$W$ limit with fixed $N$, as
\be 
\beta(N,W) = \frac{N^2}{2} -  W \frac{ \sqrt{2}}{3 \sqrt{\pi}} \frac{1}{2^{2N} } \frac{N (2N+1)!}{(N!)^2}  + W^2 \left(\sum_{k=0}^{N-1} \Delta E_{2k+1}^{(2)}   + \frac{N}{8}\right) + \mathcal{O}(W^3) \;.
\ee
Note that we recover the limit result of \myeqref{eq:2:DecayExponentViciousWalkersFixedWall}. Specialising this formula to the case $N=1$, we obtain 
\be  
\beta(N=1,W) = \frac{1}{2} - \frac{W}{\sqrt{2 \pi}} + \frac{1-\ln 2}{\pi}  W^2 +  \mathcal{O}(W^3)   
\ee
which is in agreement with the first order result $\beta(1,W) =\frac{1}{2} - \frac{W}{\sqrt{2 \pi}} + {\cal O}(W^2)$ obtained in \cite{Krapivsky1996}. Note that the constant coefficient $\frac{1}{2}$ is the universal exponent discussed in length in section \ref{subsec:2.2.1}. Details on the Taylor coefficients for general $N$ are given in \pubref{publication:NonCrossingBrownianDBM}.

\paragraph{Semi-classical scaling} We consider now the limit when $N$ and $\abs{W}=\bigO(\sqrt{N})$ are simultaneously large. In this regime, the quantum system can be analyzed in a semi-classical approximation, which is asymptotically exact.

In this large-$N$ limit, the key quantity is the density $\rho(\epsilon)$ of single-particle energy levels of the system with Hamiltonian $\hat{H}^W$ in \myeqref{eq:2:HamiltonianQuadraticHardWall}. Formally, it is defined by
\be
\rho(\epsilon) = \sum_{k \geqslant 0} \delta(\epsilon - \epsilon_k^W ) \; .
\ee
 Denoting $\epsilon_F$ the highest occupied level in the $N$-particle ground-state (or Fermi energy), we have
\be 
N = \int_0^{\epsilon_F} \dd \epsilon \ \rho(\epsilon) \; .
\label{eq:2:ConditionFermiEnergyIntegral}
\ee
The decay exponent $\beta(N,W)$ is the ground-state energy, given by
\be 
\beta(N,W) = \int_0^{\epsilon_F} \dd \epsilon \ \epsilon \ \rho(\epsilon) \; .
\label{eq:2:BetaFromLevelDensity}
\ee
Hence $\beta(N,W)$ is obtained from the knowledge of the level density $\rho$, by eliminating $\epsilon_F$ from these two equations \eqref{eq:2:ConditionFermiEnergyIntegral} and \eqref{eq:2:BetaFromLevelDensity}.

The density of states can be approximated semi-classically in the large-$N$ limit \cite{Basdevant2002,Voros1976,Abe1984,Eckerta2014}. Under the semi-classical approximation, $\epsilon_k$, the $k$\textsuperscript{th} energy level of a quantum particle under potential $V$, satisfies the Bohr-Sommerfeld quantization condition
$ \int_{X_1}^{X_2} p(X) \dd X = k \pi $ with the classical momentum $p(X) = \sqrt{2 (\epsilon_k^W - V(X))}$ and where $X_1<X_2$ are the turning points of the classical trajectories. 

In our problem, the potential can be approximated to $V(X) \simeq \frac{X^2}{8}$ since the integrals are dominated by large values of $X \sim \sqrt{N}$ and we have $X_1=- \sqrt{8 \epsilon}$ and $X_2= \min( \sqrt{ 8 \epsilon} ,W)$, as illustrated in Fig.~ \ref{fig:SemiClassicalPotentialW}. For the $k$\textsuperscript{th} level, this leads to the condition
\be 
\int _ {- \sqrt{ 8 \epsilon_k^W}}^ { \min( \sqrt{ 8 \epsilon_k^W} ,W) } \sqrt { 2 \left( \epsilon_k^W - \frac { X ^ { 2 } } { 8 } \right) } \  \dd X = k \pi  \;.
\label{eq:2:SemiClassicalCondition}
\ee 
\begin{figure}[t!]
    \centering
    \begin{subfigure}[t]{0.5\textwidth}
        \centering
    \includegraphics[height=5.5cm]{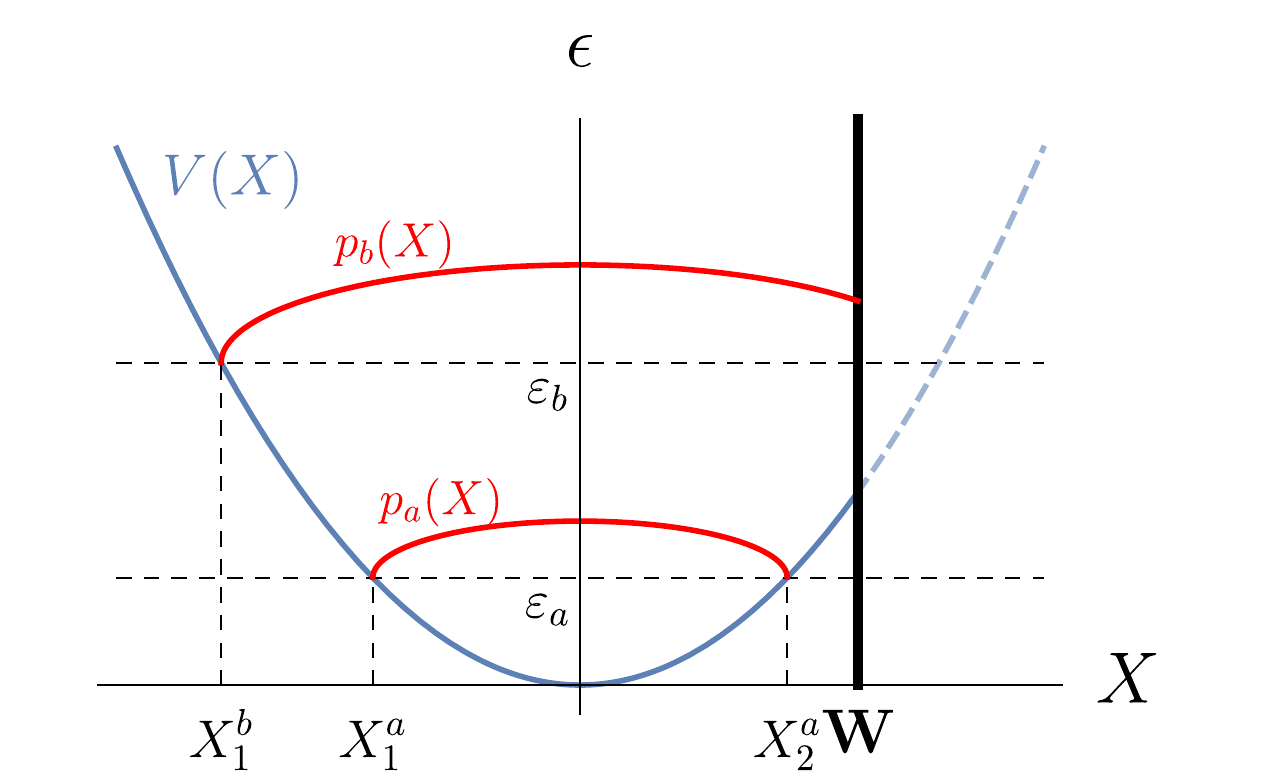} 
    \caption{$W>0$}
    \end{subfigure}%
    ~ 
    \begin{subfigure}[t]{0.5\textwidth}
        \centering
    \includegraphics[height=5.5cm]{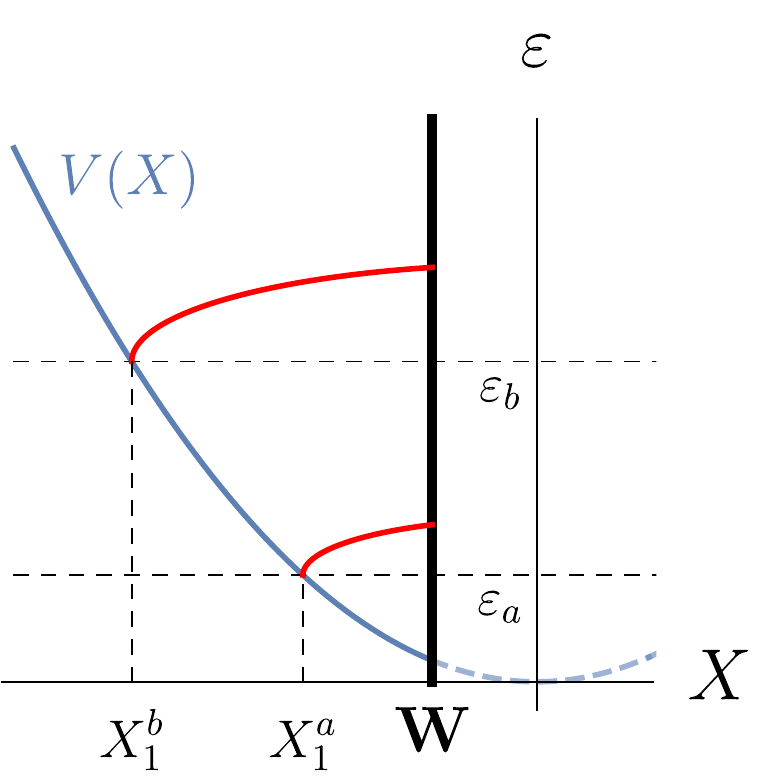} 
    \caption{$W<0$}
    \end{subfigure}
    \caption{Quadratic potential $V(X)$ with a hard-wall in $W$ for \emph{(a)} $W>0$ and \emph{(b)} $W<0$. For a given energy $\epsilon$, the red line shows the semi-classical momentum $p(X)$. This function is supported on $[X_1, X_2]$ with $X_1 = - \sqrt{8 \epsilon} $ and $ X_2 = \min( \sqrt{8 \epsilon},W)$.}
    \label{fig:SemiClassicalPotentialW}
\end{figure}

From this condition, we obtain in the continuum limit the approximate semi-classical density of states as
\be
\rho( \epsilon) \simeq  \rho^{\mathrm{sc}}( \epsilon) =  \frac{1}{\sqrt{2} \pi} \int _ {- \sqrt{ 8 \epsilon} }^ { \min( \sqrt{ 8 \epsilon} ,W) } \frac{{d}X}{ \sqrt {  \epsilon - \frac { X ^ { 2 } } { 8 }  } } \; .
\ee
This integral can be performed explicitly. Denoting $\Theta = \1_{\R^+}$ the Heaviside function, one obtains
\be
\rho^{\rm sc}(\epsilon) =
\left\{
    \begin{array}{ll}
        \displaystyle 2 \, 
        \Theta(W) & \mbox{for } \epsilon < \frac{W^2}{8} \; , \\[5pt]
        \displaystyle 1 + \frac{2}{\pi} \arcsin \left( \frac{W}{\sqrt{8 \epsilon}} \right)   & \mbox{for } \epsilon > \frac{W^2}{8} \; . 
    \end{array}
\right.
\ee
The semi-classical density depends on the sign of $W$ in the first line, for small $\epsilon$. It is the density of the harmonic oscillator $\rho_{W=+ \infty}=2$ if $W$ is positive; and zero if $W$ is negative such that the low-lying states are forbidden by the hard wall. Note that the density is continuous at $\epsilon = \frac{W^2}{8}$ in both cases, and that it converges as $\epsilon \to \infty$ to the density of the harmonic oscillator with a centered hard-wall $\rho_{W=0}=1$, which is half of the regular harmonic oscillator density. The semi-classical density of states is illustrated in Fig.~\ref{fig:SemiClassicalDensity}.

 \begin{figure}[ht!]
    \centering 
    \includegraphics[width=.6 \textwidth]{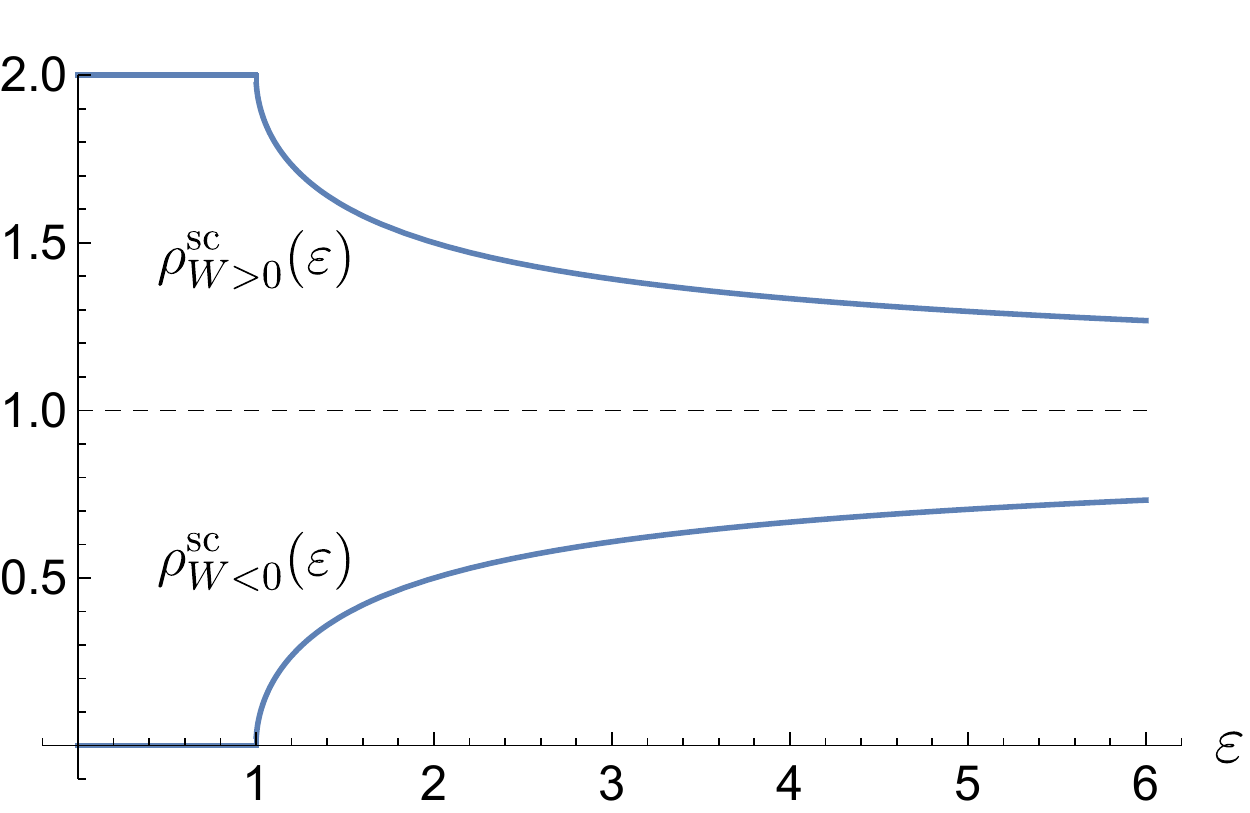} 
    \caption{Plot of the semi-classical density of states $\rho^{\rm sc}_W(\epsilon)$ versus the energy $\epsilon$ in units of $\frac{W^2}{8}$, in both cases $W >0$ and $W<0$. For large $\epsilon$, both branches converge to $\rho_{W=0} = 1$.}
    \label{fig:SemiClassicalDensity}
\end{figure}

The decay exponent $\beta(N,W)$ can be characterized from relations \eqref{eq:2:ConditionFermiEnergyIntegral} and \eqref{eq:2:BetaFromLevelDensity}:
\begin{itemize}
\item If $ W \geqslant \sqrt{4 N} $, we obtain $\epsilon_F=\frac{N}{2} \leqslant \frac{W^2}{8}$ such that the system does not feel the wall. The decay exponent is then
\bea 
\beta(N,W) \simeq \epsilon_F^2 = \frac{N^2}{4} \; .
\eea 
\item If $W < \sqrt{4 N}$, either positive or negative, then $\epsilon_F > \frac{W^2}{8}$ and the wall has an influence on the system. One then obtains the following system
\be
\begin{split}
&N =\frac{W \sqrt{8 \epsilon_F -W^2}}{4 \pi }+\frac{2 \epsilon_F  \arcsin \left(\frac{W}{2 \sqrt{2 \epsilon_F
   }}\right)}{\pi }+\epsilon_F \;,  \\[3pt]
& \beta(N, W) \simeq \frac{N \epsilon_{F}}{2}-\frac{W}{96 \pi}\left(8 \epsilon_{F}-W^{2}\right)^{\frac{3}{2}} \;.
\end{split}
\ee 
\end{itemize}
In conclusion, the decay exponent $\beta(N,W)$ takes the following scaling form in the limit 
$(N,W) \to +\infty$ with $y=\frac{W}{\sqrt{4 N}}$ fixed:
\emphbe 
\beta(N,W) \simeq \frac{N^2}{4} \ {\sf b}\left( \frac{W}{\sqrt{4 N}}\right)  
\end{empheq}
with the scaling function $\sf b$ is given by
\be
\left\{
    \begin{array}{ll}
        \displaystyle {\sf b}(y)=1 & \mbox{for }  y \geqslant 1\; , \\[5pt]
         {\sf b}(y) = y^2 \left( \frac{1}{u^2} - \frac{2}{3 \pi} y^2  \frac{(1-u^2)^{3/2}}{u^3}  \right)  &  \mbox{for }  y < 1\;   \; ,
    \end{array}
\right.
\ee
where in the second line $u= \frac{W}{\sqrt{8 \epsilon_F}} \in [-1,1]$ is obtained in terms of $y$ by solving 
\be 
 \frac{1}{y^2} = \frac{1}{2u^2}+ \frac{\sqrt{1-u^2}}{\pi u} +  \frac{\arcsin(u)}{\pi u^2}  \;. 
\ee 
The semi-classical scaling function $\sf b$ is plotted numerically in Fig.~\ref{fig:SemiClassicalScalingFunction}, along with its asymptote $2 y^2 + \frac{3}{5} \sqrt[3]{2} (3 \pi )^{2/3} y^{2/3}$ for large negative argument. Details on the asymptotics of ${\sf b}(y)$ as $y \to 1^-$, $y\to 0$ and $y \to -\infty$ can be found in \pubref{publication:NonCrossingBrownianDBM}. In particular, there is a $\frac{5}{2}$\textsuperscript{th} order non-analyticity at $y=1$, where the wall starts to be felt by the system. 

 \begin{figure}[ht!]
    \centering 
    \includegraphics[width=.6 \textwidth]{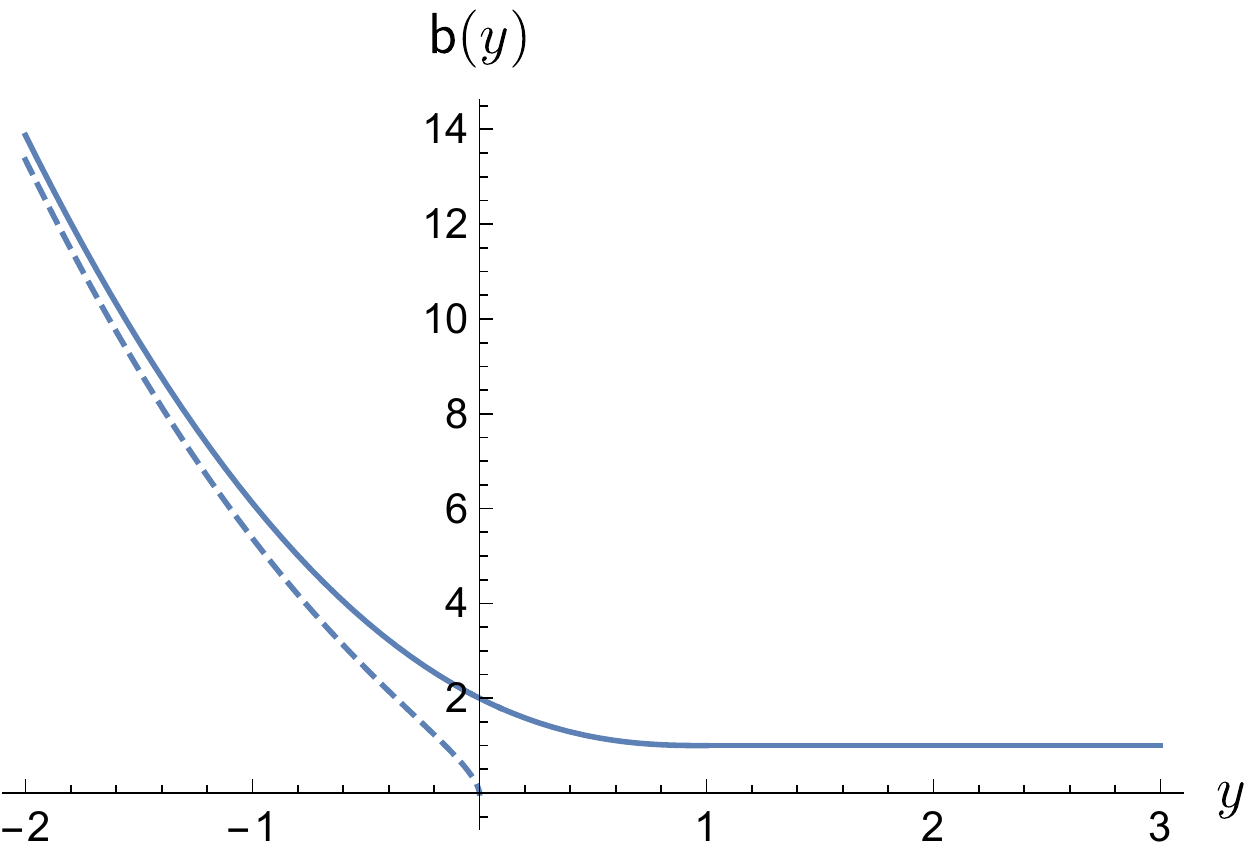} 
    \caption{Plot of the scaling function ${\sf b}(y)$ versus $y=\frac{W}{\sqrt{4N}}$. The dashed line is the asymptote curve as $y \to - \infty$.}
    \label{fig:SemiClassicalScalingFunction}
\end{figure}
 
In the last sections, we have shown how the knowledge of the decay exponent can be refined in the cases of slow barrier and large-$N$ limit with $W = \bigO(\sqrt{N})$. Other limits can also be investigated from the asymptotics of roots of parabolic cylinder functions, as the large-$N$ limit for fixed $W$, see \pubref{publication:NonCrossingBrownianDBM}.

\paragraph{Algebraic prefactor} 
The prefactor in the algebraic decay of the survival probability is the function $A(\vec{x}, t_0)$ defined in \myeqref{eq:2:PrefactorSurvivalProbability}. We can re-write it, from the Slater determinant form of $\Psi$, as a function of the single-particle wavefunctions:
\be
A(\vec x ,t_0) =  \det_{1 \leqslant i,j \leqslant N}  \left[ {\phi_{i-1}^W} ^* (\frac{x_j}{\sqrt{t_0}})  e^{\frac{{x}_j^2}{4 t_0}  } \right]     t_0^{ \epsilon^W_{\vec{0}}  }  \left(  \int\limits_{\Weyl_N^W}  {d}\vec{Y}  \   \det_{1 \leqslant i,j \leqslant N} \left[   \phi_{i-1}^W (Y_j) e^{-\frac{Y_j^2}{4} }  \right] \right)
\ee
A multiple integral involving a single determinant such as the one appearing above was computed by de Bruijn in \cite{DeBruijn1955}, in terms of the Pfaffian of an $N\times N$ matrix. Assuming $N$ even, and with the sign function $\mathrm{sgn}(z)= \1_{z>0} - \1_{z<0}$, de Bruijn's formula gives  
\begin{gather}
 \int\limits_{\Weyl_N^W}  \dd \vec{Y}  \   \det_{1 \leqslant i,j \leqslant N} \left[   \phi_{i-1}^W (Y_j) \, e^{-\frac{Y_j^2}{4} }  \right]  =  \underset{1 \leqslant i,j \leqslant N}{\Pf}  \AAA_{i,j} \\[5pt]
 \text{where} \quad \AAA_{i,j} =    \iint\limits_{\substack{Y<W \\ Z<W }}   \dd Y \dd Z \    \phi^W_{i-1} (Y) \phi_{j-1}^W (Z)  \ e^{-\frac{Y^2+ Z^2}{4} } \ \mathrm{sgn}(Y-Z)   \; .
 \label{eq:2:PfaffianPrefactorSurvivalProba}
\end{gather} 
If $N$ is odd, this can also be written as a Pfaffian, with the subtlety that the last column and row are different from the general term, see \cite{DeBruijn1955}. For explicit computations of the matrix $\AAA$ and its Pfaffian in the simple cases $W=+\infty$ and $0$, see \pubref{publication:NonCrossingBrownianDBM}.

The prefactor $A(\vec{x}, t_0)$ can thus be expressed in terms of a determinant, which depends on the initial condition at time $t_0$, and a Pfaffian, which depends on the eigenfunctions below $W$. As a consequence, the initial condition details disappear in the decay exponent $\beta(N,W)$ but remain in this prefactor.

\subsubsection{Joint distribution at large time}

Further from the survival probability results discussed in the previous section, the quantum point of view on the problem of vicious walkers under a square-root boundary provides the joint distribution of the surviving walkers, at large time.

\paragraph{Arbitrary $W$}

At large time, from approximation \eqref{eq:2:LargeTimePropagatorApproximation}, the stochastic propagator is obtained from the quantum propagator in \myeqref{eq:2:SummaryStochasticQuantumPropagators} such that
\be
P\left( \vec{y}, t , \NC_{[t_0,t]} , \Bevent_{[t_0,t]} \mid \vec{x}, t_0 \right)  
\simeq 
\frac{ N! \,  t_0^{\epsilon_{\vec{0}}^W } }{t^{\epsilon_{\vec{0}}^W +\frac{N}{2}  }} 
\ \Psi_{\vec{0}}^W(\frac{\vec{y}}{\sqrt{t} }) \
{\Psi_{\vec{0}}^W}^*(\frac{\vec{x}}{\sqrt{t_0}})  \
e^{ \sum_{i=1}^{N} \frac{x_i^2}{4t_0}  - \frac{y_i^2}{4 t}  }
\ee
where the additional factor of $t^{N/2}$ arises from the Jacobian of the transformation from $Y_i$ to $y_i/\sqrt{t}$. With the expression of the single-particle wavefunctions for general $W$ in \myeqref{eq:2:GeneralWSingleParticleWavefunction}, this propagator can be expressed formally in terms of parabolic cylinder functions as
\be
\begin{split}
& P\left( \vec{y}, t , \NC_{[t_0,t]} , \Bevent_{[t_0,t]} \mid \vec{x}, t_0 \right)  \\
& \quad \simeq 
\frac{C}{t^{\frac{N}{2}}}
\left(\frac{t_0}{t}\right)^{\beta(N,W)} 
 \det_{1 \leqslant i,j \leqslant N} D_{2 \epsilon_{i-1}^W}( -\frac{y_j}{\sqrt{t}}) \
 \det_{1 \leqslant i,j \leqslant N} {D_{2 \epsilon_{i-1}^W}}^*(- \frac{x_j}{\sqrt{t_0}}) \
e^{ \sum_{i=1}^{N} \frac{x_i^2}{4t_0}  - \frac{y_i^2}{4 t}  }
\end{split}
\label{eq:2:StochasticPropagatorLargeTime}
\ee
with $C= \prod_{i=0}^N C_k^2$ from the normalization constants defined in \myeqref{eq:2:GeneralWSingleParticleWavefunction}. As a consequence, keeping only terms that depend on $\vec{y}$, the joint distribution of surviving walkers at time $t$ is given by
\be 
 P\left( \vec{y}, t  \mid \NC_{[t_0,t]} , \Bevent_{[t_0,t]} , \vec{x}, t_0 \right) =
 K \
  e^{ -  \frac{1}{4t}\sum_{i=1}^{N}  y_i^2  } \
 \det_{1 \leqslant i,j \leqslant N} D_{2 \epsilon_{i-1}^W}( -\frac{y_j}{\sqrt{t}})  \; .
 \label{eq:2:JPDFSurvivingViciousWalkersLargeTime}
\ee
The normalization constant $K$ is the ratio of the terms in \myeqref{eq:2:StochasticPropagatorLargeTime} that do not depend on $\vec{y}$ over the survival probability $S\left(t \mid \vec{x}, t_0 \right)$, and decays as $t^{-\frac{N}{2}}$ such that the joint distribution is normalized.

The JPDF of surviving vicious walkers in \myeqref{eq:2:JPDFSurvivingViciousWalkersLargeTime} can be evaluated in the cases where the wavefunctions are simple.

\paragraph{$W \to +\infty$} In this case, the wavefunctions are simply those of the harmonic oscillator, see \myeqref{eq:1:HarmonicWavefunctions}, such that the JPDF is
\bea
P\left( \vec{y}, t  \mid \NC_{[t_0,t]} , \Bevent_{[t_0,t]} , \vec{x}, t_0 \right)  &\propto &
  e^{ -  \frac{1}{4t}\sum_{i=1}^{N}  y_i^2  } \
 \det_{1 \leqslant i,j \leqslant N} \left( H_{i-1}( -\frac{y_j}{\sqrt{t}})  e^{-\frac{y_j^2}{4t}} \right) \\
  &\propto & 
   \abs{\Delta(\vec{y}) }   \   e^{ -  \frac{1}{2t}\sum_{i=1}^{N}  y_i^2  }  \; .
\eea
In the boundary-less setting $W \to + \infty$, we recover the GOE eigenvalue JPDF as discussed in section \subsubref{subsubsec:II.1.2.b}{II.1.2} and proven in \myeqref{eq:2:ProofNonCrossingGOE}.

\paragraph{$W = 0$} In this case, the wavefunctions are the odd ones of the harmonic oscillator, such that the JPDF is
\be
P\left( \vec{y}, t  \mid \NC_{[t_0,t]} , \Bevent_{[t_0,t]} , \vec{x}, t_0 \right)  \propto 
  e^{ -  \frac{1}{4t}\sum_{i=1}^{N}  y_i^2  } \
 \det_{1 \leqslant i,j \leqslant N} \left( H_{2i-1}( -\frac{y_j}{\sqrt{t}})  e^{-\frac{y_j^2}{4t}} \right)
\ee
and can be shown, see \pubref{publication:NonCrossingBrownianDBM}, to be proportional to
\be  
P\left( \vec{y}, t  \mid \NC_{[t_0,t]} , \Bevent_{[t_0,t]} , \vec{x}, t_0 \right) 
 \propto 
 \prod_{i=1}^N \abs{ y_i } \ 
   \abs{\Delta\left(\overrightarrow{y^2}\right) }   \   e^{ -  \frac{1}{2t}\sum_{i=1}^{N}  y_i^2  } 
   \label{eq:2:JPDFSurvivingViciousWalkersWZero}
\ee
where 
\be 
\abs{\Delta\left(\overrightarrow{y^2}\right) } = \prod_{1 \leqslant i < j \leqslant N} \left( y_j^2 - y_i^2 \right) \; .
\ee
Note that this JPDF \eqref{eq:2:JPDFSurvivingViciousWalkersWZero} is identical to the one
for the eigenvalues $\lambda_i$ of the LOE, see \myeqref{eq:1:JPDFEigenvaluesWishartEnsemble}, with the correspondence $\lambda_i = y_i^2$ and by fixing $T= N+1$. In the large-$N$ limit, this yields $q=1$ such that the limit density in the bulk is given by the quarter-circle distribution \eqref{eq:1:QuarterCircleDistribution}, with $\sigma = \sqrt{t}$. Similar results were obtained in
\cite{Katori2003}, where vicious walkers were studied under a fixed wall. 
 
\subsection[Application: Dyson Brownian motion]{Application: Dyson Brownian motion under an absorbing boundary}
\label{subsec:2.2.3}

As discussed at the beginning of this chapter, non-crossing walkers share deep links with the Dyson Brownian motion of index $\beta=2$. The results presented above for the behaviour of non-crossing walkers under a moving boundary can, as a consequence, be applied to the DBM-2 under the same boundary $g(t) = W \sqrt{t}$.

The process of interest is now $(\lambda_{i,t})_{1 \leqslant i \leqslant N, t \geqslant 0}$, the eigenvalues of $\HH_t^2$ in \myeqref{eq:2:DefDBMGUE}. Since the DBM eigenvalues almost-surely never touch, the only source of danger for the walkers is the vicious moving boundary. We denote the survival probability of this system as
\be 
S_{DBM}
(t \mid \vec{x} , t_0) = \Pr_{DBM} \left( \Bevent_{[t_0,t]}  \mid \vec{\lambda}_{t_0} = \vec{x} \right)
\label{eq:2:SurvivalProbabilityDBM}
\ee
where the notations are similar to those of the previous sections.
 
\subsubsection{Square-root boundary}

Recall the mapping between the propagators of vicious walkers and the DBM-2 in \myeqref{eq:2:PropagatorRelationViciousWalkerDBM2}. One can show that this propagator relation holds in the presence of a moving boundary as well, see Appendix F of \pubref{publication:NonCrossingBrownianDBM}, such that 
\be
P_{DBM}( \vec{y},t , \Bevent_{[t_0,t]}  \mid \vec{x}, t_0)  = \frac{\Delta(\vec{y})}{\Delta(\vec{x})}  \  P(\vec{y},t,\NC_{[t_0,t]}, \Bevent_{[t_0,t]}  \mid \vec{x}, t_0)  \; .
\label{eq:2:relationPropagatorViciousWalkerDBMwithMovingBoundary}
\ee
The large-time expression of the non-crossing propagator in \myeqref{eq:2:StochasticPropagatorLargeTime} directly gives the large-time expression of the DBM propagator under the boundary. This can be used to characterize the decay of the survival probability $S_{DBM}(t \mid \vec{x} , t_0)$ defined in \myeqref{eq:2:SurvivalProbabilityDBM}, since
\be 
S_{DBM}(t \mid \vec{x} , t_0) = \int_{\Weyl_N^{W\sqrt{t}}} \dd \vec{y}  \ P_{DBM}( \vec{y},t , \Bevent_{[t_0,t]}  \mid \vec{x}, t_0) \; .
\ee
One finds from the results of the previous section that the large-time decay of the DBM-2 survival probability is
\begin{empheq}[box=\setlength{\fboxsep}{8pt}\fbox]{equation}
S_{DBM}(t \mid \vec{x} , t_0)  \  \propto \  t^{  -  \beta(N,W) + \frac{N(N-1)}{4} } 
\end{empheq}
The decay exponent
\be \beta_{DBM}(N,W) =  \beta(N,W)- \frac{N(N-1)}{4} \ee is expressed from $\beta(N,W)$, the exponent studied in the previous section and defined in \myeqref{eq:2:ResultLargeTimeSurvivalProbabilityDecay}, by removing the Fisher exponent \eqref{eq:2:DecayExponentWithoutWall}. We note that the large-$W$ limit is $ \beta_{DBM}(N,W)  \xrightarrow{W \to + \infty} 0$ such that the survival probability does not decay in this limit, as expected.

The expression of the large-time propagator is useful to characterize the joint distribution of surviving DBM particles under the boundary, as was done for vicious walkers in the previous section. In the large $W$-limit, one recovers easily the usual GUE distribution for the DBM-2 without a barrier. In the next section, we dwell on the more interesting case $W=0$.
 
\subsubsection{Fixed boundary}

\paragraph{Biorthogonal ensemble}

In the special case $W=0$, i.e. with fixed boundary at 0, the DBM propagator at large time is obtained from the computations of the previous section as
\be 
\begin{split}
P_{DBM}( \vec{y},t , \Bevent_{[t_0,t]}  \mid \vec{x}, t_0)   \simeq
 C \ e^{-\frac{1}{2t} \sum\limits_{i=1}^{N}  y_i^2    } \  
 \prod_{1 \leqslant i \leqslant N} \abs{y_i}  \  \prod _ { 1 \leqslant i < j \leqslant N } (  y _ { j } - y _ { i } ) \ ( y_j^2 - y_i^2)  \\ \times
  \prod_{1 \leqslant i \leqslant N} \abs{x_i}   \prod\limits_{1\leqslant i < j \leqslant N} \abs{ x_{j} + x_{i} }  \  \ t^{-N(N + \frac{1}{2})} 
\end{split}
\ee
with $C$ the constant appearing in \myeqref{eq:2:StochasticPropagatorLargeTime}. 

As a consequence, the JPDF of surviving particles is given by
\be 
P_{DBM}( \vec{\lambda},t  \mid  \Bevent_{[t_0,t]} , \vec{x}, t_0)  = K \
e^{-\frac{1}{2t} \sum\limits_{i=1}^{N}  \lambda_i^2    } \  
 \prod_{1 \leqslant i \leqslant N} \lambda_i  \  \prod _ { 1 \leqslant i < j \leqslant N } (  \lambda _ { j } - \lambda _ { i } ) \ ( \lambda_j^2 - \lambda_i^2)
 \label{eq:2:JPDFSurvivingDBMWZero}
\ee
with $K$ a normalization constant. The process is restricted to $(\R^-)^N$ by the boundary, but we choose for simplicity to reflect the process in the following and consider $\vec{\lambda} \in (\R^+)^N$, i.e. DBM-2 eigenvalues over the fixed boundary. We have accordingly removed the absolute values in \myeqref{eq:2:JPDFSurvivingDBMWZero}.

In order to shed light on this process, we change variables to $z_i = \lambda_i^2 / 2 t$, with corresponding JPDF
\be  
P_{\mathrm{biorthL}}(\vec{z}) =K' \ e^{- \sum_{i=1}^{N}  z_i    }   \prod _ { 1 \leqslant i < j \leqslant N } \left( z_ { j } ^{\frac{1}{2}} - z_ { i }^{\frac{1}{2}} \right) (z_j - z_i)  \;,
\ee 
supported on $ ( \R^+)^N  $. Joint distributions of the type of $P_{\mathrm{biorthL}}$ are known to belong to the so-called \emph{biorthogonal ensembles} \cite{Muttalib1995,Borodin1998,Dolivet2007}, which are a generalization of the orthogonal ensembles, i.e. $\beta$-ensembles with $\beta=2$ solvable in terms of orthogonal polynomials as presented in chapter \ref{chap:1}, with different exponents in the two Vandermonde terms.

The specific joint probability distribution $P_{\mathrm{biorthL}}$ coincides with the one of the \emph{biorthogonal Laguerre ensemble} defined in \cite{Borodin1998}, with weight  function $\omega(z) = e^{-z}$ and parameters $\alpha = 0$ and $\theta=\frac{1}{2}$ in the notations of \cite{Borodin1998}. As a consequence, $\vec{z}$ forms a determinantal point process with a kernel $K_N$ given by 
\be  
K_{N}(z, z')= \frac{1}{2} \sum_{k, i=0}^{N-1} \sum_{r=i}^{N-1} \frac{\Gamma\left(N+2(i+1)\right)}{\Gamma(\frac{k}{2}+1) \Gamma\left(2 (i+1)\right)} \,  \frac{(-1)^{i+k}}{k !(N-k-1) ! i !(r-i) !} \frac{z^{\frac{k}{2}} (z')^{r}}{\frac{k}{2}+i+1} e^{-\frac{z+z'}{2}} \;.
\ee
All characteristics of surviving DBM-2 particles over a fixed absorbing boundary can be obtained from this kernel. In the next section, we study in detail the density of the system, with a large number of particles.

\paragraph{Large-$N$ density}

This kernel provides the exact average density of DBM-particles over the fixed absorbing boundary as 
\be
\rho( \lambda ) = \frac{ \lambda}{t} \  \rho_z \left( \frac{\lambda^2}{2t} \right) 
\label{eq:2:DefDensityDBMOverBoundary}
\ee
from the density of the $\vec{z}$-process
\be 
\rho_z(z) = \frac{1}{N} \ K_N(z,z) \; .
\ee
The density $\rho$ is plotted for a few values of the number of particles $N$ in Fig.~\ref{fig:ExactDensityDBMFixedBoundary}.

 \begin{figure}[ht!]
    \centering 
    \includegraphics[width=.8 \textwidth]{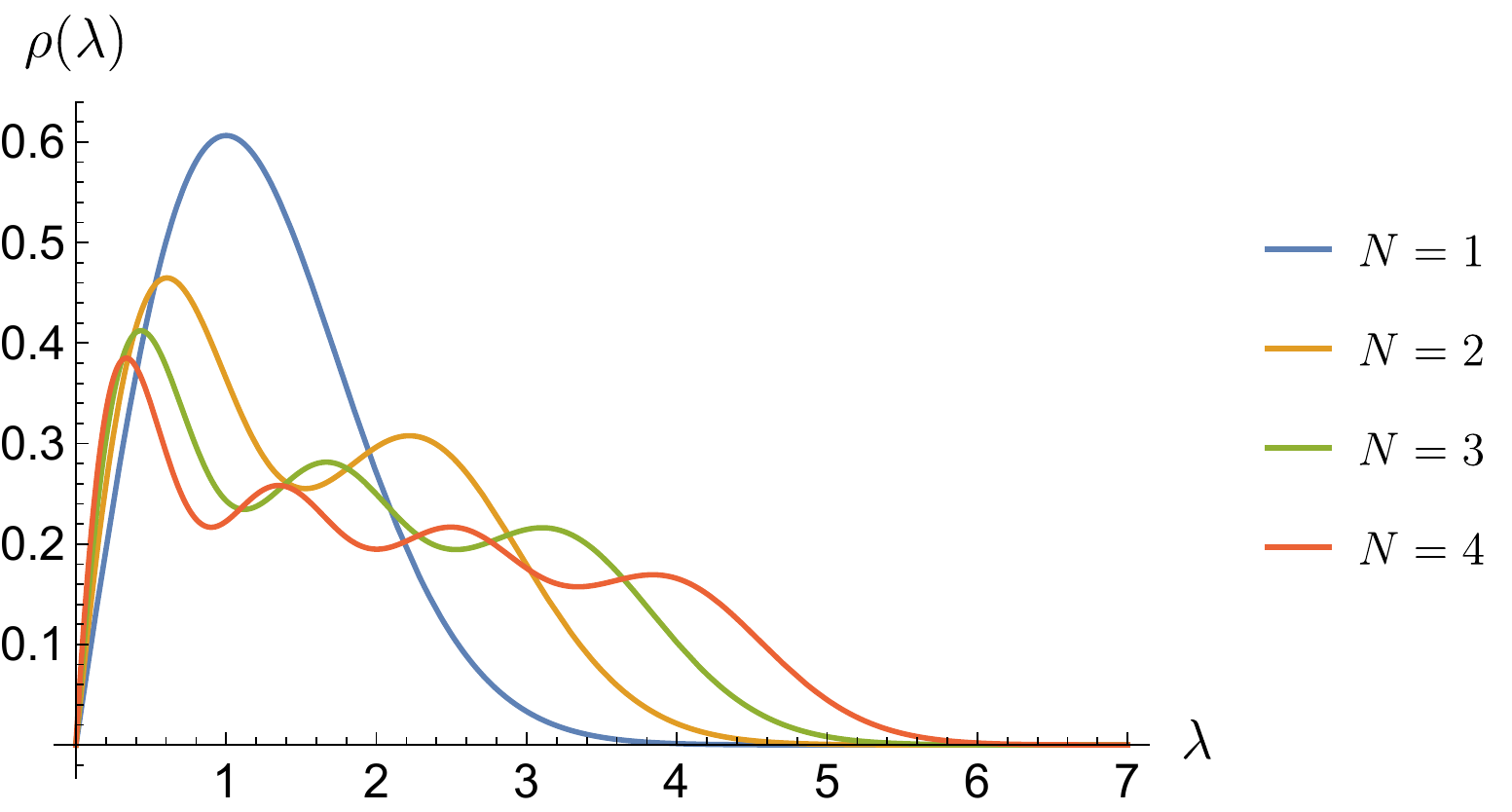} 
    \caption{Density $\rho(\lambda)$ of the surviving DBM particles over a fixed absorbing boundary, defined in \myeqref{eq:2:DefDensityDBMOverBoundary}, for various values of $N$ and at $t=1$.}
    \label{fig:ExactDensityDBMFixedBoundary}
\end{figure}

In the large-$N$ limit, the biorthogonal Laguerre ensemble has the same order-$N$ scaling as the LUE and the $\lambda$ particles thus scale as $\sqrt{N}$. More precisely, the DBM particles are typically found on the interval $[0, \sqrt{2Nt}L]$, between the hard edge at $0$ and the soft edge at $\sqrt{2Nt}L$ with $L = \left( \frac{3}{2} \right)^{\frac{3}{2}}$. We give in the following the scaling form of the density in the bulk and at the hard edge, pointing to \pubref{publication:NonCrossingBrownianDBM} for further details on the derivations and the matching of the two regimes as a power-law with exponent $-\frac{1}{3}$.

\subparagraph{In the bulk} At large $N$, with $\lambda = \bigO( \sqrt{2Nt}) $, the average density in the bulk takes the following scaling form:
\be \
\rho(\lambda) = \frac{1}{\sqrt{2 N t}} \  r_{\rm bulk} \left( \frac{1}{\sqrt{2 N t} } \ \lambda \right)
\ee
where the scaling function $r_{\rm bulk} $ has a finite support $[0,L]$ with $L = \left( \frac{3}{2} \right)^{\frac{3}{2}}$, and such that $\int_{0}^{\sqrt{2Nt}L}\rho(\lambda) \dd \lambda=1$. The scaling function can be computed by a Coulomb gas method, see \pubref{publication:NonCrossingBrownianDBM} and \cite{Borot2012,Claeys2014,Nadal2011}, as
\be
 r_{\rm bulk} (\lambda) =
\frac{1}{2 \pi \sqrt{2} }   \left(  g_-(\lambda) - g_+(\lambda) + 3 \sqrt{1 - \frac{\lambda^2}{L^2} } \, \bigg( g_-(\lambda) + g_+(\lambda) \bigg) \right) 
\label{eq:2:rBulkDefinition}
\ee
with
\be
g_\pm(\lambda) = \left( \frac{L}{\lambda} \pm \sqrt{\frac{L^2}{\lambda^2} -1 }\right) ^{1 /3}   \;.
\ee
The bulk density scaling function  $r_{\rm bulk}$ for the DBM-2 over an absorbing boundary is plotted in Fig.~\ref{fig:ScalingBulkDBMBoundary}.

 \begin{figure}[ht!]
    \centering 
    \includegraphics[width=.6 \textwidth]{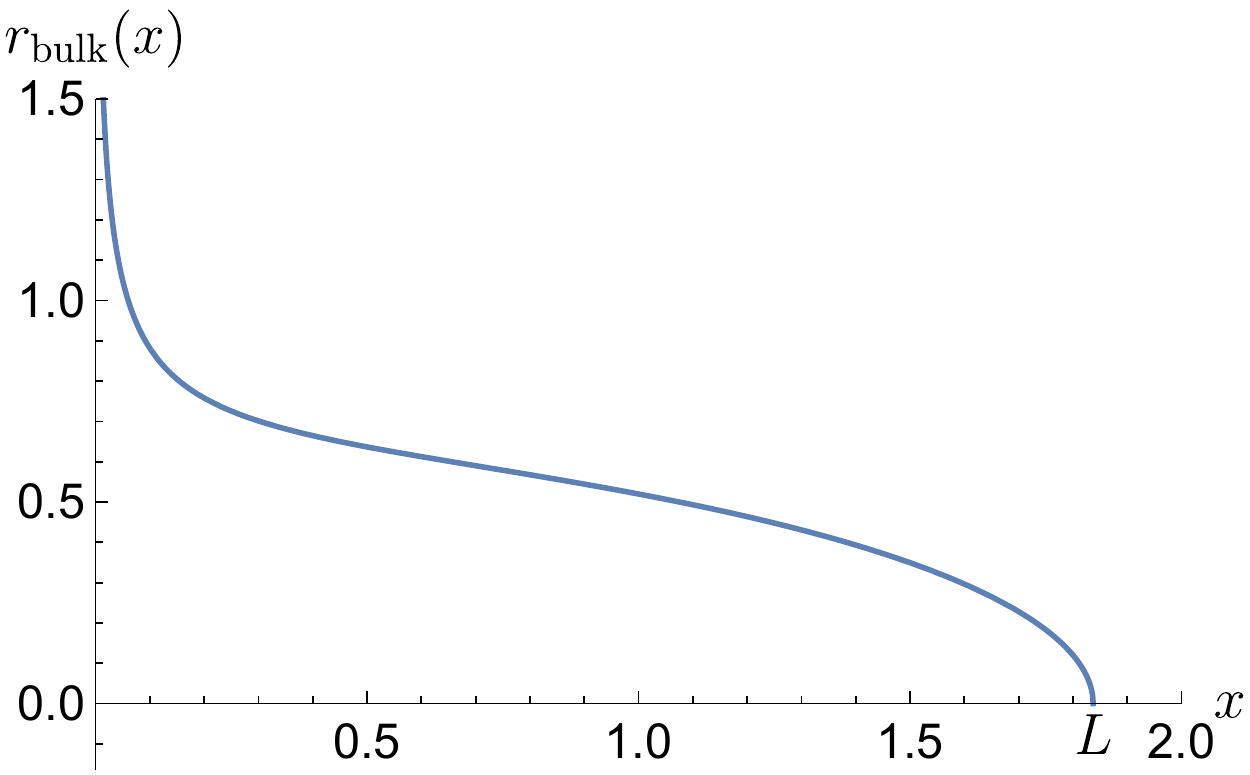} 
    \caption{Large-$N$ bulk density of the surviving DBM particles over an absorbing boundary at $0$, given by the scaling function $r_{\rm bulk}$ in \myeqref{eq:2:rBulkDefinition}.}
    \label{fig:ScalingBulkDBMBoundary}
\end{figure}

\subparagraph{At the hard edge}
 
Close to the hard edge, the density can be obtained from the limiting kernel of the biorthogonal Laguerre ensemble \cite{Borodin1998}. In contrast with the LUE behaviour discussed in \myeqref{eq:1:LimitBesselKernelHardEdgeLUE}, the limiting $\vec{z}$-process is obtained after rescaling by $N^2$, such that the hard-edge regime concerns the DBM particles at scale $\lambda \sim \frac{\sqrt{2t}}{N}$.
 
At this scale such that $\lambda \sim \sqrt{2t}/N$, the density is given by
\be 
\rho(\lambda) = \frac{1}{\sqrt{2t}} \ r_{\rm edge} \left( \frac{ N}{\sqrt{2t}} \ \lambda \right) 
\label{eq:2:DensityHardEdgeDBMWall}
\ee
where $\ r_{\rm edge} (x) = 2 x \, \rho_{{\rm edge} , z} (x^2)$ depends on the density of the rescaled $\vec{z}$-process, expressed in the $\tilde z = z / N^2$ variables as
\be 
\rho_{{\rm edge} , z} (\tilde z) =  \mathcal{K} ^{\rm edge}(\tilde z, \tilde z)
\ee
with the hard-edge limiting kernel \cite{Borodin1998} 
 \be  
\mathcal{K}^{\rm edge} (\tilde z, \tilde z') =
\sum_{j,k=0}^{\infty} \frac{(-1)^{k} \tilde z^{k}}{k ! \Gamma\left(2(1+k)\right)} \frac{(-1)^{j} (\tilde z')^{\frac{j}{2}}}{j ! \Gamma(1+\frac{j}{2})} \frac{1}{2(1+k)+ j} \; .
 \ee
 The scaling function $r_{\rm edge}$ and the hard-edge density $\rho_{{\rm edge},z}$ of the rescaled $\vec{z}$ are plotted in Fig.~\ref{fig:ScalingHardEdgeDBMBoundary}. Note that the scaling form \myeqref{eq:2:DensityHardEdgeDBMWall} ensures that a fraction $\bigO(\frac{1}{N})$ of particles are found close to the hard edge, as can be expected.

\begin{figure}[ht!]
    \centering
    \begin{subfigure}[t]{0.5\textwidth}
        \centering
    \includegraphics[height=5cm]{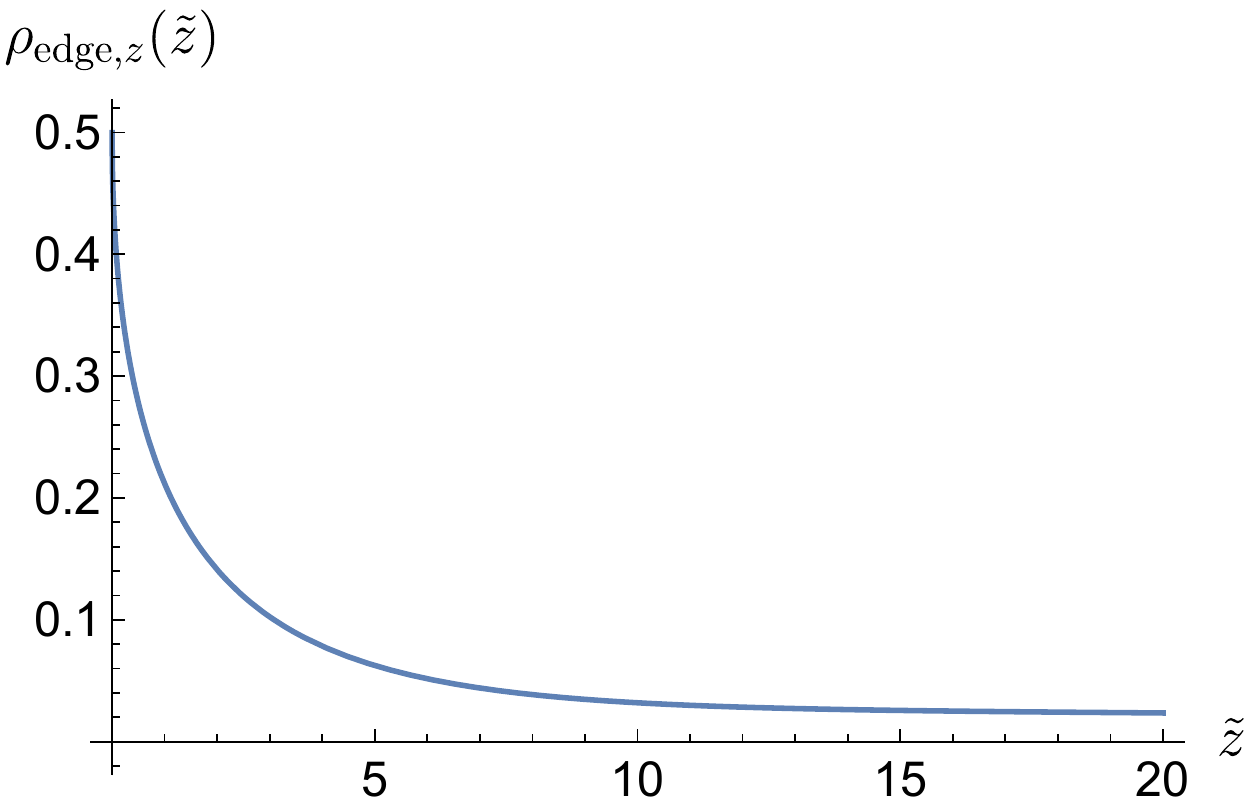}  
    \end{subfigure}%
    ~ 
    \begin{subfigure}[t]{0.5\textwidth}
        \centering
    \includegraphics[height=5cm]{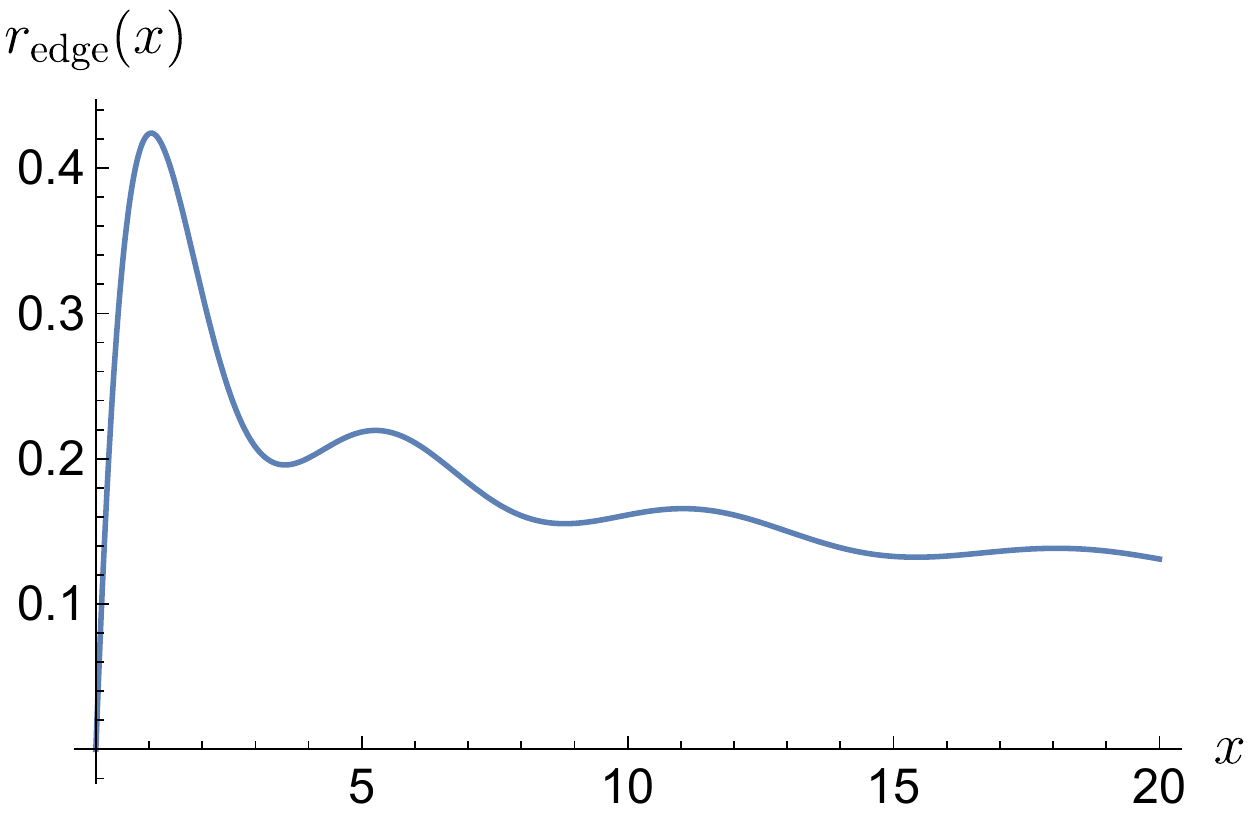}  
    \end{subfigure}
    \caption{\emph{Left:} Plot of the scaling function $\rho_{{\rm edge} , z} (\tilde z)$. \emph{Right:} Plot of the scaling function $ r_{\rm edge} (x) = 2 x \, \rho_{{\rm edge} , z} (x^2)$ which describes the hard edge scaling limit of $\rho(\lambda)$ for $\lambda \sim \sqrt{2t}/N$. At this scale, individual particles appear as oscillations in $r_{\rm edge} $, which cannot be seen in $\rho_{{\rm edge} , z}$ due to the overall decay.}
    \label{fig:ScalingHardEdgeDBMBoundary}
\end{figure}

\sectionmark{Gaussian ensemble interpolation and non-crossing condition}
\section{Gaussian ensemble interpolation and partial non-crossing condition}
\sectionmark{Gaussian ensemble interpolation and non-crossing condition}
\label{sec:2.3}

As a conclusion to this chapter, we present an interesting perspective on the links between non-crossing walkers and random matrix theory. Recall that, as proven in section \subsubref{subsubsec:II.1.2.b}{II.1.2}, the JPDF of non-crossing walkers at time $t$ is given by that of the GOE or the GUE if the non-crossing condition is imposed on $[0,t]$ or $[0, \infty[$ respectively. A natural question is then the following:

\begin{itquote}
What is the position JPDF at time $t$ of $N$ independent Brownian motions conditioned to remain non-crossing on $[0,T]$, with $t \leqslant T \leqslant \infty$ ?
\end{itquote}
This distribution should of course interpolate between the GOE and GUE eigenvalue distributions. Can it then be reduced to a $\beta$-ensemble with $\beta \in [1,2]$ ?

We compute the interpolating JPDF in the first section, and show that the latter question can be answered negatively. Instead of a $\beta$-ensemble, we show in the second section that the distribution is that of the eigenvalues in the interpolating Pandey-Mehta random matrix ensemble \cite{Mehta1983}. As a consequence, the position distribution of non-crossing walkers retains a random matrix interpretation in the intermediate case $t < T < \infty$. 

This problem and its solution constituted a project of this thesis, before it was realized that this was already known in the literature, as it had been studied by Katori and Tanemura in \cite{Katori2002,Katori2003a,Katori2003b,Nagao2003}.
 
\subsection{Solution}

Let $(B_{i,t})_{1 \leqslant i \leqslant N, 0 \leqslant t}$ be $N$ independent Brownian motions, started at initial position $\vec{x} \in \Weyl_N$ at $0$. With $T \geqslant t$, the distribution of interest is
\be 
P( \vec{y} , t \mid \NC_{[0,T]} , \vec{x}, 0 )  \; .
\ee
Rewriting this conditional distribution and by Markov property, this is equal to
\bea
P( \vec{y} , t \mid \NC_{[0,T]} , \vec{x}, 0 ) &=& 
\frac{P( \vec{y} , t , \NC_{[0,T]} \mid \vec{x}, 0 )}{\Pr( \NC_{[0,T]} \mid \vec{x}, 0 )}
\\ 
&=&  
P( \vec{y} , t , \NC_{[0,t]} \mid \vec{x}, 0 )  \ \frac{\Pr(  \NC_{[t,T]} \mid \vec{y}, t )  }{\Pr( \NC_{[0,T]} \mid \vec{x}, 0 )} 
\eea
The first term can be written as a determinant by Karlin-McGregor theorem, and converges to \myeqref{eq:2:ProofNonCrossingGOE} in the $\vec{x} \to \vec{0}$ limit. Both the terms in the fraction are computed as
\bea
\Pr( \NC_{[0,\tau]} \mid \vec{x}, 0 )  &=& \int_{\Weyl_N} \dd \vec{z}  \ \ P( \vec{z}, \tau, \NC_{[0,\tau]} \mid \vec{x}, 0 )  \\
&=&  \int_{\Weyl_N} \dd \vec{z}  \  \det_{1 \leqslant i,j \leqslant N} \   P( z_i, \tau \mid  x_j, 0 ) 
\eea
By de Bruijn's formula \cite{DeBruijn1955}, assuming $N$ is even for simplicity as in \myeqref{eq:2:PfaffianPrefactorSurvivalProba}, this is: 
\begin{gather}
\Pr( \NC_{[0,\tau]} \mid \vec{x}, 0 )   =  \underset{1 \leqslant i,j \leqslant N}{\Pf}  I(x_i,x_j,\tau )\\[5pt]
 \text{where} \quad
I(x_i,x_j, \tau ) = 
\iint_{\R^2} P\left( z_1, \tau  \mid  x_i, 0 \right)  \ P\left( z_2, \tau  \mid  x_j, 0 \right)  \ \mathrm{sgn}(z_1-z_2) \  \dd z_1 \dd z_2 \; . 
\end{gather}
It is shown in App.~\ref{app:InterpolationComputation} that, in the Brownian setting with Gaussian propagator \eqref{eq:2:PropagatorBM}, this integral is
\be 
I(x_i,x_j, \tau )  = \erf \left( \frac{x_j - x_i}{2\sqrt{\tau}} \right)
\ee
where the error function $\erf$ is defined in \myeqref{eq:2:DefErrorFunction}.

The JPDF of positions at time $t$ with the intermediate condition can be written in terms of the single-particle propagator as
\be 
P( \vec{y} , t \mid \NC_{[0,T]} , \vec{x}, 0 ) = \det_{1 \leqslant i,j \leqslant N} P(y_i, t \mid x_j , 0) \ \frac{\underset{1 \leqslant i,j \leqslant N}{\Pf}  I(y_i,y_j,T-t)}{\underset{1 \leqslant i,j \leqslant N}{\Pf}  I(x_i,x_j,T)}
\ee
such that finally, in our Brownian setting:
\begin{empheq}[box=\setlength{\fboxsep}{5pt}\fbox]{equation}  
P( \vec{y} , t \mid \NC_{[0,T]} , \vec{x}, 0 ) =
\frac{1}{(2 \pi t)^{\frac{N}{2}}} \  \det_{1 \leqslant i,j \leqslant N} e^{-\frac{(y_j-x_i)^2}{2t}} 
\ \ \frac{\underset{1 \leqslant i,j \leqslant N}{\Pf} \erf \left( \frac{y_j - y_i}{2\sqrt{T-t}} \right)  }{\underset{1 \leqslant i,j \leqslant N}{\Pf}  \erf \left( \frac{x_j - x_i}{2\sqrt{T}} \right)   }  
\end{empheq}
Note that the Pfaffian terms are well-defined since they concern antisymmetric matrices from the odd symetry of the $\erf$ function.

In order to obtain the solution for standard Brownian motions, let us now consider the limit $\vec{x} \to \vec{0}$, for example as $\vec{x} = \epsilon \,  t \, \vec{z}$ with fixed $\vec{z}$ and vanishing $\epsilon$. From the computations of App.~\ref{app:VandermondeLimit}, the determinant term is equivalent, as $\epsilon \to 0$, to
\be
\det_{1 \leqslant i,j \leqslant N} e^{-\frac{(y_j-x_i)^2}{2t}} \  \simeq  \
\epsilon^{\frac{N(N-1)}{2}} \frac{ \Delta \left( \vec{z} \right) } {\prod_{i=1}^{N-1} i!}
\ \Delta \left( \vec{y} \right)   \ e^{- \frac{1}{2t} \sum_{i=1}^N y_i^2 } \; .
\ee
The Pfaffian term, see App.~\ref{app:InterpolationComputation}, behaves as 
\be 
\underset{1 \leqslant i,j \leqslant N}{\Pf}  \erf \left( \epsilon \frac{t}{2\sqrt{T}}   (z_j - z_i)  \right)
\ \simeq \ A \ \epsilon^{\frac{N(N-1)}{2}}  \
\left(  \frac{t}{\sqrt{T}} \right)^{\frac{N(N-1)}{2}}  \Delta(\vec{z}) 
\label{eq:2:SmallEpsilonPfaffianTermEquivalent}
\ee
with a constant $A$. The sought position density of non-crossing standard Brownian motions, under partial non-crossing condition is finally
\begin{empheq}[box=\setlength{\fboxsep}{10pt}\fbox]{equation}  
P( \vec{y} , t \mid \NC_{[0,T]} , \vec{0}, 0 ) = K  \
\Delta(\vec{y})   \ e^{- \frac{1}{2t} \sum_{i=1}^N y_i^2 }   \
\underset{1 \leqslant i,j \leqslant N}{\Pf} \erf \left( \frac{y_j - y_i}{2\sqrt{T-t}} \right)
\label{eq:2:JPDFResultNonCrossingPartialCondition} 
\end{empheq}

Some remarks can be made on this distribution result. First and foremost, it can be seen to converge to the GOE and GUE measures in the limit cases $T \to t$ and $T \to \infty$ respectively, as expected. Indeed, the limits of the Pfaffian term are
\be 
\underset{1 \leqslant i,j \leqslant N}{\Pf} \erf \left( \frac{y_j - y_i}{2\sqrt{T-t}} \right) 
\xrightarrow{T\to t} 
\underset{1 \leqslant i,j \leqslant N}{\Pf}\mathrm{sgn}(y_j - y_i)  = \prod_{1 \leqslant i < j \leqslant N} \mathrm{sgn}(y_j - y_i)  = 1
\ee
such that $P( \vec{y} , t \mid \NC_{[0,t]} , \vec{0}, 0 ) $ then coincides with the JPDF of GOE eigenvalues, see \cite{DeBruijn1955} for the sign Pfaffian formula ; and
\be 
\underset{1 \leqslant i,j \leqslant N}{\Pf} \erf \left( \frac{y_j - y_i}{2\sqrt{T-t}} \right)  \
\overset{T \to \infty}{\sim} \ T^{- \frac{N(N-1)}{4} }  \Delta(\vec{y}) 
\ee
such that $P( \vec{y} , t \mid \NC_{[0, \infty[} , \vec{0}, 0 ) $ then coincides with the JPDF of GUE eigenvalues, where this line holds from \myeqref{eq:2:SmallEpsilonPfaffianTermEquivalent}. 

The interpolation is thus operated by the Pfaffian term which depends on $\vec{y}$ as $\Delta(\vec{y})^0$ for $T\to t$ and as $\Delta(\vec{y})^1$ for $T \to \infty$. However, it cannot be reduced to $\Delta(\vec{y})^\alpha$ with $\alpha \in ]0, 1[$ for intermediate $T$, as seen from the analytical expression.
As a consequence, the interpolation uncovered here between the GOE and the GUE measures is different from the one given by the Hermite $\beta$-ensemble of \myeqref{eq:1:JPDFBeta} with $1 < \beta <2$. We see in the next section that it coincides in fact with the eigenvalue measure of the Pandey-Mehta matrix ensemble.

\subsection{Links with the Pandey-Mehta matrix ensemble}

Pandey and Mehta have defined in \cite{Mehta1983} the random matrix ensemble defined by matrices
\be 
\HH = \frac{\gamma}{\sqrt{1 + \alpha^2}} \left(  \sqrt{\frac{1-\alpha^2}{2}} \, \HH_{\rm GOE} + \alpha \, \HH_{\rm GUE} \right) 
\label{eq:2:DefPandeyMehtaEnsemble}
\ee
where $\HH_{\rm GOE} $ and $\HH_{\rm GUE} $ are samples of the GOE and GUE with parameter $1$, $\alpha \in [0,1]$ is an interpolation parameter and $\gamma >0$ is a normalization parameter. The interest of this construction is that it provides a rotation-invariant ensemble of Hermitian matrices, interpolating between the GOE, as $\alpha \to 0$, and the GUE, as $\alpha \to 1$. It is of particular interest in the context of quantum chaos \cite{Haake2000}, where it allows to study the transition from the time-reversal symmetric Hamiltonian $\HH_{\rm GOE}$ to the unrestricted Hamiltonian $\HH_{\rm GUE}$.

As for the Gaussian ensembles presented in chapter \ref{chap:1}, it is possible to obtain the eigenvalue distribution by integrating out the eigenvector-related variables. This is done in a slightly more technical way in this case, involving the Harish-Chandra-Itzykson-Zuber (\textit{HCIZ}) integral. We give the final result for the eigenvalue JPDF \cite{Mehta1983}:
\be 
P( \vec{\lambda})  = K \ e^{-\frac{1+\alpha^2}{2\gamma^2} \sum_{i=1}^N \lambda_i^2 } \ \abs{\Delta(\vec{\lambda} )} \ 
\underset{1 \leqslant i,j \leqslant N}{\Pf} \erf \left( \sqrt{\frac{1-\alpha^4}{4\alpha^2\gamma^2}} (\lambda_j - \lambda_i) \right) 
\ee
It can be seen directly that this matches our JPDF \eqref{eq:2:JPDFResultNonCrossingPartialCondition} under the replacement
\be 
\begin{cases}
\displaystyle t   = & \displaystyle  \frac{\gamma^2}{1+\alpha^2} \; , \\[8 pt]
\displaystyle T   = & \displaystyle  \frac{\gamma^2}{1-\alpha^4} \; .
\end{cases} 
\ee
The limits match naturally as $T \to t \Leftrightarrow \alpha \to 0$ and $T \to \infty \Leftrightarrow \alpha \to 1$. The ratio of $T$ and $t$ is thus in direct correspondence with $\alpha$ as
\be
\frac{T}{t} = \frac{1}{1-\alpha^2} \  \in \R^+ \; .
\ee

The models are thus equivalent, such that non-crossing walkers observed at $t$ while conditioned to be non-crossing on $[0,T]$ can be interpreted as the eigenvalues of (arguably) the simplest interpolating matrix model \eqref{eq:2:DefPandeyMehtaEnsemble}. This completes our presentation of the links between non-crossing walkers and random matrix theory, in a generalization of the results presented at the beginning of the chapter.

As explained above, the results presented here were a rediscovery of findings by Katori, Tanemura and co-workers.
In \cite{Katori2002}, they identified and solved the problem studied here, and also noticed the connection with the Pandey-Mehta interpolating ensemble. Further from the single-time distribution, they also studied multi-dimensional marginals in \cite{Nagao2003}, before the law of the full process in \cite{Katori2003a,Katori2003b}. In these latter works, they identified a random matrix equivalent for the full process, as a natural generalization of the static case explained above, whose entries are sums of Brownian motions (providing the GOE ingredient) and Brownian bridges (providing the GUE ingredient).

Finally, we note that a different interpolation between GOE and GUE statistics was observed in the soft edge region in \cite{Borodin2008}. This interpolation is provided by the transition between flat and curved regions in the TASEP model of the KPZ universality class, which are respectively related to the $\beta=1$ and $\beta=2$ Gaussian edge statistics through the Airy$_1$ and Airy$_2$ processes.

\thispagestyle{empty}
\chapter{Stochastic matrix evolutions}
\label{chap:3}
 
The stochastic evolutions of random matrices, and their eigenvalue processes, are a dynamical extension of the static theory of random matrix ensembles. They are deeply connected to fermion systems and non-crossing walkers, as illustrated at length in the context of the Dyson Brownian motion, in chapter \ref{chap:2}. This Gaussian matrix diffusion is indeed in correspondence with fermions in the Calogero-Moser Hamiltonian, as proven in section \subsubref{subsubsec:2.1.1.b}{II.1.1}, and its connections with non-crossing walkers are established in section \ref{subsec:2.1.2}. The DBM occupies a central space because it is arguably the simplest stochastic matrix process: it is structure-less as an extension of the Gaussian ensembles, and its stochastic evolution is driftless, Markovian and stationary.

Besides this pillar of stochastic matrix evolutions, there exist many noteworthy processes. Firstly, there similarly exist stochastic matrix extensions for the Wishart, Jacobi and circular ensembles \cite{Forrester2010,Collins2017}. Secondly, generalized Dyson Brownian motions in an arbitrary potential are useful in the study of line ensembles and growth models \cite{Akemann2011,Johansson2006,Spohn2006,Sasamoto2007}. Modified additive processes with alternate free addition, as in the DBM, and co-diagonalizable addition, such that eigenvalues evolve independently, have been designed in order to provide asymptotically a rotation-invariant interpolating $\beta$-ensemble for the Gaussian and Wishart classes \cite{Allez2012,Allez2013}. Multiplicative versions of the DBM have also been constructed and investigated \cite{Hall2019,Driver2019,Biane1997}, with a particular interest in the corresponding Lyapunov exponents which drive the behaviour of chaotic systems \cite{Akemann2019,Monthus2021,Son2009}. Significantly, other stochastic matrix models have been introduced for their physical significance in the problems of wave-diffusion in random media and quantum transport in disordered wires \cite{Beenakker1997,Grabsch2018,Texier2016,Delande2015}. In these problems of quasi-1D wave scattering, the Dorokhov-Mello-Pereyra-Kumar approach (\textit{DMPK}) introduced in the 1980's \cite{Mello1988,Dorokhov1988} consists in studying the random evolution of the scattering matrix along the disordered multichannel wave-conductor. 
This field of study has thus provided interest in the study of different stochastic matrix models \cite{Brouwer2003,Brouwer1998,Brouwer2000}, especially in the last 20 years after it was realized that multichannel disordered wires were a convenient framework to study topological phase transitions in quasi-1D systems \cite{Kitaev2001,Oreg2010}. In discrete space-time, Chalker-Coddington type of matrix models on graphs are useful for the study of localization in two dimensions \cite{Cardy2010}, or in lattice QCD \cite{Gopakumar1995}.

In this chapter, we turn to other models of stochastic evolution for a random matrix process. Mainly, we present the process defined and studied in publication \pubref{publication:MatrixKesten}, the \emph{matrix Kesten recursion}. This model is a matricial realization of the renowned stochastic recursion equation introduced by Harry Kesten in the 1970's \cite{Kesten1973,Kesten1975} which asymptotically exhibits a heavy tail distribution. The interest for a matrix version of the Kesten recursion is then to study this heavy-tail mechanism in a multivariate setting. As we present below, the matrix Kesten recursion converges, in the continuum limit of this recursion, to the distribution of the inverse-Wishart ensemble. The full dynamics can be solved by a mapping to fermions in a Morse potential, which are non-interacting for $\beta=2$. At finite $N$, the distribution of eigenvalues exhibits heavy tails, generalizing Kesten's results in the scalar case. These results display a correspondence between the inverse-Wishart ensemble and fermions in a Morse potential, and thus add the Morse potential to the list of potentials in one dimension, presented in chapter \ref{chap:2}, for which the quantum correlations can be computed using their connection to random matrices. 
For the discrete matrix recursion, the problem can be studied using free probability in the large-$N$ limit, and translated in a recursion equation for the $\ST$ and Stieltjes transforms of the process.

In section \ref{sec:3.1}, we present the original Kesten recursion and its matricial extension studied in publication \pubref{publication:MatrixKesten}, both in the continuous- and discrete-time setting. In section \ref{sec:3.2}, we present briefly other interesting matrix evolutions. In section \ref{sec:3.3}, we present an interesting perspective on stochastic matrix evolutions, from the recent work of Grela, Nowak and Tarnowski \cite{Grela2020}.

\section{Matrix Kesten recursion}
\label{sec:3.1}

\subsection{The Kesten recursion}
\label{subsec:3.1.1}

The Kesten recursion is an example of the fascinating phenomenon of dynamical generation of random variables with heavy power-law tails. Such power-law distributions occur in problems such as diffusion in random media \cite{Bouchaud1990,Bouchaud1990a}, directed polymers in random environment \cite{Derrida1988}, growing networks \cite{Dorogovtsev2013} and Anderson localization in an external field \cite{Bellaistre2018,Delyon1984,Prigodin1980}. Power-laws are also essential in avalanches of driven elastic systems \cite{Alessandro1998,LePriol2020} and in self-organized critical systems such as sandpiles \cite{Bak1987}. Finally, they are ubiquitous in economics and finance \cite{Bouchaud2001,Gabaix2009,Bouchaud2000} and in social systems \cite{Gabaix1999,Verbavatz2020,Axtell2001,Robichet2021}, see \cite{Bouchaud2021} for a recent joint exposition.

After defining the Kesten recursion and giving the first details on its power-law behaviour, we explain in the following some application fields in which the Kesten recursion appears explicitly, and finally present some useful technical properties.

\subsubsection{Definition}

Let $(Z_n, \xi_n)_{n\geqslant 0}$ be two sequences of scalar random variables, where the $\xi_n$ are i.i.d.~positive. The Kesten recursion is defined by the following system of equations:
\begin{empheq}[box=\setlength{\fboxsep}{8pt}\fbox]{equation}
 \left\{
    \begin{array}{ll}
        Z_{0} = 0 \\[5pt]
Z_{n} = \xi_{n}\left(1+Z_{n-1}\right)
    \end{array}
\right.
\label{eq:3:KestenRecursionDef}
\end{empheq}
The solution of this recursion is formally:
\be 
Z_{n} \  = \ \sum_{j=0}^{n-1} \xi_{n} \cdots \xi_{n-j} \ = \ \sum_{j=0}^{n-1} \prod_{k=n-j}^n \xi_k\; .
\label{eq:3:KestenRecursionFormalSolution}
\ee

Harry Kesten studied the limit as $n \to \infty$ of such recursions in \cite{Kesten1973,Kesten1975,Kesten1984}, showing that the behaviour of the model depends on the sign of $\E[ \log \xi ]$. The sum in \myeqref{eq:3:KestenRecursionFormalSolution} may either grow unboundedly with $n$ if $\E[ \log \xi ] \geqslant 0$, or converge to a positive random variable, denoted $Z_\infty$, if $\E[ \log \xi ] < 0$. 

Kesten further showed that in the latter case, the tail of the distribution of $Z_\infty$ exhibits a power-law behaviour with exponent $1+\nu$, such that for $z \gg 1$
\be  
P_{\infty} (z)  \ \propto  \  \frac{1}{z^{1+\nu}}
\label{eq:3:PowerLawKestenVariable}
\ee
where $P_{\infty}$ is the PDF of $Z_\infty$. The power-law index $\nu$ is characterized by the distribution of $\xi$ as
\be 
\E \left[  \xi^\nu \right] =1 \; .
\label{eq:3:ExponentConditionKesten}
\ee
This condition can be understood as follows. From the definition \eqref{eq:3:KestenRecursionDef} one must have the equality in law $Z_\infty \overset{\mathrm{law}}{=} \xi(1+ Z_\infty)$, where $\xi$ and $Z_\infty$ are taken as independent random variables in the r.h.s. Denoting $P_{\xi}$ the PDF of $\xi$, we see that the PDF $P_{Z_\infty}$ of $Z_\infty$ must then obey the integral equation 
\be
P_{\infty}(z) = \int \dd z' P_{\infty}(z') \int \dd x P_{\xi}(x) \delta(z- x(1+ z')) = \int \frac{\dd x}{x} P_\xi(x) P_{\infty}(\frac{z}{x}-1)  \; .
\ee 
If one assumes that $P_\infty(z) \sim z^{-(1+\nu)}$ at large $z$, the relation \eqref{eq:3:ExponentConditionKesten} follows.

\subsubsection{Applications}

As presented above, the Kesten recursion is a simple random recursion which elegantly leads to power-law tail distributions. It appears explicitly in a variety of problems, which exhibit a power-law behaviour as a consequence. We present below a selection of three of these settings.

\paragraph{Random walk in a random environment}
 
The Kesten recursion has a close connection to the problem of a random walk in a random environment (\textit{RWRE}) \cite{Solomon1975,Kesten1986,Sinai1983,LeDoussal1999}. The RWRE is a random walk where the hopping probabilities are random variables, and thus constitute a random environment. Let us explain how the Kesten recursion occurs from RWRE in one dimension, following an argument by Solomon \cite{Solomon1975}. 

We denote by $\alpha_n = \mathbb{P}_{n,n+1}$ and $\beta_n =\mathbb{P}_{n,n-1}=1- \alpha_n$ the random hopping probabilities from site $n$ of the random walk on $\mathbb{Z}$, as illustrated in Fig. \ref{fig:SchemaRandomWalkRandomEnvironment}. Let these random variables be sampled once and fixed, such that they are a source of \emph{quenched} disorder for the system. With $\mu_n$ the mean first-passage time from site $n$ to site $n+1$ for a random walk in this environment, we have:
\be
\mu_{n}=\alpha_{n}+\beta_{n}\left(1+\mu_{n-1}+\mu_{n}\right) \; .
\ee
Indeed, the first-passage time is 1 with probability $\alpha_n$, if the first step from site $n$ is taken to the right; or $(1+\mu_{n-1} + \mu_n)$ in average with probability $\beta_n$, if the first step is taken to the left.

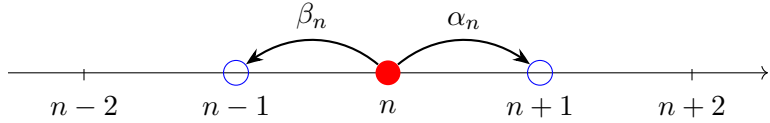
\begin{figure}[ht!]
    \centering
\begin{tikzpicture}[bullet/.style={circle,inner sep=0.7ex},x=2cm,auto,bend angle=40]
 \draw[->] (-2.5,0) -- (2.5,0);
 \path (-1,0) node[bullet,draw=blue] (-a) {}
  (0,0) node[bullet,fill=red] (0) {}
  (1,0) node[bullet,draw=blue] (a) {};
 \foreach \Y [count=\X starting from -2] in {n-2,n-1,n,n+1,n+2} 
  {\node at (\X,-0.45) {$\Y$};}
  \draw (-2,0.05) -- (-2,-0.1);
  \draw (2,0.05) -- (2,-0.1);
 \draw[-{Stealth[bend]},thick] (0) to[bend left] node{$\alpha_n$} (a);
 \draw[-{Stealth[bend]},thick] (0) to[bend right] node[above]{$\beta_n$} (-a);
\end{tikzpicture}
\caption{One-dimensional random walk in a random environment, where the transition probabilities from site $n$, $\alpha_n$ and $\beta_n=1- \alpha_n$, are random variables.}
\label{fig:SchemaRandomWalkRandomEnvironment}
\end{figure}

Defining $\xi_n = \frac{\beta_n}{\alpha_n}$ and $Z_{n}=\frac{1}{2}\left(\mu_{n}-1\right)$, we obtain directly the Kesten recursion \eqref{eq:3:KestenRecursionDef}
\be 
Z_n = \xi_n (1+ Z_{n-1})
\ee
with the initialization $Z_0= 0$ if we choose $\alpha_0 =1$ such that the random walk is constrained to the positive sites. The mean first-passage time $\mu_n$ from site $n$ to site $n+1$ is thus distributed as the Kesten random variable of \myeqref{eq:3:KestenRecursionFormalSolution}. The quantity $- \E[ \log \xi ]$ thus represents the effective bias of the environment: if it is positive, the random walker typically moves to the right, and $\mu_\infty$ is a finite random variable. Its distribution then has a power-law tail, such that it is not rare that $\mu_n$ should reach very large values for a typical random environment.

\paragraph{Directed polymers in random media and wealth condensation}
 
The Kesten recursion also appears in a simplified model of directed polymers in random media \cite{Derrida1988}, where the polymer lives on a complete graph \cite{Bouchaud2000}. The time evolution of the partition function of polymers ending on site $a = 1, \cdots , N$ at time $n$ is given by:
\be
    z_{n+1}^a = \eta_n^a \left((1-\varepsilon) z_n^a + \frac{\varepsilon}{N} \sum_{b=1}^N z_n^b \right),
\ee
where $N$ is the total number of sites and $\varepsilon$ the hopping rate of the polymer. On the complete graph, the Laplacian takes a mean-field form that allows one to simplify the analysis considerably. Introducing $\overline{z}_n = \sum_b z_n^b/N$, one can indeed show that for large $N$, the rescaled partition function $Z^a_n:=(1 - \varepsilon) z^a_n/\varepsilon \overline{z}_n$ obeys a Kesten recursion:
\be
    Z_{n+1}^a = \xi_n^a (1 + Z_n^a), 
\ee
with $\xi_n = \frac{ (1- \varepsilon) \eta_n}{\E[\eta]}$, see \cite{Bouchaud2000}. In this work, the polymer is a model of capital distribution in a population, explaining in a simplified economy the appearance of a power-law condensation for the wealth.

%

\paragraph{Investment problem}

In \cite{Dufresne1990}, Dufresne exhibited the Kesten recursion in an investment management problem, seeking the asset statistics of an investor after he has saved consistently for many years. Consider that a savings account is credited with a unit currency at the beginning of every year. Denoting $R_n$ the random rate of return during the $n$\textsuperscript{th} year, let us define the random variable $\xi_n = 1 + R_n$. The accumulated value $S_n$ is then given as a function of that of the previous year $S_{n-1}$ through the Kesten recursion:
\be 
S_n = \xi_n (1+ S_{n-1}) \; .
\ee   
This shows that the statistics of the investor's assets follows a power-law distribution after many years, such that it exhibits large variability.

Beyond these examples, the Kesten recursion has applications in the physics of the random-field Ising chain problem, as well as in biosciences and other fields \cite{Cavalli-Sforza1973}.

\subsubsection{Properties}

\paragraph{Continuous version}

A continuous version of the scalar recursion was studied in physics, as a Langevin equation describing the diffusion of a particle in a one-dimensional white noise Gaussian random field \cite{Bouchaud1990a}, and in mathematics \cite{Dufresne1990,Dufresne2001}. In this continuous setting, the exact limit distribution can be obtained as we present in the following, while only the power-law behaviour is accessible in the discrete case, apart from special cases \cite{DeCalan2012}.

\subparagraph{Vanishing time increment}
A path to a meaningful continuous version is to consider the Kesten variable multiplied by a vanishing time interval, as
\be
U_t= Z_{n=\frac{t}{ \dd t}} \,  \dd  t 
\ee 
where $\dd t$ is the infinitesimal time step. Note that for a given integer $n$, the variable $Z_n$ in \myeqref{eq:3:KestenRecursionFormalSolution} has the same distribution, by permuting the $\xi_k$ variables according to ${(0,1,\cdots\!, n)} \to {(n, \cdots\!, 1,0)}$, as 
\be 
Z_n \ \overset{\mathrm{law}}{=}  \   \sum_{j=0}^{n-1} \prod_{k=0}^j \xi_k  \; .
\label{eq:3:KestenFormalSumReordered}
\ee
Writing the random variable $\xi$ in exponential form as $\xi_n = e^{\eta_n}$, the reordered equivalent form \eqref{eq:3:KestenFormalSumReordered} gives the following convergence for the process $U$ at time $t = n  \dd t$ :
\be 
U_t  \ \overset{\mathrm{law}}{=} \ \sum_{j=0}^{n-1} \ e^{\; \sum\limits_{k = 0}^j \eta_k}  \dd t   \quad \overset{ \dd t \to 0 }{\longrightarrow}  \quad \int_0^t e^{A_s} \dd s
\label{eq:3:KestenContinuumReordered}
\ee
where the process $A_s$ is the continuous limit of the random walk $\sum\limits_{k = 0}^j \eta_k$. This process is well-defined provided the i.i.d. $\eta$ random variables scale as:
\be 
\eta = \gamma \dd t + \alpha \sqrt{\dd t} X \quad \quad \text{with} \quad X \sim \mathcal{N}(0,1) \; ,
\label{eq:3:EtaRandomScaling}
\ee
such that $A_s$ is a drifted Brownian motion
\be 
A_s = \gamma s + \alpha B_s \; .
\ee
The parameters depend on the distribution of $\xi$ as
\be 
\begin{cases}
\gamma= \frac{\E \left[ \log \xi \right]}{ \dd t} \; ; \\[5 pt]
\alpha^{2}= \frac{\mathrm{Var}\left( \log \xi \right)}{\dd t} \; .
\end{cases}
\ee
The infinite-time limit of the $U$ variable can then finally be written as
\be
U_\infty \overset{\mathrm{law}}{=} \int_{0}^{\infty} e^{\gamma t+\alpha B_{t}}\dd  t \; .
\label{eq:3:UInfinityBrownianFunctional}
\ee  
This last equation characterizes the law of $U_\infty$, in terms of the distribution of the widely studied Brownian exponential functional on the r.h.s., see \cite{Yor2001} for a review, as well as \cite{Monthus1994,Comtet1998}.

\subparagraph{Large-time distribution}
In order to exhibit the large-time distribution of the process $U_t$, let us derive its stochastic evolution from the Kesten recursion obeyed by $Z_n$. Exponentiating the scaling form \eqref{eq:3:EtaRandomScaling}, and denoting $m = \gamma +\frac{\alpha^2}{2}$ and $\sigma= \frac{\alpha}{\sqrt{2}}$, the $\xi$ variables are given by 
\be
\xi = 1 + m \dd t +  \sigma \sqrt{2  \dd t} X \quad \quad \text{with} \quad X \sim \mathcal{N}(0,1)
\ee
Injecting this in the Kesten recursion \eqref{eq:3:KestenRecursionDef} yields the following SDE for the process $U_t$
\be
\dd U_t = \left( 1+ m \; U_t \right) \dd t + \sqrt{2}\sigma U_t \dd B_t \quad \quad \quad \quad \text{(It\^o)}
\label{eq:3:StochasticEvolution1DContinuousKesten}
\ee
Since the prefactor of the Brownian term $\dd B_t$ depends on the process $U_t$ itself, a choice of prescription must be made and this SDE is understood \emph{in the sense of Itô}. We postpone the detailed discussion of this issue to section \ref{subsec:3.2.1}.
The Fokker-Planck equation for the associated PDF $P(x,t)$ is
\be 
\frac{\partial P(x,t)}{\partial t} = \sigma^2 \frac{\partial^2 }{\partial x^2} \left(  x^2 P(x,t)\right) - \frac{\partial }{\partial x}  \left(  \left( 1+ m x \right)  P(x,t) \right)  \; .
\ee
When $\gamma<0$, that is to say
\be 
m < \sigma^2 \; ,
\label{eq:3:ConditionScalarCase}
\ee
the PDF converges to a stationary limit and $U_\infty$ then follows an inverse-gamma law distributed as
\be 
P_\infty(x)=\frac{ \sigma ^{-2 + 2\frac{m}{\sigma^2}}}{\Gamma\left(1 - \frac{m}{\sigma^2} \right) } \ x^{-2 + \frac{m}{\sigma^2}}  \ e^{-\frac{1}{\sigma^{2} x}}  \; .
\label{eq:3:InverseGammaDistribution}
\ee
In other words, the inverse of $U_\infty$ follows the gamma law 
\be
U_\infty^{-1} \ \sim  \ \Gamma\left( 1 - \frac{m}{\sigma^2}, \sigma^2 \right) \; .
\ee
Note that this result confirms, in the continuous limit, the power-law behaviour of \myeqref{eq:3:PowerLawKestenVariable} with the heavy tail exponent $1+\nu=2 - \frac{m}{\sigma^2}$.

\paragraph{Vector recursion}


The vector case for the discrete recursion \eqref{eq:3:KestenRecursionDef} was also studied by Kesten \cite{Kesten1973}, promoting $Z_n$ to $d$-dimensional vectors and $\xi_n$ to $d \times d$ matrices. Under suitable hypothesis, there also exists $\nu$ such that for any unit $d$-dimensional vector $\bfx$ the following limit exists and is nonzero: 
\begin{equation}
0 < \lim _{t \rightarrow \infty} t^{\nu} \, \mathbb{P}\left( \bfx^T . \ Z_\infty >  t \right) < \infty \; .
\end{equation}
In this case however, $\nu$ is not characterized by a simple moment equation, in contrast to \myeqref{eq:3:ExponentConditionKesten} in the scalar case.

\subsection{Matrix Kesten evolution in continuous time}
\label{subsec:3.1.2}

The focus of publication \pubref{publication:MatrixKesten} was the introduction of a matrix realization of the Kesten recursion. In the present section, we define this matrix model and present the properties of its continuous limit, which can be solved exactly by a remarkable connection with the quantum mechanics of fermions in a Morse potential.

\subsubsection{A matricial version of the Kesten recursion}

\paragraph{Definition}

For $N \in \mathbb{N}^*$, let us consider $(\mathbf{\bxi}_n)_{n \geqslant 0}$ a sequence of i.i.d. symmetric positive semidefinite $N \times N$ matrices. We define the Kesten recursion for a sequence of symmetric positive semidefinite $N \times N$ matrices $(\ZZ_n)_{n \geqslant 0}$ in a manifestly self-adjoint way as:
\begin{empheq}[box=\setlength{\fboxsep}{8pt}\fbox]{equation}
\label{eq:3:DefMatrixKestenRecursion}
\ZZ_{n+1}=\sqrt{\epsilon \ID +\ZZ_{n}}  \ \bxi_{n} \ \sqrt{ \epsilon  \ID +\ZZ_{n}} 
\end{empheq}

The regularizing real parameter $\epsilon >0$ will be given a value of order $1$ in the discrete setting, and taken to $0$ in the continuous limit. The symmetric positive semidefinite property of the sequence $\bxi$ implies by recursion the same property for the sequence $(\ZZ_n)_{n \geqslant 0}$, as we denote for each $n$ by $\sqrt{\epsilon \ID +\ZZ_{n}}$ the principal square-root of $\epsilon \ID +\ZZ_{n}$. The properties of $\bxi_n$ and $\ZZ_n$ can be summed up by stating that they are symmetric matrices with only nonnegative eigenvalues. 

In parallel, we also consider a model where all matrices are Hermitian positive semidefinite. We will define as usual a Dyson index $\beta =1$ in the real symmetric case, and $\beta =2$ in the complex Hermitian case. Note that the study could be generalized to the case $\beta =4$, but we restrict to the first two cases. 

\paragraph{Continuous limit}

\subparagraph{Matrix evolution}
Let us derive the continuous limit of the matrix Kesten recursion \eqref{eq:3:DefMatrixKestenRecursion}. Remembering the picture laid out in the previous section for the scalar case, see the scaling \eqref{eq:3:EtaRandomScaling}, we choose the following structure for the matrices $\bxi_n$
\be 
\bxi_n = \exp \left(  \gamma \epsilon \ID + \sigma \sqrt{\epsilon } \HH_n \right)
\ee
where $\epsilon$ is the parameter introduced in \myeqref{eq:3:DefMatrixKestenRecursion} and where $\HH$ is a unit-variance GOE or GUE matrix. Note that this ensures positive definiteness of $\bxi$. A small-$\epsilon$ expansion of the matrix exponential definition for $\bxi_n$ gives to order $\epsilon$:
\be   
\bxi_{n}=(1+m \, \epsilon) \ID +\sigma \, \sqrt{\epsilon} \HH_{n}
\label{eq:3:bxiRandomScaling}
\ee 
where parameters $m=\gamma + \frac{1}{2} \sigma^2(N+\delta_{\beta=1})$ and $\sigma$ 
respectively tune the mean and the noise in $\bxi_n$. Indeed, for the sake of the
continuum limit, we can replace $\epsilon (\HH_n)^2$ by its expectation $\epsilon \ID (N+\delta_{\beta=1})$, where the Kronecker delta appears as correlations are different in the $\beta =1,2$ cases.
 
With this choice for the i.i.d. sequence $\bxi$
and expanding the square-root matrix, the evolution of $\ZZ$ reads at order $\epsilon$:
\be 
\ZZ_{n+1}=\ZZ_{n}+\epsilon I+m \epsilon \ZZ_{n}+ \sigma \sqrt{\epsilon} \sqrt{\ZZ_{n}} \HH_{n} \sqrt{\ZZ_{n}} + \mathcal{O}(\epsilon^{3/2})
\label{eq:3:RecursionBeforeContinousLimit}
\ee
In the $\epsilon \to 0$ limit, we define a continuous time matrix process $\VV_t$, where the time variable is $t = n \epsilon$, simply as
\be 
\VV_t = \ZZ_{t/\epsilon} \;.
\ee
We have avoided here the scaling between the discrete and continuous processes, which appeared in the scalar case, from the insertion of $\epsilon$ in the definition of the recursion. From the recursion \eqref{eq:3:RecursionBeforeContinousLimit}, we obtain the time evolution for $\VV_t$ as the following It\^o matrix SDE, where $\HH_t$ is the DBM: 
\emphbe 
\dd \VV_t = \left( \ID +m \VV_t\right) \dd t +\sigma \sqrt{\VV_t} \dd \HH_{t} \sqrt{\VV_t}  
\label{eq:3:ContinuousKestenSDE}
\end{empheq} 
Similarly as in the scalar case, the multiplicative noise imposes the choice of a prescription for this SDE, which is understood here in the sense of Itô, see section \ref{subsec:3.2.1}. Taking $N=1$ in the real case $\beta=1$ recovers exactly the continuous Kesten evolution \eqref{eq:3:StochasticEvolution1DContinuousKesten} as expected, because the diagonal terms in the DBM-1 are $\sqrt{2}$ times a standard Brownian motion. 

\subparagraph{Joint eigenvalue evolution}

The stochastic evolution \eqref{eq:3:ContinuousKestenSDE} is rotation-invariant, as conjugation by a unitary matrix, as $\VV \to \UU \VV \UU^{-1}$, preserves the SDE. As a consequence, the eigenvectors of $\VV_t$ are uniformly distributed on the unit sphere as soon as they are in the initial condition, for example with $\VV_0=0$. The model can then be crucially reduced to the study of the eigenvalues.

Let us denote $\{\lambda_{i,t} \}_{1 \leqslant i \leqslant N}$, the set of positive eigenvalues of $\VV_t$. Their evolution can be derived by perturbation theory, as presented for the DBM in section \ref{sec:2.1}. This procedure yields, see publication \pubref{publication:MatrixKesten}: 
\emphbe 
\label{eq:3:EigenContinuousKestenSDE}
\dd \lambda_{i,t}=\left(1+m \lambda_{i,t}\right) \dd t+  \sigma^2\sum\limits_{\substack{1 \leqslant j \leqslant N \\j\neq i}} \frac{\lambda_{i,t} \lambda_{j,t}}{\lambda_{i,t}-\lambda_{j,t}} \dd t+\sqrt{\frac{2}{\beta}} \sigma \lambda_{i} \dd B_{i,t}
\end{empheq}
where the $(B_{i,t})_{1 \leqslant i \leqslant N}$ are $N$ independent standard Brownian motions. This evolution can be compared with that of the DBM in \myeqref{eq:2:DBMEigenvalueSDE}: the evolution of the eigenvalues is similarly obtained with no a priori knowledge of the evolution of
the eigenvectors, and display a repulsive interaction term. However a
difference is that, because of the multiplicative noise, the repulsion depends also multiplicatively on the eigenvalues.

From this stochastic evolution, one can obtain the evolution of the eigenvalue JPDF $P(\vec{\lambda},t )$. The Fokker-Planck equation associated with the eigenvalue evolution of the It\^o process $\VV_t$ is:
\be
\frac{\partial P(\vec{\lambda},t)}{\partial t}=\frac{\sigma^2}{\beta} \sum_{i=1}^N \frac{\partial^{2}}{\partial \lambda_{i}^{2}}\left(  \lambda_{i} ^{2} P(\vec{\lambda},t)\right)-\sum_{i=1}^N \frac{\partial}{\partial \lambda_{i}}\left(\left(1+m \lambda_{i}+  \sigma^2\sum\limits_{\substack{1 \leqslant j \leqslant N \\j\neq i}}\frac{\lambda_{i} \lambda_{j}}{\lambda_{i}-\lambda_{j}}\right) P(\vec{\lambda},t)\right) 
\ee
where the initial condition is specified by the eigenvalues of $\VV_0$.

\subparagraph{Stationary solution}

The stationary solution $P_\infty$ of this diffusion can be found as a zero-flux solution as
\be 
0 = \frac{\partial}{\partial \lambda_{i}}\left( \frac{1}{\beta}\left(  \sigma \lambda_{i}\right)^{2} P_\infty \right)-   \left(1+m \lambda_{i}+ \sigma^{2} \sum\limits_{\substack{1\leqslant j\leqslant N\\ j\neq i}}    \frac{\lambda_{i} \lambda_{j}}{\lambda_{i}-\lambda_{j}}\right) P_\infty \; .
\ee
This equation is found to be solved by the inverse-Wishart eigenvalue distribution \eqref{eq:1:EigPDFInverseWishart} with parameters
\be 
\left\{
    \begin{array}{ll}
        T = N  - \frac{2 m }{\sigma^2 } + \frac{2}{\beta }  -1  \\[10pt]
    \sigma_W = \frac{\sigma}{\sqrt{2}}
    \end{array}
\right. 
\label{eq:3:InverseWishartParametersStationary}
\ee
where we rename $\sigma_W$ the parameter in \eqref{eq:1:EigPDFInverseWishart} in order to avoid ambiguities; such that the stationary JPDF is
\emphbe 
\label{eq:3:EigenJPDFStationarySolution}
P_\infty (\vec{\lambda})=K_\beta^N \ \abs{ \Delta(\vec{\lambda}) } ^\beta  \ \prod_{i=1}^N \lambda_{i}^{ \beta \left( \frac{m}{\sigma^{2}} - N +1 \right) -2 } \  e^{-\frac{\beta}{\sigma^{2}} \sum_{i=1}^N \frac{1}{\lambda_{i}}} 
\end{empheq}
We note that for $N=1$ the Vandermonde factor is absent 
and setting $\beta=1$ one recovers exactly the inverse gamma law of \myeqref{eq:3:InverseGammaDistribution}.

The eigenvector-basis rotation invariance then allows to conclude that the inverse-Wishart matrix distribution \eqref{eq:1:MatPDFInverseWishart} is a stationary distribution for this model. Note that the normalizability condition translates here to
\be
m < \frac{\sigma^2}{\beta}
\label{eq:3:StochasticNormCondition}
\ee
which coincides, as expected for $N=1$ and $\beta=1$, with the scalar-case condition \eqref{eq:3:ConditionScalarCase}. For more details on the marginal density for the largest eigenvalue and its matching with the scalar case, see \pubref{publication:MatrixKesten}.

\subsubsection{Quantum Morse-Sutherland mapping}

Further from the large-time solution, the full finite-time dynamics can be solved exactly by a quantum mapping to fermions in a Morse potential, in the spirit of section \subsubref{subsubsec:2.1.1.b}{II.1.1}.

\paragraph{Stochastic-to-quantum connection}

To elicit this mapping we first perform a change of variables by defining 
\be 
x_{i,t} = \log \lambda_{i,t}  \; . 
\label{eq:3:ChangeVariableMorseFermionMapping}
\ee
From It\^o rules, the stochastic evolution of the $x$ variables has additive noise following:
\begin{equation} 
\dd x_{i,t} = \mathcal{F}_i \dd t + \sqrt{\frac{2}{\beta}} \sigma \dd B_{i,t}  \qquad \mathcal{F}_i = e^{-x_{i,t}} + m - \frac{\sigma^2}{\beta} +  \sigma^2 \sum\limits_{\substack{1\leqslant j\leqslant N\\ j\neq i}}  \frac{1}{e^{ x_{i,t} - x_{j,t}} -1},
\label{eq:3:TransformedEigenContinuousKestenSDE}
\end{equation}
where $\mathcal{F}_i$ is the force felt by particle $x_i$. The initial condition is fixed as $x_{i,t=0}=x_0$ a large negative constant, for all $i$. 

\subparagraph{Operator mapping}
Let us denote $\bar{P}(\vec{x},t)$ the PDF of the $x$-process. In the following, $\bar{P}$ indicates a distribution function on the $x$ variables. Denoting consequently $\bar{P}_\infty$ the distribution obtained from \myeqref{eq:3:EigenJPDFStationarySolution} through the change of variables \eqref{eq:3:ChangeVariableMorseFermionMapping}, we have 
\be 
\bar{P}_\infty (\vec{x}) = e^{\sum_{i=1}^N x_i }  \ P_\infty (\exp \vec{x})  \; .
\label{eq:3:PtildeinftyFromPinfty}
\ee
The time-dependent PDF $\bar P(\vec{x},t)$ verifies the following Fokker-Planck equation 
\be  
\frac{\partial \bar P}{\partial t } =  \sum_{i=1}^N   \left( \frac{\sigma^2}{\beta} \frac{\partial^2 \bar P }{\partial x_i^2} - \frac{\partial}{\partial x_i} ( \mathcal{F}_i \bar P)  
\right)
= - \Ham_{FP} \bar P 
\ee
where we have introduced the generator $ \Ham_{FP}$ of this diffusion. The auxiliary function
\be
\psi(\vec{x},t) = \frac{ \bar P(\vec{x},t)} {  \sqrt{ \bar P_\infty(\vec{x}) } }
\ee
is then a solution of the following imaginary-time Schrödinger equation, which we rewrite by introducing the operator $\Ham$ and the constant $E_0$:
\be 
\frac{\partial \psi }{\partial t} =  \sum_{i=1}^N \left( \frac{\sigma^2}{\beta} \frac{\partial^2 \psi}{\partial x_i^2} 
- \left(  \beta\frac{\mathcal{F}_i^2}{4 \sigma^2} + \frac{1}{2}\frac{\partial \mathcal{F}_i}{\partial x_i}  \right) \ \psi
\right)
= - \frac{\sigma^2}{\beta} (  \Ham - E_0) \psi  \; .
\label{eq:3:KestenContinuousSchrodingerEquation}
\ee
The quantum Hamiltonian $\Ham$ is 
\emphbe 
\Ham =  \sum_{i=1}^N  \left( - \frac{\partial^2}{\partial x_i^2} + V_\mathrm{M}(x_i) +\sum\limits_{\substack{1\leqslant j\leqslant N\\ j\neq i}}  V_\mathrm{int}(x_i,x_j )   \right)
\label{eq:3:HamiltonianMorseSutherland}
\end{empheq}
with a one-particle potential of the Morse form $V_\mathrm{M}(x)=A e^{-2 x} - B e^{-x}$ \cite{Morse1929} and with Sutherland $\sinh^{-2}$ interaction potential $V_\mathrm{int}$ \cite{Sutherland2004}:
\emphbe
\begin{split} 
V_\mathrm{M}(x_i) &= \beta^2 \frac{e^{-2 x_i}}{4 \sigma^4}  - \frac{\beta}{\sigma^2}
 \left(  \frac{\beta}{2}(N-1-  \frac{m}{ \sigma^2} )+1\right) e^{- x_i}  
  \\[5pt]
  V_\mathrm{int}(x_i,x_j ) &= 
   \frac{\beta (\beta -2) }{16} 
  \frac{1 }{\sinh(\frac{x_i-x_j}{2})^2} 
\end{split} 
\label{eq:3:PotentialMorseSutherland}
\end{empheq}
Notice the translation by the constant $E_0$ and the rescaling by $\sigma^2/\beta$ in \myeqref{eq:3:KestenContinuousSchrodingerEquation}, such that the stochastic-to-quantum operator mapping can be formally written as 
\be
\frac{\sigma^2}{\beta} (\Ham - E_0) = {\bar{P}_\infty \, \! }^{-\frac{1}{2}} \  \Ham_{FP} \  {\bar{P}_\infty \, \! }^{\frac{1}{2}} \; . 
\ee
The lowest Fokker-Planck eigenvalue is equal to $0$, the eigenvalue of the stationary state, such that the constant $E_0$ is in fact the ground-state energy of $\Ham$, and is given by
\be  
\label{eq:3:ConstantEZeroTranslationHamiltonian}
E_0 =
- N   \frac{( \beta m- \sigma^2 )^2}{4\sigma^4} 
 +   \frac{N(N-1)}{4} \left(\frac{\beta^2}{\sigma^2} m - \beta  \frac{\beta+2}{2}  \right)  
 - N(N-1)(N-2)\frac{\beta^2}{12}  \; .
\ee 

\subparagraph{Fermionic symmetry}
As in the DBM case, it can be shown that the non-crossing statistics of the $x$-process implies the fermionic symmetry of the quantum system, see \pubref{publication:MatrixKesten}. As a consequence, we are led to the study of $N$ fermions in the Hamiltonian $\Ham$.

In summary, we have detailed in the previous lines a connection of the eigenvalue process of the continuous Kesten evolution \eqref{eq:3:ContinuousKestenSDE}, through the change of variables \eqref{eq:3:ChangeVariableMorseFermionMapping}, with a system of fermions with Hamiltonian $\Ham$ in \myeqref{eq:3:HamiltonianMorseSutherland}. For both $\beta=1,2$, this gives at large time a direct connection between the inverse-Wishart matrix ensemble in \myeqref{eq:3:EigenJPDFStationarySolution} and the ground-state of this many-body Morse-Sutherland quantum system. This result is thus an addition to the literature of random matrix theory, through the lens of its connections to many-body quantum systems \cite{Mehta2004,Forrester2010,Dean2019,Lacroix-A-Chez-Toine2018,Lacroix-A-Chez-Toine2019,Cunden2018}. As presented below, a specificity of the present case is that the Morse potential can only accomodate a finite number of fermions in its quantized bound states.

\paragraph{The single-particle Morse potential}

We present in this paragraph the properties of the Morse potential
\be 
V_\mathrm{M}(x) = g^2\left( e^{-2(x-x_0)} -  2 e^{-(x-x_0)}\right)
\label{eq:3:MorsePotentialgxzero}
\ee 
which appears as the external potential in \myeqref{eq:3:PotentialMorseSutherland}. This potential is well-known in the literature as it is a convenient model for the potential energy of a diatomic molecule. Indeed, it is fully tractable as we detail in the following, and has the expected properties of an interatomic interaction: it displays a minimum $V_\mathrm{M}(x_0)=-g^2$ at position $x_0$, diverges exponentially fast for $x \to - \infty$ and tends to zero for $x \to +\infty$. See Fig. \ref{fig:MorsePotential} for a plot of the Morse potential. In the Kesten setting, the parameters of the potential are fixed per \myeqref{eq:3:PotentialMorseSutherland} as 
\be 
\begin{cases}
g= \frac{\beta}{2}(N -\frac{m}{\sigma^2}-1)+1 \; ; \\[5pt]
e^{-x_0} = \sigma^2 ( N - \frac{m}{\sigma^2}-1 + \frac{2}{\beta} ) \; .
\end{cases}
\label{eq:3:Defgxzero}
\ee 
Notice that the parameters depend explicitly on the matrix size $N$, which has important consequences. 

Let us review the properties of a single quantum particle in Hamiltonian $V_\mathrm{M}$ \cite{Morse1929,Dong2007}. The spectrum of this Hamiltonian is composed of two branches: a finite branch of discrete bound states with negative eigenenergies, and a continuum of positive-eigenenergy diffusion states.

Let us denote $K = g- \frac{1}{2} $.
The discrete states, indexed by $k \geqslant 0$, exist under the condition
\be 
k <   K
\ee  
such that the maximum number of fermions that can be stacked in the discrete spectrum of the Morse potential is the largest integer $m$ such that $ m < K+1  $, denoted as the floor of an infinitesimally smaller value $N_{\mathrm{max}}= \lfloor \left(  K+1 \right)^- \rfloor$. For $0 \leqslant k <  K$, the bound states have energy level and corresponding eigenfunction
\begin{gather} 
\varepsilon_k = -\left(K - k \right)^2 \; ;    \\[5pt]
    \psi_{k}(x) = N_{k} \left( 2ge^{-(x-x_0) }\right)^{g-k-\frac{1}{2}} e^{- ge^{-(x-x_0) } } L_{k}^{(2g-2 k-1)}\left(2ge^{-(x-x_0) } \right) \; .
\end{gather}
The eigenfunction is expressed in terms of the Laguerre polynomials $L^{(\alpha)}_k$, see chapter \ref{chap:1}, with 
\be 
N_k  =  \sqrt{ \frac{k !(2 g-2 k-1)}{\Gamma(2 g-k)}}
\ee a normalization constant. The admissible bound states described above are plotted in Fig. \ref{fig:MorsePotential} along with the Morse potential.

 \begin{figure}[ht!]
    \centering 
    \includegraphics[width=.95 \textwidth]{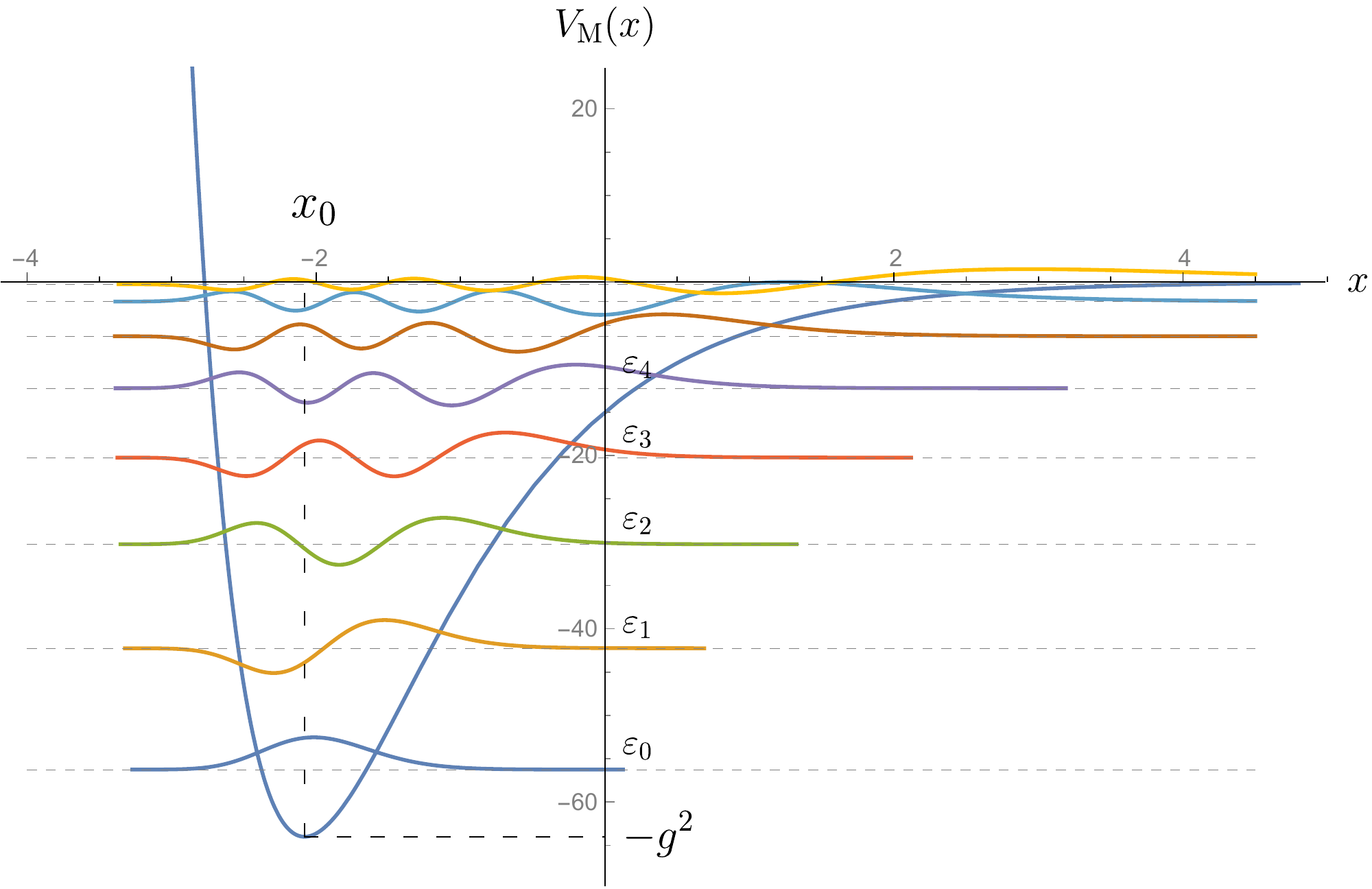} 
    \caption{The Morse potential $V_\mathrm{M}(x)$ in \myeqref{eq:3:MorsePotentialgxzero} is plotted in the blue line, with parameters $g=8$ and $x_0 \simeq -2.079$ which corresponds for example to $N=5$, $\sigma=1$ and $m=-3$ in the $\beta=2$ Kesten model. The eight admissible bound-state eigenfunctions $\psi_k$ are plotted at altitude $\epsilon_k$, in coloured lines.}
    \label{fig:MorsePotential}
\end{figure}

In addition to this discrete spectrum of bound states, there is a continuum of positive energy eigenstates $\epsilon=p^2 >0$. The associated eigenfunctions are of the form 
\be 
\psi_p(x) \sim e^{ \frac{x-x_0}{2} } \ W_{g, \ii p}\left(2 g e^{-(x-x_0)}\right)
\ee 
in terms of the Whittaker $W_{k,\mu}$ function, see App.~\ref{app:specialfunctions}, such that $\psi_p$ is a real function decreasing fast for $x<0$ and oscillating for $x>0$.

In the following section, we turn back to the $N$-particle case and study the many-body fermionic system with Hamiltonian $\Ham$ in \myeqref{eq:3:HamiltonianMorseSutherland}, with a restriction to $\beta =2$.
 
\subsubsection{Full solution for the non-interacting case $\beta=2$}

The single-particle results of the previous paragraph can be directly applied to the case $\beta =2$. Indeed, since the interaction potential $V_\mathrm{int}$ vanishes, the particles are simply $N$ non-interacting fermions in the Morse potential.

\paragraph{Ground state of the non-interacting case}
 
In the present case $\beta =2$, the parameters of the Morse potential are $g = N-\frac{m}{\sigma^2}$ and $e^{-x_0} = \sigma^2 \left(N-\frac{m}{\sigma^2}\right)$ such that the energy levels and corresponding eigenfunctions of the Morse potential are
\begin{gather} 
\varepsilon_k =  - \left( N-\frac{m}{\sigma^2} - k - \frac{1}{2}\right)^2  \; ;    \label{eq:3:EnergyLevelMorsePotentialBeta2} \\[5pt]
    \psi_{k}(x) = N_{k} \left( \frac{2}{\sigma^2} e^{-x}\right)^{N-\frac{m}{\sigma^2}-k-\frac{1}{2}} e^{-  \frac{1}{\sigma^2} e^{- x }} L_{k}^{(2N-\frac{2m}{\sigma^2}-2 k-1)}\left(\frac{2}{\sigma^2} e^{- x } \right)  \; .
     \label{eq:3:EigenfunctionMorsePotentialBeta2} 
\end{gather}

\subparagraph{Ground-state wavefunction and energy}

The ground-state wavefunction of the $N$-fermion model is expressed as a Slater determinant with corresponding energy $\epsilon_{\vec{0}}$:
\bea
          \Psi_{\vec{0}} ( \vec{x} ) &=&  \frac{1}{\sqrt{N!}} \det_{1\leqslant i,j \leqslant N} \left( \psi_{i-1}(x_j)\right) \; , \\
        \epsilon_{\vec{0}} &=&  \sum_{k=0}^{N-1} \epsilon_k  \; .
\eea
Two things can be verified algebraically: the coincidence of the ground-state energy $\epsilon_{\vec{0}}$ with the constant $E_0$ in \myeqref{eq:3:ConstantEZeroTranslationHamiltonian}; and the coincidence of the fermionic JPDF with the large-time stochastic distribution as $ \abs{ \Psi_{\vec{0}} ( \vec{x} ) }^2 = \bar{P}_\infty (\vec{x})$. Furthermore, as mentioned above, the Morse potential has a finite number of bound states $N_\mathrm{max}= \lfloor \left(K+1\right)^- \rfloor$. In order for the ground-state of $N$ fermions to be well-defined, the condition $ N \leqslant N_\mathrm{max}$ must hold, i.e.
 \be 
 m < \frac{\sigma^2}{2} 
 \label{eq:3:NormalizabilityConditionBeta2}
 \ee
 which coincides perfectly with the stochastic normalizability condition \eqref{eq:3:StochasticNormCondition} in the present case $\beta= 2$.
 
Finally, we define the Fermi energy $\epsilon_F$ as the last occupied level
\be 
     \epsilon_F   =  \epsilon_{N-1} = - \left( \frac{1}{2} - \frac{m}{\sigma^2} \right)^2 \; .
     \label{eq:3:FermiEnergyKesten}
\ee 
We note that it is independent of the number of particles, such that the potential with growing depth, given from $g$ in \myeqref{eq:3:Defgxzero}, is filled by a growing number of $N$ particles up to a fixed energy level $\epsilon_F$. This is not true any more if $\sigma$ depends on $N$, as we discuss below.

\subparagraph{Determinantal structure of the ground state}
The fermionic setting suggests a determinantal rewriting of the stationary solution:
\be 
\bar P_\infty(\vec{x}) =
 \frac{1}{N!}
\det_{1 \leqslant i,j \leqslant N}  K_{\mathrm{M}}(x_i,x_j) 
\ee
where the kernel $K_{\mathrm{M}}$, which respects the trace-class, positivity and reproducibility properties, is given formally by the weighted sum of the first $N$ Morse eigenfunctions as
\be
\begin{split}
K_{\mathrm{M}}(x,y) = (2/\sigma^2)^{2(N-\frac{m}{\sigma^2}-\frac{1}{2} ) }  
   \sum_{k=0}^{N-1}  \frac{ \sigma^{4k} N_k^2}{4^{k}}     e^{-(x+y)(N-\frac{m}{\sigma^2}-k-\frac{1}{2})} e^{-  \frac{1}{\sigma^2}(  e^{- x }+e^{-y})}  \\ 
       L_k^{(2N-\frac{2m}{\sigma^2}-2k-1)}\left(\frac{2}{\sigma^2}e^{-x}\right)
    L_k^{(2N-\frac{2m}{\sigma^2}-2k-1)}\left(\frac{2}{\sigma^2}e^{-y}\right) \; .
    \end{split}
\ee 
All statistical properties of the fermion positions in the ground state are characterized by this kernel, such as their average density
\be 
\label{eq:3:ExactDensityFermiGas}
\bar P (x)= \frac{1}{N}K_{\mathrm{M}}(x,x) \; .
\ee 
This coincides with the infinite-time average density of the $\beta=2$ Kesten-matrix eigenvalues $\lambda_{i}$ as $ P(\lambda) = \frac{1}{\lambda} \bar P(\log \lambda) $. We recall from chapter \ref{chap:1} that a similar kernel can be obtained on the inverse-Wishart side from the orthogonal polynomial method for $\beta=2$.

\paragraph{Large-$N$ behaviour}

For large $N$, the density $\bar{P}(x)$ of the Fermi gas in the Morse potential without interactions can be well described by approximations in the bulk region and close to the edges. The results are presented briefly in the following. We point to \pubref{publication:MatrixKesten} for the full computations concerning the Morse potential in \myeqref{eq:3:MorsePotentialgxzero} with arbitrary parameters $g$ and $x_0$, and their application to the Kesten setting.

\subparagraph{Density in the bulk}

When $N$ is large, the density is well described in the bulk by a semi-classical approximation for the density of a Fermi gas, also called the local density approximation (\textit{LDA}) \cite{Castin2007}. The bulk density approximation is supported on $[x_-,x_+]$ and given by:
\be 
\label{eq:3:BulkDensityKestenFermigas}
\bar{P}_\mathrm{bulk}(x)=  \frac{1}{\sigma^2 \pi N } \sqrt{( e^{-x}- e^{-x_+})(e^{-x_-} - e^{-x})  } 
\ee
where the positions of the edges $x_\pm$ are
\be 
x_\pm = x_0 - \log \left( 1 \mp \sqrt{\frac{1}{1 - \frac{m}{N\sigma^2}}} \sqrt{2- \frac{1}{1 - \frac{m}{N\sigma^2}}} \right)   
\ee
expressed in relation to the bottom of the Morse potential at position $x_0 = - \log \left( \sigma^2 (N -\frac{m}{\sigma^2}) \right)$.

We can rewrite this density from a random matrix point of view. Let us recall $T= N - \frac{2m}{\sigma^2}$ the inverse-Wishart parameter obtained for the large time distribution of the Kesten model, from \myeqref{eq:3:InverseWishartParametersStationary} in the present case $\beta=2$. Introducing the rescaled $1/\lambda$ variable
\be 
\Lambda = \frac{1}{T} \frac{2}{\sigma^2} \ e^{-x} 
\label{eq:3:DefinitionLambda}
\ee
the previous asymptotic density becomes, under the change of variables: 
\be 
\begin{split}
\bar{P}_\mathrm{bulk}(x)\dd x
&=  \frac{\sqrt{ (\Lambda - \Lambda_-) (\Lambda_+ - \Lambda) }}{2 q \pi \Lambda}  \dd  \Lambda  =  \rho_{\rm MP}(\Lambda)\dd  \Lambda  \\
& \text{with} \quad \Lambda_\pm =\frac{1}{T} \frac{2}{\sigma^2} e^{-x_{\mp}} 
\end{split}
\ee 
where $ \rho_{\rm MP}$ is the Mar\v{c}enko-Pastur distribution \eqref{eq:1:MarcenkoPasturRewritten} with parameters
\be 
\begin{cases}
\displaystyle q = \frac{N}{T}=  \frac{N}{N- \frac{2m}{\sigma^2}}   \; ; \\[10pt]
\displaystyle \sigma_{W} = 1 \; .
\end{cases} 
\label{eq:3:MPParametersBulkDensity}
\ee 
Here, we have renamed the Wishart variance $\sigma_W$ as in \myeqref{eq:3:InverseWishartParametersStationary}.

We recover in this asymptotic Mar\v{c}enko-Pastur result the fact that the $\Lambda_i \propto 1/\lambda_i$ are distributed according to the eigenvalues of the Wishart ensemble, since the $\lambda_i$ are distributed as the eigenvalues of the inverse-Wishart ensemble. The parameters $q$ and $\sigma_{W} =1$, coincide as expected with the parameters obtained from the inverse-Wishart correspondence \myeqref{eq:3:InverseWishartParametersStationary} after the rescaling implemented in the definition \eqref{eq:3:DefinitionLambda}.

In addition to giving a RMT perspective on the Fermi gas asymptotic result at large-$N$, this obtention of the Mar\v{c}enko-Pastur distribution indicates that the system displays two different regimes depending on parameter $\sigma$:
\begin{itemize}[itemsep=.3em,topsep=10pt]

\item Either $\sigma$ scales as $1/\sqrt{N}$ such that
\be \sigma \simeq  \frac{ \tilde \sigma}{\sqrt{N}} \ee
with $\tilde \sigma$ constant, such that 
\be 
q=  \frac{1}{1- \frac{2m}{\tilde \sigma ^2}} <1
\label{eq:3:qValueSigmaTildeRegime}
\ee
where $m$ is necessarily negative per condition \eqref{eq:3:NormalizabilityConditionBeta2}. In this case, asymptotics of the Wishart ensemble presented in chapter \ref{chap:1} show that there are two soft edges at the boundaries of the support, where the statistics are described by the universal Airy kernel. 

The Fermi gas and the Kesten eigenvalues are then, in turn, distributed between two soft edges at fixed positions. This can be understood intuitively as follows: at large $N$, the depth of the trapping potential grows as $-g^2 \sim  N^2$. In this regime, the Fermi energy in \myeqref{eq:3:FermiEnergyKesten} grows at the same speed $\epsilon_F \sim N^2$, and the fermions thus remain confined between two soft edges.

\item Or $\sigma$ remains constant of order $1$, such that asymptotically
\be 
q=1 \; .
\ee 
In this case, the lower edge of the Wishart ensemble at $0$ is a hard edge, where the statistics are described by the universal Bessel kernel. The Fermi gas and the Kesten eigenvalues then lose their right edge and extend infinitely to the right for $N = + \infty$.

In this case, the Fermi energy remains of order 1, $\epsilon_F = \bigO(1)$, see \myeqref{eq:3:FermiEnergyKesten}. The depth of the potential however still grows as $-g^2 \sim  N^2$, such that the support of the density extends very far to the right, as explained above.
\end{itemize}
The difference between these two regimes is illustrated in Fig. \ref{fig:FermiEnergyScaling}. In the following, we present the edge densities for both cases.

\begin{figure}[ht!]
    \centering
    \begin{subfigure}[t]{0.45\textwidth}
        \centering
    \includegraphics[height=5cm]{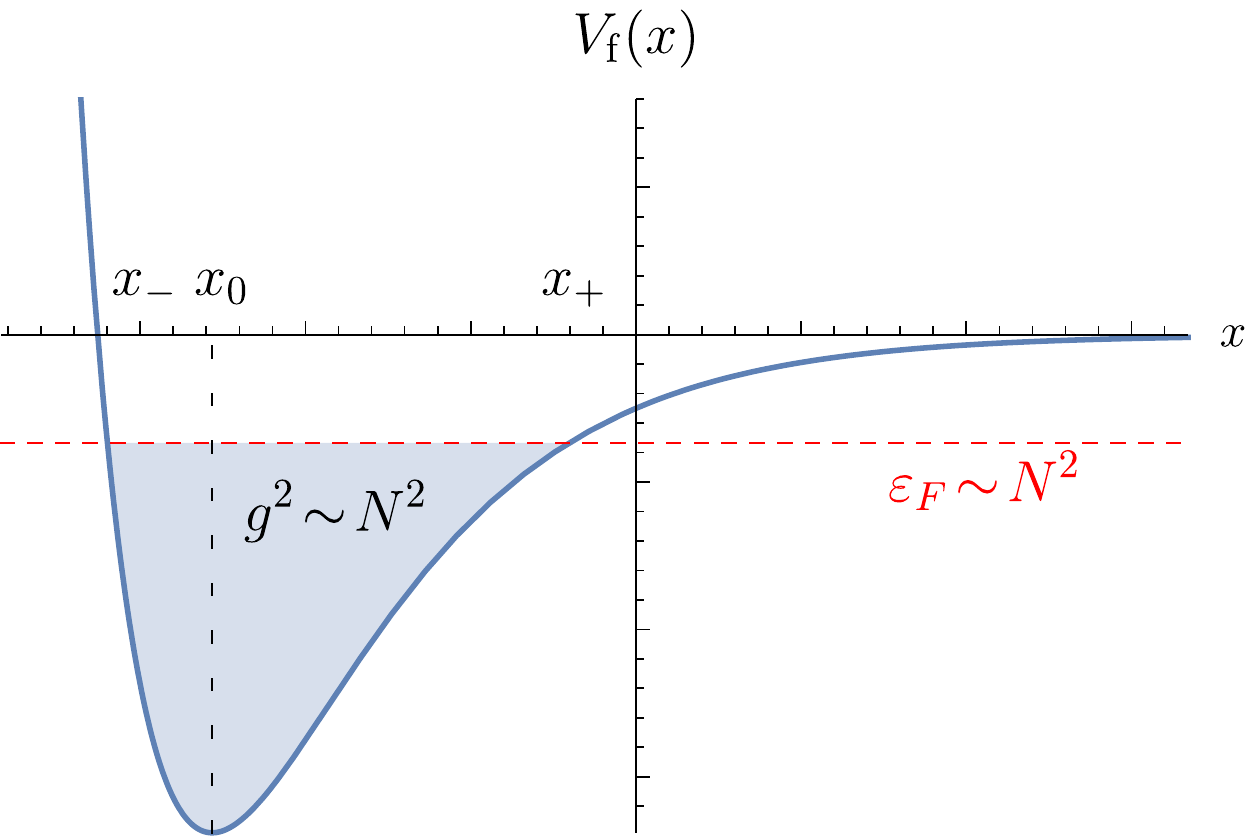}  
    \end{subfigure}%
    \hfill
    \begin{subfigure}[t]{0.45\textwidth}
        \centering
    \includegraphics[height=5cm]{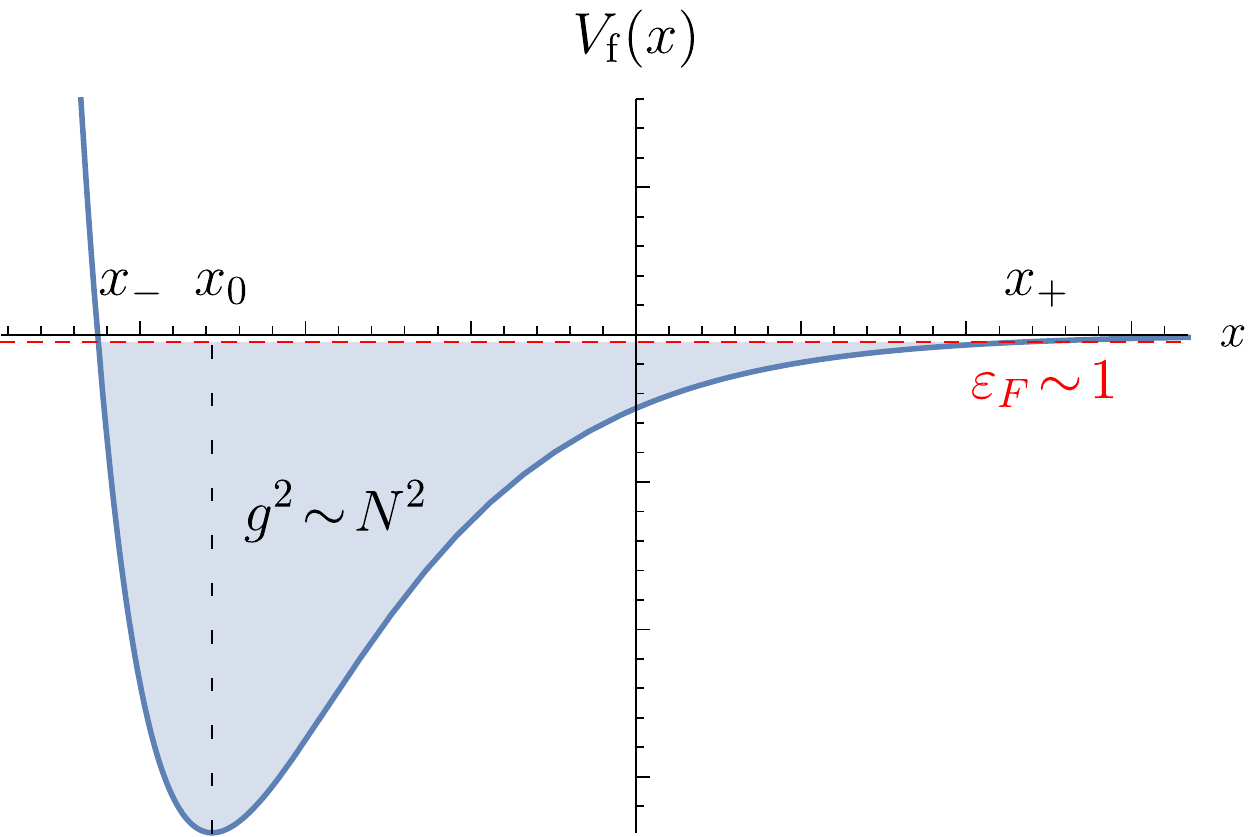}  
    \end{subfigure}
    \caption{Fermi sea of non-interacting fermions in the Morse potential, for large number $N$ of particles, in the frame where $x_0$ is fixed. \emph{Left:} Regime $\sigma \sim \tilde \sigma /\sqrt{N}$. \ \emph{Right:} Regime of constant $\sigma$.}
    \label{fig:FermiEnergyScaling}
\end{figure}

\subparagraph{Density at the edges with $\sigma \sim \tilde{\sigma}/\sqrt{N}$} With vanishing $\sigma$ at speed $1/\sqrt{N}$, the edges of the Fermi gas converge to the fixed positions
\be 
x_\pm = -  \log \left( \tilde{\sigma}^2 - m  \mp  \tilde{\sigma}^2 \sqrt{1- \frac{2m}{\tilde{\sigma}^2 }} \right) \; .
\ee
Between these two edges, the density is asymptotically given by \myeqref{eq:3:BulkDensityKestenFermigas}.

Near the two edges, denoting $ w_\pm =  V_\mathrm{M}'(x_\pm) ^ {-1/3} = \mathcal{O} \left( N^{-2/3}\right) $ the length scale which characterizes the width of the edge regions, the density is correctly described by the one obtained from the Airy kernel such that:
\be   
\bar{P}^\mathrm{edge}_\pm(x) \simeq \frac{1}{N \abs{w_\pm}} F_{1}\left(\frac{x-x_\pm}{w_\pm}\right)
, \quad
 F_{1}(z)=\left[\mathrm{Ai}^{\prime}(z)\right]^{2}-z[\mathrm{Ai}(z)]^{2}
 \label{eq:3:FermiGasDensitySigmaTildeFromAiryKernel}
\ee 
The agreement of the 1-particle density with the approximations in the bulk and at the edges, in the case of the $\sigma \sim 1/\sqrt{N}$ scaling, is illustrated in Fig. \ref{fig:KestenDensitySigmaTilde}. In this figure, the parameters are fixed as $\tilde \sigma=2$ and $m=-3$. For these parameters, we have $x_-\simeq -2.59$, $x_0\simeq -1.95$ and $x_+ \simeq 0.39$. The number of fermions is $N=10$ in the main plot and $N=50$ in the inset, where we see that $\bar{P}^\mathrm{edge}_+(x)$ is closer to the exact density $\bar{P}(x)$ for this higher number of particles. 

 \begin{figure}[ht!]
    \centering 
    \includegraphics[width=.85 \textwidth]{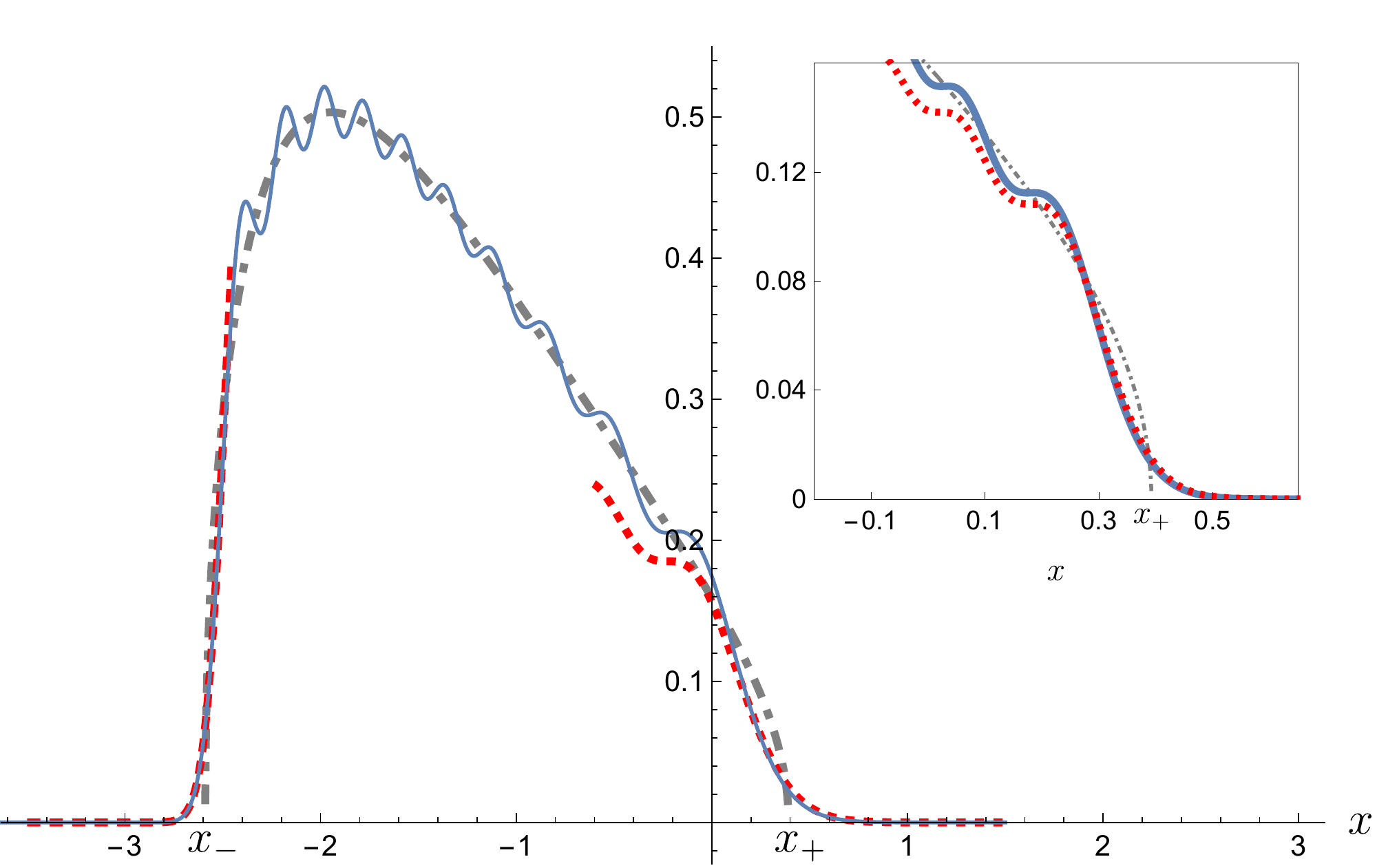} 
    \caption{\emph{Scaling $\sigma \sim 1/\sqrt{N}$ :\ } Plot of the exact average density $\bar{P}(x)$ \eqref{eq:3:ExactDensityFermiGas} for $N=10$ particles (solid blue line). The bulk density $\bar{P}_{\rm bulk}(x)$ \eqref{eq:3:BulkDensityKestenFermigas} is the dashed gray line, and the Airy-kernel edge densities $\bar{P}^\mathrm{edge}_\pm(x)$ \eqref{eq:3:FermiGasDensitySigmaTildeFromAiryKernel} are the dashed red lines. The inset shows the region near the right edge for $N=50$.}
    \label{fig:KestenDensitySigmaTilde}
\end{figure}

The density of the Kesten variables in the regime $\sigma \sim 1/ \sqrt{N}$ is then deduced straightforwardly from the above results.

\subparagraph{Density at the edges with constant $\sigma$}

In the case where $\sigma$ is a constant of order 1, the position of the minimum of the potential is asymptotically ${x_0 = - \ln(\sigma^2 N)}+ \mathcal{O}(\frac{1}{N})$ and the left edge is at a fixed distance of the minimum $x_- = x_0 - \ln 2 + \mathcal{O}(\frac{1}{N^2})$, while the right edge is at diverging distance from the minimum ${x_+ = x_0 + 2 \ln (\sigma^2 N) - \ln (\frac{m^2}{2}) + \mathcal{O}(\frac{1}{N})}$. In the translated frame where the position of the minimum is fixed, the Fermi gas only has one edge to the left, while extending infinitely to the right for $N = + \infty$.  

In the vicinity of the left edge, the Airy kernel characterizes the behaviour of the Fermi gas. With $w =V_\mathrm{M}'(x_-)^{-1/3} = \mathcal{O}(N^{-2/3})$, the density is then described as above by:
\be   
\bar{P}^\mathrm{edge}_-(x) \simeq \frac{1}{N \abs{w}} F_{1}\left(\frac{x-x_-}{w}\right) \; .
\label{eq:3:LeftEdgeDensityFermiGas}
\ee  

To the far right, the density of the Fermi gas can be obtained from the Bessel kernel \eqref{eq:1:BesselKernel}, which describes the hard edge behaviour of the Wishart ensemble, as
\be 
\bar{P}^\mathrm{far \, right}(x) \simeq \frac{8}{\sigma^2}e^{-x} K^{\mathrm{Bessel}}_{-\frac{2m}{\sigma^2}}
\left( \frac{8N}{\sigma^2}e^{-x} , \frac{8N}{\sigma^2}e^{-x} \right) \; .
\ee 
This can be written as a function of the distance between $x$ and the diverging edge $x_+ \simeq \log \left( \frac{2 N \sigma^2}{m^2} \right)$, with the expression of the Bessel kernel at the singular point in terms of Bessel functions: 
\be 
\begin{split}
\bar{P}^\mathrm{far \, right}(x) =
\frac{2}{\sigma^2}e^{-x} \bigg( &
J_{-\frac{2m}{\sigma^2}} \left( \frac{2\abs{m}}{\sigma^2}e^{-\frac{x-x_+}{2}} \right)^2 \\
&- J_{-\frac{2m}{\sigma^2}+1} \left( \frac{2\abs{m}}{\sigma^2}e^{-\frac{x-x_+}{2}}\right)
J_{-\frac{2m}{\sigma^2}-1} \left( \frac{2\abs{m}}{\sigma^2}e^{-\frac{x-x_+}{2}}\right)
\bigg)  \; .
\end{split}
\label{eq:3:FarRightDensityFermiGas}
\ee
The agreement of the 1-particle density with the approximations in the bulk and at the edges, in the case of constant $\sigma$, is illustrated in Fig. \ref{fig:KestenDensityConstantSigma}. In this figure, the parameters are fixed as $\sigma=2$ and $m=-3$. The main plot is drawn for $N=10$ fermions, with $x_-\simeq -4.45$, $x_0\simeq -3.76$ and $x_+ \simeq 2.26$. The inset has $N=50$, where $x_+ \simeq 3.81$. We see that the exact density is very closely approximated by $\bar{P}^\mathrm{far \, right}(x)$ in that region.

 \begin{figure}[ht!]
    \centering 
    \includegraphics[width=.85 \textwidth]{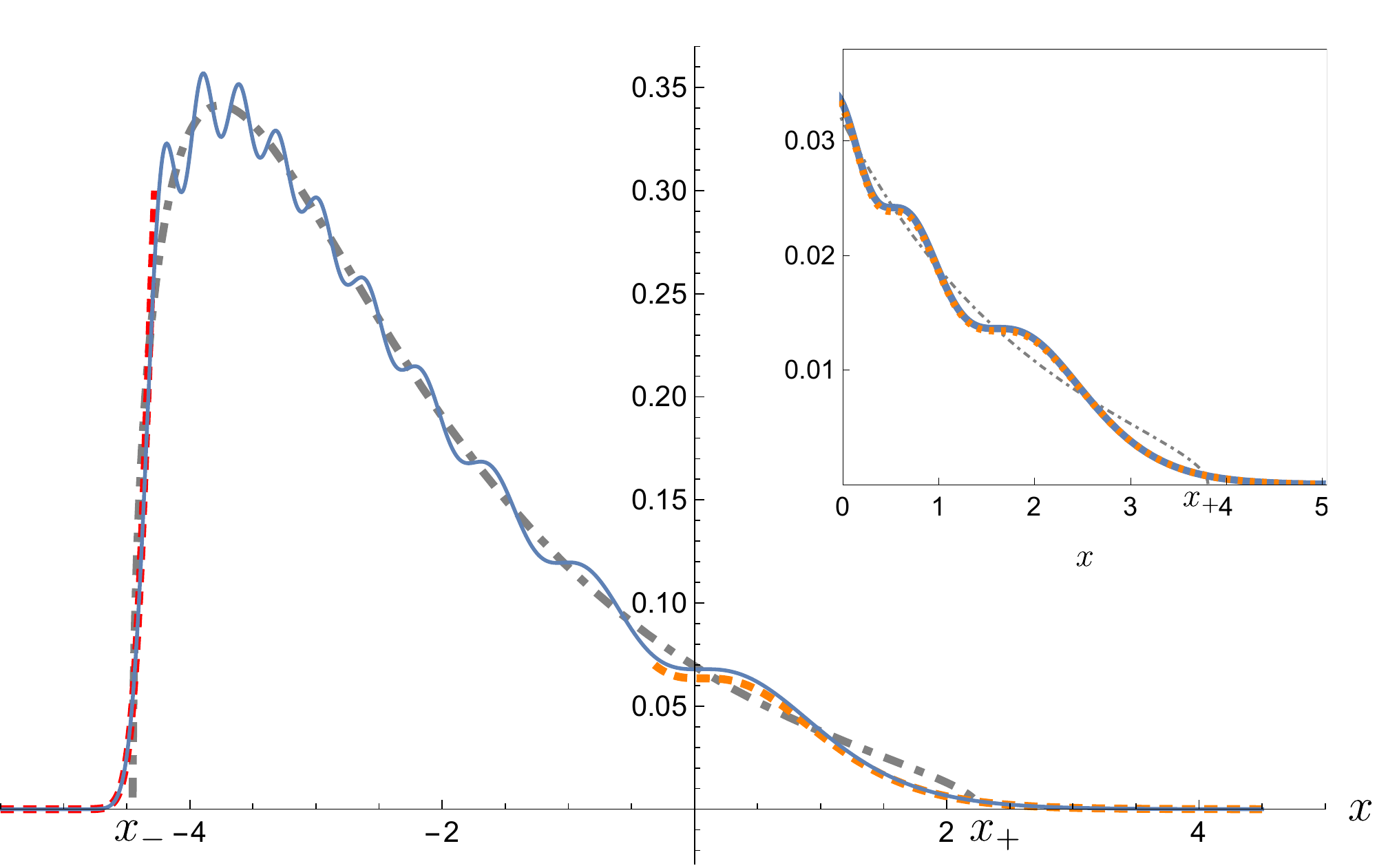} 
    \caption{\emph{Constant $\sigma$ :\ } Plot of the average density $\bar{P}(x)$ \eqref{eq:3:ExactDensityFermiGas} for $N=10$ particles (solid blue line), with the approximations in the bulk $\bar{P}_{\rm bulk}(x)$ \eqref{eq:3:BulkDensityKestenFermigas} (dashed gray line), at the left edge $\bar{P}^\mathrm{edge}_-(x)$ \eqref{eq:3:LeftEdgeDensityFermiGas} (red dashed line) and on the far right $\bar{P}^\mathrm{far \, right}(x)$ near $x_+ \sim \log N$ \eqref{eq:3:FarRightDensityFermiGas} (orange dashed line). The inset shows this latter region for $N=50$.}
    \label{fig:KestenDensityConstantSigma}
\end{figure}

The density of the Kesten variables in the regime of constant $\sigma$ is then deduced straightforwardly from the above results. For more details on the tail of the distribution of the rightmost eigenvalue from asymptotics of the solution of the Painlevé III equation, see \pubref{publication:MatrixKesten}.

\paragraph{Finite-time solution}

Further from the large-time stationary distribution detailed in the previous sections through the ground state of the related Fermi gas, one can solve analytically the full dynamics of the Kesten eigenvalue process in the present case $\beta=2$. In this section, we give the analytical expression for the PDF $P(\vec{\lambda},t)$ at arbitrary finite time.

Taking into account the bound states and the continuous branch of the spectrum, the Euclidean propagator, or Green's function in imaginary time, of a one-particle system in the Morse potential \eqref{eq:3:MorsePotentialgxzero} can be expressed in terms of the Whittaker $W$ function as:
\be 
\begin{split}
G(y,t \mid x,0)  &=  \sum_{k=0}^{\lfloor \left(g - \frac{1}{2} \right)^- \rfloor} \psi_k(x) \psi_k(y) e^{- (g - k - \frac{1}{2})^2 t}  \\
& + \int_0^{+\infty} \frac{ \dd p}{2 g \pi^2} \ p \ \sinh(2 \pi p)  \ \abs{\Gamma(\ii p - g + \frac{1}{2})}^2  \ e^{\frac{x+y}{2} - x_0} \\ 
& \qquad W_{g, \ii p}(2 g e^{-(x-x_0)}) \ W_{g, \ii p}(2 g e^{-(y-x_0)}) \ e^{- p^2 t} \; .
\end{split}
\ee
From completeness of the eigenbasis, this propagator satisfies $G(y,0 \mid x,0) = \delta(x-y)$. The propagator of the $N$-particle system of non-interacting fermions in the Morse potential is then simply:
\be
G(\vec{y},t \mid \vec{x},0)= \bra{\vec{y}} e^{-t \Ham } \ket{\vec{x}}= \det_{1 \leqslant i,j \leqslant N } G(y_i, t \mid x_j,0)
\ee
which satisfies $G(\vec{y},0 \mid \vec{x},0) = \delta(\vec{x}-\vec{y})$. 
From this quantum result, the finite time solution of the transformed Kesten stochastic diffusion in the $x= \log(\lambda)$ variables \eqref{eq:3:TransformedEigenContinuousKestenSDE} can be obtained as:
\emphbe
\bar P( \vec{x} ,t)   
=    \bra{ \vec{x} } e^{- t \Ham_{FP} } \ket{ \vec{x}_0 } =  \left(\frac{\bar P_\infty( \vec{x} )}{\bar P_\infty( \vec{x}_0 )} \right)^{\frac{1}{2} }   G(\vec{x}, t \mid \vec{x}_0,0) \ e^{E_0 t}, \quad \quad \bar P( \vec{x} ,0) = \delta( \vec{x} - \vec{x}_0)   
\end{empheq}

Upon changing back to the $\lambda = e^{x}$ variables, these computations yield an analytical formula for the finite-time solution of the Kesten eigenvalue diffusion as:
\be 
P(\vec{\lambda},t) = \bar  P( \vec{x} = \ln \vec{\lambda},t)  \; \prod_{i=1}^N \frac{1}{\lambda_i} \; .
\ee

\subsubsection{Interacting case $\beta=1$}

We have treated the non-interacting case $\beta=2$ in detail, and will only mention briefly the properties of the interacting case $\beta =1$. In this case, the Kesten process can give many informations about the ground-state statistics of the interacting fermions. Note that these statistics cannot be easily found from first principles as a mere Slater determinant, in contrast with the non-interacting case. 

In the case $\beta=1$, the large-time stationary solution for the Kesten eigenvalue process was also obtained in \myeqref{eq:3:EigenJPDFStationarySolution}. The ground-state wavefunction of the interacting fermion system is directly obtained from this result as
\be 
\abs{\Psi_{\vec{0}} (\vec{x})} = \sqrt{  \bar{P}_\infty(\vec{x})  } 
\ee 
where $\bar{P}_\infty$ is expressed in terms of $P_\infty$ in \myeqref{eq:3:PtildeinftyFromPinfty}, and where the sign of the wave-function can be deduced from fermionic symmetry. The ground-state energy is also found from the above studies, as it is given by the constant $E_0$ in \myeqref{eq:3:ConstantEZeroTranslationHamiltonian} with $\beta=1$.

Furthermore, the large-$N$ behaviour of the system can also be obtained from the inverse-Wishart statistics of the Orthogonal class $\beta=1$. For the case $\sigma=\mathcal{O}(1)$, the left edge is now described, upon the same change of variable, by the universality of the GOE soft edge. The fermions on the far right, corresponding to the largest eigenvalues of the continuous matrix Kesten recurrence, are now described, upon this change of variable, by the $\beta=1$ Bessel Pfaffian point process, characterized by the universal Bessel $2 \times 2$ matrix kernel mentioned in section \ref{subsec:1.2.2}.

The finite-time solution of the diffusion requires the knowledge of the whole spectrum of Morse-Sutherland fermions, beyond the ground state discussed in the previous lines, addressed in the literature of integrable systems \cite{Inozemtsev1996}.

\subsection{Discrete recursion in the large matrix-size limit}
\label{subsec:3.1.3}

In analogous fashion with the scalar setting, the matrix Kesten recursion defined in \myeqref{eq:3:DefMatrixKestenRecursion} is fully solvable in its continuous version \eqref{eq:3:ContinuousKestenSDE}. We have detailed this solution through a mapping to Morse-Sutherland fermions in the previous section, with a particular emphasis on the non-interacting case $\beta=2$.

In this section, we turn back to the discrete setting, where $\epsilon$ remains of order 1 in \eqref{eq:3:DefMatrixKestenRecursion}. Similarly as in the scalar setting, the discrete version is not easily solved. However, the powerful toolbox of free probability presented in section \ref{sec:1.3} gives a way to handle this problem in the limit of large matrix-size $N \to \infty$, to which we turn in the following. We note that the Dyson index $\beta$ does not have the same significance in this section as in the previous one, since, in the free case, all the information is contained in the eigenvalue density: results apply to both $\beta=1,2$ cases.

\subsubsection{Free product approach}

Free probability tools can help characterize the behaviour of the Kesten matrix $\ZZ_n$ if the large matrices $\varepsilon \ID + \ZZ_n$ and $\bxi_n$ are free such that the Kesten recursion 
\be
\ZZ_{n+1}=\sqrt{\epsilon \ID +\ZZ_{n}}  \ \bxi_{n} \ \sqrt{ \epsilon  \ID +\ZZ_{n}} 
\ee
gives $\ZZ_{n+1}$ as their free product. The $\ST$-transform of the $\ZZ$ matrices is then a useful tool, as we discuss in the following.

\paragraph{$\ST$-transform recursion} 

Let us make two assumptions on the distribution of the large i.i.d.~random matrices $\bxi_n \sim \bxi$:
\begin{itemize}[itemsep=.3em,topsep=10pt]
\item it is invariant under eigenbasis-rotation ;
\item the spectrum of $\bxi$ is compactly supported in the large-$N$ limit.
\end{itemize}
We have under the first assumption, as explained in section \ref{sec:1.3}, that $\varepsilon \ID + \ZZ_n$ and $\bxi_n$ are asymptotically free, in the large-$N$ limit, for arbitrary index $n$. $\ZZ_{n+1}$ is their free product, such that by the factorization property \eqref{eq:1:FreeMultiplicationSTransform}:
\emphbe 
\ST_{\ZZ_{n+1}} = \ST_{\xi} \times \ST_{\varepsilon \ID + \ZZ_n}
\end{empheq}
This relation characterizes the spectral distribution of $\ZZ_{n+1}$, through the transform $\ST_{\ZZ_{n+1}}$, in terms of the spectral distribution of $\ZZ_n$, through the transform $\ST_{\varepsilon \ID + \ZZ_n}$. This function can be expressed in terms of the $\ST$-transform of $\ZZ_n$ itself, see \pubref{publication:MatrixKesten}, yielding the following recursion equation on the functional sequence $(\ST_{\ZZ_n})_{n \geqslant 0}$, for all $\omega$:
\emphbe  
\frac{\ST_{\bxi}(\omega)}{\ST_{\ZZ_{n+1} } (\omega) } = \frac{1}{\ST_{\ZZ_{n} }\left( \, \omega(1- \epsilon \,\frac{\ST_{\ZZ_{n+1} }(\omega)}{\ST_{\bxi}(\omega)} )   \, \right) } + \epsilon
\label{eq:3:KestenSTransformRecursion}
\end{empheq}

\subparagraph{Stationary self-consistent equation}
At large-$n$, the $\ST$-transform, denoted $\ST_\infty$, of the stationary spectral distribution solves the following self-consistent equation, for all $\omega$:
\be  
\frac{\ST_{\bxi}(\omega)}{\ST_\infty  (\omega) } = \frac{1}{\ST_\infty \left( \, \omega(1- \epsilon \,\frac{\ST_\infty (\omega)}{\ST_{\bxi}(\omega)} )   \, \right) } + \epsilon \; .
\label{eq:3:KestenSTransformInftySelfConsistentEquation}
\ee

In the previous lines, we have presented the relation which characterizes the evolution of the spectral distribution of $\ZZ_n$ for an arbitrary $\bxi$-distribution, in \eqref{eq:3:KestenSTransformRecursion}. At late times, this characterizes the stationary spectral distribution of $\ZZ_\infty$ in \myeqref{eq:3:KestenSTransformInftySelfConsistentEquation}. Note that these two equations are functional relations, holding for all $\omega$, which are not easily solved for an arbitrary choice of the noise matrix distribution. In the continuous limit however, these discrete relations can be shown to match the results of section \ref{subsec:3.1.2}, as we show below. Further, they give access to some information in a perturbative manner, as we discuss next.

\paragraph{Continuous limit} 

We recall the suitable form for the random matrix $\bxi$ in \myeqref{eq:3:bxiRandomScaling}, in the continuous-time $\varepsilon \to 0$ limit, as:
\be   
\bxi_{n}=(1+m \, \epsilon) \ID +\sigma \, \sqrt{\epsilon} \HH_{n}
\label{eq:3:bxiRandomScalingRappel}
\ee
where $\HH$ is a unit-variance GOE or GUE matrix. The large-$N$ spectral density is the same for both versions of $\HH$, such that there is no difference in the infinite-$N$ results between the $\beta=1$ and $\beta =2$ cases. The $\sqrt{N}$-scaling in the convergence of a Gaussian matrix spectral density to Wigner's semicircle in \myeqref{eq:1:WignerSemicircleConvergence}, imposes that
\be 
\sigma = \frac{\tilde{\sigma}}{\sqrt{N}}
\ee
with $\tilde{\sigma}$ fixed, in order for $\bxi$ to have a compactly supported asymptotic spectrum. We recognize here the large-$N$ situation of the previous section: this is the regime where the spectrum remains bounded, while on the other $\sigma= \bigO(1)$ implies a blowup of the support. From the continuous setting, recall also that $m<0$. The spectral distribution of $\bxi$ is fixed, by \myeqref{eq:3:bxiRandomScalingRappel}, as a translated semicircle distribution, such that by Table \ref{tab:RMTtransforms}:
\be  
\ST_{\bxi} (\omega) =  \frac{-1-m \varepsilon + \sqrt{(1+m\epsilon)^2 +4 \tilde{\sigma}^2 \varepsilon \omega}}{2 \tilde{\sigma}^2 \varepsilon \omega}  \; .
\ee

In the small-$\varepsilon$ limit, it is then natural to introduce the following expansion for the function of interest:
\be 
\ST_{\ZZ_n} (\omega) = \ST^{(0)}_{\ZZ_n} (\omega) + \varepsilon \ST^{(1)}_{\ZZ_n} (\omega) + \varepsilon^2 \ST^{(2)}_{\ZZ_n} (\omega) + \cdots
\label{eq:3:EpsilonExpansionSTransform}
\ee
At first order in $\varepsilon$, the recursion \eqref{eq:3:KestenSTransformRecursion} implies
\be 
\ST_{\ZZ_{n+1}}(\omega)  - \ST_{\ZZ_{n}}(\omega) = - \epsilon \ST_{\ZZ_{n}}(\omega) \left( m + \tilde{\sigma}^2 \omega + \ST_{\ZZ_{n}}(\omega) + \omega \ST_{\ZZ_{n}}'(\omega) \right)  \; .
\ee
In the continuous $\epsilon \to 0$ limit, the $\ST$-transform $\ST(\omega,t)$ of the continuous matrix process $\VV_t$ thus obeys the following PDE
\emphbe 
\frac{\partial}{\partial t } \ST(\omega,t) = - \ST(\omega,t) \left( m +  \tilde{\sigma}^2 \omega + \ST(\omega,t) + \omega \frac{\partial}{\partial \omega } \ST(\omega,t)  \right)
\label{eq:3:ContinuousEvolutionKestenSTransform}
\end{empheq}

\subparagraph{Stationary distribution}
The stationary solution of this evolution is
\be 
\ST_\infty(\omega) = -m - \frac{\tilde{\sigma}^2}{2} \omega = \abs{m} \left( 1 - \frac{\tilde{\sigma}^2}{2 \abs{m} }\omega \right)  \; .
\ee
This is the $\ST$-transform of the large-time $\VV_t$ process, which is thus distributed as
\begin{empheq}[box=\setlength{\fboxsep}{8pt}\fbox]{equation}
\VV_{\infty} \   \sim \   \InverseW
\end{empheq}
where $\InverseW$ follows an inverse Mar\v{c}enko-Pastur distribution, see Table \ref{tab:RMTtransforms}, with parameters
\be 
\begin{cases}
\displaystyle \kappa = \frac{-m}{\tilde{\sigma}^2} \; ; \\[5pt]
\displaystyle \sigma_W = \sqrt{\abs{m}} \; .
\end{cases}
\label{eq:3:IMPParametersContinuousTimeLargeNKesten}
\ee
As in the previous section, we add an index ${}_W$ to the standard deviation parameter of the IMP distribution, to avoid ambiguities. 

As expected, this distribution matches the large-$N$ stationary results of the previous section in the regime $\sigma \sim 1/\sqrt{N}$, where $q$ was obtained in \myeqref{eq:3:qValueSigmaTildeRegime}. Indeed, the $q$ and $\kappa$ inverse Mar\v{c}enko-Pastur parameters are related as $q=\frac{1}{2 \kappa +1}$, see \myeqref{eq:1:DefKappaParameterIMP}. Furthermore, the variance can also be shown to agree with \myeqref{eq:3:MPParametersBulkDensity} asymptotically. This shows a perfect matching between the stationary solution in the continuous limit for the present discrete-time/large-$N$ setting ; and the one obtained in the large-$N$ limit for the continuous-time/finite-$N$ setting in \myeqref{eq:3:EigenJPDFStationarySolution}.

\subparagraph{Exact solution}

The PDE \eqref{eq:3:ContinuousEvolutionKestenSTransform} can be solved analytically through a hodograph transform \cite{Siddiqui1985}. The convergence of $\ST(\omega,t)$ to $\ST_\infty(\omega)$ can further be investigated by linearizing the evolution equation of $\Delta \ST(\omega,t) =  \ST(\omega,t) - \ST_\infty(\omega)$. For details on both these approaches, see \pubref{publication:MatrixKesten}.

\paragraph{Non-vanishing $\varepsilon$}

The introduction of the $\varepsilon$-expansion \eqref{eq:3:EpsilonExpansionSTransform} allows to reach further than the continuous-time stationary distribution. At large-time with non-zero $\varepsilon$, the stationary distribution of $\ZZ_\infty$ can be characterized by perturbative corrections to its spectral density, moments, and distribution edges. We point to \pubref{publication:MatrixKesten} for the results of this perturbative analysis. 

In addition, the first moments can be obtained for arbitrary finite $\epsilon$ from a small-$\omega$ expansion. As an example, the first moment of $\ZZ_\infty$ can be obtained as
\be 
 \phi( \ZZ_\infty) = - \frac{m \epsilon + 1}{m} \quad \xrightarrow[\epsilon \to 0]{}  \quad \frac{1}{\abs{m}}
\ee
where the continuous-time limit coincides with \myeqref{eq:3:IMPParametersContinuousTimeLargeNKesten}.

\subsubsection{Evolution of other functions}

In the previous section, we have presented how the Kesten matrix diffusion can be studied from the point of view of the $\ST$-transform, under mild assumptions in the large-matrix-size limit. Crucially, this results in the discrete \eqref{eq:3:KestenSTransformRecursion} and continuous \eqref{eq:3:ContinuousEvolutionKestenSTransform} evolution equations. The study can be generalized to other functions of the Kesten eigenvalue process in the continuous setting, as we present now.

\paragraph{Stieltjes transform}

The Stieltjes transform of the Kesten matrix, with eigenvalues $(\lambda_{i,t})_{1 \leqslant i \leqslant N}$, is
\be
\g (z,t) = \frac{1}{N} \sum_{i=1}^N \frac{1}{z- \lambda_{i,t}} \; .
\ee 
As presented in chapter \ref{chap:1}, this function coincides with its expectation in the infinite $N$ limit, by the self-averaging property. The time-evolution of $\g(z,t)$ can be obtained from the eigenvalue SDE \eqref{eq:3:EigenContinuousKestenSDE}, by an application of Itô's formula. Removing the noise term, which vanishes for infinite $N$, the evolution of $\g$ is governed by the following PDE:
\be 
\frac{\partial}{\partial t} \mathfrak{g}(z,t)  = \frac{\partial}{\partial z} \left(  
- \mathfrak{g} + (\tilde{\sigma}^2 - m )z\mathfrak{g} - \frac{1}{2}  \tilde{\sigma}^2 (z\mathfrak{g})^2
\right) \; .
\label{eq:3:EvolutionStieltjesTransformLargeN}
\ee

This evolution can be deduced from \myeqref{eq:3:ContinuousEvolutionKestenSTransform}, as expected but not directly obvious. The stationary solution is easily seen to be the following inverse Mar\v{c}enko-Pastur Stieltjes transform
\be 
\g_\infty (z) =\frac{z(\kappa+1)-\frac{\kappa}{\sigma_W^2}-\kappa \sqrt{z- \mu_-^{IMP} } \sqrt{z-\mu_+^{IMP} } }{z^{2}}
\ee
with $\sigma_W^2 = \abs{m}$, $\kappa =- \frac{m}{\tilde{\sigma}^2}$ and  $
 \mu_{\pm}^{IMP} =\frac{1}{\kappa \sigma_W^2} \left( \kappa+1 \pm \sqrt{2 \kappa+1} \right)$, in keeping with the previous results and Table \ref{tab:RMTtransforms}.

\paragraph{Expected characteristic polynomial}

Another function of the eigenvalues which displays an interesting evolution equation is the expected characteristic polynomial $\Pi (z,t)$ defined as
\be 
\Pi (z,t) = \E \left[   \,
\prod_{i=1}^N (z- \lambda_{i,t} )
\, \right]  \; .
\ee
This function was studied for the DBM and diffusing Wishart process in \cite{Blaizot2010,Blaizot2014,Blaizot2013}.

Let us fix a finite matrix-size $N$, with Dyson index equal indifferently to $\beta=1$ or $2$. The evolution of $\Pi (z,t)$ can be found in closed form from the eigenvalue SDE \eqref{eq:3:EigenContinuousKestenSDE} as the following PDE
\be 
\begin{split}
\frac{\partial}{\partial t}  \Pi(z,t) = N \left( m -  \frac{\sigma^2(N-1)}{2} \right)  \Pi(z,t) & \\  +\left(  z ( \sigma^2   (N-1)  -m) -1 \right) \frac{\partial}{\partial z}  \Pi (z,t) & - \frac{\sigma^2}{2} z^2 \frac{\partial^2}{\partial z^2}  \Pi(z,t) \; .
\end{split}
\label{eq:3:EvolutionExpectedCharacteristicPolynomial}
\ee

Remarkably, the dynamics of $\Pi(z,t)$ can be related to the Morse potential $V_\mathrm{M}(x)$ in \myeqref{eq:3:MorsePotentialgxzero}. Indeed one can perform a change of functions as 
\be
\Pi(z=e^x,t) = f(x) \; \psi(x,t) 
\ee 
with 
\be
f(x)  = e ^ {  (N-\frac{1}{2}- \frac{m}{\sigma^2} ) x   }  \, e^{\frac{1}{\sigma^2} e^{-x}} \; .
\ee
One finds that $\psi(x,t)$ evolves according to \emph{minus} the Schrödinger Hamiltonian in the Morse
potential as
\be
\frac{\partial}{\partial t} \psi 
= \frac{\sigma^2}{2} \left( - \frac{\partial^2}{\partial x^2} + V_\mathrm{M}(x) - \epsilon_N \right) \psi  
\ee
where $V_\mathrm{M}(x)$ is obtained by setting $\beta =2$ in \myeqref{eq:3:PotentialMorseSutherland} and $\epsilon_N = - \left( \frac{1}{2} + \frac{m}{\sigma^2} \right)^2 $ is the $(N+1)$\textsuperscript{th} energy level in this potential, see \myeqref{eq:3:EnergyLevelMorsePotentialBeta2}. 
 
Because the initial condition for $\Pi(z,t=0)$ is a polynomial in $z$ of degree $N$, the
initial wavefunction $\psi(x,t=0)$ belongs to the subspace spanned by the $N+1$ lowest levels of the
Morse potential. Since the time evolution is reversed, it converges to the highest energy accessible. As a consequence, $\Pi(z,t)$ converges to the following degree-$N$ polynomial
\be 
\lim_{t \to +\infty} \Pi(z,t) = a_N \; z^N \; L_N^{-1-\frac{2 m}{\sigma^2}}(\frac{2}{\sigma^2 z})
\ee
which is obtained from $\psi_{N+1}$ in \myeqref{eq:3:EigenfunctionMorsePotentialBeta2}, with a constant $a_N$. As expected, this function can be verified to be a stationary solution for the PDE \eqref{eq:3:EvolutionExpectedCharacteristicPolynomial}.

We remark that the previous analysis holds at finite $N$ in both cases $\beta=1,2$ because $\prod_{i=1}^N (z- \lambda_{i,t})$ is linear in each eigenvalue. Taking the expectation on Itô's formula for the evolution of this product then removes any $\beta$-dependent terms inherited from $\dd \lambda_{i,t}$ in SDE \eqref{eq:3:EigenContinuousKestenSDE}.

\section{Other matrix evolutions}
\label{sec:3.2}

After having detailed the matrix Kesten process in the previous section, we turn here to other interesting processes: processes already studied in the literature that bear close similarities to the Kesten process on the one hand, and generalized classes of processes inspired by it on the other hand. 

Beforehand, we give a brief mention of the difference between the Itô and Stratonovich prescriptions, which will be useful in the rest of the section. In the previous sections and chapters, every SDE was considered to be in the Itô sense. We now present the different Stratonovich prescription, and authorize Stratonovich SDEs in the next sections.

\subsection{Itô and Stratonovich stochastic prescriptions} 
\label{subsec:3.2.1}

As mentioned after \myeqref{eq:3:StochasticEvolution1DContinuousKesten}, a stochastic differential equation where the Brownian term's prefactor depends on the diffusing process itself must be specified with a prescription, and understood in the sense of either \emph{Itô} or \emph{Stratonovich}.

\subsubsection{Scalar prescriptions}

Let us consider a scalar stochastic process $(X_t)_{t \geqslant 0}$ obeying the SDE
\be 
\dd X_t = f(X_t) \dd t + g(X_t) \dd B_t \; .
\label{eq:3:SDEPrescriptionsScalar}
\ee
The two prescriptions are obtained by understanding the evolution as the continuous limit of two different discrete schemes. These two schemes, presented in the following, differ by the choice of the exact value to choose for the variable $X_t$ in the noise term $g(X_t) \dd B_t$.

\paragraph{Itô prescription}

The Itô convention assigns to \myeqref{eq:3:SDEPrescriptionsScalar} the limit of the following discrete scheme:
\be 
X_{n+1}-X_n = f(X_n)  + g(X_n)  \dd B_{n+1} 
\ee
where $\dd B_{n+1}$ is a Gaussian random variable sampled independently from the value of $X_n$. As a consequence, in the continuum, the prefactor $g(X_t)$ is independent of the Brownian increment $\dd B_t$, such that 
\be 
\E \left[ g(X_t) \dd B_t \right] = 0 \; .
\ee

This prescription leads to the following Fokker-Planck equation for the PDF $P(x,t)$ of the $X$-process, employed in the previous sections: 
\be 
\frac{\partial}{\partial t} P(x,t) = \frac{1}{2} \frac{\partial^2 }{\partial^2 x} \left(  g(x)^2 P(x,t)  \right) - \frac{\partial}{ \partial x} \left(  f(x) P(x,t) \right) \; .
\label{eq:3:ItoFPEquation}
\ee

\paragraph{Stratonovich prescription}

The Stratonovich convention assigns to \myeqref{eq:3:SDEPrescriptionsScalar} the limit of the following discrete scheme:
\be 
X_{n+1}-X_n = f(X_n)  + g\left( \frac{X_n + X_{n+1}}{2}  \right)  \dd B_{n+1}  \; .
\ee
Instead of being defined from the previous value $X_n$, the Brownian term's prefactor is defined from the half-sum of the values before and after the jump. In the continuum, the prefactor $g(X_t)$ is not independent of $\dd B_t$ any more, such that
\be
\E \left[ g(X_t) \dd B_t \right] =  \E \left[  \frac{1}{2} g(X_t) g'(X_t) \right] \; .
\ee

This prescription leads to a different Fokker-Planck equation:
\be 
\frac{\partial}{\partial t} P(x,t) = \frac{1}{2} \frac{\partial}{\partial x} \left(  
g(x) \frac{\partial}{\partial x}  \left( g(x) P(x,t) \right) 
 \right) - \frac{\partial}{ \partial x} \left(  f(x) P(x,t) \right) \; .
\label{eq:3:StratoFPEquation}
\ee

\paragraph{Itô-Stratonovich Drift}

As can be seen from the comparison of the FP equations \eqref{eq:3:ItoFPEquation} and \eqref{eq:3:StratoFPEquation}, these prescriptions are finally equivalent up to a drift term $\Delta_{\mathrm{IS}}$ given by
\be 
\Delta_{\mathrm{IS}} = \frac{1}{2} \,   g \,  g' \; .
\ee 
Considering two versions of the SDE \eqref{eq:3:SDEPrescriptionsScalar}, one in Itô prescription with drift function $f_{\mathrm{I}}$ and the other in Stratonovich prescription with drift function $f_{\mathrm{S}}$, this means that both SDEs describe the same process if:
\be
 f_{\mathrm{I}}  = f_{\mathrm{S}} + \Delta_{\mathrm{IS}} \; .
\ee 

Note that the drift term can also be obtained from the SDE \eqref{eq:3:SDEPrescriptionsScalar} as the continuous limit of the difference of the two discrete schemes as:
\bea
\Delta_{\mathrm{IS}}  \, \dd t  &=& 
 g \left( \frac{X_{t+ \dd t} + X_t }{2} \right) \dd B_{t+\dd t}  - g(X_t) \dd B_{t+\dd t}  \\
&=&  
 g \left( X_t +  \frac{X_{t+ \dd t} - X_t }{2} \right) \dd B_{t+\dd t} - g(X_t) \dd B_{t+\dd t}  \\
 &=&
 \begin{aligned}
 g \left( 
 X_t +  \frac{1}{2}
  \left(
   f(X_t) \dd t +  g \left( \frac{X_{t+ \dd t} + X_t }{2} \right) \dd B_{t+\dd t}   
    \right) 
  \right) \dd B_{t+\dd t}  \\ - g(X_t) \dd B_{t+\dd t}  
  \end{aligned}
\eea
Expanding in a Taylor expansion at order $\dd B ^2 \sim \dd t$, this is as expected
\be 
\Delta_{\mathrm{IS}}  \, \dd t =  \frac{1}{2} g(X_t) g'(X_t) \, \dd t \; .
\label{eq:3:ResultDriftSDEComputations}
\ee

We point to \cite{VanKampen1981} for more details on these stochastic prescriptions, and their different behaviour under variable change. See \cite{Pesce2013} for an experimental display of the importance of the prescription choice in physical models of randomness.

\subsubsection{Matrix prescriptions}

The same prescriptions can be defined for the stochastic evolution of a matrix process. If the noise term's prefactor depends on the matrix process itself, the SDE should be endowed with a prescription. 

Similarly as in the scalar case, the matrix Itô and Stratonovitch prescriptions denote the limits of the two discrete schemes presented above, where the value of the prefactor's variable is taken before the jump in the former case, and as the half-sum of the values before and after the jump in the latter case.

As in the scalar case, both prescriptions are identical up to a drift term which is now a matrix denoted
\be 
\DelBB_{\mathrm{IS}} \; .
\ee
This drift is conveniently computed, for an actual process, by similar computations to the ones leading to \myeqref{eq:3:ResultDriftSDEComputations}.

In the rest of this chapter, the prescription used is specified in parenthesis for each SDE.

\subsection{Processes related to the matrix Kesten evolution} 
\label{subsec:3.2.2}

Some models studied in the literature present some similarities, but also some differences, with the Kesten matrix process presented in this chapter. These models were studied by Rider and Valk\'o on the one hand, and by Grabsch and Texier on the other hand. We discuss the precise relation between these models in the following.

\subsubsection{Model studied by Rider and Valk\'o} 
 
Rider and Valk\'o define in \cite{Rider2016} the following geometric $N \times N$ drifted matrix Brownian motion $\MM$:
\be 
\dd \MM_{t}=\left(\frac{1}{2} -  \mu\right) \MM_{t} \dd t + \MM_{t} \dd \XX_{t} , \quad \MM_{0}=\ID \quad \quad \quad \quad \text{(It\^o)} 
\ee
where the elements of the Brownian matrix $\XX_t$ are $N^2$ independent real Brownian motions, in contrast with the Hermitian symmetry of the Dyson Brownian motion. We recall that they are related as in \myeqref{eq:2:DefDBMGOE}.
	
Since $\MM_t$ is not symmetric, the authors study the law of $\MM_t \MM^T_t$. To this aim, they define the process $\QQ_t =  \MM_{-t}^{-1} \left(  \int_{-t}^\infty \MM_s \MM_s^T \mathrm{d}s \right) \MM_{-t}^{-T}$ for negative time $t$, and show that it obeys the It\^o SDE:
\be 
\label{eq:3:RiderValkoSDE}
\dd \QQ_{t}=\left( (1 + \Tr  \QQ_{t} ) \ID +      (1-2 \mu) \QQ_{t} \right) \dd t +  \QQ_{t} \dd  \XX_{t} + \dd  \XX_{t}^T \QQ_{t}  \quad \quad \quad \quad \text{(It\^o)} \; .
\ee
The stationary distribution of this SDE, i.e.\ the law of $\QQ_0 = \int_{0}^\infty \MM_s \MM_s^T \dd s$, is proved to be the following inverse-Wishart distribution:
\be
\label{eq:3:RiderValkoInverseWishart}
\gamma_{2\mu}^{-1}(\QQ) \ \propto \ \operatorname{det} ( \QQ )^{ - \mu -1 - \frac{N-1}{2}} \ e^{-\frac{1}{2} \Tr \QQ^{-1}}   \; .
\ee
The motivation of this construction is to provide an extension to the matrix realm of the Dufresne identity which gives the distribution of the exponential Brownian functional \eqref{eq:3:UInfinityBrownianFunctional} as an inverse-gamma law.

\subsubsection{Models studied by Grabsch and Texier}

\paragraph{Multichannel random-mass Dirac equation}

In \cite{Grabsch2016}, Grabsch and Texier study topological phase transitions occurring in a multichannel random wire modelled by a random-mass Dirac equation. As explained in the chapter introduction, multichannel disordered wires are naturally studied through stochastic matrix evolutions. In \cite{Grabsch2016} specifically, a Riccati matrix $\ZZ_t$ is introduced, which obeys the SDE:
\be 
\dd \ZZ_t= \left( -\ZZ_t^{2} +k^2 \ID -\mu \GG \ZZ_t-\mu \ZZ_t \GG \right)\dd t  + \sigma_\beta (\ZZ_t \dd \HH_t +\dd \HH_t   \ZZ_t) \quad \quad \quad \quad \mathrm{(Stratonovich)}
\ee
where both symmetry cases $\beta=1,2$ are considered, such that the matrix $\ZZ$ is real symmetric for $\beta = 1$ and complex Hermitian for $\beta=2$. $\HH_t$ is a Hermitian matrix Brownian motion with correlations given by $\GG$, which coincides with the DBM in the case where $G=I$ provided one sets $\sigma_\beta=\sqrt{\beta/2}$. We keep here the evolution variable as $t$, but the evolution parameter, in the original setting, is in fact a position $x$ along the longitudinal direction of the disordered wire, see \cite{Grabsch2016}. Notice finally that this stochastic equation is given with the Stratonovitch prescription.

Using the Fokker-Planck operator acting on the entries of the matrix $\ZZ$, Grabsch and Texier demonstrate that the matrix diffusion for $\ZZ$ admits the following stationary distribution for $\sigma_\beta=\sqrt{\beta/2}$:
\be 
 f(\ZZ) \  \propto \  \det ( \ZZ)^{-\mu-1-\beta\frac{N-1}{2}   }  \  e^{-\frac{1}{2} \Tr \left\{ \GG^{-1}\left(\ZZ+k^{2} \ZZ^{-1}\right)\right\} } \; .
\ee
Let us restrict to the special case $\GG=\ID$. After the change of variables $\{\ZZ,k^2\} \to \{ \epsilon \ZZ, \epsilon   \}$ the SDE becomes in the $\epsilon \to 0$ limit:
\be 
\label{eq:3:LimitSDE}
\dd \ZZ_t= \left(   \ID - 2 \mu \ZZ_t \right)\dd t + \sigma_\beta( \ZZ_t \dd \HH_t +\dd \HH_t   \ZZ_t ) \quad \quad \quad \quad \mathrm{(Stratonovich)}
\ee
which admits as a consequence the following stationary distribution, restoring the parameter $\sigma_\beta$ for future comparisons:
\be 
\label{eq:3:StatDistLimit}
 \tilde{f}(\ZZ) \ \propto \ \det ( \ZZ)^{
 - \frac{\beta}{2\sigma_\beta^2}\mu -1 - \beta \frac{N-1}{2}}
 \  e^{ -\frac{\beta}{4\sigma_\beta^2} \Tr \ZZ^{-1}} \; .
 \ee

\paragraph{Wigner-Smith time delay}
 
In a recent work \cite{Grabsch2020}, Grabsch and Texier encountered the SDE \eqref{eq:3:LimitSDE} in a different problem: the evolution of the (symmetrized) Wigner-Smith time delay matrix $\tilde{\QQ}$ of a multi-channel disordered wire, where the size $L$ of the disordered region plays the role of the time parameter in the SDE. More precisely, the scattering process is defined by a Schr\"odinger equation with random potential coupling the $N$ channels. Their work generalizes a problem which is known to coincide in the case $N=1$ with the Brownian exponential functional \eqref{eq:3:UInfinityBrownianFunctional}. In the general $N$ case, they show that the symmetrized Wigner-Smith matrix $\tilde{\QQ}$ can be written as:
\be
\label{eq:3:QTildeDef}
\tilde \QQ_t = \XX_t \left( \int_0^t \dd s  \, \XX_s^{-1} \left(\XX_s^{ \dagger}\right)^{-1} \right) \XX_t^\dagger
\ee
where $\XX_t$ follows the following Stratonovitch flow:
\be 
\dd \XX_t = (-\mu + \sigma_\beta \, \dd \HH_t) \XX_t\quad \quad \quad \quad \mathrm{(Stratonovich)}
\ee 
such that $\tilde \QQ_t$ satisfies the SDE \eqref{eq:3:LimitSDE} where $\ZZ_t$ is replaced by $\tilde \QQ_t$. They rewrite \eqref{eq:3:QTildeDef} by defining $\LL^\dagger_s = \XX_t \XX_{t-s}^{-1}$, which satisfies $\dd \LL_s = (-\mu + \sigma_\beta \dd \tilde \HH_s)\LL_s$ in Stratonovitch prescription, where $\dd \tilde \HH_s = \dd \HH_{t-s}$. Upon reordering $s \to t -s$ in \eqref{eq:3:QTildeDef}, they obtain:
\be 
\label{eq:3:QTildeIntegraleLambdacroixLambda}
\tilde \QQ_t  \overset{\mathrm{law}}{=} 
\int_{0}^t \LL^\dagger_s \LL_s \, \dd s \; .
\ee
As a result they show that both $\tilde \QQ_t $ and the r.h.s.~of \myeqref{eq:3:QTildeIntegraleLambdacroixLambda} are distributed as \eqref{eq:3:StatDistLimit} in the limit of large time.

As pointed out by Grabsch and Texier, this is very close to the results of Rider and Valk\'o in the real case ($\beta =1$) as can be seen by comparing the stationary distributions \eqref{eq:3:RiderValkoInverseWishart} and \eqref{eq:3:StatDistLimit} with $\sigma_\beta = \sqrt{\beta/2}$. Fixing $\beta= 1$, the two SDEs \eqref{eq:3:RiderValkoSDE} and \eqref{eq:3:LimitSDE} differ by a drift term $ ( \QQ + \ID \Tr \QQ  ) \dd t$. This is shown below to be exactly the drift $\DelBB_{\rm IS}$ to be removed to the Itô SDE \eqref{eq:3:RiderValkoSDE} with noise term $ \QQ \, \dd \XX +  \dd \XX^T \, \QQ  $ to transform it into a Stratonovich SDE. The only difference of the two SDEs is the absence of structure in the noise matrix $\dd \XX_t$ used by Rider and Valk\'o. Grabsch and Texier show in \cite{Grabsch2020} that this is transparent at the level of the distribution. Indeed, the non-symmetric structure can be dealt with by gauging out the non-Hermitian part  of $\dd \XX_t$ in a way that leaves the law of $\int_{0}^\infty \MM_s \MM_s^T \dd s$ unaltered. In conclusion, the models introduced by both groups are finally identical for $\beta =1$. 
 
Since the expression of $\tilde{\QQ}$ as a matrix Brownian functional and the stationary distribution for SDE \eqref{eq:3:LimitSDE} were both established also in the complex Hermitian case ($\beta =2$) by Grabsch and Texier, their work is in fact an extension of the matrix Dufresne identity to the complex case.

\paragraph{Drift computation}

We compute the Itô-Stratonovich drift $\DelBB_{\rm IS}$ corresponding to the noise structure
\be 
\dd \QQ = \QQ \, \dd \XX +  \dd \XX^T \, \QQ  
\ee
where we recall that $\XX_t$ is the matrix process whose entries are $N^2$ independent Brownian motions. As explained in section \ref{subsec:3.2.1}, this drift can be obtained as the continuous limit of the Stratonovich and Itô discrete schemes, at order $\dd \XX^2$:
\bea 
\DelBB_{\rm IS} \, \dd t  &=& 
\begin{aligned}
    \frac{\QQ_t + \QQ_{t+\dd t} }{2} \dd \XX_{t+\dd t} + \dd \XX_{t+\dd t}^T \frac{\QQ_t + \QQ_{t+\dd t} }{2}  \\    - \left(  \QQ_t \dd \XX_{t+\dd t} + \dd \XX_{t+\dd t}^T \QQ_t \right) 
 \end{aligned}   
    \\
&=& \frac{1}{2} \dd \QQ_t  \dd \XX_{t+\dd t}  +\frac{1}{2} \dd \XX_{t+\dd t}^T   \dd \QQ_t   \\
&\simeq & \frac{1}{2} (    \QQ_t  \dd \XX_{t+\dd t}^2 + {\dd \XX_{t+\dd t}^T}^2 \QQ_t) + \dd \XX_{t+\dd t}^T \QQ_t \dd \XX_{t+\dd t} \; .
\eea 
In the continuous limit, the drift is given as a consequence in terms of the matrix process $\QQ$ and with $\XX$ a matrix filled with $N^2$ independent standard Gaussian random variables:
\be 
\DelBB_{\rm IS}  =  \E\left[ 
\frac{1}{2} (   \QQ \,  \XX^2 +  {\XX^T}^2  \, \QQ) +  \XX^T  \, \QQ \,   \XX
\right] \; .
\ee 
The correlations between entries of the matrix $\XX$ verify
\be 
\E  \left[ \XX_{ij} \XX_{kl} \right]  = \delta_{ik} \delta_{jl}
\ee
which yields
\be 
\DelBB_{\rm IS} = \QQ +  \ID \, \Tr(\QQ)\; .
\ee 
As explained above, this is the difference between the Stratonovich SDE of Grabsch-Texier and the Itô SDE of Rider-Val\'o, such that their models are identical. This process is closely related to the Kesten evolution presented in the previous sections.

\subsubsection{Connections with the Kesten evolution} 

The Kesten evolution is naturally related to the Brownian exponential functional, as shown in the scalar case in \myeqref{eq:3:UInfinityBrownianFunctional}. The matrix model was however not defined from this point of view, but instead from the recursion \eqref{eq:3:DefMatrixKestenRecursion}. We showed that, in the continuous limit, this recursion yields the SDE \eqref{eq:3:ContinuousKestenSDE}
\be 
\label{eq:3:ItoSDEIVC}
\dd \VV _t = \left( \ID +m \VV _t\right) \dd t +\sigma \sqrt{\VV _t} \dd \HH_{t} \sqrt{\VV _t}
 \quad \quad \quad \quad \text{(It\^o)} \; . 
 \ee
This is close to the diffusion of Rider-Valk\'o and Grabsch-Texier, as the drift terms are essentially the same, but there is a crucial difference in the noise term, which is $\sqrt{\VV } \dd \HH \sqrt{\VV }$ instead of $\VV \dd \HH + \dd \HH \VV $. 

Remarkably, although these two matrix processes are \emph{different}, the eigenvalue processes are actually \emph{identical} provided one compares Eq. \eqref{eq:3:ItoSDEIVC} in It\^o prescription and Eq. \eqref{eq:3:LimitSDE} in Stratonovitch prescription, with some correspondence between the parameters. This is shown in \pubref{publication:MatrixKesten} through the study of the joint evolution of the eigenvalues in the Grabsch-Texier model, which was not done in \cite{Grabsch2016,Grabsch2020}. We can identify the eigenvalues of $\VV _t$ and those of $\ZZ_t$ if we set $\sigma_\beta = \frac{\sigma}{2}$ and $\mu = -\frac{m}{2} + \frac{\sigma^2}{4}(N+2-\beta)$. Inserting these values in the stationary distribution \eqref{eq:3:StatDistLimit} reproduces exactly our result \eqref{eq:3:EigenJPDFStationarySolution} for $\beta=1,2$.

The evolution processes of the eigenvectors, however, have no obvious reason to be related. Since the stationary matrix measure is isotropic in both cases, the two models converge to the identical inverse-Wishart distribution.


\subsection{Generalized processes}
\label{subsec:3.2.3}

After having presented in details the Kesten matrix stochastic evolution \eqref{eq:3:ContinuousKestenSDE} as well as the related processes \eqref{eq:3:RiderValkoSDE} and \eqref{eq:3:LimitSDE}, we now give results on other processes constructed in the idea of generalizing and widening the scope of the previous sections. Some details on the computations leading to these results are given in App.~\ref{app:GeneralizedProcesses}. For simplicity, we restrict to the real symmetric $\beta =1$ case in the following. 
 
\subsubsection{Generalized Kesten process}

The Kesten process \eqref{eq:3:ContinuousKestenSDE} can be generalized by adding drift terms depending on any power of $\VV_t$, as
\be 
\dd \VV_t = \left( m \VV_t + \sum_{k\neq 1} p_k  \VV_t^k \right) \dd t + \sigma \sqrt{\VV_t} \dd \HH_t \sqrt{\VV_t}   \quad \quad \quad \quad \text{(It\^o)}  \; .
\label{eq:3:GeneralizedKesten}
\ee
We suppose, here and for the following models, that the coefficient $p_{k_{\rm{max}}}$ of the highest-power term is negative as to ensure the convergence of the process. The coefficient of the linear term $\propto \VV_t$ is kept aside as it drives the power-law behaviour of the stationary solution, as we will see below. The perturbative analysis of the eigenvalue process $(\lambda_{i,t})_{1 \leqslant i \leqslant N}$ yields  the Itô SDE
\be 
\dd \lambda_{i,t} =  ( m \lambda_{i,t} +\sum\limits_{k\neq 1} p_k \lambda_{i,t}^k) \dd t  + \sigma^2 \sum\limits_{j \neq i } \frac{\lambda_{i,t} \lambda_{j,t}}{\lambda_{i,t} - \lambda_{j,t}}   \dd t + \sqrt{2} \sigma \lambda_{i,t} \dd B_{i,t}
\ee
which is a generalization of \eqref{eq:3:EigenContinuousKestenSDE} in the presence of additional drift terms. The stationary JPDF of the eigenvalue process can be found to be
\be
P_\infty( \vec{\lambda}) \propto
\abs{\Delta(\vec{\lambda})}  \, \prod\limits_{i=1}^N \lambda_{i}^{\frac{m}{ \sigma^2} - (N+1)}    \; e^{ \frac{1}{ \sigma^2} \sum\limits_{1 \leqslant  i \leqslant N \atop k \neq 1}   \frac{p_{k}}{k-1} \lambda_{i}^{k-1} } \; .
\label{eq:3:GeneralizedKestenStationarySol}
\ee

This is an interesting generalization of the Kesten process, since the freedom gained in the drift terms allow to define the drift function as any sufficiently regular function $f$ which admits a Taylor expansion around $0$. The stationary solution of the model then depends on the Taylor coefficients as seen above, with the power-law exponent driven by the coefficient of the linear term.

\subsubsection{Generalized Grabsch-Texier-Rider-Valk\'o process}

A similar generalization, with arbitrary-power drift terms, can be defined for the processes \eqref{eq:3:RiderValkoSDE} and \eqref{eq:3:LimitSDE}. As explained in the previous section, these models are equivalent upon Itô-Stratonovich drift translation, up to the difference of symmetry in the noise terms. We deal with the two possible cases in the following, with $\dd \XX_t$ and $\dd \HH_t$ respectively a  structure-less and a symmetric noise term.

\paragraph{With asymmetric noise}

We can define a similar generalized model as \eqref{eq:3:GeneralizedKesten}, where the noise term is the one of \eqref{eq:3:RiderValkoSDE}, i.e. $\VV \dd \XX + \dd \XX^T \VV_t$, as:
\be 
\dd \VV_t =  \left(m \VV_t + \sum_{k\neq 1} p_k  \VV_t^k \right) \dd t  + \sigma \frac{ \VV_t \dd \XX_t + \dd \XX_t^T \VV_t }{\sqrt{2}}  \quad \quad \quad \quad \text{(Stratonovich)} \; .
\label{eq:3:GeneralizedGTRVAsymmetric}
\ee 
The noise term is renormalized for ease of comparison with the previous generalized model. 

The evolution of eigenvalues can be obtained from perturbative analysis, with the warning that the SDE should be taken to the Itô prescription before applying the computations described above \myeqref{eq:2:EigenvaluePerturbationTheory}. In this case, the drift term to be taken into account is the one computed in the previous section, up to a rescaling:
\be 
\DelBB_{\rm IS} = 
\frac{1}{2} \VV_t +   \frac{\Tr ( \VV_t) }{2} \ID  \; .
\label{eq:3:DriftISAsymmetricGTRV}
\ee
Taking SDE \eqref{eq:3:GeneralizedGTRVAsymmetric} to Itô prescription and applying the perturbation theory framework yields the following Itô SDE for the eigenvalues
\be 
\dd \lambda_{i,t} = \left(  \left(m + \sigma^2 \frac{N+1}{2} \right) \lambda_{i,t}  +  \sum_{k\neq 1} p_k  \lambda_{i,t}^k    \right)\dd t +\sigma^2 \sum\limits_{j \neq i } \frac{\lambda_{i,t}    \lambda_{j,t}}{\lambda_{i,t} - \lambda_{j,t}}  \dd t+  \sqrt{2} \sigma \lambda_{i,t} \dd B_{i,t} \; .
\label{eq:3:EigenvalueSDEAsymmetricGTRV}
\ee
The stationary solution for the JPDF of eigenvalues is then
\be
P_\infty( \vec{\lambda}) \propto  \abs{\Delta(\vec{\lambda})}   \, \prod_{i=1}^N \lambda_{i}^{\frac{m}{ \sigma^2} - \frac{N+1}{2 }   }   \  \; e^{ \frac{1}{ \sigma^2} \sum\limits_{1 \leqslant  i \leqslant N \atop k \neq 1}   \frac{p_{k}}{k-1} \lambda_{i}^{k-1} } \; .
\label{eq:3:GeneralizedGTRVAsymmetricStationarySol}
\ee

As in the simple case where $p_k = 0$ if $k \neq 0$ discussed in \ref{subsec:3.2.2}, the models with noise terms given by $\sqrt{\VV} \dd \HH  \sqrt{\VV} $ and $\VV \dd \XX + \dd \XX^T \VV_t$ are closely related, at the level of the eigenvalue evolution and stationary distribution. Notice indeed the proximity of \myeqref{eq:3:GeneralizedGTRVAsymmetricStationarySol} with the stationary solution of the generalized Kesten process in \myeqref{eq:3:GeneralizedKestenStationarySol}. 

The reason for this connection is a subtle interplay between the form of the noise and the prescription used, such that terms of the Stratonovitch-It\^o drift combine with the perturbative eigenvalue drift to recover the same eigenvalue flow. Under the perturbation computations, both noise terms indeed lead to different interaction terms
which finally match after taking into account the drift $\DelBB_{\rm IS}$. These manipulations are detailed in App.~\ref{app:GeneralizedProcesses}.

\paragraph{With symmetric noise}

A similar model can be defined with symmetric noise as
\be 
\dd \VV_t = \left(m \VV_t + \sum_{k\neq 1} p_k  \VV_t^k \right) \dd t  + \sigma \frac{\VV_t \dd \HH_t+ \dd \HH_t\, \VV_t }{\sqrt{2}}   \quad \quad \quad \quad \text{(Stratonovich)} 
\label{eq:3:GeneralizedGTRVSymmetric}
\ee  
which can be shown, in a similar fashion as above,
to have eigenvalues evolving as the Itô SDE
\be 
\dd \lambda_{i,t} = \left(  (m + \sigma^2 (N+1) ) \lambda_{i,t}  +  \sum_{k\neq 1} p_k  \lambda_{i,t}^k    \right)\dd t
 +  2 \sigma^2 \sum\limits_{j \neq i } \frac{\lambda_{i,t} \lambda_{j,t}}{\lambda_{i,t} - \lambda_{j,t}} \dd t  + 2 \sigma \lambda_{i,t} \dd B_{i,t}
\label{eq:3:EigenvalueSDESymmetricGTRV}
\ee
with stationary solution given by
\be
P_\infty( \vec{\lambda}) \propto
\abs{\Delta(\vec{\lambda})}  \, \prod_{i=1}^N \lambda_{i}^{\frac{m}{ 2 \sigma^2} - \frac{N+1}{2 } }   \ \; e^{ \frac{1}{2  \sigma^2} \sum\limits_{1 \leqslant  i \leqslant N \atop k \neq 1}   \frac{p_{k}}{k-1} \lambda_{i}^{k-1} } \; .
\label{eq:3:GeneralizedGTRVSymmetricStationarySol}
\ee  

In this case, the correlations of the symmetric noise yield a different drift term
\be 
\DelBB_{\rm IS} = 
(1+\frac{N}{2}) \VV_t +   \frac{\Tr ( \VV_t) }{2} \ID  \; .
\label{eq:3:DriftISSymmetricGTRV}
\ee
Note that this case, which is different on the structure of the noise matrix, recovers a very similar eigenvalue evolution, related to the previous case by a simple rescaling of the parameter $\sigma$. This simple relation is similar to the one discussed below \myeqref{eq:3:QTildeIntegraleLambdacroixLambda} about the connection between the symmetric noise of the Grabsch-Texier model and the structure-less noise of the Rider-Valk\'o model.

\subsubsection{Square-root multiplicative noise}

Beyond adding arbitrary drift terms, the models studied in this chapter can be further generalized by changing the power to which the process appears in the noise term. In each of the previous models, this power was $1$, meaning that rescaling the process $\VV_t$ by a scalar $z$ leads to a rescaling of the noise term by $z$.

Many models can be defined with other exponents in the noise term prefactor, but not so many can be easily solved in order to find the stationary solution, if we restrict to simple $\VV^k$ drift terms. A first model where a solution can be found is the following, where the process appears with exponent $\frac{1}{2}$ in the prefactor of the noise term:
\be 
\dd \VV_t  = (m\ID+ \sum\limits_{k\neq 0} p_k \VV_t^k) \dd t  + \sigma \frac{\sqrt{\VV_t} \dd \XX_t + \dd \XX_t^T \sqrt{\VV_t}}{\sqrt{2}}  \quad \quad \quad \quad \text{(Itô)}
\label{eq:3:SquareRootNoiseModel}
\ee
Note that, in this case, we put aside the term with power $k=0$. The perturbative analysis yields the following joint eigenvalue Itô SDE
 \be 
\dd \lambda_{i,t} =
    ( m+ \sum\limits_{k\neq 0} p_k \lambda_{i,t}^k) \dd t  +   \frac{\sigma^2}{2} \sum\limits_{j \neq i } \frac{\lambda_{i,t} + \lambda_{j,t}}{\lambda_{i,t} - \lambda_{j,t}} \dd t +  \sigma \sqrt{2 \lambda_{i,t}}\dd B_{i,t}
    \label{eq:3:SquareRootEigenvalueSDE}
    \ee
which admits the following stationary solution
\be
P_\infty( \vec{\lambda}) \propto
\abs{\Delta(\vec{\lambda})} \, \prod_{i=1}^N \lambda_{i}^{  \frac{m}{\sigma^2} - \frac{N+1}{2}  }   \; e^{ \frac{1}{ \sigma^2}  \sum\limits_{1 \leqslant  i \leqslant N \atop k \neq 0}  \frac{p_k}{k} \lambda_{i}^k   }  \; .
\label{eq:3:SquareRootMultNoiseStationarySolution}
\ee

The term leading the power-law behaviour is the prefactor of the identity matrix drift, while the others appear in the exponential term. In spite of the different power of the multiplicative noise, the solution is remarkably close to the models presented above.

\subsubsection{Squared multiplicative noise}
 
We present another model where a stationary solution can be found, with overall squared weight in the prefactor of the noise term. In this case, the model is defined with the Stratonovich prescription, because the drift $\DelBB_{\rm IS}$ allows to simplify the interaction term in the eigenvalue evolution, and to finally solve for the stationary solution.
 
The model in question is defined as
\be 
 \dd \VV_t = ( m \VV_t^3 + \sum\limits_{k\neq 3} p_k \VV_t^k )\dd t +    \sigma \VV_t \dd \HH_t \VV_t \quad \quad \quad \quad \text{(Stratonovich)} \; .
 \label{eq:3:SquaredModelDefProcess}
 \ee
Note that, in this case, we put aside the term with power $k=3$. Taking this SDE to the Itô prescription by taking into account the drift 
\be 
\DelBB_{\rm IS} =  \VV_t^3 + \Tr(\VV_t) \; \VV_t^2 \; ,
\label{eq:3:SquaredISDrift}
\ee
the eigenvalue process is found, by perturbative analysis, to follow the Itô SDE
\be 
\dd \lambda_{i,t} =
\left(  (m+2 \sigma^2) \lambda_{i,t}^3 + \sum\limits_{k\neq 3}  p_k \lambda_{i,t}^{k}  \right) \dd t + \sigma^2 \lambda_{i,t}^2 \sum_{j \neq i} \frac{\lambda_{i,t} \lambda_{j,t}  }{\lambda_{i,t} - \lambda_{j,t}}  \dd t + \sqrt{2}  \sigma \lambda_{i,t}^2 \dd B_{i,t} \; .
\label{eq:3:SquaredEigenvalueSDE}
\ee 

The stationary eigenvalue JPDF is then
\be
P_\infty( \vec{\lambda}) \propto \abs{\Delta(\vec{\lambda})}  \prod_{i=1}^N \lambda_{i}^{\frac{m}{\sigma^2} - (N+1)} \; e^{ \frac{1}{\sigma^2}  \sum\limits_{1 \leqslant  i \leqslant N \atop k \neq 3} \frac{p_{k}}{k-3} \lambda_{i}^{k-3}} 
\label{eq:3:SquareMultiplicativeStationarySolution}
\ee
where the term leading the power-law behaviour corresponds to $k=3$, such that it was set aside in the model definition. Once again, the solution is remarkably close to that of the other models presented above.


\section{A Hamilton-Jacobi point of view}
\label{sec:3.3} 

As a conclusion to this chapter, we present an interesting perspective on stochastic matrix evolutions, proposed recently by Grela, Nowak and Tarnowski in \cite{Grela2020}. In this work, the authors give an original formulation for the evolution of a large random matrix. As laid out extensively in this chapter, this study can be done by focusing on the stochastic evolution of the entries and eigenvalues, or on evolution equations for the standard transforms such as Burger's equation \eqref{eq:2:BurgersEquationDBM} for the Stieltjes transform of the DBM. Instead, the authors of \cite{Grela2020} build an \emph{eikonal formulation}, i.e. a Hamilton-Jacobi framework for random matrix evolutions.

From the self-averaging property of RMT, the description of the evolution of a large random matrix relies on the deterministic evolution of its spectral density. The cornerstone of the problem is then the \emph{trajectory} of the spectral density, or equivalently of the related transforms, along the evolution of the time variable. 

The proposition of \cite{Grela2020} is to take the focus away from the trajectory of the matrix, in order to gain a different point of view on the system's dynamics. Their effort can be understood in analogy with the wavefront description of light propagation, which grants one with an encompassing point of view, as opposed to the light-ray trajectory description. Another powerful analogy is the one that can be drawn with a mechanical system. One can either describe the intricacies of a mechanical system's trajectory, integrating the evolution equation under the influence of different actions. Or one can take a step back and opt for a Hamilton-Jacobi formalism, describing the system itself through its Hamiltonian operator. This latter Hamilton-Jacobi framework is the one developed in \cite{Grela2020} in the scope of large matrix random evolutions.

To this aim, the authors of \cite{Grela2020} propose to consider a canonical pair of complex variables $(q,p)$, corresponding respectively to $(z,\g)$, the complex variable which appears as the argument of the Stieltjes transform $\g(z)$ and the value of the transform itself. The correspondence is thus 
\be 
(z,\g)  \ \  \Leftrightarrow \ \ (q,p) \; .
\ee
In contrast with the trajectory description, where the pseudo-position $z$ and pseudo-moment $\g(z,t)$ are deeply intertwined along the time evolution, these variables are to be considered completely independent in the Hamilton-Jacobi formalism. In this formalism, the evolution of the variables is driven by the pair of Hamilton equations
\be 
\begin{cases}
\displaystyle \frac{\partial q }{\partial t}  = \frac{\partial H}{\partial p}  \; ;   \\[8pt]
\displaystyle \frac{\partial p }{\partial t}  = -  \frac{\partial H}{\partial q}  \; .
\end{cases}
\ee

As an illustration of the matrix Hamilton-Jacobi formalism, let us consider the DBM with large matrix-size $N$, which corresponds to successive addition of infinitesimal Gaussian matrices. As recalled above, the evolution of the spectral density is conveniently described in terms of the Stieltjes transform $\g$ as
\be 
\frac{\partial }{\partial t } \g(z,t) + \g(z,t) \frac{\partial }{\partial z } \g(z,t)  = 0 \; .
\ee
Using the method of complex characteristics, the trajectory is found from the initial condition $\g_0$ as the solution of \myeqref{eq:2:SolutionBurgersCharacteristics}, i.e.
\be 
\g(z,t) = \g_0 \bigg(z- t \, \g(z,t) \bigg)  \; .
\label{eq:3:EvolutionEquationStieltjesDBM}
\ee
We recall that the trivial initial solution $\g_0(z) = \frac{1}{z}$ yields the expanding semi-circle distribution, see \myeqref{eq:2:SolutionSemiCircle}, as expected for the DBM. It can be seen that the DBM evolution described in \myeqref{eq:3:EvolutionEquationStieltjesDBM} corresponds to 
\be
\begin{cases}
\displaystyle \frac{\partial}{\partial t} z = \g  \\[8pt]
\displaystyle \frac{\partial}{\partial t} \g = 0
\end{cases}
\ee
such that the DBM evolution is represented, in this point of view, by the free Hamiltonian
\be 
H_{\rm DBM}  = \frac{p^2}{2} \; .
\label{eq:3:HHamiltonJacobiDBM}
\ee
The simplest matrix diffusion, the Dyson Brownian motion where infinitesimal gaussian matrices are successively added, is represented in this framework by the Hamiltonian of an isolated mechanical particle, which travels ballistically in the complex plane, with constant velocity from a given initial position. 

The authors build the formalism by characterizing the Hamiltonian for all Hermitian additive processes. With $\RT$ the $\RT$-transform of the infinitesimal increment, they show that the Hamiltonian of the corresponding diffusion is
\be 
H_{\rm additive} = \int_0^p \RT(\omega) \dd \omega \; .
\ee
One can then verify that \myeqref{eq:3:HHamiltonJacobiDBM} is indeed the particular case where the increment has unit-variance semicircle distribution such that $\RT (\omega) =  \omega$, see Table \ref{tab:RMTtransforms}.

In addition to the Hermitian case, they construct a generalized resolvent function and corresponding Hamiltonian for non-Hermitian and non-normal additive problems. Moreover, their formalism is not restricted to the additive case and can accomodate processes such as the Ornstein-Uhlenbeck process with restoring coefficient $k$, for which the Hamiltonian is also a function of the pseudo-position $z$ as
\be 
H_{\mathrm{OU}}=\frac{1}{2} p^{2}+ k (1-q p) \; .
\ee

In this scope, it is interesting to note that the formalism can be applied to the Kesten process studied in this chapter. One finds by inspection that, in the Hamilton-Jacobi point of view, the large-$N$ Kesten evolution characterized by \eqref{eq:3:ContinuousEvolutionKestenSTransform} and \eqref{eq:3:EvolutionStieltjesTransformLargeN} is endowed with the following Hamiltonian:
\be 
H_{\mathrm{Kesten}}= p-\left(\tilde{\sigma}^{2}-m\right) q p+\frac{\tilde{\sigma}^{2}}{2} q^{2} p^{2} \; .
\ee
 

\thispagestyle{empty}
\chapter{Scalar and matrix bridge processes} 
\label{chap:4}

A \emph{bridge} process is a stochastic process defined on a time interval $[0,T]$ and conditioned to end in a certain configuration at final time $T$. For a (continuous) scalar or matrix bridge, the final value of the process is thus fixed, such that the process converges to this value as $t \to T$. In the previous chapters, we have presented a range of stochastic processes, with a focus on non-intersecting scalar processes in chapter \ref{chap:2}, and matricial processes in chapter \ref{chap:3}. In this chapter, we study both these constructions concurrently, as we turn to the special properties of the bridge versions of scalar and matrix stochastic processes. 

In the scalar realm, the Brownian bridge is the bridge version of the standard Brownian motion. Both processes share subtle links and can be mapped onto each other in several ways. Remarkably, although the bridge process is conditioned on an event in the future, a differential description can be constructed, such that the condition appears as a drift in a SDE. These properties will be detailed below. Most importantly, the survival properties of a bridge above a semicircle boundary have been studied in \cite{Ferrari2005}. This work unveiled the renowned \emph{Ferrari-Spohn} distribution for the position of the bridge above the barrier at mid-time, expressed in terms of Airy functions. The motivation for this problem is to understand crudely the behaviour of the top trajectory in a multi-layer PNG model \cite{Prahofer2000,Ferrari2004}, or equivalently the top walker in a group of non-intersecting Brownian bridges. Indeed, because of conditioning, the expected trajectory of the upper walker is a semicircle, such that a unique bridge conditioned to remain above this moving boundary is a toy model for the upper walker. If the correct scaling is recovered in this toy model, the full distributions however differ, and the Ferrari-Spohn (\textit{FS}) distribution is observed instead of the Tracy-Widom distribution \cite{Widom2004}, presented in the scope of random matrix extreme eigenvalues in chapter \ref{chap:1}. The object of publication \pubref{publication:ConstrainedNCFerrariSpohn} was to study a multiple-walker version of the toy model described above, and thus to investigate the generalization of the Ferrari-Spohn distribution observed for $N$ non-crossing bridges above a semicircle boundary.

In the matrix realm, bridges have a special significance as they appear in the asymptotics of the HCIZ integral, and in the large deviations of some models of random matrix theory. The chapter will be closed on this interesting perspective.

In section \ref{sec:4.1}, we present the properties of the simplest scalar bridge process, the Brownian bridge, and detail its properties and links with the standard Brownian motion. In section \ref{sec:4.2}, we present the extension to $N$ non-crossing walkers of the renowned Ferrari-Spohn distribution, and detail the connection with fermions in a linear potential. Finally, in section \ref{sec:4.3}, we turn to the importance of matrix bridge processes in the HCIZ integral, which plays a prominent role in the large deviations of extreme eigenvalues in some ensembles of random matrix theory.

\section{Scalar Brownian bridge}
\label{sec:4.1}

We open this chapter on bridge processes by reviewing the properties of the simplest such object, the scalar Brownian bridge. When conditioned to remain above a semicircle boundary, the distribution of its position obeys the Ferrari-Spohn law, presented below. 

\subsection{Definition and properties}
\subsubsection{Definition}

The Brownian bridge $(A_t)_{0 \leqslant t \leqslant T}$ is a Brownian motion, such that
\begin{itemize}[itemsep=.3em,topsep=4pt]  
\item $A_t$ is almost-surely continuous ;
\item all increments are independent and normally distributed ; 
\end{itemize}
with a final position fixed to $z \in \R$, such that
\emphbe 
A_T = z
\label{eq:4:FinalConditionBridge}
\end{empheq}
In other words, the law of $(A_t)_t$ is that of the Brownian motion, conditioned on \eqref{eq:4:FinalConditionBridge}. A realization of this process is illustrated in Fig.~\ref{fig:BrownianBridge}.

 \begin{figure}[ht!]
    \centering 
    \includegraphics[width=.45 \textwidth]{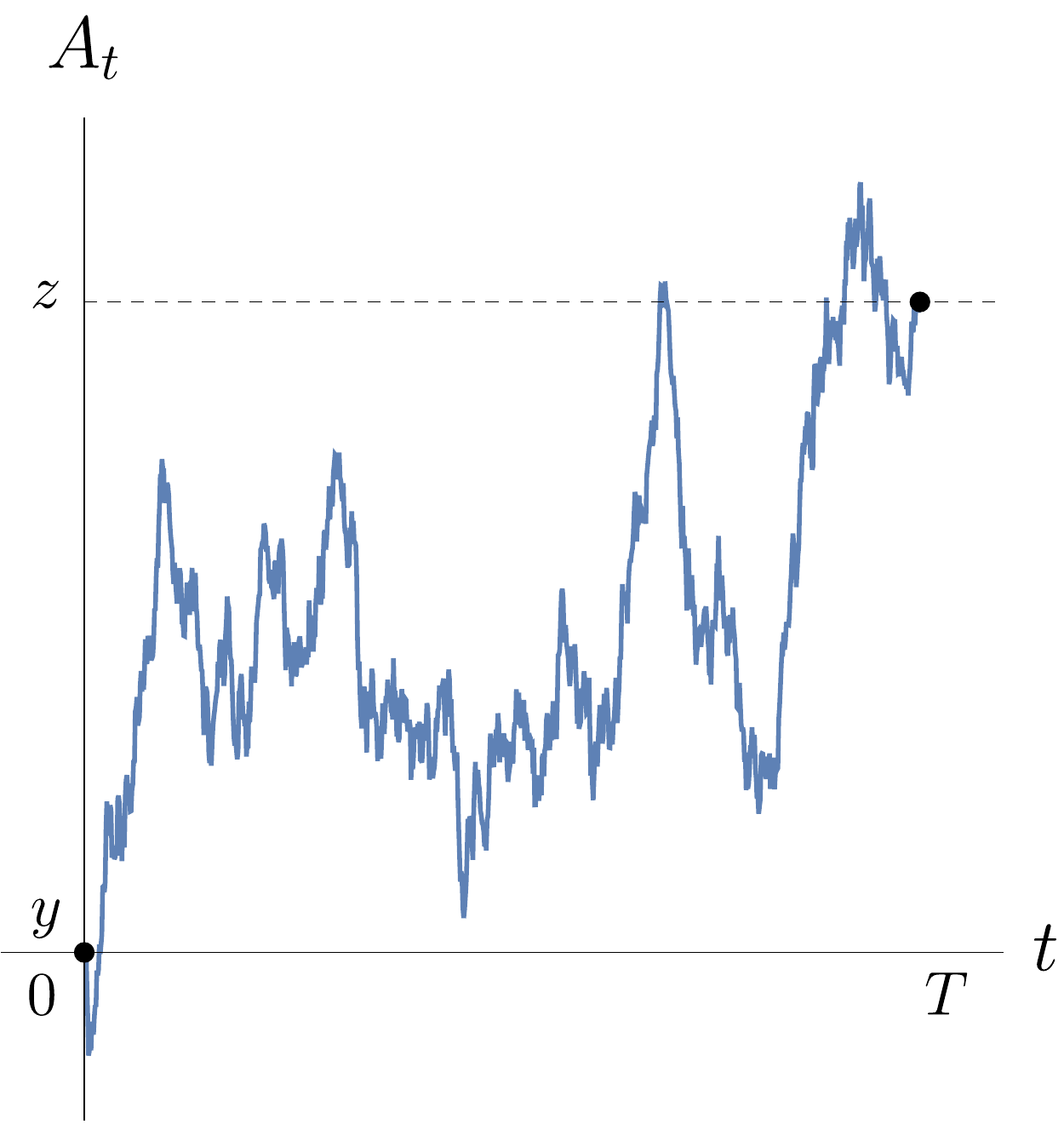} 
    \caption{Realization of the Brownian bridge $A_t$, started from position $y$ at time $t=0$ and ended at $z$ at time $T$.}
    \label{fig:BrownianBridge}
\end{figure}

\subsubsection{Properties}

We give a few properties of the Brownian bridge, which can be obtained easily from its definition \cite{Mansuy2008}. We consider that the process is started from an arbitrary initial position $y$. 

The single-time distribution of the process at time $t$, i.e. the distribution of the random variable $A_t$, can be obtained in conditional form from the Markov property of Brownian motion, as
\be 
P(x,t) =  \frac{  P(x, t \mid y,0) \,  P(z, T \mid x,t)  } { P(z,T \mid y, 0) }
\ee
with the propagator given in \myeqref{eq:2:PropagatorBM}. This yields
\bea
P( x , t  ) &=& \frac{1}{\sqrt{2\pi \frac{ t (T-t)}{T} }} e^{ - \frac{(x-y)^2}{2t } } \, e^{ - \frac{(z-x)^2}{2(T-t)} } \, e^{   \frac{(z-y)^2}{2T}  }  \\
 &= &   \frac{1 }{\sqrt{2\pi \frac{ t (T-t)}{T} }}  \, e^{  - \frac{T}{2 t (T-t) }  \left( x-  \frac{y (T-t) + z t }{T} \right)^2    }  \; .
\eea 
As a consequence, $A_t$ is distributed as a Gaussian random variable with expectation 
\be 
\E \left[  A_t  \right]  =  \frac{y (T-t) + z t }{T} =  y +  \frac{t}{T} (z-y) \; ,
\label{eq:4:ExpectationBridge}
\ee
and variance
\be 
\E \left[  \left( A_t - \E \left[  A_t  \right] \right) ^2 \right] =   \frac{t (T-t)}{T} \; .
\label{eq:4:VarianceBridge}
\ee
Let us denote the standard deviation
\be 
s(t) = \sqrt{ \frac{t (T-t)}{T} } \; .
\label{eq:4:StandardDevBridge}
\ee
The expectation and variance are illustrated for many realizations of the process in Fig.~\ref{fig:MultipleBrownianBridges}. In this plot, the linear expectation is the black dashed line, and the red dashed lines are one standard variation above and below the expectation, i.e. $\E[A_t] \pm s(t)$.

 \begin{figure}[ht!]
    \centering 
    \includegraphics[width=.7 \textwidth]{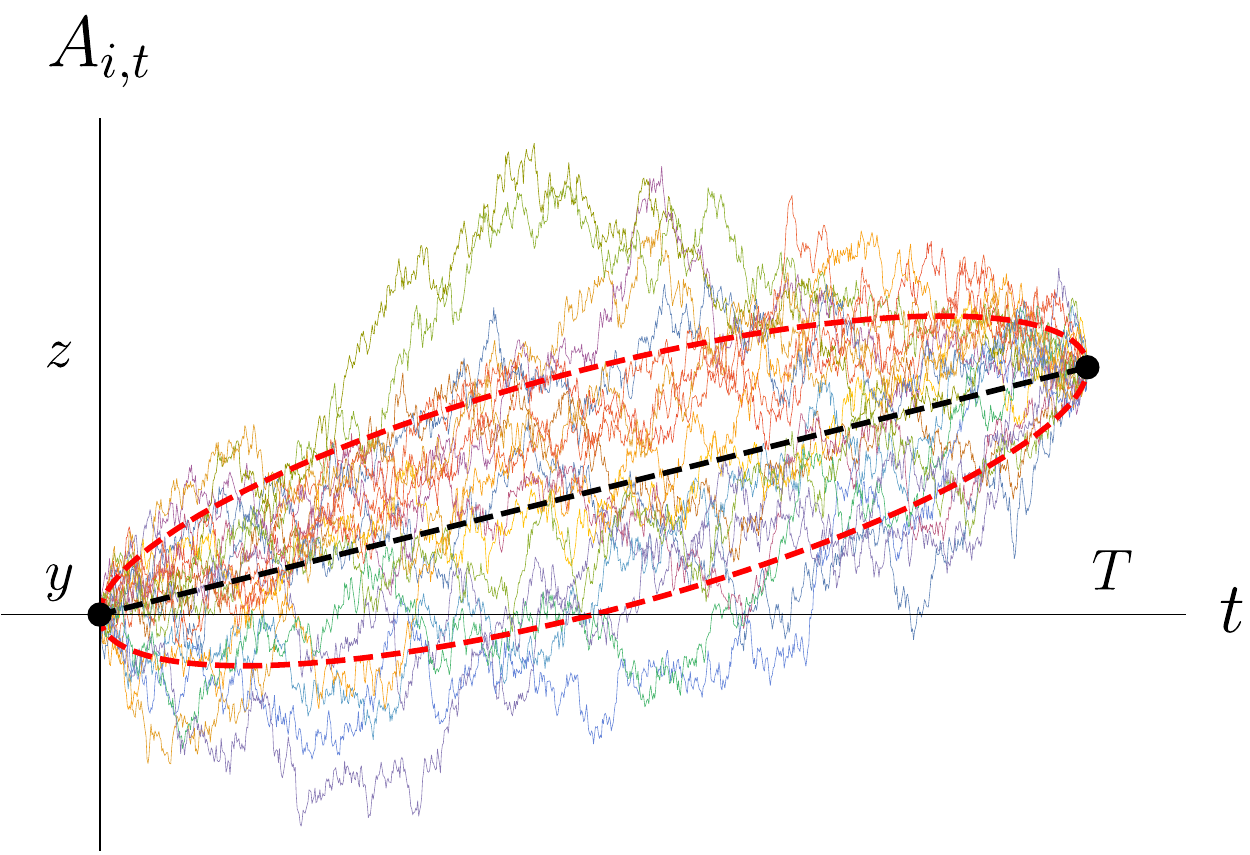} 
    \caption{Realization of $N=20$ independent Brownian bridges $A_{i,t}$ started from $a$ at time $t=0$ and ended at $b$ at time $T$. The black dashed line indicates $\E [A_t]$ and the red dashed lines indicate $\E[A_t] \pm s(t)$, see Eqs.~\eqref{eq:4:ExpectationBridge} and \eqref{eq:4:StandardDevBridge}.}
    \label{fig:MultipleBrownianBridges}
\end{figure}

More generally for two instants $t_1 < t_2$, the covariance of the process $(A_t)_{0 \leqslant t \leqslant T}$ is 
\be 
\E \left[
 \left( A_{t_1} - \E \left[  A_{t_1} \right] \right)   \left( A_{t_2}- \E \left[  A_{t_2} \right] \right) 
\right]
=
\frac{t_1(T-t_2)}{T} \; .
\ee

\subsubsection{Mappings to the standard Brownian motion}

The Brownian bridge can be constructed from a realization of the Brownian motion $(B_t)_{t \geqslant 0}$ as we detail now. Two cases can occur: either one considers that the realization is already known, or one considers that the details of this realization are still unknown and cannot be used in the construction. Without loss of generality, we consider in the following that the process is started from $y=0$.

\paragraph{With knowledge of the Brownian endpoint}

Let us consider that a realization $(B_t)_{t \geqslant 0}$ of the Brownian motion is fixed, such that, in particular, its value $B_T$ at time $T$ is known. One can construct a statistically correct realization of the Brownian bridge $(A_t)_{0 \leqslant t \leqslant T}$ ended at $z$ by defining \cite{Mansuy2008}
\be 
A_t = B_t + \frac{t}{T} ( z - B_T ) \; .
\label{eq:4:MappingWithKnowledge}
\ee
We immediately have $A_T = z$ with probability $1$, as expected. 

\paragraph{Without knowledge of the Brownian endpoint}

The previous trick however fails if one considers that a special value $B_T$ of the Brownian realization is out of reach. Another representation of the Brownian bridge from the Brownian motion, which escapes this drawback and does not require future information about the Brownian motion trajectory, is as follows.

\subparagraph{General bridge endpoint}

From the Brownian motion $ (B_t)_{t \geqslant 0}$, the Brownian bridge $(A_t)_{0 \leqslant t \leqslant T}$ ending at $A_T =z$ can be constructed as \cite{Revuz1995}
\be 
A_t = \frac{T-t}{T}  B_{\frac{Tt}{T-t}} + \frac{t}{T} z \; .
\label{eq:4:MappingBrownianBridgeAsymmetric}
\ee
One can verify that this Gaussian process respects the expectation and covariance of the Brownian bridge presented in the previous subsection, such that it is indeed equal in law to a Brownian bridge started from $0$ and ended at $z$ at time $T$.

\subparagraph{Symmetric bridge}

Let us consider the particular case where $z=y=0$, such that the bridge is returned to its initial position at the ending time $T$. The expectation of the bridge is then $\E [ A_t ] = 0$ for any $t$, and the mapping is simply a rescaling of the standard Brownian motion in both space and time directions, as 
\be 
A_t = \frac{T-t}{T}  B_{\frac{Tt}{T-t}}  \; .
\label{eq:4:MappingBrownianBridgeSymmetric}
\ee
It is remarkable that the standard deviation $s(t)$ in \myeqref{eq:4:StandardDevBridge} is then recovered, under the same mapping, from the square-root standard deviation of the Brownian motion as
\be 
s(t) = \frac{T-t}{T} \sqrt{\frac{Tt }{T-t}} =  \sqrt{\frac{t(T-t)}{T}} \; .
\ee

The equivalence of the standard Brownian motion and the symmetric Brownian bridge through mapping \eqref{eq:4:MappingBrownianBridgeSymmetric} is illustrated in Fig.~\ref{fig:MappingBrownianBridge}, along with the equivalence of their respective standard deviations under the same mapping. Notice in particular that every crossing of the Brownian trajectory $B_t$ with $\pm \sqrt{t}$ corresponds to a crossing of the bridge trajectory $A_{\tilde{t}}$ with $\pm s(\tilde{t})$, at the relevant time $\tilde{t}=\frac{Tt}{T+t}$ under the mapping. More crossings can be observed on the bridge side, since $t \in [0,T]$ in this case corresponds to $t \in [0, + \infty[$ in the Brownian case, which outflanks the plot range.

\begin{figure}[ht!]
    \centering
    \begin{subfigure}[t]{0.45\textwidth}
        \centering
    \includegraphics[width=\textwidth]{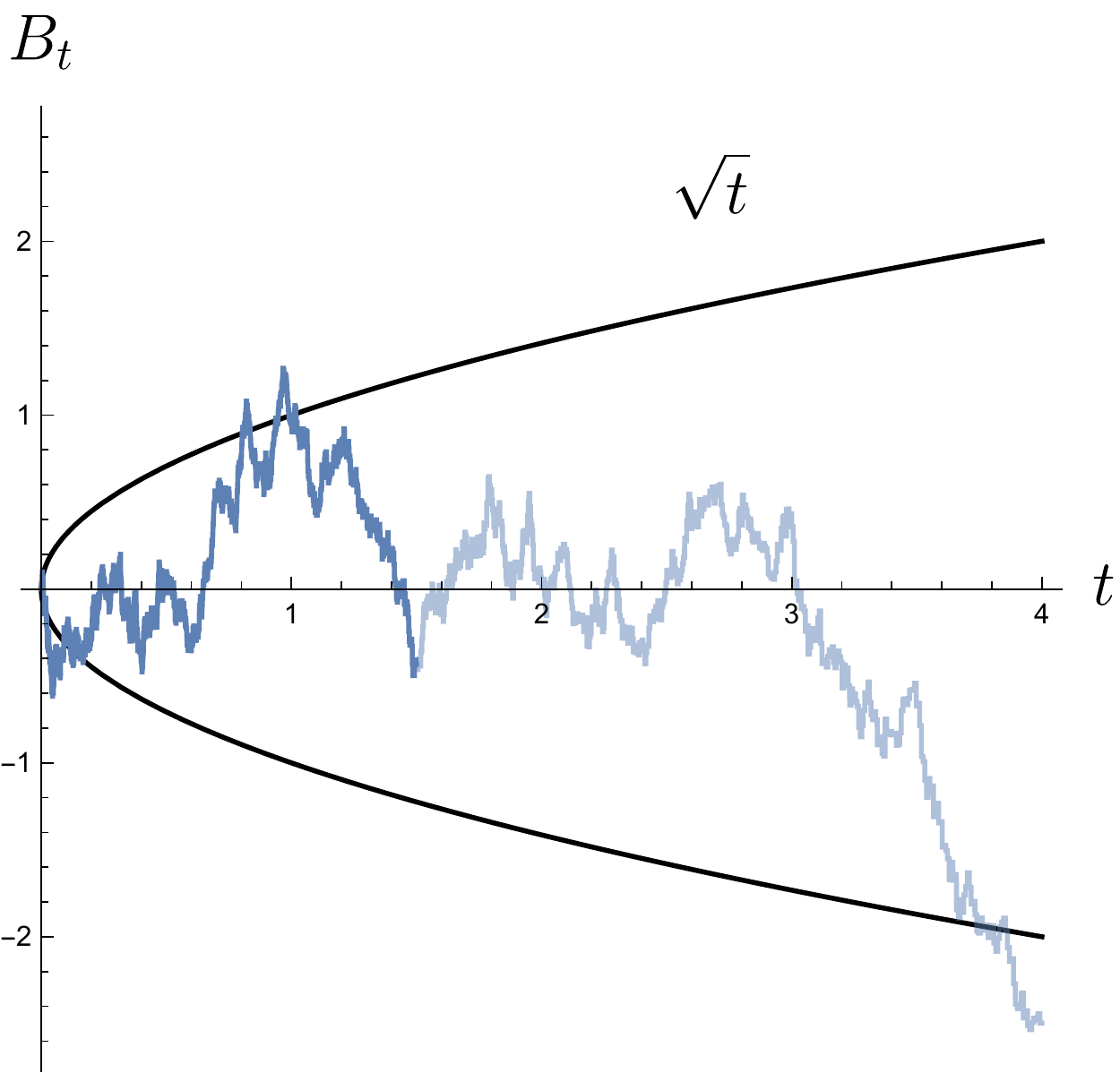}  
    \end{subfigure}%
    \hfill
    \vbox to 180pt {\vfil  \hbox to 1cm{ $\boldsymbol{\Longleftrightarrow}$ } \vfil }
    \hfill
    \begin{subfigure}[t]{0.45\textwidth}
        \centering
    \includegraphics[width=\textwidth]{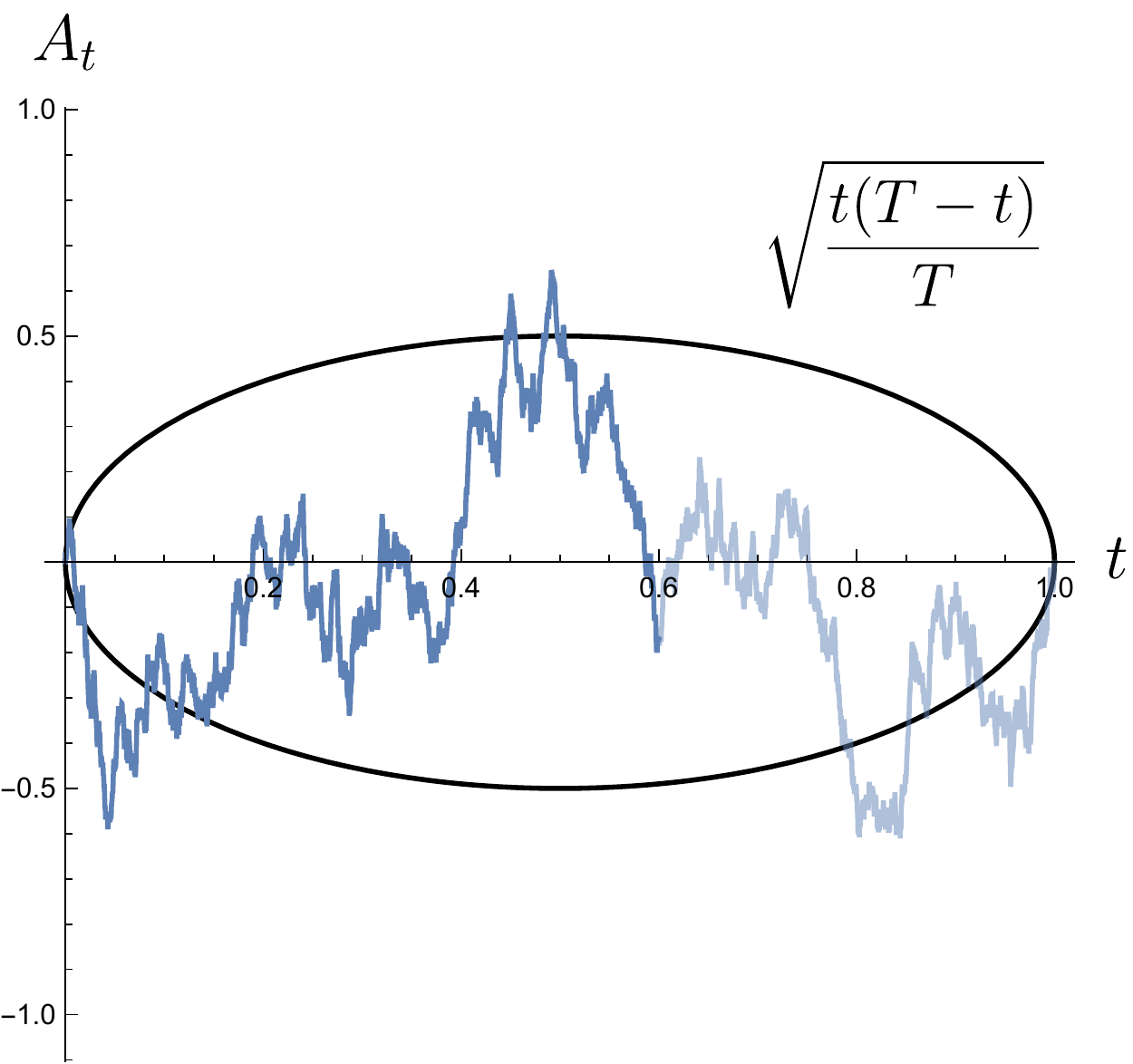}  
    \end{subfigure}
    \caption{\emph{Left}: Realization of the standard Brownian motion $(B_t)_{t \geqslant 0}$. \emph{Right:} Brownian bridge realization $(A_t)_{0 \leqslant t \leqslant T=1}$ obtained from the Brownian realization through \myeqref{eq:4:MappingBrownianBridgeSymmetric}. The running time parameter in the mapping is indicated as the end of the dark blue colour in both plots. The standard deviations of each process, which coincide under the mapping, are plotted in the black lines.}
    \label{fig:MappingBrownianBridge}
\end{figure}

\subsubsection{Conditional drift : stochastic differential equation for a bridge}

The bridge process is conditioned on its endpoint $A_T$, such that an intuitive description of $(A_t)_{0 \leqslant t \leqslant T}$ requires information about the future as in \myeqref{eq:4:MappingWithKnowledge}. The second mapping of the Brownian bridge to the standard Brownian motion, presented in the previous lines, however suggests that a description which does not break temporal locality is possible. 

We can indeed write a time-local SDE for the evolution of $A_t$, by applying Itô's formula to \myeqref{eq:4:MappingBrownianBridgeAsymmetric}:
\be 
\dd A_t = \left( - \frac{A_t - \frac{t}{T}z}{T- t } + \frac{z}{T} \right) \dd t + \frac{T-t}{T} \dd B_{ \frac{Tt}{T-t}}  =   \frac{z - A_t}{T- t} \dd t + \dd B_t \; .
\label{eq:4:SDEBridge}
\ee
The Brownian bridge diffuses as a Brownian motion, with a linear force attracting it towards its endpoint $z$ with a prefactor $1/(T-t)$ which increases with $t$ and diverges as $t \to T$. We emphasize that \myeqref{eq:4:SDEBridge} is a time-local differential equation for the evolution of the process.

This can be seen as a particular case of a broader problem of obtaining a Langevin equation for a conditioned process, and an application of Doob's $h$-transform theory. Let us consider an unconstrained stochastic process obeying a given SDE and constrain this process such that we add a condition on an event concerning the future. Let us denote $Q(x,t)$ the probability under which the unconstrained process respects this very condition. For example, in the case of the bridge, $Q(x,t)$ is the probability that the unconstrained Brownian motion ends in $z$ at time $T$, i.e.
\be 
Q(x,t) = P( z , T \mid x, t )  = \frac{1}{\sqrt{2 \pi (T-t) }} e^{- \frac{(x-z)^2}{2(T-t)}} \; .
\label{eq:4:ConditionVerificationProbability}
\ee 
It can be shown, see \cite{Majumdar2015}, that the constrained process respects the same SDE as the unconstrained process, with an additional drift term given by
\be 
\frac{\partial}{\partial x} \log Q(x,t)  \; .
\label{eq:4:LogQConditionalDrift}
\ee
This conditional drift enforces the condition through infinitesimal time evolution. As expected in the particular bridge case, it can be verified from \myeqref{eq:4:ConditionVerificationProbability} that
\be 
\left. \left( \frac{\partial}{\partial x} \log Q(x,t) \right) \right|_{x=A_t} = \frac{z-A_t}{T-t}    \; .
\ee

We finally remark that this conditional drift gives a proof of the well-known coincidence of the three-dimensional Bessel process, evolving according to \myeqref{eq:2:BesselSDE} in the special case $D=3$, with the Brownian motion conditioned to never reach the origin \cite{Pitman1975}. Indeed for large time $t$, the survival probability $S(t \mid x,0)$ of a Brownian motion above a fixed vicious boundary at $g=0$ is asymptotically, per \myeqref{eq:2:LargeTimeSurvivalProbaFixedBoundary}:
\be 
S(t \mid x,0) \simeq x  \sqrt{\frac{2 }{\pi t}}
\ee
such that 
\be 
\left. \left( \frac{\partial}{\partial x} \log S(t \mid x,0) \right) \right|_{x=R_t}  = \frac{1}{R_t}
\ee
which recovers the drift in SDE \eqref{eq:2:BesselSDE} with $D=3$. A comparison with \myeqref{eq:2:DBMSDECloseToACollision} explains, with the framework of this paragraph, why the DBM-2 coincides with the process of independent Brownian motions conditioned to never intersect, as explained in detail in section \subsubref{subsubsec:II.1.2.b}{II.1.2}.

\subsection{Moving boundary: the Ferrari-Spohn distribution}

In \cite{Ferrari2005}, Ferrari and Spohn considered a Brownian bridge $(A_t)_{0 \leqslant t \leqslant T}$ started and ended at $0$, and conditioned to stay above a semicircle wall defined with $W>0$ as
\be 
g(t) = W \sqrt{ \frac{t (T-t)}{T}} \; .
\label{eq:4:SemicircleBoundary}
\ee
They studied the typical fluctuations of the distance $x = A_{\frac{T}{2}} - g\left( \frac{T}{2} \right)$ between the Brownian particle and the wall, which we specialize in this presentation to the mid-time $\frac{T}{2}$. Dropping the dependence on $T$, we denote the corresponding Brownian probability density for the mid-time distance $x$ as
\be 
R(x,W)= P\left( x+ g\left( \frac{T}{2} \right)  , \frac{T}{2} \mid \Bevent_{[0,T]} , \{ 0, 0 \} , \{ 0, T \} \right) \; .
\label{eq:4:DefSingleParticleDensity}
\ee
This PDF gives the probabilistic weight associated with finding the Brownian motion at $x + g\left( \frac{T}{2} \right)$ at time $\frac{T}{2}$ knowing that it is started and ended at $0$, and remains above the boundary for $t \in [0,T]$. Abiding by the notations of chapter \ref{chap:2}, $\Bevent_{[0,T]}$ is the event where the particle remains away from the boundary $g$, and we add brackets when several position-time pairs appear on the same line, in order to avoid ambiguities. The bridge under consideration in this problem is illustrated in Fig.~\ref{fig:FerrariSpohnSingleBridge}.

 \begin{figure}[ht!]
    \centering 
    \includegraphics[width=.85 \textwidth]{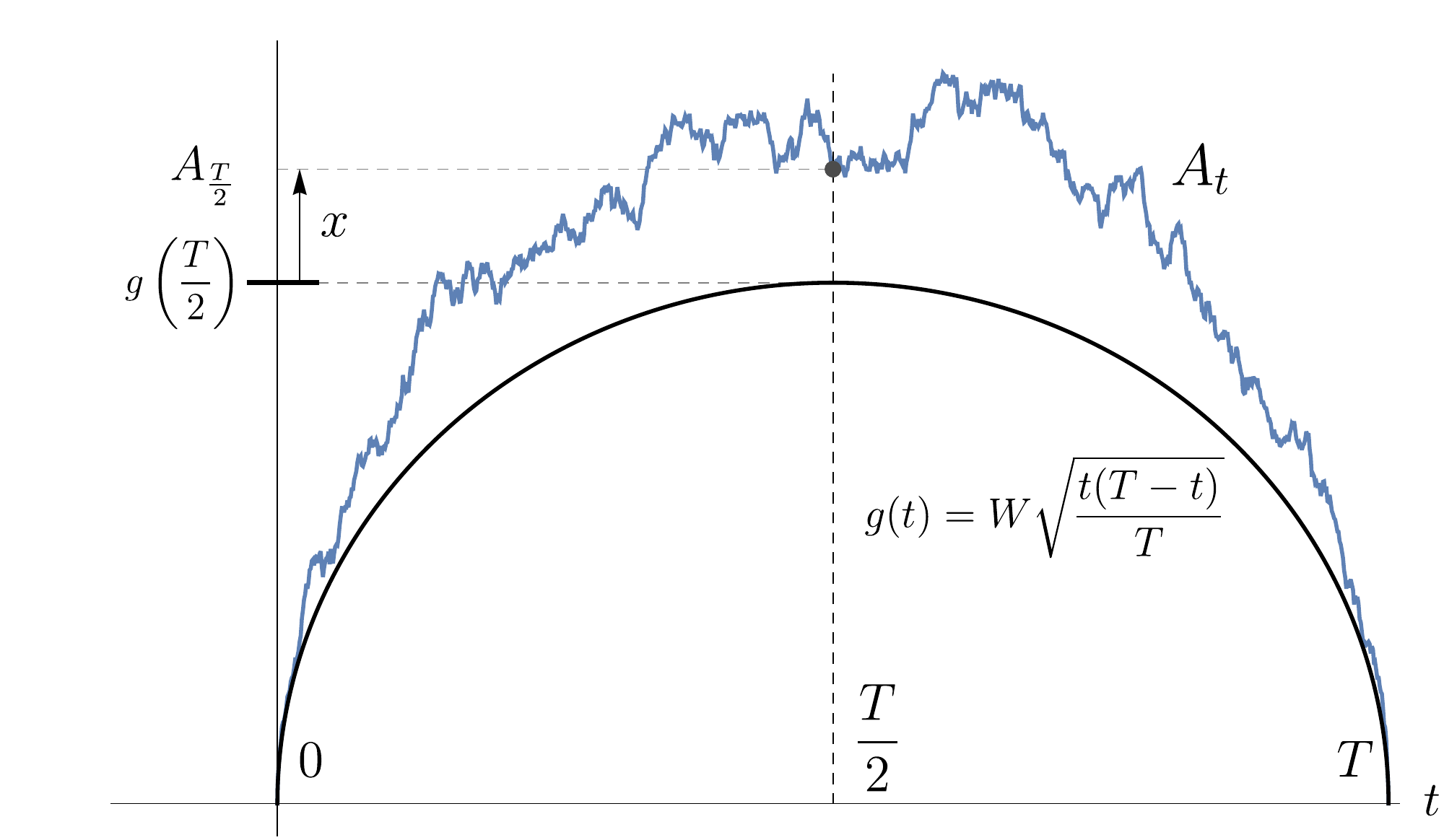} 
    \caption{Brownian bridge conditioned to remain above a semicircle boundary $g(t)$ and found at distance $x$ of the wall at the mid-time $\frac{T}{2}$.}
    \label{fig:FerrariSpohnSingleBridge}
\end{figure}

In the large-$W$ limit, Ferrari and Spohn found that the typical fluctuations of the mid-time distance are described by a universal distribution given in terms of the square of the Airy function $\Ai$. Under the assumption $W \gg 1$ on this dimensionless parameter, the distribution \eqref{eq:4:DefSingleParticleDensity} is the \emph{Ferrari-Spohn distribution}:
\emphbe
R(x,W) \simeq 
P_{\rm FS}(x)=\frac{\left(4W\right)^{1/3}}{\sqrt{T}\Ai'(\alpha_{1})^{2}}
\Ai\left(\alpha_{1}+\frac{\left(4W\right)^{1/3}}{\sqrt{T}}\ x\right)^{2} 
\Theta (x)
\label{eq:4:FerrariSpohnDistribution}
\end{empheq}
The constant $\alpha_1=-2.33811\ldots$ is the smallest zero, in absolute value, of the Airy function such that $P_{\rm FS}(x) \to 0$ as $x \to 0$, as expected. As introduced earlier in this thesis, $\Theta$ is the Heaviside function such that $P_{\rm FS}$ is supported on $\R^+$, on which it is normalized. This distribution is plotted in Fig.~\ref{fig:FerrariSpohnDistribution}, in the case where $4W$ and $T$ are both fixed to unity.

 \begin{figure}[ht!]
    \centering 
    \includegraphics[width=.65 \textwidth]{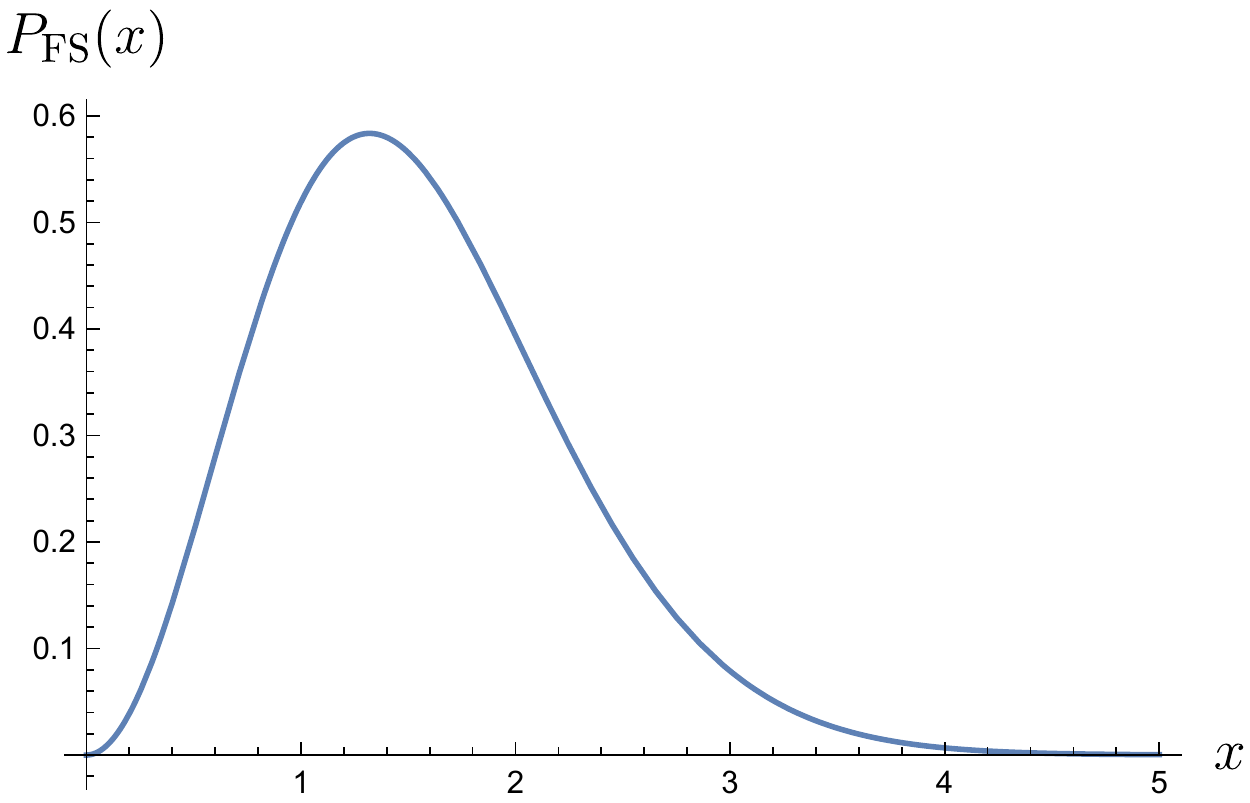} 
    \caption{Plot of the Ferrari-Spohn distribution $P_{\rm FS}$ in \myeqref{eq:4:FerrariSpohnDistribution}, for $T=4W=1$.}
    \label{fig:FerrariSpohnDistribution}
\end{figure}

The limit holds when the distance reached by the wall, $g\left(T/2\right)\sim W\sqrt{T}$, is much larger than the typical diffusion length $\sim \sqrt{T}$, such that the wall pushes the particles into a rare-event regime. The characteristic scale of fluctuations of the distance between the bridge's position and the wall is then given, in terms of $T$, by the prefactor in \myeqref{eq:4:FerrariSpohnDistribution} as $\frac{\sqrt{T}}{W^{1/3}}$.

The exact setting of \cite{Ferrari2005} is $W= \sqrt{T}$ such that the semi-circle is not deformed, and the problem is studied in the large-$T$ limit. The position fluctuations are then observed
on scale $T^{1/3}$, given by the characteristic scale $\frac{\sqrt{T}}{W^{1/3}}$ upon fixing $W= \sqrt{T}$. 

More generally, the results of \cite{Ferrari2005} allow for a general point of observation further from the mid-point $\frac{T}{2}$, and a range of parabolic boundaries further from the semicircle presented here. In each case, the Ferrari-Spohn distribution holds, with a prefactor of the variable fixed by
\be 
\ell = \left( - 2 \, g''(t_{\rm obs}) \right)^{1/3} 
\ee
where $g''(t_{\rm obs})$ is the second derivative of the boundary $g$ at the point of observation $t_{\rm obs}$. One sees that, for the semicircle boundary in \myeqref{eq:4:SemicircleBoundary}, this gives indeed:
\be 
\ell = \frac{(4W)^{1/3}}{ \sqrt{T}} \; .
\ee
Finally, the results of \cite{Ferrari2005} extend to a description of the temporal evolution on a time scale $T^{2/3}$.
 
Further from the typical regime described by the Ferrari-Spohn distribution, large deviations in this problem and its extensions were studied in \cite{Smith2019,Meerson2019}.
In addition, the Ferrari-Spohn distribution has also been encountered in other settings. It has been obtained for the position of a Brownian excursion conditioned to have very small total area \cite{Agranov2020}, where we recall that a Brownian excursion is a Brownian bridge started and ended at the origin, and conditioned to remain strictly positive in the meantime. The FS distribution has also been encountered in models of two-dimensional Brownian motion conditioned on non-absorption by a stationary wall and pushed into a large-deviation regime through further constraints \cite{Nechaev2019,Vladimirov2020}.

\section[Generalized Ferrari-Spohn distribution for non-crossing bridges]{Generalized Ferrari-Spohn distribution for multiple non-crossing bridges}
\label{sec:4.2}

In publication \pubref{publication:ConstrainedNCFerrariSpohn}, a generalization of the Ferrari-Spohn model was considered for $N$ Brownian bridges conditioned not to cross each other. In a direct extension of the single-particle case presented in the previous paragraph, the statistics of the positions at mid-time were investigated, for $N$ non-crossing bridges $(\vec{A}_{t})_{  0 \leqslant t \leqslant T}$ started and ended at $\vec{0}$ and conditioned to avoid the semicircle wall $g(t)$ in \myeqref{eq:4:SemicircleBoundary}. The statistical quantity of interest is thus the following joint distribution, denoted $R(\vec{x},W)$ and expressed in terms of the set of distances to the wall $\vec{x}$ at mid-time as
\be
R(\vec{x},W) = P\left( \vec{x}+ g\left( \frac{T}{2} \right)  , \frac{T}{2} \mid \Bevent_{[0,T]} ,  \NC_{[0,T]} , \{ \vec{0}, 0 \} , \{ \vec{0}, T \} \right) \; .
\label{eq:4:DefMultipleParticleDensity}
\ee
As in chapter \ref{chap:2}, $ \NC_{[0,T]}$ denotes the event where the walkers remain non-crossing, and $\Bevent_{[0,T]}$ the one where the lowest particle remains above the boundary. Note that $R(\vec{x},W)$ also depends on $T$ but this dependence is not made explicit for simplicity. The situation is illustrated in Fig.~\ref{fig:FerrariSpohnMultipleBridges}, where the distance $x_k = A_{k , \frac{T}{2}} - g\left( \frac{T}{2}\right)$ of the $k$\textsuperscript{th} particle with the wall at mid-time is highlighted.

 \begin{figure}[ht!]
    \centering 
    \includegraphics[width=.85 \textwidth]{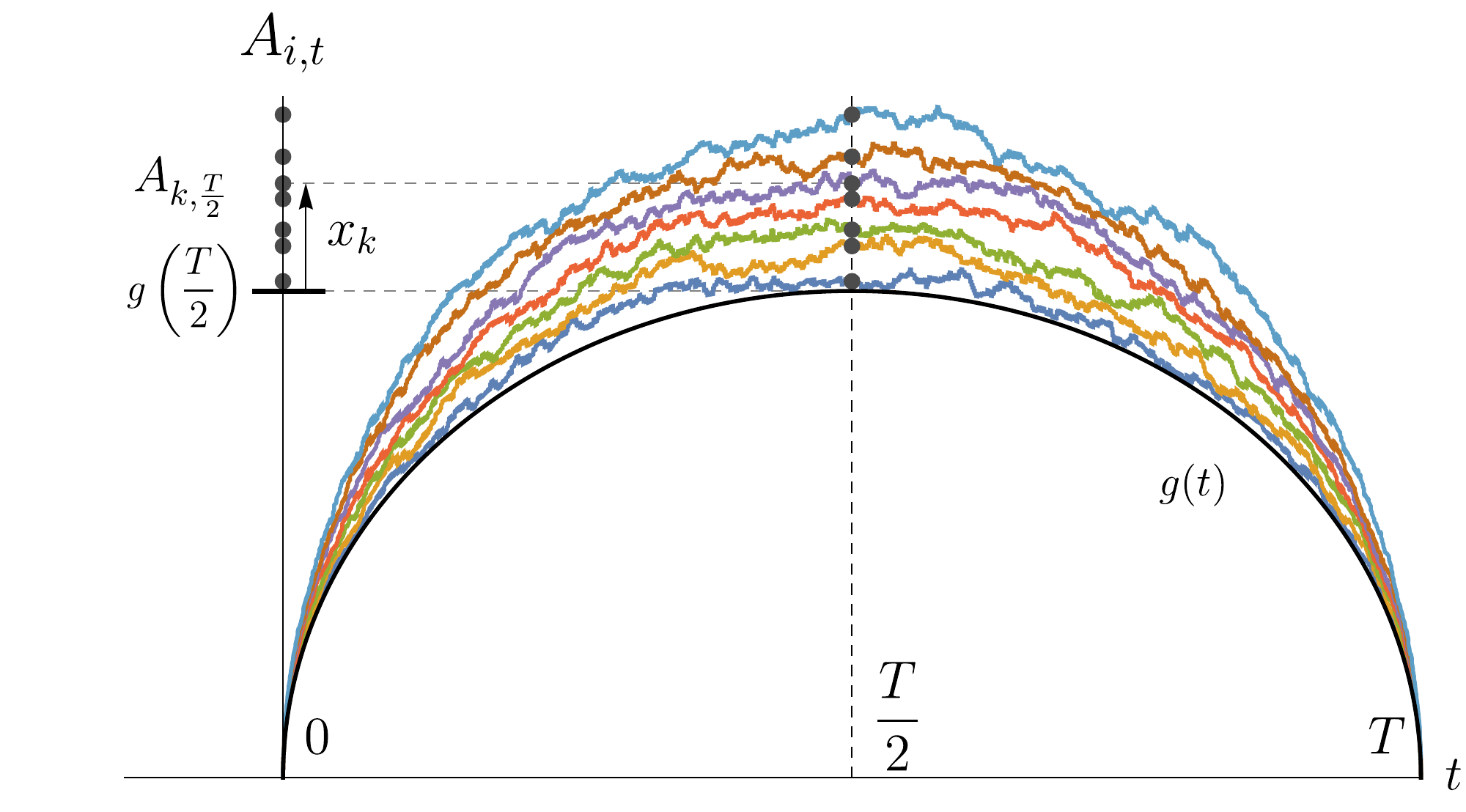} 
    \caption{Non-crossing Brownian bridges conditioned to remain above a semicircle boundary $g(t)$ and found at distances $\vec{x}$ of the wall at the mid-time $\frac{T}{2}$.}
    \label{fig:FerrariSpohnMultipleBridges}
\end{figure}

In the limit of large $W$, the distribution \eqref{eq:4:DefMultipleParticleDensity} is found to converge to a multi-particle generalization of the Ferrari-Spohn distribution $P_{\rm FS}$ in \myeqref{eq:4:FerrariSpohnDistribution}, as we detail below. The resulting distribution displays a determinantal structure, such that it is in direct equivalence with the position statistics of a fermionic system. More precisely, the corresponding system is subject to a linear potential above a hard wall. This mapping allows one to undertake a large-$N$ analysis of the system. Finally, we mention below links with other systems such as area-tilted random walks.

\subsection{Generalized Ferrari-Spohn distribution} 

The problem defined above can be solved
 by a transformation into a problem concerning a collection of non-crossing Brownian motions over a barrier with a square-root profile, such that results from \pubref{publication:NonCrossingBrownianDBM} presented in chapter \ref{chap:2} apply directly. In order to do this, the mapping \eqref{eq:4:MappingBrownianBridgeSymmetric} between Brownian motions with a square-root barrier and Brownian bridges with a semicircle barrier is employed. The problem needs to be slightly modified in the first place, as presented in the next lines.

\subsubsection{Problem redefinition}

\paragraph{Boundary condition regularization}

We begin by regularizing the initial and final conditions, in order to escape the singularity caused by the null probability of the conditioning event $\{  \Bevent_{[0,T]} ,  \NC_{[0,T]} , \{ \vec{0}, 0 \} , \{ \vec{0}, T \}  \}$. This singular conditioning is not problematic in the definition \eqref{eq:4:DefMultipleParticleDensity} but will be in the next operations on the joint distribution, such that the regularization is necessary.

The initial and final positions of the $N$ Brownian bridges $\vec{A}_t$ are thus taken as a fixed set of positions $\vec{s}$, such that $0 < s_1 < \cdots < s_N$. Furthermore, the results from chapter \ref{chap:2} impose a starting time that is not exactly zero, therefore we introduce a time $t_0$ at which the bridges are started, and the final position is now observed at $T-t_0$ by symmetry.  The initial and final conditions are thus given by:
\be
\vec{A}_{t_0} =\vec{A}_{T-t_0}= \vec{s} .
\label{eq:4:BoundaryConditionRegularization}
\ee
With these precise conditions, the joint density is redefined as
\be 
R ( \vec{x}, W ) = P\left( \vec{x}+ g\left( \frac{T}{2} \right)  , \frac{T}{2} \mid \Bevent_{[t_0,T-t_0]} ,  \NC_{[t_0,T-t_0]} , \{ \vec{s}, t_0 \} , \{ \vec{s}, T-t_0 \} \right) \; .
\ee 
The joint limit $( t_0 \to 0 , \vec{s} \to \vec{0} )$ will be taken in the last step to recover \myeqref{eq:4:DefMultipleParticleDensity}.

\paragraph{Conditional manipulations}

After this regularization, the distribution $R(\vec{x},W)$ can be reshaped in order to obtain quantities upon which the results of chapter \ref{chap:2} may be applied. This is done by massaging the conditional density, as follows.

We begin the by writing the conditional density $R\left(\vec{x},W\right)$ as
\be 
R\left(\vec{x},W\right) = \frac{P\left( \left\{ \vec{x}+ g\left( \frac{T}{2} \right) , \frac{T}{2} \right\}  , \{ \vec{s} , T-t_0 \} , \Bevent_{[t_0,T-t_0]}  ,  \NC_{[t_0,T-t_0]}  \mid    \vec{s}, t_0 \right)  }{P\left(   \{ \vec{s} , T-t_0 \},  \Bevent_{[t_0,T-t_0]}  ,  \NC_{[t_0,T-t_0]}  \mid \vec{s}, t_0  \right)} \; .
\ee
By Markov's property, the numerator can be cut in two terms, which are equal by symmetry, such that
\be 
R\left(\vec{x},W\right) = \frac{
P\left(  \left\{ \vec{x}+ g\left( \frac{T}{2} \right) , \frac{T}{2} \right\}   ,   \Bevent_{[t_0,\frac{T}{2}]}  ,  \NC_{[t_0,\frac{T}{2}]}  \mid   \vec{s}, t_0  \right) ^2
}{P\left(  \{ \vec{s} , T-t_0 \} ,  \Bevent_{[t_0,T-t_0]}  ,  \NC_{[t_0,T-t_0]}  \mid   \vec{s}, t_0  \right)} \; .
\label{eq:4:IntermediateComputation}
\ee

We now introduce back a conditioning on the final position in order to recover bridge processes, modifying the numerator term as
\begin{align}
 & P\left(  \left\{ \vec{x}+ g\left( \frac{T}{2} \right) , \frac{T}{2} \right\}   ,   \Bevent_{[t_0,\frac{T}{2}]}  ,  \NC_{[t_0,\frac{T}{2}]}  \mid   \vec{s}, t_0  \right)      \\
&  \qquad \qquad \qquad  =  \frac{
P\left(  \left\{ \vec{x}+ g\left( \frac{T}{2} \right) , \frac{T}{2} \right\}   ,   \Bevent_{[t_0,\frac{T}{2}]}  ,  \NC_{[t_0,\frac{T}{2}]}  , \{ \vec{0} , T \} \mid   \vec{s}, t_0  \right) 
}{ P \left( \vec{0} , T \mid  \vec{x}+ g\left( \frac{T}{2} \right) , \frac{T}{2}  \right)} 
 \\
= & P\left(  \left\{ \vec{x}+ g\left( \frac{T}{2} \right) , \frac{T}{2} \right\}   ,   \Bevent_{[t_0,\frac{T}{2}]}  ,  \NC_{[t_0,\frac{T}{2}]}   \mid   \{ \vec{s}, t_0 \} , \{ \vec{0} , T \}  \right) 
\frac{
 P \left( \vec{0} , T \mid  \vec{s}  , t_0 \right)} 
{P \left( \vec{0} , T \mid  \vec{x}+ g\left( \frac{T}{2} \right) , \frac{T}{2}  \right)}   \; .
\end{align}
Injecting this relation in \eqref{eq:4:IntermediateComputation} yields the following reshaped expression for $R(\vec{x},W)$:
\be
R\left(\vec{x},W\right) = 
\frac{
 P\left(  \left\{ \vec{x}+ g\left( \frac{T}{2} \right) , \frac{T}{2} \right\}   ,   \Bevent_{[t_0,\frac{T}{2}]}  ,  \NC_{[t_0,\frac{T}{2}]}   \mid   \{ \vec{s}, t_0 \} , \{ \vec{0} , T \}  \right) ^2 }
{P\left(  \{ \vec{s} , T-t_0 \} ,  \Bevent_{[t_0,T-t_0]}  ,  \NC_{[t_0,T-t_0]}  \mid   \vec{s}, t_0  \right)}
\frac{
 P \left( \vec{0} , T \mid  \vec{s}  , t_0 \right)^2} 
{P \left( \vec{0} , T \mid  \vec{x}+ g\left( \frac{T}{2} \right) , \frac{T}{2}  \right)^2} 
 \, .
\label{eq:4:ReshapedProblemDefDetailedDerivation}
\ee 

The interest of this calculation is to express the desired JPDF in terms of a survival probability density concerning $N$ Brownian bridges over a semicircle boundary. More precisely the term $Q(\vec{x},W)$ defined by
\be 
\label{eq:4:QdefRegularized}
Q(\vec{x},W) =  P\left(  \left\{ \vec{x}+ g\left( \frac{T}{2} \right) , \frac{T}{2} \right\}   ,   \Bevent_{[t_0,\frac{T}{2}]}  ,  \NC_{[t_0,\frac{T}{2}]}   \mid   \{ \vec{s}, t_0 \} , \{ \vec{0} , T \}  \right)  
\ee
is the probability density that $N$ Brownian bridges on $[0,T]$ started and ended from $\vec{0}$ and known to be in $\vec{s}$ at $t_0$, are found at positions $ \vec{x} + g(\frac{T}{2}) $ at mid-time, and that they stay non-crossing and above the semicircle barrier on $[t_0,\frac{T}{2}]$. We now turn to the study of $Q(\vec{x},W)$, as the other terms in \myeqref{eq:4:ReshapedProblemDefDetailedDerivation} depend trivially on the positions $\vec{x}$.

\subsubsection{Brownian motions over a square-root boundary}
 
As seen in the previous paragraph, the problem has been transformed into the study of $Q( \vec{x}, W)$, which concerns symmetric bridges on $[0,T]$ over a semicircle boundary on $[t_0, \frac{T}{2}]$. This object can be transformed into quantities presented in chapter \ref{chap:2}. Indeed, \myeqref{eq:4:MappingBrownianBridgeSymmetric} gives a mapping of the present Brownian bridges over a semicircle boundary towards standard Brownian motions over a square-root boundary. 
 
By inverting this change of variables, each Brownian bridge $A_{i,t}$ defined for $t \in [0,T]$ and conditioned to end back at $0$ at $t = T$, is mapped to a standard Brownian motion $B_{i,u}$ with a time variable $u =\frac{T t}{T-t} \in [0,\infty[$ as
\be 
A_{i,t}
\quad  \xrightarrow{ \ \;  t \ \;  \to \ \; u=\frac{T t}{T-t} \ \;  } \quad
B_{i,u} =   \frac{T+u}{T} \; A_{ \frac{T u}{T+u} } \; .
\ee
The semicircle barrier $g(t)$ is mapped under this transformation to the square root barrier $h(u)=W \sqrt{u}$:
\be 
g(t) =   W \sqrt{\frac{t}{T}  (T- t ) }
\quad  \xrightarrow{ \quad \quad \quad \quad} \quad
h(u) = W \sqrt{u} \; .
\ee 
Under this transformation, the initial condition $\vec{A}_{t_0} = \vec{s}$ is mapped to
\be  
\vec{B}_{u_{0}}=\frac{T+u_0}{T}\vec{s}\quad \text{where} \quad u_{0}=\frac{Tt_{0}}{T-t_{0}} \; .
\ee   
We are particularly interested in the mid-time $t = \frac{T}{2}$, where the following relations hold:
\be 
t = \frac{T}{2} \quad \quad  \implies \quad \quad 
\left\{
    \begin{array}{ll}
        u= T \; , \\[3pt]
        A_{i, \frac{T}{2}} = \frac{1}{2} \, B_{i,T} \; , \\[3pt]
        g(\frac{T}{2}) = \frac{1}{2} \,   h(T) =   W\frac{\sqrt{T}}{2} \; .
    \end{array}
\right.
\ee

As can be seen easily, this mapping preserves the non-crossing properties among walkers and with the boundary: the $\vec{B}$-paths and the boundary $h$ do not cross on $[u_0,T]$ if and only if the $\vec{A}$-paths and boundary $g$ do not cross on $[t_0, \frac{T}{2}]$. As a consequence, $Q(\vec{x},W)$ can be expressed in terms of the following probability density for the $(B_{i,u})$ process, where we denote $\Bevent^h$ the event where the square-root boundary $h$ is not touched by the walkers in order to avoid ambiguities with the previous lines:
\emphbe 
Q(\vec{x},W)  =
 2^N  P\left( \left\{ 2 \vec{x}  + h(T) , T  \right\} , \Bevent^h_{[u_0,T]}   ,   \NC_{[u_0,T]}  \mid  \frac{T + u_0  }{T} \vec{s}  , u_0 \right)
\end{empheq}

This quantity is precisely given in \myeqref{eq:2:SummaryStochasticQuantumPropagators} by the computations on Brownian motions over a square-root boundary presented in chapter \ref{chap:2}, with the slight modification that the walkers are \emph{above} the boundary in this section. Let us denote $(\phi_k, \varepsilon_k)_{k\geqslant 0}$ the sequence of 1-particle eigenfunctions and eigenenergies of the 1-particle Hamiltonian $\hat{H}$ defined as in \myeqref{eq:2:HamiltonianQuadraticHardWall} with reversed inequalities. In particular
\be 
\phi_k(X)=C_{k} D_{2  \varepsilon_k }(X) 
\label{eq:4:EigenfunctionsPositiveSign}
\ee
is defined on $[W,+\infty[$, with a modified sign from \myeqref{eq:2:GeneralWSingleParticleWavefunction}. The JPDF $Q(\vec{x},W)$ can then be expressed analytically as
\emphbe 
\begin{split}
Q(\vec{x},W)  &=  \frac{2^{N}}{ T^{N/2}   } \,
e^{-\frac{1}{4}\sum\limits_{k=1}^N\left[  \left(W+\frac{2x_{k}}{\sqrt{T}}\right)^{2} -  \frac{T}{(T-t_{0})t_{0}}{s}_{k}^{2}\right]} 
 \\
& \times \sum_{\vec{k}\in\Omega_{N}}
\Bigg[ 
\det\limits_{1 \leqslant i,j \leqslant N}\phi_{k_{i}}\left(W+2\frac{x_{j}}{\sqrt{T}}\right) \
\det\limits_{1 \leqslant i,j \leqslant N}\phi_{k_{i}}^{*}\left(\sqrt{\frac{T  }{(T-t_{0}) t_0 }} \, {s}_{j}\right) \
\\
 & \qquad \qquad \qquad \qquad \qquad \qquad \qquad \qquad \left(\frac{T-t_{0}}{t_{0}}  \right) ^{- \epsilon_{\vec{k}}}
\Bigg]  
\end{split}
\label{eq:4:QFromChap2}
\end{empheq}
where the initial conditions are expressed in terms of $t_0$. We recall that $\Omega_{N}$ is defined in \myeqref{eq:1:DefOmegaNManyBodyEigenstates} and that the multiple-fermion energies are given by
\be \epsilon_{\vec{k}}= \sum\limits_{i=1}^N \epsilon_{k_i} \; . \ee

\subsubsection{Limit results}

Before stating the results, we remove the regularization of the boundary condition, and then turn to the large-$W$ limit.

\paragraph{Initial condition limit $\{ t_0, \vec{s} \} \to \{ 0 , \vec{0}\}$}

A non-zero initial condition was necessary for the conditional manipulations and the use of the results of chapter \ref{chap:2}. We may remove it by scaling $\vec{s}$ with $t_0$ as follows:
\be
\label{eq:4:Sdef}
\vec{s} =g\left(t_{0}\right)\vec{S}=W \sqrt{\frac{t_{0}}{T}(T-t_{0})}\,\vec{S},
\ee
where $\vec{S}$ satisfies $1<S_{1}<\cdots<S_{N}$, and taking the limit $t_0 \to 0$ with $\vec{S}$ held constant. The dominant contribution in \eqref{eq:4:QFromChap2} then corresponds to the ground state such that
\be
\begin{split}
Q(\vec{x},W) 
\simeq \frac{2^{N}}{ T^{N/2}   } \,
e^{-\sum\limits_{k=1}^N  \left[ \frac{1}{T}  \left( \frac{W \sqrt{T}}{2}+ x_{k}   \right)^{2} - \frac{W^2 S_k^2}{4} \right]} 
\det\limits_{1 \leqslant i,j \leqslant N}\phi_{i-1}\left(W+2\frac{x_{j}}{\sqrt{T}} \right)  \\
\det\limits_{1 \leqslant i,j \leqslant N}\phi_{i-1}^{*}\left( W S_{j}  \right) \ \left(\frac{T}{t_{0}}  \right) ^{- \varepsilon_{\vec{0}}}  \, .
\end{split}
\ee
We note that the eigenfunctions $\phi_i$ are defined on $[W, + \infty[$ such that $  \det\limits_{1 \leqslant i,j \leqslant N}\phi_{i-1}^{*}\left(W S_{j}  \right) $ is a well-defined constant.
  
The JPDF of interest, $R(\vec{x},W)$, is obtained in \myeqref{eq:4:ReshapedProblemDefDetailedDerivation} from the expression of $Q(\vec{x},W)$. Evaluating the simple propagator terms which cancel the exponential term, we have in the limit $t_0 \to 0$:
 \be  
R\left(\vec{x},W\right) \simeq  \frac{\left(\frac{2}{T}\right)^{N}\left(\frac{t_{0}}{T}\right)^{2\varepsilon_{\vec{0}}}\left(\det\limits _{1\leqslant i,j\leqslant N}\phi_{i-1}^{*}\left(W S_{j}\right)\right)^{2} e^{\frac{1}{2} \sum\limits_{k=1}^N W^2 S_k^2   }}
{P\left(  \{ \vec{s} , T-t_0 \} ,  \Bevent_{[t_0,T-t_0]}  ,  \NC_{[t_0,T-t_0]}  \mid   \vec{s}, t_0  \right)}
\left[\det\limits _{1\leqslant i,j\leqslant N}\phi_{i-1}\left(W+2\frac{x_{j}}{\sqrt{T}}\right)\right]^{2}  \, .
\ee
We can directly read, from this equivalent expression, the limiting joint distribution function of the positions of the bridges above the semicircle boundary by discarding the terms which do not depend on variables $\vec{x}$. We conclude that, at $t_0=0$ and for arbitrary $W$, the JPDF $ R(\vec{x},W) $ defined in \myeqref{eq:4:DefMultipleParticleDensity} is given by
\emphbe 
 R(\vec{x},W) =
C  \
\left[ \det\limits_{1 \leqslant i,j \leqslant N}\phi_{i-1}\left(W+2\frac{x_{j}}{\sqrt{T}}\right) \
  \right]^2 
  \label{eq:4:ResultRDistributionZeroInitialCondition}
\end{empheq}
where we recall that the expression of the eigenfunctions $\phi$ of the Schrödinger operator with a quadratic potential and a hard wall at $W$ is given in \myeqref{eq:4:EigenfunctionsPositiveSign}. We point to \pubref{publication:ConstrainedNCFerrariSpohn} for additional details on the convergence, for the value of the prefactor $C$ and for the conclusions which can be drawn on the vanishing speed of the denominator of \myeqref{eq:4:ResultRDistributionZeroInitialCondition}.
 
\paragraph{Large-$W$ limit}

In order to achieve a generalization of the Ferrari-Spohn distribution, we finally specialize this expression to the large-$W$ case, where the wall pushes the particles in a large-deviation regime.

In the large-$W$ limit, the Hamiltonian of the underlying fermionic system $\hat{H}$ is nearly linear, neglecting a $(X-W)^2$ term, as
\be
\hat{H} \simeq
 \left\{
    \begin{array}{ll}
        \displaystyle  - \frac{1}{2} \frac{\partial^2}{\partial X^2}+ \frac{1}{8}W^2 + \frac{1}{4}(X-W) - \frac{1}{4}    & \mbox{for } X  > W \; ,  \\[8 pt]
        \displaystyle   + \infty    & \mbox{for } X \leqslant W \;  . \\
    \end{array}
\right. 
\label{eq:4:HamiltonianLinear}
\ee
As detailed in \pubref{publication:ConstrainedNCFerrariSpohn}, the eigenfunctions $\phi_{k}$ and eigenvalues $\epsilon_{k}$ can then be shown to be given in terms of the Airy function by the following expressions
\begin{equation}
\label{eq:4:AiryEigenfunction}
\left\{ \begin{array}{ll}
\displaystyle \phi_{k}(X)=D_{k,W}\,\Ai \left(\alpha_{k+1}+\left(\frac{W}{2}\right)^{1/3}(X-W)\right)\\[8pt]
\displaystyle \epsilon_{k}=-\frac{\alpha_{k+1}}{2}(\frac{W}{2})^{2/3}+\frac{1}{8}W^{2}-\frac{1}{4}
\end{array}\right.
\end{equation}
where $(\alpha_k)_{k \geqslant 1}$ is the sequence of zeros of the Airy function. The normalization constant $D_{k,W} $ is such that $\int_W^\infty \phi_k^2 (X) \dd X = 1$. It is found from the integral $\int_{\alpha_k}^\infty \Ai(x)^2 \dd x = \Ai'(\alpha_k)^2$, see App.~\ref{app:specialfunctions}, as
\begin{equation}
D_{k,W}= \frac{(W/2)^{\frac{1}{6}}}{\abs{ \operatorname{Ai}^\prime ( \alpha_{k+1} ) }} \; .
\end{equation}
    
Let us give some details about the validity of the approximations made above. Neglecting the quadratic term in the potential as we did in \myeqref{eq:4:HamiltonianLinear} requires $\Delta X = X-W \ll W$. On the other hand, the eigenfunctions of \myeqref{eq:4:AiryEigenfunction} are spread over a spatial scale $\Delta X \sim W^{-1/3}$. We therefore find that the requirement is simply $W \gg 1$, as assumed in the large-$W$ regime. Recalling that the wall's position is given by \myeqref{eq:4:SemicircleBoundary}, this requirement simply means that the distance reached by the wall, $g\left(T/2\right)\sim W\sqrt{T}$, is much larger than the typical diffusion length $\sim \sqrt{T}$, such that the wall pushes the particles into a large-deviation regime. 

In the large $W$ regime, the limiting distribution for the distance of the bridges with the wall is obtained by injecting the eigenfunctions \eqref{eq:4:AiryEigenfunction} in \eqref{eq:4:ResultRDistributionZeroInitialCondition}. We can then give the final result of our analysis, which gives $R(\vec{x},W)$ as the following generalization of the Ferrri-Spohn distribution:
\emphbe
\label{eq:4:RNWithNormalization}
R(\vec{x},W)=C_{N,W} \ \left[\det\limits _{1\leqslant i,j\leqslant N}\operatorname{Ai}\left(\alpha_{i}+\frac{(4W)^{1/3}}{\sqrt{T}}x_{j}\right)\right]^{2} 
\end{empheq}
The normalization constant, such that $\int_{{\R^+}^N} R( \vec{x},W) \, \dd \vec{x} =1$, is given by
\be 
C_{N,W}=  
\frac{ (4W)^{\frac{N}{3}}
}{ N! \
T^{\frac{N}{2}} \
\prod_{k=1}^N \operatorname{Ai}^\prime (\alpha_k)^2
} \; .
\ee 
This distribution can be seen to coincide with \myeqref{eq:4:FerrariSpohnDistribution} in the single-bridge case $N=1$.

\subsubsection{General observation time}

The goal of the previous computations was to extend the mid-time result given in \myeqref{eq:4:FerrariSpohnDistribution} for a single bridge, but we emphasize that the manipulations can be extended to any observation time $t \in ] 0 , T [$, further from the special case $t = \frac{T}{2}$ detailed above.

Crucially, for a general observation time $t$, the term $Q(\vec{x}, W)^2$ in the expression of $R( \vec{x}, W)$ is replaced by the product of two different terms
\be 
\begin{split}
& P\left(  \left\{ \vec{x}+ g\left( t \right) , t \right\}   ,   \Bevent_{[t_0,t]}  ,  \NC_{[t_0,t]}   \mid   \{ \vec{s}, t_0 \} , \{ \vec{0} , T \}  \right) 
 \\ & \times
  P\left(  \left\{ \vec{x}+ g\left( T-t \right) , T-t \right\}   ,   \Bevent_{[t_0, T-t ]}  ,  \NC_{[t_0,T-t]}   \mid   \{ \vec{s}, t_0 \} , \{ \vec{0} , T \}  \right) \; .
\end{split}
\ee 
The mapping to Brownian motions over a square-root boundary can be applied to both these terms separately. The procedure goes through and yields the following result
\be 
 R_t(\vec{x},W) =
C  \
\left[ \det\limits_{1 \leqslant i,j \leqslant N}\phi_{i-1}\left(W+x_j \sqrt{ \frac{T}{t(T-t)}}  \right) \
  \right]^2  
\ee 
which gives back \myeqref{eq:4:ResultRDistributionZeroInitialCondition} for $t=\frac{T}{2}$. The large-$W$ limit then yields the general-observation-time generalization of the Ferrari-Spohn distribution
\emphbe 
R_t(\vec{x},W)=C \ \left[\det\limits _{1\leqslant i,j\leqslant N}\operatorname{Ai}\left(\alpha_{i}+ \left(\frac{W}{2} \right)^{\frac{1}{3}} \sqrt{\frac{T}{t(T-t)}} \, x_{j}\right)\right]^{2} 
\end{empheq}

Similarly as in the single-bridge case, the prefactor of the $x$ variable is given by the second derivative of the boundary at the time of observation as
\be 
\ell = \left( - 2 \, g''(t) \right)^{1/3}  =\left(\frac{W}{2} \right)^{\frac{1}{3}} \sqrt{\frac{T}{t(T-t)}} \; .
\ee
This suggests that the $N$-particle setting extends to general concave boundary shapes, although we are limited to the semi-circle boundary with the method employed in this derivation.

In the generalized Ferrari-Spohn problem, the distribution of positions can thus be derived for a general observation time, as shown in this section. Notice that the situation is different in other non-crossing bridge problems where analytical solutions can only be found for particular times, as in \cite{Grela2021}.

\subsection{Density of particles above the boundary} 

In the previous section, we have derived the joint statistics of the generalized Ferrari-Spohn problem. In this section, we characterize the density of these non-crossing bridges above the semicircle boundary at mid-time, i.e. the system described by JPDF \eqref{eq:4:RNWithNormalization}. We start by giving details on the manifest determinantal structure of this distribution.

\subsubsection{Fermions in a linear potential above a hard wall}

The crux of the derivation presented in the previous pages relies on bootstrapping the results of chapter \ref{chap:2}, which themselves rely upon a mapping to a fermionic system, with Hamiltonian given in \myeqref{eq:4:HamiltonianLinear} in the large-$W$ limit. 

After the various changes of variables, the final JPDF \eqref{eq:4:RNWithNormalization} itself can be seen as the joint distribution of the ground-state positions of $N$ non-interacting fermions, in the single-particle Hamiltonian
\be 
\hat{H}_{\rm lin}=
 \left\{
    \begin{array}{ll}
        \displaystyle  - \frac{1}{2} \frac{\partial^2}{\partial x^2}+ \frac{2 W}{T^{3/2}} x & \mbox{for } x > 0 \; ,  \\[8 pt]
        \displaystyle   + \infty    & \mbox{for } x \leqslant 0 \;  . \\
    \end{array}
\right. 
\ee
The probabilistic system derived in the previous section is thus in correspondence with fermions in a linear potential $ V_{\rm lin}(x)= \frac{2 W}{T^{3/2}} x$, above a hard wall placed at the origin. 

The eigenfunctions $\psi_k$ of $\tilde{\hat{H}}$ are given from the $\phi_k$ in \myeqref{eq:4:AiryEigenfunction} through the change of variables $X \to W + \frac{2x}{\sqrt{T}}$ as
\be 
\psi_k(x)  = \frac{(4 W)^{1 / 6}}{T^{1 / 4}\left|\Ai^{\prime}\left(\alpha_{k+1}\right)\right|} \Ai \left(\alpha_{k+1}+\frac{(4 W)^{1 / 3}}{\sqrt{T}} x\right)
\ee
such that the ground state statistics match \myeqref{eq:4:RNWithNormalization} as announced, following
\be 
\frac{1}{N!}  \, \abs{  \det_{1 \leqslant i, j \leqslant N}  \psi_{i-1}( x_j)       }^2  = R(\vec{x},W) \; .
\ee
The corresponding energy levels are $\xi_k = - \alpha_{k+1}  \frac{(2W^2)^{1/3}}{T}$.
The linear potential above a hard-wall and its first few eigenfunctions $\psi_k$ are plotted in Fig.~\ref{fig:AiryEigenfunctions}. 

 \begin{figure}[ht!]
    \centering 
    \includegraphics[width=.65 \textwidth]{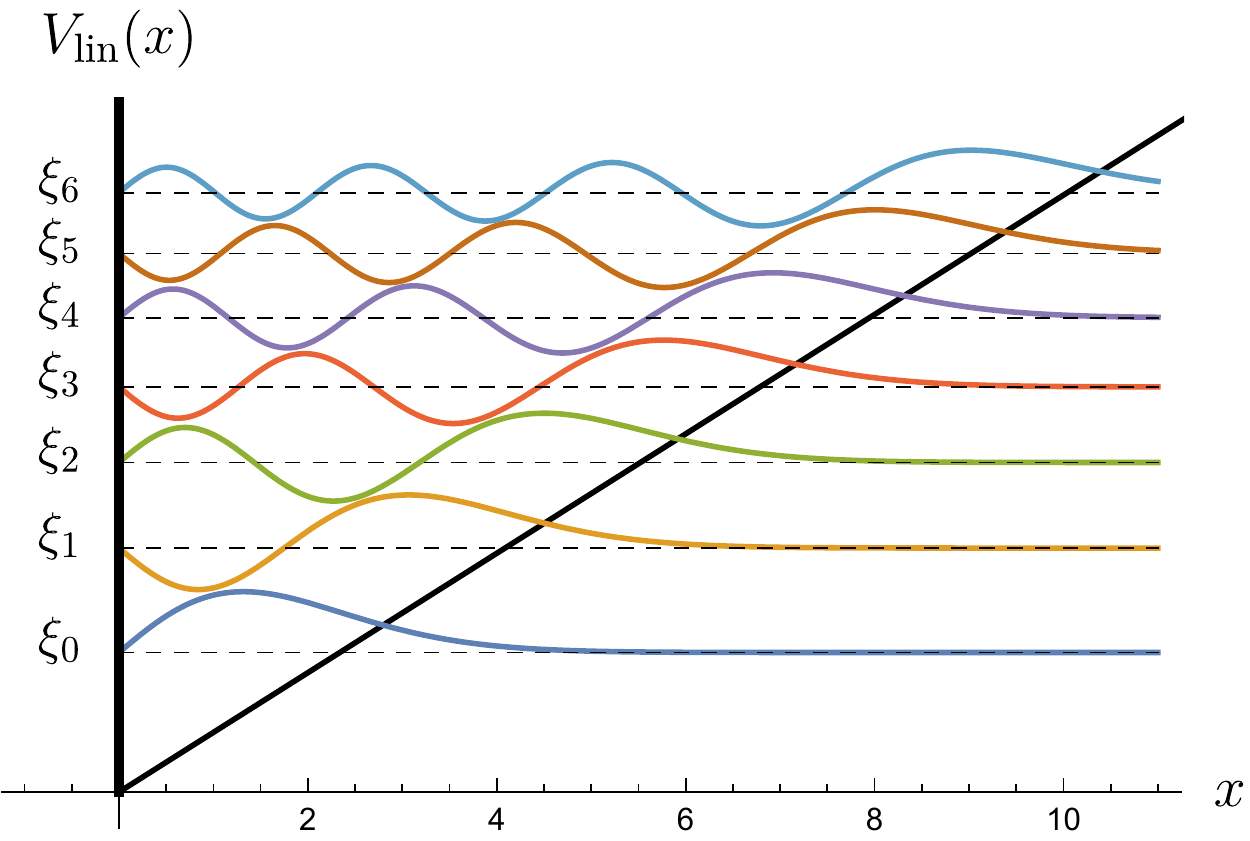} 
    \caption{Plot of the linear potential $V_{\rm lin}(x)$ above a hard-wall at the origin with $T=4W=1$ (black line). The first seven eigenfunctions $(\psi_k)_{0 \leqslant k \leqslant 6}$ are plotted at the corresponding energy level $\xi_k$ (coloured lines). Notice that $\abs{\psi_0}^2$ recovers the PDF plotted in Fig.~\ref{fig:FerrariSpohnDistribution}.}
    \label{fig:AiryEigenfunctions}
\end{figure}

As presented in section \ref{subsec:1.2.2} the ground state of non-interacting fermions in the ground state is endowed with a determinantal structure, such that the positions of non-crossing bridges above the semicircle boundary constitute a DPP with kernel
\be
K( x, y ) = \sum_{k=0}^{N-1} \psi_k(x) \, \psi_k(y) \; .
\ee
In particular, the density of the system is then given analytically by
\be 
P(x)= \frac{1}{N} K(x,x)   \; .
\label{eq:4:DensityParticles}
\ee
This density is plotted for $N= 20$ particles in Fig.~\ref{fig:DensityBridgesSemiCircle}. In the next section, we give precisions on the asymptotic behaviour of the density in the large-$N$ limit.

\subsubsection{Large-$N$ behaviour}

In the limit where the number of Brownian bridges is very large, the behaviour of the density \eqref{eq:4:DensityParticles} can be obtained from the universal properties of noninteracting fermions, related to the universal properties of RMT presented in chapter \ref{chap:1}. The fermion system of interest here presents three different regimes in the bulk, at the soft edge around the top particle and close to the hard edge. We present briefly the exposition of \pubref{publication:ConstrainedNCFerrariSpohn}.

\paragraph{In the bulk}

In the bulk regime, the density is well described by the LDA \cite{Castin2007}, which yields the following approximation
\be 
P^{\rm{bulk}}\left(x\right)=\frac{\sqrt{2}}{\pi N}\sqrt{\frac{\left(3\pi WN\right)^{2/3}}{2^{1/3}T}-\frac{2W}{T^{3/2}}x } \qquad , \qquad x \in \left[ 0 , x_{\rm{edge}} \right] 
\label{eq:4:PBulkBridges}
\ee
where 
\be 
x_{\rm{edge}}=\frac{\left(3\pi N\right)^{2/3}T^{1/2}}{2^{4/3}W^{1/3}} \; .
\ee
The effective Fermi energy of the fermion system is given by
\be 
\mu= \frac{ \left(3\pi WN\right)^{2/3} }{ 2^{1/3}T} \; .
\ee

More precisely, the bulk statistics of the particles on a scale $\left( N P^{\rm{bulk}}(x) \right)^{-1}$ are described by the sine kernel $K^{\mathrm{sine}}$ \eqref{eq:1:DefSineKernel}.

\paragraph{At the soft edge}

Near the soft edge at position $x_{\rm{edge}}$, the statistics are described by the Airy kernel $K^{\mathrm{Airy}}$ \eqref{eq:1:AiryKernel}. The width of the edge regime is
\be
w_{N}= \left[2 \, V_{\rm lin}(x_{\rm{edge}})\right]^{-1/3}  =\frac{\sqrt{T}}{\left(4W\right)^{1/3}} \; .
\ee
The density is then correctly described by 
\be  
P^{\rm{edge}}(x) \simeq \frac{1}{N w_{N}}F_{1}\left(\frac{x-x_{\rm{edge}}}{w_{N}}\right),\qquad
 F_{1}\left(z\right)=\left[\Ai '\left(z\right)\right]^{2}-z\left[\Ai \left(z\right)\right]^{2}
 \label{eq:4:PEdgeBridges}
\ee
as discussed in \myeqref{eq:3:FermiGasDensitySigmaTildeFromAiryKernel}.

As a result of the Airy statistics in the soft edge region, the fluctuations of the position of the highest particle, i.e. the one furthest from the wall, can be characterized as seen in chapter \ref{chap:1}. This position can be written as $x_{N}=x_{\rm{edge}}+w_{N}\chi$ where $\chi$ converges, for large $N$, to the GUE Tracy-Widom distribution \eqref{eq:1:TracyWidom2}.

As explained in the chapter introduction, a motivation for the Ferrari-Spohn model \cite{Ferrari2005} is to study a single bridge above a sea of non-crossing bridges, whose expected envelope is the semicircular boundary. The semicircle toy model then yields the Ferrari-Spohn distribution instead of the Tracy-Widom distribution. As shown above, the study of \pubref{publication:ConstrainedNCFerrariSpohn} sheds light on the intermediate regime between these two distributions, as the Ferrari-Spohn distribution is obtained for $N=1$ and the Tracy-Widom distribution is recovered for $N \gg 1$. For intermediate values of $N$, the distribution of the top particle can be obtained as a marginal of \eqref{eq:4:RNWithNormalization}.

\paragraph{At the hard edge}

Close to the hard-edge, the kernel converges to the hard-wall kernel 
\be 
K_{{\rm Hb}}\left(x,y\right)=\frac{\sin\left(x-y\right)}{\pi\left(x-y\right)}-\frac{\sin\left(x+y\right)}{\pi\left(x+y\right)}
\label{eq:4:HardWallKernel}
\ee
as 
\be
\frac{1}{k_N}   K\left( \frac{x}{k_N}, \frac{y}{k_N}\right)  \quad  \xrightarrow[N \to \infty] \quad  K_{{\rm Hb}}\left( x, y \right)
\ee
where the scaling is fixed by
\be 
k_{N}=\pi N P^{\rm bulk}(0)=\sqrt{2\mu}=  \frac{\left(6\pi WN\right)^{1/3} }{\sqrt{T}} \; .
\ee

$K_{{\rm Hb}} $ is the universal kernel obtained for fermions close to a hard-wall \cite{Dean2019,Lacroix-A-Chez-Toine2017,Lacroix-A-Chez-Toine2018,Lacroix-A-Chez-Toine2019}. It can be seen to hold in particular for the hard-box fermionic potential, through the relation to the JUE mentioned in chapter \ref{chap:1}. Note that $K_{{\rm Hb}}$ is related to the Bessel kernel $K^{\mathrm{Bessel}}_\alpha$ \eqref{eq:1:BesselKernel} with parameter $\alpha=\frac{1}{2}$ as
\be 
K_{{\rm Hb}}\left(x,y\right) = 2 \sqrt{x y} \ K^{\mathrm{Bessel}}_{1/2}\left(x^2, y^2\right) \; .
\ee

All statistics close to the hard-wall, such as the PDF of the position of the particle closest to the wall, can be expressed in terms of the kernel $K_{{\rm Hb}}$ \eqref{eq:4:HardWallKernel}. The density near the wall is obtained in the limit $y \to x$ as
\be
P^{\rm hard}(x) \simeq  \frac{k_{N}}{N} F_{{\rm Hb}}\left(k_{N} \, x\right) \quad , \quad F_{{\rm Hb}}(z)=\frac{1}{\pi}\left[1-\frac{\sin\left(2z\right)}{2z}\right] \; .
\label{eq:4:PHardBridges}
\ee 

The exact density $P(x)$ of the Fermi gas, given in \myeqref{eq:4:DensityParticles}, is plotted in Fig.~\ref{fig:DensityBridgesSemiCircle} for $N=20$ particles, along with the approximations $P^{\rm bulk}(x)$ in the bulk, given in \myeqref{eq:4:PBulkBridges}, $P^{\rm edge}(x)$ at the soft edge, given in \myeqref{eq:4:PEdgeBridges}, and $P^{\rm hard}(x)$ at the hard edge, given in \myeqref{eq:4:PHardBridges}. As in previous figures, the parameters are fixed as $T=4W=1$.

This concludes our presentation of the properties of non-crossing bridges above a semicircular boundary, in the limit where the boundary pushes the particles in a large-deviation regime, following \pubref{publication:ConstrainedNCFerrariSpohn}. Fig.~\ref{fig:DensityBridgesSemiCircle} shows the density of this ensemble of non-crossing particles above the boundary, in direct extension of the single-particle Ferrari-Spohn setting illustrated in Fig.~\ref{fig:FerrariSpohnDistribution}.

 \begin{figure}[ht!]
    \centering  
    \includegraphics[width=.85 \textwidth]{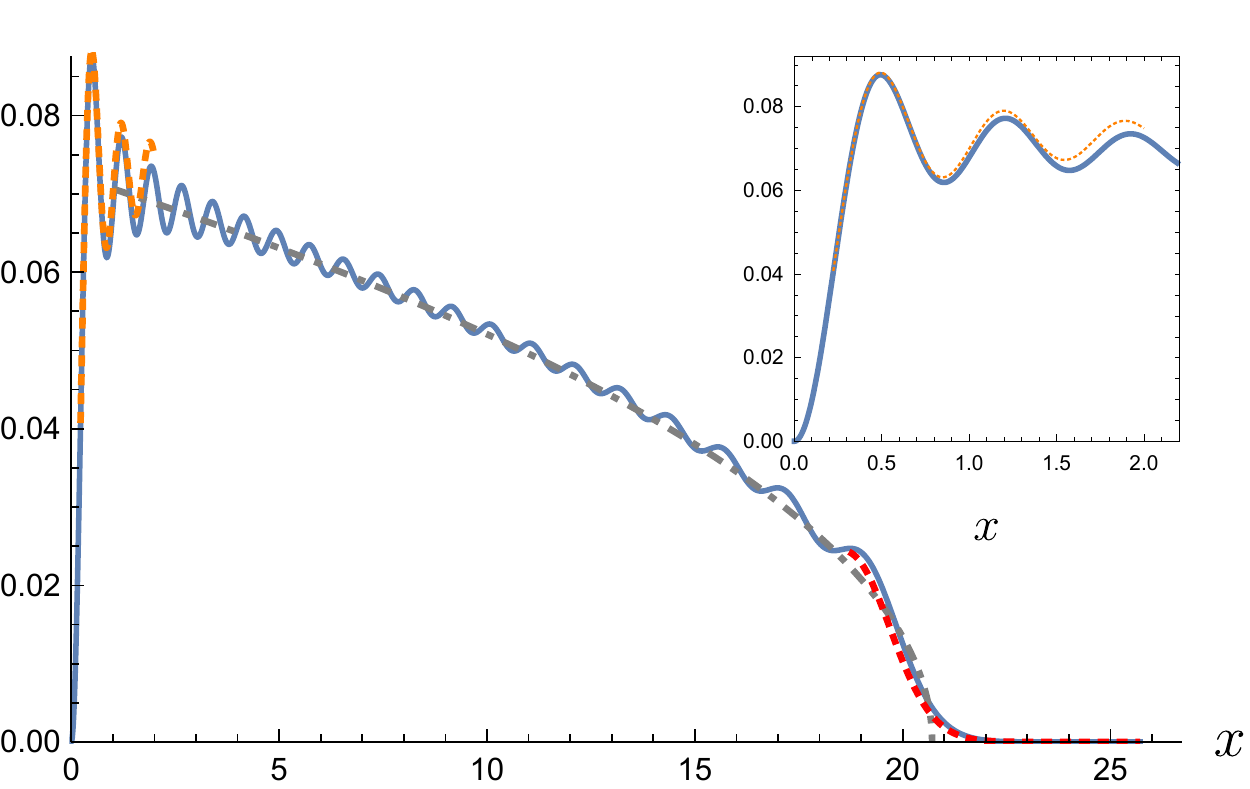} 
    \caption{Plot of the exact average density $P(x)$ for $N=20$ particles, in the blue line. The bulk density $P^{\rm bulk}(x)$ is the dashed gray line, the soft edge density $P^{\rm edge}(x)$ is the dashed red line and the hard-edge density $P^{\rm hard}(x)$ is the dashed orange line. The inset shows the region near the hard edge.}
    \label{fig:DensityBridgesSemiCircle}
\end{figure}

\subsection{Links with area-tilted excursions}

Let us consider a single Brownian excursion $(X_t)_{0\leqslant t \leqslant T}$. The fluctuations of the area $\int_0^T X_t \, \dd t$ covered by this process are described by the Airy distribution, which also applies to the maximal height of a fluctuating interface of Edwards-Wilkinson or KPZ type \cite{Majumdar2005}. As mentioned earlier, when conditioning this Brownian excursion to cover only a very small area, the fluctuations of its position at a given time are described by the Ferrari-Spohn distribution. We point in particular to Appendix C of \cite{Agranov2020} for a detailed derivation of the connection between the Brownian excursion and the Ferrari-Spohn model, in a path integral formalism.

Accordingly, it is natural to expect the results presented in the previous pages to be valid for $N$ non-crossing Brownian excursions $X_{i,t}$ conditioned on the sum of areas $\sum_{i=1}^{N}\int_0^T X_{i,t} \, \dd t$ being small, such that their joint position distribution over the axis is given by a JPDF similar to \eqref{eq:4:RNWithNormalization}, with a squared determinant involving Airy functions.

This is reminiscent of models of non-crossing diffusing particles near a hard wall subject to generalized area tilts, recently studied in \cite{Ioffe2018, Caputo2019}. In these works, the discrete random walks considered are subject to a self-potential of tilted area type, such that the probabilistic weight of each walker's path configuration $(X_i)_{1 \leqslant i \leqslant N}$ is multiplied by
\be
e^{- \sum_{i=1}^N V( X_i) }
\ee
with a general potential $V$. In the particular case $V(x) = x$, this effectively recovers the situation mentioned above since large areas are prevented by the tilt, expressed in the continuum limit as
\be 
e^{- \int_{0}^T X_{t} \,  \dd t} \; .
\ee 
In this particular case, and in the continuum limit, a distribution similar to \myeqref{eq:4:RNWithNormalization} is obtained in \cite{Ioffe2018} as the stationary JPDF of the non-crossing Brownian excursions' positions above the hard wall.

\section{The HCIZ integral and the Dyson Brownian bridge}
\label{sec:4.3}


As a conclusion to this chapter on bridge processes, we present an interesting perspective offered by the special case of matrix bridge processes. They are tightly connected to a non-commutative Fourier-analytic approach of random matrix theory, which is based on the Harish-Chandra-Itzykson-Zuber (\textit{HCIZ}) integral. This matrix integral, which we define in the following, can indeed be related to the bridge version of the DBM, with fixed initial and final conditions. The large-matrix-size limit of this integral exhibits an asymptotic behaviour which involves generalized hydrodynamical bridges in the full-rank setting, and recovers free probability results in the low-rank setting. 

As presented below, the applications of the HCIZ integral include large deviation computations which describe the behaviour of extreme eigenvalues in some random matrix ensembles. Finally, bridge problems from a given initial condition to a given final distribution, such as the ones appearing in the HCIZ integral, have interesting connections to the Schrödinger bridge problem and optimal transport.

\subsection{The HCIZ integral}
\label{subsec:4.3.1}

Twenty years after the work of Harish-Chandra on integrals over Lie groups \cite{Harish-Chandra1956,Harish-Chandra1957}, Claude Itzykson and Jean-Bernard Zuber studied a similar construction in the context of random matrices \cite{Itzykson1979}, which is now called the HCIZ integral.

\subsubsection{Definition and bridge interpretation}

\paragraph{Definition}
 
The HCIZ integral can be defined for the three values of the Dyson index $\beta =1,2,4$. Let us denote the corresponding $N\times N$ matrix group $\grG_\beta (N)$, which is $\mathrm{O}(N)$,  $\mathrm{U}(N)$  or $\mathrm{Sp}(N)$ respectively. Let us further denote $\Vol_\beta = \Vol(\mathrm{G}_\beta (N) ) $ the volume of these compact groups, given in \myeqref{eq:app:VolumeGroups} of App.~\ref{app:normalization}. 

For $\AAA$ and $\BB$ two $N \times N$ self-adjoint matrices, with real, complex or quaternionic entries according to $\beta$, the HCIZ integral is defined as 
\emphbe 
\mathcal{I}_{N}^{(\beta)}(\AAA, \BB)  = \frac{1}{\Vol_\beta}  \int_{\grG_{\beta}(N)} \dd \UU \ e^{ \Tr \left( \AAA \UU \BB \UU^{-1} \right)}
\label{eq:4:HCIZDef}
\end{empheq}
This object is also called a spherical integral in mathematics. Because of the volume term, the integral can also be defined directly over the normalized Haar measure of the corresponding compact group.

Note that, for an arbitrary matrix $\overline{\UU} \in \grG_\beta(N)$, the translation invariance of the Haar measure in the HCIZ integral ensures that
\be 
\mathcal{I}_{N}^{(\beta)}\left(\overline{\UU} \AAA {\overline{\UU}}^{-1} ,\BB\right)  = \mathcal{I}_{N}^{(\beta)}\left(\AAA,  \overline{\UU} \BB  {\overline{\UU}}^{-1}\right)  =  \mathcal{I}_{N}^{(\beta)}(\AAA, \BB)  \; .
\ee
As a consequence, $\mathcal{I}_{N}^{(\beta)}(\AAA, \BB)$ only depends on the eigenvalues of $\AAA$ and $\BB$.

We also remark that the expected HCIZ integral factorizes if $\AAA$ is the free addition of two matrices as $\AAA = \AAA_1 +  \overline{\UU} \AAA_2 {\overline{\UU}}^{-1}$, with the Haar-distributed random matrix $\overline{\UU}  \in \grG_\beta(N)$. Integrating over the random matrix $\overline{\UU}$ indeed yields
\bea
\E \left[ \mathcal{I}_{N}^{(\beta)}(\AAA, \BB) \right] &=& 
\int_{\grG_{\beta}(N)} \frac{\dd \overline{\UU} }{\Vol_\beta}   \ \mathcal{I}_{N}^{(\beta)}(\AAA, \BB) \\
&=&  \int_{\grG_{\beta}(N)^2} \frac{\dd \UU \dd \overline{\UU} }{\Vol_\beta^2}  \ e^{ \Tr   \AAA_1  \UU \BB \UU^{-1} }  \  e^{ \Tr   \overline{\UU} \AAA_2 {\overline{\UU}}^{-1} \UU \BB \UU^{-1} }    \\
&=&  \mathcal{I}_{N}^{(\beta)}(\AAA_1 , \BB) \ \mathcal{I}_{N}^{(\beta)}(\AAA_2 , \BB) 
\label{eq:4:FactorizationHCIZFreeSum}
\eea
where the second integral simplifies by the translation invariance of the Haar measure.

\paragraph{Exact expression from the Dyson bridge for $\beta=2$}

The central result related to the HCIZ integral \eqref{eq:4:HCIZDef} is the exact formula provided by Harish-Chandra for a general connected and semisimple compact Lie group, and rediscovered by Itzykson and Zuber in the unitary case  $\grG_2 (N) = \mathrm{U}(N)$ as we now state. 

\subparagraph{HCIZ integral formula}
For $\AAA$ and $\BB$ two $N \times N$ Hermitian matrices with ordered eigenvalues given by $\vec{a}$ and $\vec{b}$ respectively, the HCIZ integral is equal to
\emphbe
\mathcal{I}_{N}^{(2)}(\AAA, \BB)  =
\left(\prod_{p=1}^{N-1} p !\right) \ \frac{\det\limits_{1 \leqslant i,j \leqslant N} e^{a_{i} b_{j}} }{\Delta(\vec{a}) \, \Delta(\vec{b})}
\label{eq:2:HCIZExactFormula}
\end{empheq} 
In the fully degenerate limit where $\vec{a} \to \vec{0}$, the computations of App.~\ref{app:VandermondeLimit} show that this exact formula recovers $\mathcal{I}_{N}^{(2)}\to 1$, as expected from the definition \eqref{eq:4:HCIZDef}.

\subparagraph{Dyson bridge proof} This formula can be proved in an insightful way by turning to a bridge interpretation, as we detail now. A slight remodelling of the  $\beta=2$ HCIZ integral of $\AAA$ and $\BB$ gives
\be 
C_2 \ \mathcal{I}_{N}^{(2)}(\AAA, \BB) \  e^{- \frac{\Tr (\AAA^2 + \BB^2)}{2}} = \frac{C_2}{\Vol_\beta}
 \int_{\mathrm{U}(N)} \dd \UU \ e^{ - \frac{1}{2} \Tr \left( \AAA -  \UU \BB \UU^{-1} \right)^2} \; ,
 \label{eq:4:RemodeledHCIZ}
\ee
where we have included $C_2$, the constant defined in \myeqref{eq:1:invariantdefGUEC2Def} with $\sigma=1$.

Let us now recall that $\HH_1^2$, the DBM with Dyson index $\beta=2$ at time $t=1$ defined in \myeqref{eq:2:DefDBMGUE}, is distributed as the $\sigma=1$ GUE matrix distribution \eqref{eq:1:invariantdefGUE}. As a consequence, the r.h.s. of \myeqref{eq:4:RemodeledHCIZ} is the exact probabilistic weight for the DBM-2 started at $\HH_0^2=\AAA$ to end, at time $t=1$, in any rotation $\UU \BB \UU^{-1}$ of the matrix $\BB$. We deduce
\be 
C_2 \ \mathcal{I}_{N}^{(2)}(\AAA, \BB) \  e^{- \frac{\Tr (\AAA^2 + \BB^2)}{2}} = \E \left[ \1_{ \HH_1^2 = \UU \BB \UU^{-1}} \mid  \HH_0^2=\AAA \right] 
\label{eq:4:ExpectationDysonBridgeHCIZ}
\ee 
where the conditional expectation is taken over both the value $\HH_1^2$ of the DBM at $t=1$ and a normalized-Haar-measure distributed random matrix $\UU$.

At the eigenvalue level, this expectation is the probabilistic weight of the $N$-particle DBM-$2$ bridge configurations starting at $\vec{a}$ and ending at $\vec{b}$, between $t=0$ and $t=1$, after taking into account the Jacobian term $\Delta(\vec{b})^2$ and the proportionality factor $N! \, \Xi_2^N$ given in \myeqref{eq:app:ProportionalityFactoDistributions}, both needed to recover the (ordered) eigenvalue distribution from the matrix distribution. We furthermore recall per chapter \ref{chap:2} that the DBM-2 eigenvalue process is identical to the process of never-crossing Brownian motions.  The expectation in \myeqref{eq:4:ExpectationDysonBridgeHCIZ} can then be written simply in terms of the following Brownian joint probability density
\be 
C_2 \ \mathcal{I}_{N}^{(2)}(\AAA, \BB) \  e^{- \frac{\Tr (\AAA^2 + \BB^2)}{2}} = \frac{1}{N! \, \Xi_2^N \Delta(\vec{b})^2} \ P \left( \vec{b}, 1 \mid \vec{a}, 0 , \NC_{[0, \infty [}  
\right)\; .
\ee 
The application of relation \eqref{eq:2:NeverCrossingInTermsofIntermediatePropagator} and of the Karlin-McGregor theorem \eqref{eq:2:KarlinMcGregorTheorem} gives
\be 
 P \left( \vec{b}, 1 \mid \vec{a}, 0 , \NC_{[0, \infty [}   \right)  =  \frac{\Delta(\vec{b})}{\Delta ( \vec{a} )} \det_{1 \leqslant i,j \leqslant N} P(b_i, 1  \mid a_j, 0)
\ee
in terms of the single-particle Brownian propagator, such that
\bea
C_2 \ \mathcal{I}_{N}^{(2)}(\AAA, \BB) \  e^{- \frac{\Tr (\AAA^2 + \BB^2)}{2}} &=& \frac{1}{N!\, \Xi_2^N \sqrt{2\pi}^N} \ \frac{\det\limits_{1 \leqslant i,j \leqslant N}  e^{-\frac{(a_i - b_j)^2}{2}}}{ \Delta(\vec{a}) \, \Delta(\vec{b})}   \\
&=& 
 \frac{e^{- \frac{1}{2} \sum\limits_{i=1}^N (a_i^2 + b_i^2 ) } }{N!\, \Xi_2^N \sqrt{2\pi}^N} \ \frac{\det\limits_{1 \leqslant i,j \leqslant N}  e^{a_i  b_j}}{ \Delta(\vec{a}) \, \Delta(\vec{b})} 
 \; .
\eea
Simplifying the squared terms finally recovers the celebrated HCIZ integral formula \eqref{eq:2:HCIZExactFormula}.

This derivation shows that the HCIZ integral is fundamentally related to a probabilistic weight concerning a DBM bridge, with bounds given in terms of the two arguments $\AAA$ and $\BB$ of $\mathcal{I}_{N}^{(2)}(\AAA, \BB)$. The Dyson bridge matrix process is, as a consequence, connected to many aspects of RMT through the HCIZ integral, as we develop below. Note that the Dyson bridge connection can be extended to the $\beta=1,4$ cases. However, we have not included them in the derivation above, because the absence of a simple expression for the DBM propagator for $\beta \neq 2$ prevents one from obtaining an expression similar to \myeqref{eq:2:HCIZExactFormula}. We recall that this is directly linked to the presence of interactions in the corresponding Calogero-Moser Hamiltonian's potential \eqref{eq:2:PotentialCalogeroMoser} when $\beta \neq 2$.

For more details on the above definition of the HCIZ integral and for algebraic proofs of the integral formula, see \cite{Zuber2012,McSwiggen2018,Potters2020,Mergny2021}.

\subsubsection{Large-$N$ limit}

We present briefly the asymptotics of the HCIZ integral in the large-$N$ limit, which can be investigated in the general full-rank setting, where $\AAA$ and $\BB$ both have many non-zero eigenvalues, and in the particular low-rank setting, where one of them has most eigenvalues equal to zero.

\paragraph{Full-rank setting}

The large-$N$ asymptotics of the HCIZ integral were first obtained by Matytsin in the unitary case \cite{Matytsin1994}, before being derived in full mathematical rigour \cite{Guionnet2002a} and extended to the orthogonal case \cite{Zuber2008}. These asymptotics can be useful in many applications, such as matrix models where the HCIZ integral \eqref{eq:4:HCIZDef} appears as a consequence of interaction, and thus controls the large-$N$ behaviour of the system and the asymptotic spectral distributions \cite{Itzykson1979}.

Let us state the asymptotic result in the $\beta=2$ case. Assuming that the eigenvalues of $\AAA$ and $\BB$ scale as $\sqrt{N}$ such that the occupation densities of $\vec{a} / \sqrt{N}$ and $\vec{b}/ \sqrt{N}$ are described in the limit by two distribution functions $\rho_a$ and $\rho_b$, we have
\emphbe 
\mathcal{I}_{N}^{(2)}(\AAA, \BB) \  \simeq \ e^{ N^2  \mathcal{F}(\rho_a, \rho_b) } 
\label{eq:4:FullRankAsymptoticsHCIZ}
\end{empheq}
The leading term grows exponentially at speed $N^2$ with a rate function $\mathcal{F}$ related to the action corresponding to the Euler system
\be 
\frac{\partial \rho}{\partial t} + \frac{ \partial (\rho v) }{\partial x} =0 \quad , \quad \frac{\partial v}{\partial t} +v \frac{\partial v}{\partial x} -\pi^{2} \rho \frac{ \partial \rho}{\partial x} =0 
\label{eq:4:EulerSystem}
\ee
describing the one-dimensional evolution of a fluid with density $\rho(x,t)$ and velocity $v(x,t)$, with boundary conditions fixed by
\be 
\rho(x,t=0) = \rho_a(x) \quad , \quad \rho(x,t=1) = \rho_b(x) \; .
\ee
The leading term in the large-$N$ limit of $\mathcal{I}_{N}^{(2)}(\AAA, \BB)$ is thus related to a hydrodynamical bridge problem, and is characterized by the dynamics of a fluid evolving as \eqref{eq:4:EulerSystem} from the eigenvalues of $\AAA/\sqrt{N}$ at initial time to those of $\BB/\sqrt{N}$ at final time. This hydrodynamical bridge is reminiscent of the Dyson bridge, explained in the previous section to be at the center of the HCIZ construction. Remarkably, it was shown in \cite{Bun2014} that this hydrodynamical description can be obtained directly from the finite-$N$ Dyson bridge setting, through an instanton approach to the large deviations of the DBM bridge.

Finally, in order to stress the connections with the Dyson bridge, we remark that the complex field $f(x,t) = v(x,t) + \ii \pi \rho(x,t)$ evolves, per \myeqref{eq:4:EulerSystem}, as the complex Burgers equation which characterizes the Stieltjes transform evolution for the DBM, see \myeqref{eq:2:BurgersEquationDBM}, as
\be 
\frac{\partial f(x,t)}{\partial t}+f(x,t) \frac{\partial f(x,t)}{\partial x} =0 \; .
\ee  
For more details on the hydrodynamical system and its complex Burgers version, see \cite{Menon2017,Matytsin1994,Menon2012}.

\paragraph{Low-rank setting} 

The analysis is simpler in the case where one of the matrices has low rank, or even rank one as we turn to now. The result was obtained in the context of spin glasses in \cite{Marinari1994} for $\beta =1,2$, and derived in full mathematical rigour in \cite{Guionnet2005}. In this regime, the HCIZ integral is tightly related to the partition function of the spherical Sherrington-Kirkpatrick model \cite{Kosterlitz1976,Baik2016}.

With $\bfx $ be a unit $N$-dimensional vector, we consider that the second argument of the HCIZ integral is
\be 
\BB= N  t \,  \bfx \bfx^\dagger 
\ee
such that its eigenvalues are $\vec{b} = \{ N t , 0 , \cdots, 0 \}$. The corresponding HCIZ integral then depends on the eigenvalues $\vec{a}$ of $\AAA$ and on the parameter $t$, such that we denote it
\be 
 \mathcal{I}_{\vec{a}}^{(\beta)}( t ) = \mathcal{I}_{N}^{(\beta)}(\AAA, N  t \,  \bfx \bfx^\dagger )    \; .
\ee
For small enough $t$, the large-$N$ behaviour of the rank-one HCIZ integral is found to grow exponentially at speed $N$, according to
\emphbe
\mathcal{I}_{\vec{a}}^{(\beta)}( t )  \ \simeq \ e^{ N \int_0^t \RT_\AAA\left( \frac{2}{\beta} u\right) \, \dd u }
\label{eq:4:LowRankHCIZAsymptotics}
\end{empheq}
where $\RT_\AAA$ is the $\RT$-transform of $\AAA$, defined in \myeqref{eq:1:DefRTransform}. For higher values of $t$, a phase transition is encountered at a given $t^*$ and the rate function is different thereafter, see \cite{Guionnet2005}. Notice the difference of the $N$ scaling in this low-rank setting with the $N^2$ scaling which characterizes the full-rank asymptotics in \myeqref{eq:4:FullRankAsymptoticsHCIZ}.

The appearance of $\RT_\AAA$ in the large-$N$ HCIZ integral of $\AAA$ and a rank-one matrix shows that the HCIZ construction has profound ties to free probability. These ties can be illustrated by applying \myeqref{eq:4:FactorizationHCIZFreeSum} to the current setting. Indeed, taking $\vec{c}$ to be the eigenvalues of the free sum of two matrices $\AAA$ and $\BB$, we have in the self-averaging large-$N$ limit
\be
 \mathcal{I}_{\vec{c}}^{(\beta)}( t ) 
=  \mathcal{I}_{\vec{a}}^{(\beta)}( t ) \times   \mathcal{I}_{\vec{b}}^{(\beta)}( t )   \; .
\ee
Therefore, the HCIZ integral is, in the large-$N$ limit, the exponential of a function which is additive under the free sum. Recalling \myeqref{eq:1:FreeAdditionRTransform}, it is therefore not surprising that the $\RT$-transform should appear in the asymptotics of the log of the rank-one HCIZ integral.

Finally, we remark that the nature of $\mathcal{I}_{\vec{a}}^{(\beta)}(t)$ as a function of the random matrix $\AAA$ and the external parameter $t$ which is multiplicative under the asymptotically free sum makes it a natural matrix counterpart of the characteristic function for a scalar random variable $X$, defined as $ \E [ e^{i t X} ]$. This illustrates Harish-Chandra's initial motivation, when introducing the HCIZ integral, to develop a non-commutative Fourier analysis on Lie algebras \cite{Harish-Chandra1957a}; we point to \cite{Guionnet2005} for a discussion on the validity of upgrading the parameter $t$ to a complex number.
 
For more details on the relation of the HCIZ construction with free probability, derivations of \myeqref{eq:4:LowRankHCIZAsymptotics}, replica computations surrounding the HCIZ integral and results in the setting where the rank is small compared to $N$ but higher than one, we point to \cite{Potters2020,Mergny2021}. Interestingly, a multiplicative version of the HCIZ integral can be constructed, see \cite{Mergny2020}, which displays a rank-one large-$N$ behaviour similar to \myeqref{eq:4:LowRankHCIZAsymptotics} with the $\ST$-transform taking the place of the $\RT$-transform.

\subsection{Large deviations of extreme eigenvalues in some RMT ensembles}
\label{subsec:4.3.2}

The HCIZ integral has applications in many areas of physics and mathematics. It appears naturally in harmonic analysis, representation theory and symplectic geometry, as well as in the combinatorics of Hurwitz number, see \cite{McSwiggen2018}. In physics, the HCIZ integral appears naturally in matrix models presenting interactions: it was initially studied in relation to QCD and map enumeration \cite{Itzykson1979,Matytsin1994,Zvonkin1997}, but it also appears in spin glass models \cite{Potters2020} and in quantum ergodic transport \cite{Bauer2020,Fyodorov1999}.

Remarkably, the HCIZ integral also plays a central role in the large deviations of some RMT ensembles which have an external source \cite{Brezin2016}, as noticed in the formal derivation of the full-rank large-$N$ asymptotics \cite{Guionnet2002a}. In order to probe the largest eigenvalue's right-tail large deviations in some ensembles, the idea has recently emerged in \cite{Guionnet2020a} to tilt the matrix distribution by the rank-one HCIZ integral. As explained in \cite{Belinschi2020}, this procedure amounts to tilt the law of the matrix in a random direction, in order to push the top eigenvalue in a large-deviation regime. This yields large-deviation results which are, as a consequence, directly expressed in terms of the rank-one asymptotics presented in the previous section. Because of the phase transition mentioned below \myeqref{eq:4:LowRankHCIZAsymptotics}, the ensuing rate functions can be intricate. Consequently, we will not present them in full detail here, but instead mention results that were recently obtained with this technique, for some ensembles of RMT.

Let $\AAA$ and $\BB$ be two large deterministic Hermitian matrices, and $\UU$ a unitary Haar-distributed matrix. In \cite{Guionnet2020}, large-deviation results were obtained for the largest eigenvalue of 
\be 
\AAA + \UU \BB \UU^{-1} \; .
\label{eq:4:APlusUBUMinus}
\ee 

With $\AAA$ a large deterministic matrix as above, and $\HH$ a Wigner matrix (GOE or GUE for example), the large-deviation principle for the largest eigenvalue of 
\be 
\AAA + \HH
\ee
was established in \cite{McKenna2021}. Notice the difference with the previous model, as the eigenvalues of $\HH$ are not fixed here, unlike those of $\BB$ in \myeqref{eq:4:APlusUBUMinus}.

For $\HH$ a GOE or GUE matrix and $\bfx$ a fixed vector, the largest eigenvalue of the spiked model
\be 
\theta \, \bfx \bfx^\dagger  + \HH
\ee
was described in the large-deviation regime in \cite{Biroli2020}. The transition in the behaviour of the largest eigenvalue of this rank-one perturbation of a Gaussian matrix was famously described by Baik-Ben Arous-Péché (\textit{BBP}) in \cite{Baik2005}; the investigation of \cite{Biroli2020} thus explores the large-deviation regime beyond the BBP mechanism. We remark that the results are also generalized to the Wishart ensemble. 

In a distinctly more accessible derivation, the generalized Wishart ensemble was studied in \cite{Maillard2021}. The model in question is
\be 
\XX^\dagger  \, \GGamma \, \XX
\ee
defined as in \myeqref{eq:1:DefWishart} with an arbitrary covariance matrix $\GGamma$. Specifically, the top eigenvalue's large deviations were described. Remarkably, the rate function is shown in this case to take a very simple form as the integral of the difference of the two admissible branches of the Stieltjes transform, see Eq.~(3) of \cite{Maillard2021}.

We insist that the rank-one HCIZ tilt technique, used to obtain the largest eigenvalue large-deviation regime in these ensembles, limits all the above results to the \emph{right tail}. Indeed, the tilt pushes one eigenvalue in a rare configuration, but is incapable of pushing a large portion of eigenvalues away from their typical position, which is required to reach a \emph{left-tail} large-deviation event. As a consequence, the development of a measure-tilt by the full-rank HCIZ integral is necessary in the left-tail setting. This is illustrated in \cite{Belinschi2020} where large deviations are obtained for the empirical eigenvalue distribution for the model defined in \myeqref{eq:4:APlusUBUMinus} through such a full-rank tilt, further from the largest-eigenvalue study of \cite{Guionnet2020}. Notice in this regard that the respective scaling behaviours of the low-rank and full-rank cases coincide exactly with the Coulomb-gas energy cost of pulling a single particle ($\propto N$) or pushing a large portion ($\propto N^2$), as discussed in subsection \ref{subsec:1.2.3}.

\subsection{Schrödinger bridges and optimal transport}
\label{subsec:4.3.3}

In subsection \ref{subsec:4.3.1}, we have shown how the HCIZ integral is intimately connected to the transportation of particles from a given distribution at time $t=0$ to another one at time $t=1$. In a wider scope and relaxing interactions between particles, this is reminiscent of a bridge problem introduced by Erwin Schrödinger in the 1930's. This problem has deep connections to the theory of optimal transport, on which we close the perspectives laid out in this section.

\subsubsection{Schrödinger bridge problem}

In \cite{Schrodinger1931,Schrodinger1932}, Schrödinger introduced the following thought experiment concerning the statistical physics of hot gas particles. The problem is to find the most likely trajectory taken by a cloud of $N$ independent Brownian particles, started according to a distribution $\rho_0$ at $t=0$ and known to have ended in an arbitrary distribution $\rho_1$ at $t=1$, potentially very different from the distribution expected from the convolution of $\rho_0$ with the Brownian propagator over a unit-time evolution. 

This maximum-entropy problem can be recast and solved in the modern framework of stochastic processes and large deviations, but recall that these did not yet exist at the time. The remarkable aspect of this problem lies in the analogy that can be drawn with the laws of quantum mechanics. Indeed, the solution $\rho(x,t)$, describing the most probable distribution of the particles at intermediate time $t \in [0,1]$, can be written as a product
\be 
\rho(x,t) = \psi(x,t) \ \overline{\psi}(x,t) 
\ee
where $\psi$ is the forward solution of the stochastic evolution from the initial condition $\rho_0$, and $\overline{\psi}$ is the backward solution from the final condition $\rho_1$. This is evocative of the quantum framework where the probability distribution is the modulus square of the wavefunction, see \myeqref{eq:1:PDFFromWavefunction}, and where the conjugate of the wavefunction is a solution of the time-reversed Schrödinger equation. 

We point to the survey \cite{Leonard2014} for more details on the solution of the bridge problem and its quantum analogy. For financial applications of the Schrödinger bridge construction, see \cite{Henry-Labordere2019}. As mentioned in these works, the Schrödinger bridge solution can be shown to respect a SDE with drift term similar to \eqref{eq:4:LogQConditionalDrift}, as expected from the explanations of subsection \ref{subsec:4.3.1}.

\subsubsection{Optimal transport}

Beyond quantum mechanics, the ties of this problem extend to the Monge-Kantorovich theory of optimal transport. This field focuses on the study of the optimization problem of finding the transport map, from an initial mass distribution to a final mass distribution, that minimizes a quadratic cost-function. This problem induces the Wasserstein distance between probability distributions and is solved, in some practical applications, by an implementation of Sinkhorn's algorithm, which relies on the observation that the doubly-stochastic matrix defining the transport map can be obtained as the result of alternatively normalizing the rows and columns of a matrix. 

The Schrödinger bridge problem is an entropy regularization of optimal transport, adding a noise ingredient to the particles' trajectories in a blurry approximation of the Monge-Kantorovich problem. Conversely, optimal transport can be viewed as a zero-temperature limit of Schrödinger bridges. Interestingly, Sinkhorn's algorithm is itself a special case of a numerical scheme developed in the study of Schrödinger bridges. For a thorough review of these profuse connections from the point of view of stochastic control, see \cite{Chen2021}.

We point to \cite{Sejourne2019} for a recent exposition of optimal transport and some results on its entropy regularization in the unbalanced case where the initial and final distributions have different mass.


\thispagestyle{empty}
\backmatter
\titlespacing*{\chapter}{0pt}{0pt}{40pt}
\chapter{Conclusion}
\vspace{1cm}

In this manuscript, we have given a joint presentation of random matrix theory, non-crossing walkers and trapped fermions, highlighting the remarkable connections shared by these systems. After a broad introduction to RMT, we focused on non-crossing scalar processes before turning to stochastic matrix processes, underlining in both cases their links to quantum fermionic systems. Finally, we turned to a joint treatment of scalar and matrix models through the particular case of their bridge processes. In each chapter, we took care to address perspectives offered by these themes, detailing further questions and connections to other fields.\\[10pt]

The results presented in the realm of this thesis can be summarized as follows.\\ 
Moving boundary problems were studied for non-crossing walkers, in the critical square-root boundary case. The exponent driving the decay of the survival probability was related to the ground-state energy of a system of non-interacting fermions in a harmonic potential with a hard-wall, and studied in various limits. The distribution of walkers above the boundary was also studied, and extensions of these results were given for the Dyson Brownian motion with a boundary.\\
Links between non-crossing walkers and the Gaussian ensembles were expanded, recovering results known in the literature, in a notable connection between partially non-crossing walkers and the Pandey-Mehta interpolating random matrix ensemble.\\
Stochastic matrix processes were introduced and studied, with a particular interest in the matrix Kesten recursion, a matrix model inspired from the Kesten random recursion. In the continuous-time limit, a mapping was established between this model and time-evolving fermions in the Morse potential with Sutherland $\sinh^{-2}$ interactions. At large time, this mapping yields a new connection between the inverse-Wishart ensemble of random matrices and the ground-state of this fermionic system. Further, we introduced generalized models and solved for their stationary distributions.\\
A generalization of the Ferrari-Spohn problem was considered for multiple non-crossing bridges above a semi-circle distribution. The joint distribution of the bridge positions above the boundary was obtained in a direct determinantal expansion of the renowned Ferrari-Spohn distribution, and studied in relation with a fermionic system in a linear potential with a hard wall.\\[10pt]

In the scope of the links between random matrices and quantum systems, our presentation could have been extended in several directions.\\
Such connections were recently observed in second-quantization fermionic systems. In particular, the transport properties of the $XX$ spin chain, equivalent to a system of free fermions on the $\Z$ lattice through a Jordan-Wigner transformation, were studied in a quench protocol with an initially filled $\Z^-$ half-space and empty $\Z^+$ half-space. We point for instance to \cite{Antal1999} for the obtention of the magnetization profile. The local statistics of this fermionic system at the edge of the front, propagating from $\Z^-$ into $\Z^+$, were expressed in terms of the Airy kernel, and were thus related to the edge spectrum of GUE matrices, in \cite{Eisler2013}. On the global scale, a richer connection was established with the Bessel kernel of RMT through a remarkable determinant expression in \cite{Moriya2019}, allowing the obtention of large deviations for the current from the large-deviation literature on the Wishart ensemble, see \cite{Marino2015}. In publication \pubref{publication:TransportFreeFermions}, we studied a generalization of this problem where the filled half-space in the initial condition presents non-trivial quantum correlations, in contrast with the uncorrelated initial condition chosen in the works previously cited. In this scope, an extension of the determinant expression of \cite{Moriya2019} was obtained. This correlated quench for the free fermions on the lattice showed, for this particular example, that the effect of correlations was to slow down the transport. This observation was then detailed in a general framework for the continuous case, and discussed in the interacting case.\\
In addition to fermions on the lattice as detailed above, RMT also has connections with models of Majorana fermions relevant to topological superconductors, as recent research indicates \cite{Beenakker2015,Grabsch2019}. A complete review of these themes is beyond the scope of the present manuscript.\\[10pt]

Finally, we hope that the results and the many connections outlined in this thesis will instil enthusiasm for future research in the field. In this respect, we give in the following a few tracks which are worthy of investigation.\\ 
In the same fashion as the matrix generalization of a random recursion undertaken in this work, developing matrix extensions for other disordered systems would be of great interest. An example, in the spirit of \cite{Kardar1996,OConnell2012,OConnell2021}, would be to investigate a matrix extension of the KPZ equation and the directed polymer model, and to examine the universality class associated to such models. The study of the Kesten recursion was intended as a first step in this direction.\\
Beyond the few known cases, many connections between one-dimensional trapped fermion systems and random matrix theory ensembles certainly remain to be discovered. Their unveiling would be valuable for the application of random matrix tools to a broader range of fermion systems, and for a deeper understanding of these links. Furthermore, the study of cold trapped fermions presents, on its own, a collection of interesting open questions in higher-dimensional, non-zero-temperature or interacting cases.\\
As presented in the last chapter, the large-$N$ asymptotics of the HCIZ integral, for which only some special cases are well understood, remain a promising research track. The richness of the HCIZ theory is hinted by its connections to integrable systems, fluid dynamics and large deviations of random matrix models, such that other links can certainly be expected.  \\
Lastly, improving our understanding of the depth of random matrix universality, and its ubiquitous appearances both in the physical and mathematical realms, is a long-term ambition which promises many new interesting discoveries.

\newpage 
\chaptermark{}
\null \newpage  
\thispagestyle{empty}
\begin{appendices} 
\setcounter{subsection}{0}
\renewcommand{\thesubsection}{\Alph{subsection}}  
\numberwithin{equation}{subsection}
\numberwithin{table}{subsection}
\titleformat{\subsection}[hang]{\LARGE\sffamily\bfseries}{\rlap{\thesubsection}}{2em}{}

\subsection{Normalization for \texorpdfstring{$\beta$}{Beta}-ensembles} 
\label{app:normalization}
\vspace{0.5cm}

We give a few details about the normalization constants appearing in the matrix PDF and eigenvalue JPDF of the $\beta$-ensembles, and especially the connection between the two normalization constants. As presented in chapter \ref{chap:1}, these density functions with respect to the appropriate Lebesgue measures can be written as
\bea
\mathcal{P}(\HH)  &=& C_{\beta,V} \ \ e^{-\Tr \, V(\HH) }   \\
P(\vec{\lambda} )  &=&  C_{\beta,V}^N  \ \  \abs{\Delta(\vec{\lambda})}^\beta e^{- \sum_{i=1}^N V( \lambda_i ) }
\eea
where the normalization constants are such that
\be
\int_{\OOmega_\beta} \mathcal{P}(\HH) \ \dd \HH \ = \  \int_{\R^N} P(\vec{\lambda} ) \ \dd \vec{\lambda} \ = \ 1 \; .
\ee
The matrix integral is done on the space $\OOmega_\beta$ of real symmetric, complex Hermitian or quaternionic self-dual matrices, for $\beta=1,2,4$ respectively. Note that the JPDF of eigenvalues is sometimes defined only on the Weyl chamber $ \Weyl_N = \{ \vec{\lambda} \in \R^N  \mid  \lambda_1 \! \geqslant \!  \cdots \! \geqslant \!  \lambda_N \}$ such that the normalization constant $C_{\beta,V}^N $ then differs by a factor $N!$.

Let us introduce the three compact groups defining the symmetries of these three ensembles as
\be 
\mathrm{G}_\beta (N) =
 \left\{
    \begin{array}{ll}
        \mathrm{O}(N)  & \mbox{ if } \beta =1 \; ,\\
        \mathrm{U}(N) & \mbox{ if } \beta =2 \; ,\\
        \mathrm{Sp}(N) & \mbox{ if } \beta =4 \; .
    \end{array}
\right.
\ee
The volume of these compact groups is given by \cite{Zhang2015}:
\be 
\Vol(\mathrm{G}_\beta (N) )=
\int_{\mathrm{G}_\beta (N)}  \ \dd \UU \  = 
\frac{2^{N} \pi^{ \frac{\beta}{2}\frac{N(N+1)}{2}}}{\prod_{k=1}^{N} \Gamma\left(\frac{\beta}{2} k \right)}
\; .
\label{eq:app:VolumeGroups}
\ee
Let us also recall the surface $A_\beta$ of a unit-sphere in $\beta$ dimensions, i.e. the volume of modulus-one numbers in $\R, \C$ or $\Quaternions$, is:
\be 
A_\beta = \frac{2 \pi^{ \frac{\beta}{2}} }{ \Gamma( \frac{\beta}{2}) } =
\left\{
    \begin{array}{ll}
        2 &\mbox{ if }\beta =1 \; , \\
        2\pi &\mbox{ if }\beta =2 \; ,\\
        2\pi^2 &\mbox{ if }\beta =4 \; .
    \end{array}
\right.
\ee
Note that this coincides, as expected, with the volume of the symmetry groups for $N=1$, as $A_\beta = \Vol(\mathrm{G}_\beta (1) )$.

The connection between normalization constants can be obtained as follows. The eigenvalue JPDF is obtained by integrating out the eigenvectors, which the density does not depend on, and the normalization $ C_{\beta,V}^N $ is thus equal to $C_{\beta,V}$ times the volume of $\mathrm{G}_\beta (N)$. However, symmetry factors should be taken into account: multiplying an eigenvector by a modulus-one number in $\R, \C$ or $\Quaternions$ does not change the matrix, such that there is an overcount factor $A_\beta$ for each eigenvector ; and jointly permuting the $N$ eigenvalue-eigenvector pairs does not change the matrix either, such that there is also an overcount factor $N!$. This gives
\be 
C_{\beta,V}^N =  \frac{\Vol( \mathrm{G}_\beta(N) ) }{N! \ (A_\beta)^N}  \  C_{\beta,V} 
\ee
where the proportionality factor can be expressed precisely as 
\be 
\Xi_\beta^N =
\frac{\Vol( \mathrm{G}_\beta(N) ) }{N! \ (A_\beta)^N}  = 
\pi^{ \frac{\beta}{2} \frac{N(N-1)}{2}} \prod_{j=1}^N \frac{\Gamma (1 + \frac{\beta}{2}) }{\Gamma (1 + \frac{\beta}{2}j)}
\; .
\label{eq:app:ProportionalityFactoDistributions}
\ee 

The specific values of these normalization constants for the main ensembles are obtained via \emph{Selberg integrals} \cite{Mehta2004,Forrester2010} and are given in the main text.

\subsection{Special functions} 
\label{app:specialfunctions}
\vspace{0.5cm}

We define the special functions appearing in the thesis, and give their main properties \cite{DLMF}.

\subsubsection*{Gamma}

For complex argument such that $\mathrm{Re}(z) >0$, the Gamma function $\Gamma$ is defined as
\be 
\Gamma (z)  =  \int_{0}^{+\infty} t^{z-1} e^{-t}  \dd t \; 
\ee 
such that it is the Mellin transform of $z \to e^{-z}$. It verifies 
\be
\Gamma(z+1) = z \Gamma (z) 
\ee
and $\Gamma(1) = 1$ such that, for $n \in \N$
\be 
\Gamma(n+1)=n ! \; .
\ee

\subsubsection*{Beta}

For complex arguments $(x,y)$ such that $\mathrm{Re}(x) >0$ and $\mathrm{Re}(y) >0$, the Beta function $\mathrm{B}$ is defined as
\be 
\mathrm{B}(x,y) = \int_0^1 t^{x-1}(1-t)^{y-1} \dd t \; .
\ee
It is symmetric under the exchange $x \leftrightarrow y$ and is naturally related to the Gamma function as
\be 
\mathrm{B}(x,y) = \frac{\Gamma(x) \Gamma(y)}{\Gamma(x+y)} \; .
\ee

\subsubsection*{Airy} 
The Airy function of the first kind $\Ai$ is defined as the solution of the differential equation
\be
\Ai''(x) = x \,  \Ai(x)
\ee
with the boundary condition
\be 
\lim_{x\to \infty} \Ai(x)  = 0 \; .
\ee
For real arguments, it has the following integral representation
\be 
\Ai(x)= \frac{1}{\pi} \int_{0}^{\infty} \cos \left(\frac{t^{3}}{3}+x t\right) \dd t \; .
\ee
The Airy function can be extended to the complex plane. For complex arguments, it has the following integral representation
\be 
\Ai(z)  = \frac{1}{2 \pi \mathrm{i}} \int_{\infty \mathrm{e}^{- \ii \pi  / 3}}^{\infty \mathrm{e}^{ \ii \pi / 3}} \exp \left(\frac{1}{3} t^{3}-z t\right) \dd t \; .
\ee
By integrating $\Ai^2 = \left( \Ai''/x \right)^2$ on $[\alpha_k, +\infty[$ with $\alpha_k$ a zero of the Airy function as $\Ai(\alpha_k)=0$, we obtain the following integral
\be
\int_{\alpha_k}^\infty \Ai(x)^2 \ \dd x = \Ai'(\alpha_k)^2 \; .
\ee
More generally, the following equalities hold:
\begin{eqnarray}
\int_{y}^{\infty} \mathrm{Ai}(x)^2 \ \dd x &=& \mathrm{Ai}^{\prime}(y)^2-y \mathrm{Ai}(y)^2  \\
\int_{y}^{\infty} x \mathrm{Ai}(x)^2 \  \dd x &=& \frac{1}{3}\left(y \mathrm{Ai}^{\prime}(y)^2-y^{2} \mathrm{Ai}(y)^{2}-\operatorname{Ai}(y) \mathrm{Ai}^{\prime}(y)\right)  \\
\int_{y}^{\infty} \mathrm{Ai}^{\prime}(x)^2 \ \dd x &=&  \frac{1}{3}\left(y^{2} \mathrm{Ai}(y)^2-y \mathrm{Ai}^{\prime}(y)^2-2 \operatorname{Ai}(y) \mathrm{Ai}^{\prime}(y)\right)  \\
\int_{y}^{\infty} x^{2} \mathrm{Ai}(x)^2 \ \dd x &=& \frac{1}{5}\left(y^{2} \mathrm{Ai}^{\prime}(y)^2-y^{3} \mathrm{Ai}(y)^2-2 y \operatorname{Ai}(y) \mathrm{Ai}^{\prime}(y)+\mathrm{Ai}(y)^2\right)  \\
\int_{y}^{\infty} x \mathrm{Ai}^{\prime}(x)^2 \ \dd x &=& \frac{1}{5}\left(y^{3} \mathrm{Ai}(y)^2-y^{2} \mathrm{Ai}^{\prime}(y)^2-3 y \operatorname{Ai}(y) \operatorname{Ai}^{\prime}(y)+\frac{3}{2} \mathrm{Ai}(y)^2\right)  \quad 
\end{eqnarray}
The first zeros of the Airy function are given in Table \ref{tab:ZerosAiry}.

\begin{table}[ht!]
\centering
\vspace{.5cm}
\begin{tabular}{| c | c | c | c | c | c |}
  \hline
 $k$  & $1$ & $2$ & $3$ & $4$ & $5$ \\ \hhline{|=|=|=|=|=|=|}
 & & & & & \\[-8pt] 
 $\alpha_k$  & \ $-2.33811\ldots$ \ &\  $-4.08795\ldots$ \ & \ $-5.52056\ldots$ \ & \ $-6.78671\ldots$ \ & \ $-7.94413\ldots$ \ \\[8pt] 
  \hline
\end{tabular} 
\caption{First five zeros of the Airy function.} 
\label{tab:ZerosAiry}
\end{table}

\subsubsection*{Bessel}

The Bessel function of the first kind of order $\alpha$,  denoted $J_\alpha$, is defined as the solution of the differential equation
\be 
x^2 J_\alpha''(x) + x J_\alpha'(x) + (x^2- \alpha^2) J_\alpha(x) = 0
\ee
with the boundary condition that $\lim\limits_{x\to 0} J_\alpha(x)$ is finite. This defines an analytic function $z \to J_\alpha(z)$ on the complex plane, except for a branch point at $z=0$ when $\alpha \notin \N$. For $\alpha = n \in \N$, it has the following integral representation
\be 
J_n(z) =
\frac{1}{\pi} \int_{0}^{\pi} \cos \left( z \sin \theta-n \theta \right) \dd \theta=\frac{\ii^{-n}}{\pi} \int_{0}^{\pi} \mathrm{e}^{\ii z \cos \theta} \cos (n \theta) \dd  \theta \; .
\ee
It also has a contour integral representation
\be 
J_n(z) = 
\oint_{|x|=1} \frac{\mathrm{d} x}{2 \ii \pi x} \ e^{\frac{z}{2} \left(x-\frac{1}{x}\right)} \ (-x)^{n} \; .
\ee

\subsubsection*{Parabolic cylinder}

The parabolic cylinder function $D_\nu$ is the solution of the differential equation
\be 
D_\nu '' (x) + (\nu + \frac{1}{2} - \frac{x^2}{4}) D_\nu(x) = 0 
\ee
with the boundary condition
\be 
\lim_{ x \to \infty }  \ \frac{D_\nu(x)}{ x^\nu \, e^{-\frac{x^2}{4}} }  \ = 1 \; .
\ee
For $\nu = n \in \N$, the parabolic cylinder functions reduce to
\be 
D_n(x) = e^{-\frac{x^2}{4}} H_n(x) 
\ee
in terms of the Hermite polynomial defined above \myeqref{eq:1:HermiteGaussianWeight}. For general $\nu \in \R$, $D_\nu$ admits the integral representation
\be 
D_\nu (x) = 
\frac{\mathrm{e}^{-\frac{1}{4} x^{2}}}{\Gamma\left(- \nu \right)} \int_{0}^{\infty} t^{- \nu -1 } \mathrm{e}^{-\frac{1}{2} t^{2}-x t} \dd t \; .
\ee

\subsubsection*{Whittaker $W$}

The Whittaker $W_{k,\mu}$ function is the solution of Whittaker's equation
\be 
W_{k,\mu}''(x) + \left(- \frac{1}{4} + \frac{k}{x} + \frac{1/4-\mu^2}{x^2} \right) W_{k,\mu}(x)  = 0
\ee
with the boundary condition
\be 
\lim_{ x \to \infty }  \ \frac{W_{k,\mu}(x)}{ x^k \, e^{-\frac{x}{2}} }  \ = 1 \; .
\ee
The Whittaker function is related to the parabolic cylinder function as
\be 
D_\nu(x) = \frac{W_{\frac{1}{4}+ \frac{\nu}{2}, \pm \frac{1}{4} } \left( \frac{x^2}{2} \right)}{\sqrt{x} \, 2^{ - \frac{1}{4} - \frac{\nu}{2}   }} \; .
\ee

\subsection{Vandermonde determinant limit} 
\label{app:VandermondeLimit}
\vspace{0.5cm}

In this appendix, we prove the following limit
\emphbe 
\lim_{\epsilon \to 0} \ \epsilon^{-\frac{N(N-1)}{2}} \prod_{i=1}^{N-1} i! \  \det_{1 \leqslant i,j \leqslant N} e^{ \epsilon \,  x_{i} \,  y_{j} }  \    = \  \Delta ( \vec{x} )  \,   \Delta ( \vec{y} )   
\label{eq:appA:VandermondeLimit}
\end{empheq}
Keeping the first $N$ terms in the exponential expansion $e^{ x} =  \sum_{k\geqslant0} \frac{ x^k}{k!}$, the determinant of interest can be expressed as
\be 
\det_{1 \leqslant i,j \leqslant N} e^{ \epsilon \, x_{i} \, y_{j} } 
\ \simeq  \ 
\det_{1 \leqslant i,j \leqslant N} \left(  \sum_{k=0}^{N-1}  \frac{(\epsilon x_{i} y_{j} )^k}{k!}  \right)   \; .
\ee
This can be rewritten by a matrix product as
\be 
\det_{1 \leqslant i,j \leqslant N}  e^{ \epsilon \,  x_{i} \, y_{j} }  
\ \simeq \
\left|\begin{array}{cccc}
1 & x_{1} & \cdots & x_{1}^{N-1} \\
 1 & \cdots & \cdots & \cdots \\ 
 1 & x_{N} & \cdots & x_{N}^{N-1}
 \end{array}\right|
 \times
\left|\begin{array}{ccc}
1 & 1 & 1 \\ 
\left(\epsilon y_{1} \right) & \cdots & \left(\epsilon y_{N} \right) \\ 
\cdots & \cdots & \cdots \\ 
\frac{\left(\epsilon y_{1} \right)^{N-1}}{(N-1) !} & \cdots & \frac{\left( \epsilon y_{N} \right)^{N-1}}{(N-1) !}\end{array}\right|
\ee
which gives the result when bringing the constants out of the rightmost determinant, as
\be 
\det_{1 \leqslant i,j \leqslant N}  e^{ \epsilon \,  x_{i} \, y_{j} }  
\ \simeq \
\frac{
\epsilon^{\frac{N(N-1)}{2}}}
{\prod_{i=1}^{N-1} i!}
\
\Delta \left( \vec{x} \right)
\,
\Delta \left( \vec{y} \right) 
\ee
such that the limit \eqref{eq:appA:VandermondeLimit} holds.

\subsection{An integral computation}
\label{app:InterpolationComputation}
\vspace{0.5cm}

We prove in this appendix that the integral
\be 
I(x_i,x_j, \tau ) = 
\iint_{\R^2} P\left( z_1, \tau  \mid  x_i, 0 \right)  \ P\left( z_2, \tau  \mid  x_j, 0 \right)  \ \mathrm{sgn}(z_1-z_2) \  \dd z_1 \dd z_2  
\ee 
defined in section \ref{sec:2.3}, is equal to 
\emphbe 
I(x_i,x_j, \tau )  = \erf \left( \frac{x_j - x_i}{2\sqrt{\tau}} \right)
\end{empheq}

From the expression of the Brownian propagator \eqref{eq:2:PropagatorBM}, the integral is
\be 
I(x_i,x_j, \tau ) =  \frac{1}{2 \pi \tau}
\int_{\R^2} e^{ - \frac{(z_1-x_i)^2 + (z_2-x_j)^2}{2 \tau} } \ \mathrm{sgn}(z_1-z_2) \  \dd z_1 \dd z_2  \; .
\ee
It can be computed as follows
\be
I(x_i,x_j, \tau ) = \frac{1}{2 \pi \tau}
\int_{\R} \dd z_1 \,  e^{ - \frac{(z_1-x_i)^2 }{2 \tau} }  \int_\R \dd z_2 \, e^{- \frac{(z_2-x_j)^2}{2 \tau}} \mathrm{sgn}(z_1-z_2)  
\ee
where 
\bea
 \int_\R \dd z_2 \, e^{- \frac{(z_2-x_j)^2}{2 \tau}} \mathrm{sgn}(z_1-z_2)  &=& \int_{-\infty}^{z_1}  \dd z_2 \,  e^{- \frac{(z_2-x_j)^2}{2 \tau}} -       \int_{z_1}^\infty \dd z_2 \,  e^{- \frac{(z_2-x_j)^2}{2 \tau}} \\
 &=&  \sqrt{2\tau} \left( 
 \int_{-\frac{z_1-x_j}{\sqrt{2 \tau}}}^\infty \dd u \, e^{-u^2} - \int_{\frac{z_1-x_j}{\sqrt{2 \tau}}}^\infty  \dd u \, e^{-u^2}   
 \right)\\
 &=& 2 \sqrt{2\tau} \int_0^{\frac{z_1-x_j}{\sqrt{2\tau}}} \dd u \, e^{-u^2} \\
 &=&  \sqrt{2 \pi \tau } \  \erf \left( \frac{z_1-x_j}{\sqrt{2 \tau}} \right)
\eea
in terms of the $\erf$ function defined in \myeqref{eq:2:DefErrorFunction}. As a consequence, we have
\be 
I(x_i,x_j, \tau ) = \frac{1}{\sqrt{2 \pi \tau}} \int_\R \dd z_1 \, e^{ - \frac{(z_1-x_i)^2}{2 \tau} }  \erf \left( \frac{z_1-x_j}{\sqrt{2 \tau}} \right) \; .
\ee
By virtue of the following $\erf$ integration formula \cite{DLMF}
\be 
\int_{-\infty}^{\infty} e^{-b^{2}(x-c)^{2}} \erf(a(x-d)) \dd x=\frac{\sqrt{\pi}}{b} \erf\left(\frac{a b(c-d)}{\sqrt{a^{2}+b^{2}}}\right)
\ee
we have the desired result
\be 
I(x_i,x_j, \tau )  = \erf \left( \frac{x_j - x_i}{2\sqrt{\tau}} \right) \; .
\ee

%

\subsection{Details on generalized processes}
\label{app:GeneralizedProcesses}
\vspace{0.5cm}

In this appendix, we give a few details on the results of subsection \ref{subsec:3.2.3} where some processes inspired by the matrix Kesten recursion are introduced and discussed.

\subsubsection{Generalized Kesten process}

The stationary solution of the Generalized Kesten process \eqref{eq:3:GeneralizedKesten} can be found as the solution $P_\infty$ of the following multivariate stationary Fokker-Planck equation
\bea
0 &=& - \sum_{i=1}^N \frac{\partial}{\partial \lambda_i} \left(  \left(  m \lambda_i  + \sum_{k \neq 1} p_k \lambda_i^k + \sigma^2 \sum_{j \neq i } \frac{\lambda_i \lambda_j}{\lambda_i - \lambda_j}    \right)   P_\infty (\vec{\lambda}) \right)  + \sigma^2 \sum_{i=1}^N \frac{\partial^2}{\partial \lambda_i^2} \left(  \lambda_i^2 P_\infty(\vec{\lambda}) \right) \nonumber \\
&=& \sum_{i=1}^N \frac{\partial}{\partial \lambda_i} \left[
- \left(  m \lambda_i  + \sum_{k \neq 1} p_k \lambda_i^k + \sigma^2 \sum_{j \neq i } \frac{\lambda_i \lambda_j}{\lambda_i - \lambda_j}    \right)   P_\infty (\vec{\lambda})
+ \sigma^2 \frac{\partial}{\partial \lambda_i} ( \lambda_i^2 P_\infty (\vec{\lambda} ) ) 
\right] 
\eea
As given in \myeqref{eq:3:GeneralizedKestenStationarySol}, one finds the following stationary JPDF
\be
P_\infty( \vec{\lambda}) \propto
\abs{\Delta(\vec{\lambda})}  \, \prod\limits_{i=1}^N \lambda_{i}^{\frac{m}{ \sigma^2} - (N+1)}    \; e^{ \frac{1}{ \sigma^2} \sum\limits_{1 \leqslant  i \leqslant N \atop k \neq 1}   \frac{p_{k}}{k-1} \lambda_{i}^{k-1} } \; . 
\ee

\subsubsection{Generalized Grabsch-Texier-Rider-Valk\'o process}

\paragraph{Perturbative evolution}

As detailed in the main text, the perturbative evolution of eigenvalues in models \eqref{eq:3:GeneralizedGTRVAsymmetric} and \eqref{eq:3:GeneralizedGTRVSymmetric} is obtained by first taking the matrix SDE to the Itô prescription. This is done by adding $\sigma^2$ times the drift $\DelBB_{\rm IS}$, respectively given in \myeqref{eq:3:DriftISAsymmetricGTRV} and \myeqref{eq:3:DriftISSymmetricGTRV}.

\subparagraph{Asymmetric noise}
The first model, with asymmetric noise, is then equivalent to the following Itô matrix SDE
\be 
\dd \VV_t =  \left( \left( m+\frac{\sigma^2}{2}\right) \VV_t + \sum_{k\neq 1} p_k  \VV_t^k +  \frac{ \sigma^2 }{2} \Tr ( \VV_t) \ID  \right) \dd t  + \sigma \frac{ \VV_t \dd \XX_t + \dd \XX_t^T \VV_t }{\sqrt{2}}   \quad \quad \text{(Itô)} \; .
\label{eq:app:ItoGrabschTexierAsymmetric}
\ee
Leaving the trivial drift terms aside and following the perturbative framework described below \myeqref{eq:2:PerturbationTheoryBeginning}, the evolution of the eigenvalues $\vec{\lambda}_t$ induced by the noise term follows
\be 
\dd \lambda_{i,t} = \sqrt{2} \sigma \lambda_{i,t} \dd B_{i,t}  +\frac{ \sigma^2 }{2}
\sum\limits_{j \neq i } \frac{\lambda_{i,t}^2 + \lambda_{j,t}^2}{\lambda_{i,t} - \lambda_{j,t}} \dd t \; .
\label{eq:app:AsymmetricPartialSDE}
\ee
Adding the drift terms of \myeqref{eq:app:ItoGrabschTexierAsymmetric}, that were overlooked in the above equation, then yields the eigenvalue evolution given in SDE \eqref{eq:3:EigenvalueSDEAsymmetricGTRV} as
\be 
\dd \lambda_{i,t} = \left(  \left(m + \sigma^2 \frac{N+1}{2} \right) \lambda_{i,t}  +  \sum_{k\neq 1} p_k  \lambda_{i,t}^k    \right)\dd t +\sigma^2 \sum\limits_{j \neq i } \frac{\lambda_{i,t}    \lambda_{j,t}}{\lambda_{i,t} - \lambda_{j,t}}  \dd t+  \sqrt{2} \sigma \lambda_{i,t} \dd B_{i,t} 
\; .
\ee

\subparagraph{Symmetric noise}
By the explanations given above, the second model, with symmetric noise, is equivalent to the following Itô matrix SDE
\be 
\dd \VV_t = \left( \left( m + \sigma^2(1+\frac{N}{2})  \right)  \VV_t + \sum_{k\neq 1} p_k  \VV_t^k + \frac{\sigma^2 }{2} \Tr ( \VV_t)  \ID   \right) \dd t  + \sigma \frac{\VV_t \dd \HH_t+ \dd \HH_t\, \VV_t }{\sqrt{2}}  \quad \quad  \text{(Itô)}  \; .
\label{eq:app:ItoGrabschTexierSymmetric}
\ee
Leaving the trivial drift terms aside, the evolution of the eigenvalues $\vec{\lambda}_t$ is different in this symmetric case, and follows
\be 
\dd \lambda_{i,t} =  2  \sigma \lambda_{i,t} \dd B_{i,t}  +\frac{ \sigma^2 }{2}
\sum\limits_{j \neq i } \frac{(\lambda_{i,t} + \lambda_{j,t})^2}{\lambda_{i,t} - \lambda_{j,t}} \dd t \; .
\ee
Adding the drift terms of \myeqref{eq:app:ItoGrabschTexierSymmetric}, that were overlooked in the above equation, yields the eigenvalue evolution given in SDE \eqref{eq:3:EigenvalueSDESymmetricGTRV} as
\be 
\dd \lambda_{i,t} = \left(  (m + \sigma^2 (N+1) ) \lambda_{i,t}  +  \sum_{k\neq 1} p_k  \lambda_{i,t}^k    \right)\dd t
 +  2 \sigma^2 \sum\limits_{j \neq i } \frac{\lambda_{i,t} \lambda_{j,t}}{\lambda_{i,t} - \lambda_{j,t}} \dd t  + 2 \sigma \lambda_{i,t} \dd B_{i,t}
\; .
\ee

\paragraph{Stationary solutions}

The stationary distribution \eqref{eq:3:GeneralizedGTRVAsymmetricStationarySol} is found as the solution of the stationary Fokker-Planck equation
\be 
0 =  \sum_{i=1}^N \frac{\partial}{\partial \lambda_i} \left[
 - \left(  \left(m + \sigma^2 \frac{N+1}{2} \right) \lambda_{i}  +  \sum_{k\neq 1} p_k  \lambda_{i}^k  + \sigma^2 \sum\limits_{j \neq i } \frac{\lambda_{i}    \lambda_{j}}{\lambda_{i} - \lambda_{j}}   \right) P_\infty(\vec{\lambda}) 
+ \sigma^2 \frac{\partial}{\partial \lambda_i}  \lambda_i^2  P_\infty(\vec{\lambda}) 
\right]
\ee
and \eqref{eq:3:GeneralizedGTRVSymmetricStationarySol} is found simply by rescaling $\sigma \to \sqrt{2} \sigma$.

\subsubsection{Square-root multiplicative noise}
In the study of model \eqref{eq:3:SquareRootNoiseModel}, the derivation of the eigenvalue evolution is very similar to the one of the asymmetric model above. Leaving drift terms aside, the eigenvalue SDE is obtained from \myeqref{eq:app:AsymmetricPartialSDE} under replacement $\lambda \to \sqrt{\lambda}$ in all terms except the denominator $\propto \frac{1}{\lambda_i - \lambda_j}$ which is inherent to perturbation theory. It thus reads
\be 
\dd \lambda_{i,t} = \sqrt{2 \lambda_{i,t} } \sigma \dd B_{i,t}  +\frac{ \sigma^2 }{2} 
\sum\limits_{j \neq i } \frac{\lambda_{i,t} + \lambda_{j,t}}{\lambda_{i,t} - \lambda_{j,t}} \dd t \; .
\ee
Adding drift terms, we recover the full SDE \eqref{eq:3:SquareRootEigenvalueSDE}, as
 \be 
\dd \lambda_{i,t} =
    \left( m+ \sum\limits_{k\neq 0} p_k \lambda_{i,t}^k \right) \dd t  +   \frac{\sigma^2}{2} \sum\limits_{j \neq i } \frac{\lambda_{i,t} + \lambda_{j,t}}{\lambda_{i,t} - \lambda_{j,t}} \dd t +  \sigma \sqrt{2 \lambda_{i,t}}\dd B_{i,t} \; .
    \ee
The stationary solution \eqref{eq:3:SquareRootMultNoiseStationarySolution} is then obtained from the stationary Fokker-Planck equation
\be 
0 = 
 \sum_{i=1}^N \frac{\partial}{\partial \lambda_i} \left[
 - \left(   
  m+ \sum\limits_{k\neq 0} p_k \lambda_{i,t}^k +   \frac{\sigma^2}{2} \sum\limits_{j \neq i } \frac{\lambda_{i,t} + \lambda_{j,t}}{\lambda_{i,t} - \lambda_{j,t}} \right)
 P_\infty(\vec{\lambda}) 
+ \sigma^2 \frac{\partial}{\partial \lambda_i} 
 \lambda_i
 P_\infty(\vec{\lambda})  \right] \; . 
\ee 
 
\subsubsection{Squared multiplicative noise}
The model \eqref{eq:3:SquaredModelDefProcess} is equivalent, by addition of $\sigma^2$ times the drift \eqref{eq:3:SquaredISDrift}, to the Itô matrix SDE
\be 
 \dd \VV_t = \left( (m+ \sigma^2) \VV_t^3 + \sum\limits_{k\neq 3} p_k \VV_t^k   + \sigma^2 \Tr (\VV_t) \VV_t^2 \right) \dd t +    \sigma \VV_t \dd \HH_t \VV_t  \quad \quad \text{(Itô)} \; .
\ee
The perturbation evolution of the eigenvalues under this noise term is obtained, leaving drift terms aside, as
\be 
\dd \lambda_{i,t} =  \sqrt{2}  \sigma \lambda_{i,t}^2 \dd B_{i,t}  + \sigma^2  
\sum\limits_{j \neq i } \frac{\lambda_{i,t}^2 \lambda_{j,t}^2 }{\lambda_{i,t} - \lambda_{j,t}} \dd t
\; .
\ee
Adding the drift terms recovers the full eigenvalue SDE given in \myeqref{eq:3:SquaredEigenvalueSDE} as
\be 
\dd \lambda_{i,t} =
\left(  (m+2 \sigma^2) \lambda_{i,t}^3 + \sum\limits_{k\neq 3}  p_k \lambda_{i,t}^{k}  \right) \dd t + \sigma^2 \lambda_{i,t}^2 \sum_{j \neq i} \frac{\lambda_{i,t} \lambda_{j,t}  }{\lambda_{i,t} - \lambda_{j,t}}  \dd t + \sqrt{2}  \sigma \lambda_{i,t}^2 \dd B_{i,t} \; .
\ee
The stationary solution \eqref{eq:3:SquareMultiplicativeStationarySolution} is finally obtained as the solution of the following stationary Fokker-Planck equation
\be  
0 = 
 \sum_{i=1}^N \frac{\partial}{\partial \lambda_i} \left[
 - \left(   
 (m+ 2 \sigma^2) \lambda_{i}^3 + \sum\limits_{k\neq 3}  p_k \lambda_{i}^{k} + \sigma^2 \lambda_i^2 \sum_{j \neq i} \frac{\lambda_{i} \lambda_{j}  }{\lambda_{i} - \lambda_{j}}  
  \right)
 P_\infty(\vec{\lambda}) 
+ \sigma^2 \frac{\partial}{\partial \lambda_i} 
 \lambda_i^4
 P_\infty(\vec{\lambda})  \right] \; .  
\ee

\end{appendices} 
\titleformat{\subsection}[hang]{\Large\sffamily\bfseries}{\rlap{\thesubsection}}{3em}{}


\cleardoublepage
\null \newpage
\thispagestyle{empty}
\null  \newpage
\phantomsection
\addcontentsline{toc}{chapter}{Bibliography}
\bibliographystyle{./bib/theseen-href-bruno-modif}
\bibliography{bib/bibliography}


\end{document}